\documentclass[12pt,preprint,aps,showpacs,color]{revtex4}
\usepackage{epsfig}

\begin{document}

\newcommand{\npa}[1]{Nucl.~Phys.~{A#1}}
\newcommand{\etal}{{\em et al.}}
\newcommand{\neglect}[1]{  }
\newcommand{\Label}[1]{\label{#1} \message{FIG \thefigure :label #1 fig.} }
\newcommand{\blabla}{{\em Dies \\ ist \\ nur \\ ein \\ sinnloser \\ Text \\ zum \\ Auff\"ullen  !}}
\newcommand{\rfl}{red full line}
\newcommand{\bdl}{blue dashed line}
\newcommand{\bpl}{black dotted line}
\newcommand{\gml}{green dash-dotted line}
\newcommand{\gfl}{green full line}
\newcommand{\lhs}{left hand side}
\newcommand{\rhs}{right hand  side}
\newcommand{\Figref}[1]{Fig.\ \ref{#1}}
\newcommand{\figref}[1]{fig.\ \ref{#1}}  
\newcommand{\lhsref}[1]{left hand side of fig.\ \ref{#1}}  
\newcommand{\rhsref}[1]{right hand side of fig.\ \ref{#1}}  

\title{Dynamics of $K^+$  
Production in Heavy Ion Collisions close to Threshold\footnote{Extract from the habilitation thesis}}

\author{Christoph~Hartnack\footnote{email:hartnack@subatech.in2p3.fr}}

\address{SUBATECH, Laboratoire de Physique Subatomique et des
Technologies Associ\'ees \\ UMR 6457, University of Nantes - IN2P3/CNRS -
Ecole des Mines
de Nantes \\
4 rue Alfred Kastler, F-44072 Nantes, Cedex 03, France\\
 }

\date{December 15, 2004}
%

\begin{abstract}

In this article the production of $K^+$ 
at energies close to the threshold is studied in detail. 
The production mechanisms, the influence of 
in-medium effects, cross sections, the nuclear equation of state 
and the dynamics of the nucleons  
on the kaon dynamics are discussed. A special regard will be taken
on the collision of  Au+Au at 1.5 GeV, a reaction that has recently been analyzed 
in detail by experiments performed by the KaoS and FOPI collaborations
at the SIS accelerator at GSI.


\end{abstract}


\maketitle


\newpage

\tableofcontents

\newpage

\section{Introduction}
Kaons are pseudoscalar mesons containing strangeness.
Strangeness is a property of some baryons and mesons which
at their discovery had `strange' long lifetime compared to 
the nuclear resonances known at that time. The quark model
explained that property by the content of a strange (anti)quark.
Normal nuclear matter - protons, neutrons and (following the old
Yukawa idea) pions - are build up by two types of quarks, the so-called
{\it up-} and {\it down} quarks. A further quark the so-called {\it strange}
quark allows the description of the novel particles, as it can be
seen in fig. \ref{multiplett}. 
\begin{figure}[hbt]
\epsfig{file=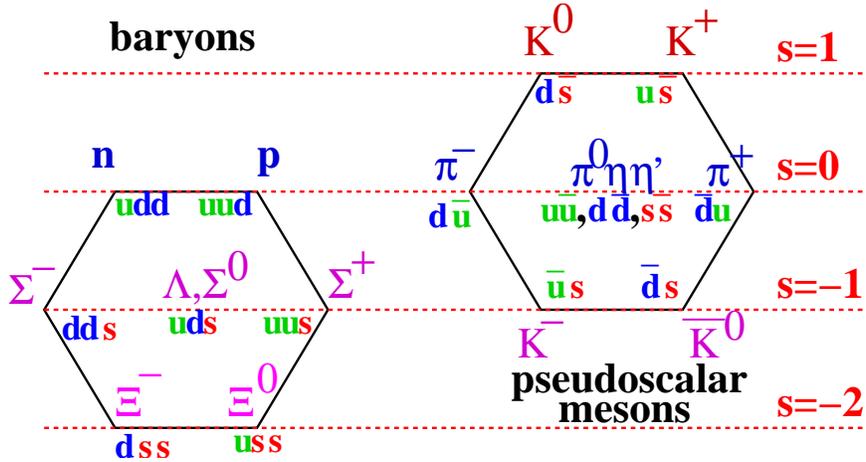,width=0.7\textwidth} 
 \caption{The basic SU(3) multipletts for baryons and pseudoscalar mesons}
\Label{multiplett}
\end{figure}

In the following time further particles giving need for new quark flavors
were discovered. Today six quark flavors are assumed: up, down, strange, charm,
bottom and top. These quarks are stable concerning the strong interaction, so that
the whole net number of quarks of each flavor is conserved. Anti-quarks are
counted with an opposite sign, so that the production of new particles via
the production of quark-antiquark-pairs and rearrangement of the other quarks
is possible. 
The conservation of the net quark numbers led to several conservation 
numbers like the strangeness content $S$ 
\begin{equation}
S = \sum \bar{s} - \sum s
\end{equation}
Similarly, the charm quantum number describes the number of charm quarks etc.
The net numbers of up and down quarks did not give rise to special conservation
quantity since their conservation is implicitly assured by the conservation
of the net baryon number and the charge.

However, the quark numbers  are not stable against the weak interaction.
Thus, a kaon may decay within about $10^{-8}$s into lepton pairs or mesons, 
e.g. $K^+ \to \mu^+ \bar{\nu}$. 
This effect is, however, important for the experimental detection but will
not touch our theoretical considerations.

The interest of the kaon itself in a heavy ion collision is that during
this short reaction time of about  $10^{-22}$s strangeness is rigidly conserved
and a produced $K^+$ can effectively not be destroyed. Thus, strangeness is
a direct signal from a heavy ion reaction. 
This property has triggered for a long time a full spectrum of theoretical and
experimental activities, whose exhaustive description would be 
quite impossible. 
For first ideas see for example \cite{dowa,aik,RanKo,Barth,Ahle,schaffi,cassing,koli}

In this article the production of kaons 
in relativistic heavy ion collisions at energies close to the threshold
of elementary production is studied.
This energy domain corresponds to experiments performed at the
SIS accelerator at GSI (Darmstadt, Germany). 
This article focus on the positively charged $K^+$ which is better 
accessible by experiment.
The $K^0$ is assumed to have similar properties. However, there is no 
unique relation to experiment, since the $K^0$ and its antiparticle, the
$\bar{K^0}$, mix together into the short living $K^0_S$ and the long-living
$K^0_L$. The antikaons ($K^-$ and $\bar{K^0}$) are not discussed neither, although
their production is strongly coupled to the kaon production. Its exhaustive
discussion would drastically enhance the size of this article.     
A special attention will be paid to the reaction Au+Au at 1.5 AGeV incident energy
which has been recently investigated in detail by the KaoS and FOPI
collaborations at GSI.

\section{Microscopic description of heavy ion collisions within IQMD}

The Isospin Quantum molecular dynamics model (IQMD) \cite{ha89,hart,baprc,iqmd} 
is a semi-classical model which describes heavy ion collisions on an
event by event basis. 
Only a brief sketch of the model will be given here.  
For a detailed description see \cite{hart,iqmd}.
For microscopic models of heavy ion collisions in general see 
\cite{st86,cas90,ai91}. For some review dedicated to strangeness
production see \cite{urqmd}.

\subsection{Potentials in IQMD}

In IQMD particles are represented by the 1 particle Wigner density.
\begin{equation} \Label{fdefinition}
 f_i (\vec{r},\vec{p},t) = \frac{1}{\pi^3 \hbar^3 }
 {\rm e}^{-(\vec{r} - \vec{r}_{i} (t) )^2  \frac{2}{L} }
 {\rm e}^{-(\vec{p} - \vec{p}_{i} (t) )^2  \frac{L}{2\hbar^2}  }
\end{equation} 
The total 1 particle Wigner density is the sum of all nucleons.
The expectation value of the total Hamiltonian is
\begin{eqnarray} 
\langle H \rangle &=& \langle T \rangle + \langle V \rangle 
\nonumber \\ \Label{hamiltdef}
&=& \sum_i \frac{p_i^2}{2m_i} +
\sum_{i} \sum_{j>i}
 \int f_i(\vec{r},\vec{p},t) \,
V^{ij}  f_j(\vec{r}\,',\vec{p}\,',t)\,
d\vec{r}\, d\vec{r}\,'
d\vec{p}\, d\vec{p}\,' \quad.
\end{eqnarray}
The baryon-potential consists of the real part of the 
$G$-Matrix which is supplemented by the Coulomb interaction
between the charged particles. The former can be further subdivided in 
a part containing the contact Skyrme-type interaction only, a contribution
due to a finite range Yukawa-potential, and a momentum dependent part.
%
\begin{eqnarray}
V^{ij} &=& G^{ij} + V^{ij}_{\rm Coul} \nonumber \\
       &=& V^{ij}_{\rm Skyrme} + V^{ij}_{\rm Yuk} + V^{ij}_{\rm mdi} + 
           V^{ij}_{\rm Coul} + V^{ij}_{sym}\nonumber \\
       &=& t_1 \delta (\vec{x}_i - \vec{x}_j) + 
           t_2 \delta (\vec{x}_i - \vec{x}_j) \rho^{\gamma-1}(\vec{x}_{i}) +
           t_3 \frac{\hbox{exp}\{-|\vec{x}_i-\vec{x}_j|/\mu\}}
               {|\vec{x}_i-\vec{x}_j|/\mu} + \Label{vijdef}  \\
       & & t_4\hbox{ln}^2 (1+t_5(\vec{p}_i-\vec{p}_j)^2)
               \delta (\vec{x}_i -\vec{x}_j) +
           \frac{Z_i Z_j e^2}{|\vec{x}_i-\vec{x}_j|} + \nonumber \\
       & & t_6 \frac{1}{\varrho_0}
 T_{3}^i T_{3}^j \delta(\vec{r}_i - \vec{r}_j) 	\nonumber   
\end{eqnarray}
In the description of the Coulomb interaction $V^{ij}_{\rm Coul} $, 
$Z_i,Z_j$ are the charges of the baryons $i$ and $j$. 

The momentum dependence $V_{\rm mdi}^{ij}$ of the $N$--$N$ interaction,
which may optionally be used in QMD, is fitted to
experimental data \cite{ar82,pa67} on
the real part of the nucleon optical potential \cite{ai87b,bert88b},
which yields
\begin{equation}
\Label{mdipar}
U_{mdi} =        \delta \cdot \mbox{ln}^2 \left( \varepsilon \cdot  
                \left( \Delta \vec{p} \right)^2 +1 \right) \cdot
                        \left(\frac{\rho_{int}}{\rho_0}\right)
\end{equation}
%

The asymmetry energy is taken into account by the term 
\begin{equation}
V^{ij}_{sym}= t_6 \frac{1}{\varrho_0}
 T_{3}^i T_{3}^j \delta(\vec{r}_i - \vec{r}_j) \quad t_6 = 100\, \rm MeV
\end{equation}
where $T_{3}^i$ and $T_{3}^j$ denote the isospin $T_3$ 
of the particles $i$ and $j$, i.e. 1/2 for protons and -1/2 for neutrons.

The potential part of the equation of state 
resulting from the convolution of 
the distribution
functions $f_i$ and $f_j$ with the interactions  
$V_{\rm Skyrme}^{ij}+ V_{\rm mdi}^{i,j}$ 
(local interactions including
momentum dependence) reads:
\begin{equation} \Label{eosinf}
U \,=\, \alpha \cdot \left(\frac{\rho_{int}}{\rho_0}\right) +
        \beta \cdot \left(\frac{\rho_{int}}{\rho_0}\right)^{\gamma} +
        \delta \cdot \mbox{ln}^2 \left( \varepsilon \cdot
                \left( \Delta \vec{p} \right)^2 +1 \right) \cdot
                        \left(\frac{\rho_{int}}{\rho_0}\right)
\end{equation}

The parameters $t_1 ... t_5$ are uniquely related to
the corresponding values of $\alpha, \beta, \gamma, \delta$ and $\epsilon$ 
which serve as input. The standard values of these parameters can be found in
table \ref{eostab}.

\begin{table}[hbt]
\begin{tabular}{lcccccc}
 &$\alpha$ (MeV)  &$\beta$ (MeV) & $\gamma$ & $\delta$ (MeV) &$\varepsilon \, 
 \left(\frac{c^2}{\mbox{GeV}^2}\right) $ & $\kappa$ (MeV) \\
\hline
 S  & -356 & 303 & 1.17 & ---  & ---  & 200   \\
 SM & -390 & 320 & 1.14 & 1.57 & 500  & 200 \\
 H  & -124 & 71  & 2.00 & ---  & ---  & 376  \\
 HM & -130 & 59  & 2.09 & 1.57 & 500  & 376 \\
 INT& -157 & 103 & 1.58 & ---  & ---  & 284    \\
 VH & -110 & 56  & 2.40 & ---  & ---  & 456 \\ 
\end{tabular}
\caption{Parameter sets for the nuclear equation of 
state used in the
QMD model}
\Label{eostab}
\end{table}
In the calculations presented in this article the parametrization SM
is used as standard. It is 
a combination of Skyrme type and momentum dependent potential with
a low compressibility.

\subsection{Collisions}

Two particles collide if their minimum distance $d$, 
i.e.\ the minimum relative 
distance of the centroids of the Gaussians during their motion, 
in their CM frame
fulfills the requirement: 
\begin{equation}
 d \le d_0 = \sqrt{ \frac { \sigma_{\rm tot} } {\pi}  }  , \qquad
 \sigma_{\rm tot} = \sigma(\sqrt{s},\hbox{ type} ).
\end{equation}
where the cross section is assumed to be the free cross section of the
regarded collision type ($N-N$, $N-\Delta$, \ldots).

The total cross section is the sum of the elastic cross section and all
inelastic cross sections.
\begin{equation}
 \sigma_{\rm tot} = \sigma_{\rm el}+\sigma_{\rm inel}  = \sigma_{\rm el}+\sum_{\rm channels} \sigma_i
\Label{tot-el-inel}
\end{equation}

For instance for a pp collision we may have 
\begin{equation}
 \sigma_{\rm tot} = \sigma_{\rm el}+ \sigma(pp \to p \Delta^+) + \sigma(pp \to n \Delta^{++}) 
\Label{pp-example}
\end{equation}

The cross sections for the different channels are given by experiment or 
by spin/isospin coefficients. For the pp case for example we have
\begin{equation}
\sigma(pp \to n \Delta^{++}) = 3 \sigma(pp \to p \Delta^+) =\frac{3}{4}\sigma_{\rm inelastic}
\Label{pp-delta}
\end{equation}

Inaccessible reactions like $\Delta N \to NN$ are calculated 
from their reverse reactions  (here $NN \to\Delta N$) using  
detailed balance.

The possibility of reaching a channel in a collision is given by its
contribution to the total cross section:
\begin{equation}
 P_{\rm channel} = \frac{\sigma_{\rm channel}}{\sigma_{\rm tot}} \qquad {\rm e.g.}
 \quad P_{pp \to p \Delta^+}= \frac{1}{4} \frac{\sigma_{\rm tot}-\sigma_{\rm el}}{\sigma_{\rm tot}}
\Label{proba}
\end{equation}
In the example we took use of eqns. \ref{tot-el-inel}, \ref{pp-example} and 
\ref{pp-delta}.
 
In the numerical simulation the choice of the channel is done randomly 
with the weight of the probability of the channel.

\subsection{Virtual particles}
The production of kaons in this energy domain is a very rare process.
Their production cross section is only a few nanobarn. In comparison of
a total cross section of about 20-40 mb the possibility of producing
strangeness is very small. Therefore, the method presented a few lines above
will cause severe statistical limitations to the description of kaons in this
way. However, simulation codes oriented toward higher energies, like UrQMD 
\cite{urqmd} apply successfully this method for reactions nearby the threshold. 
Their results (without optical potential) are quite comparable to IQMD
results \cite{bleicher}.
A method to overcome this problem is the way of ``perturbative production''
of kaons, as it was for example done in \cite{kaon94}. 
In this method one only looks for the probabilities of producing a 
kaon in a collision (see eq. \ref{proba}) and notes these reactions with their
probability $P$, their cm-momentum $\vec{p}$ and their invariant mass
$\sqrt{s}$. The collision itself continues normally (without kaon production)
and later on the reactions are analyzed to estimate the properties of the 
kaons.

However, this method has the disadvantage that further
interactions of the particles can only be roughly estimated. 
To overcome this we use the method
of virtual particles:
\begin{itemize}
\item Each particle $i$ has a probability $P_i$. Protons, neutrons, deltas
and pions start with $P_i=1$.
 
\item After production of a new particle with a reaction probability $P_r$
the parents have a probability $(1-P_r)$ for continuing undisturbedly.
The produced particle has the probability $P_n=P_1\cdot P_2 \cdot P_r$, where
$P_1$ and $P_2$ are the probabilities of the parents.

\item 
In the collision of two particles $i$ and $k$ with the probabilities 
$P_i$ and $P_k$, the final state of the collision will be calculated.
Each particle will join this final state with the probability of his 
reaction partner:
\begin{equation}
P(\mbox{\rm $i$ into final state})=P_k \qquad P(\mbox{\rm $i$ remains in prev. state})=1-P_k
\end{equation}

\item
The interactions potentials are the interaction potentials of free
particles multiplied by the probabilities of the interacting particles.

\begin{equation}
V_{i}=\sum_k P_k V_{ik}
\end{equation}

\end{itemize}

\begin{figure}[hbt]
\epsfig{file=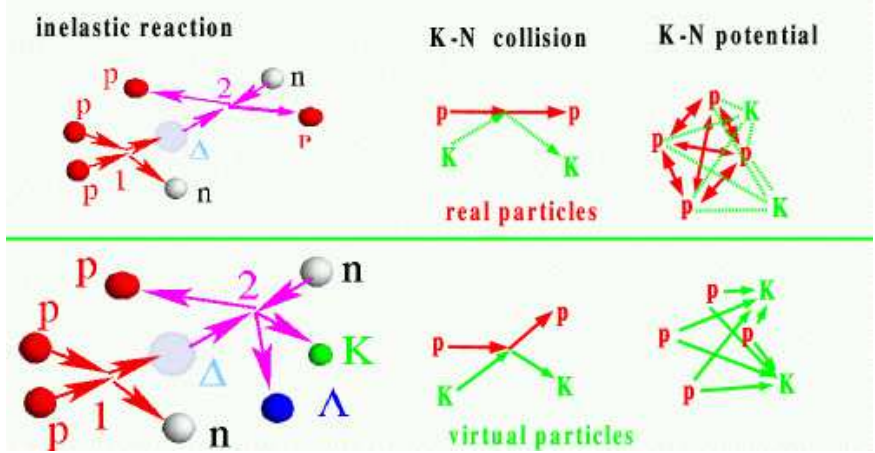,width=0.7\textwidth} 
 \caption{The major differences of ``real'' and ``virtual'' particles.}
\Label{virtual-particles}
\end{figure}

Since $P \ll 1$, $ 1-P \approx 1$, we can simplify the scheme in the following way.

\begin{itemize}

\item
Nucleons, deltas and pions are {\it real} particles with $P=1$. Strange 
particles are {\it virtual} particles who have a very small probability $P_i$.

\item
In a collision virtual particles are produced with a reaction
probability $P_r=\sigma({\rm K prod})/\sigma(tot)$. The parent particles
neglect this production and follow another channel of the collision according
to its probability. (Fig. \ref{virtual-particles} , top-left).

The produced particles act as if the production reaction had taken place
and carry a probability of $P_r$ (Fig. \ref{virtual-particles} , bottom-left).

\item
In a collision of a real and a virtual particle, the real particle will
ignore this reaction (Fig. \ref{virtual-particles} , top-middle). 
The virtual particle will act, as if this collision had taken place
(Fig. \ref{virtual-particles} , bottom-middle).

\item
The real particles do not see a potential interaction with the 
virtual particles  (Fig. \ref{virtual-particles}, top-right). 
The virtual particles feel a potential with the nuclear matter
(Fig. \ref{virtual-particles}, bottom-right).

\end{itemize}

This method has the advantage to allow for high statistic calculation of
kaon one-body observables including all effects of the medium like potential
propagation, rescattering, absorption etc. 
However, there are some major drawbacks of this method

\begin{enumerate}
\item
The energy conservation is no more assured on an exact level. Thus, the
event characteristics of kaon producing events are identical to the 
characteristics of events without kaons. Questions like `{\it Do events with
kaons produce less high energetic pions than other events}' (which would be
interesting for analogies between kaons and high energy pions) 
cannot be addressed.

\item
KN- correlations cannot be performed since the event characteristic is not
correct for kaon producing events.

\item
Higher order processes might be described incorrectly. To give an example
for a problematic description:
\begin{enumerate}
\item A $NN$ collision produces virtually $N\Lambda K^+$. In the `real world'
it produces a $N\Delta$-pair.

\item The virtual $\Lambda$ rescatters with another nucleon while the real $\Delta$
decays into $N \pi$.

\item The virtual $\Lambda$ and the real $\pi$ resulting from the delta decay
scatter and produce a $N K^-$-pair. 
\end{enumerate} 
The latter process should not be allowed since a real production of the  $\Lambda$ 
would have avoided the  $\Delta$-production and therefore the $\pi$ should not
exist if the  $\Lambda$ is there.
This process is explicitly forbidden (by triggering on different parents
of the collision partners) but it could be hidden by intermediate rescattering
of a pion $\pi N \to \Delta \to \pi N$.
However, its probability is rather low.

\end{enumerate}

\subsection{The KN-optical potential}
The description of the  KN-optical potential is a subject of vivid discussion.
We use in our calculation a parametrization calculated by J\"urgen
Schaffner-Bielich \cite{schaffi} which is derived from relativistic mean field
(RMF) calculations and transformed into a Schr\"odinger-type potential
of the following form:

\begin{equation} 
 U_{opt}^K = \sqrt{(\vec{k}-g_v\vec{\Sigma_v})^2 + m_K^2 + m_Kg_s\Sigma_s }+
  g_v\Sigma_v^0     - \sqrt{k^2 + m_K^2 }
\Label{uopt-eq}
\end{equation} 

The parametrization contains scalar fields ($g_s\Sigma_s$) which couple
to the mass term and vector fields $g_v(\Sigma_v^0,\vec{\Sigma_v})$
which couple to the momenta.
The used parametrization is depicted 
on the l.h.s. of fig.\ \ref{opt-pot-1}.

\begin{figure}[hbt]
\begin{tabular}{cc}
\epsfig{file=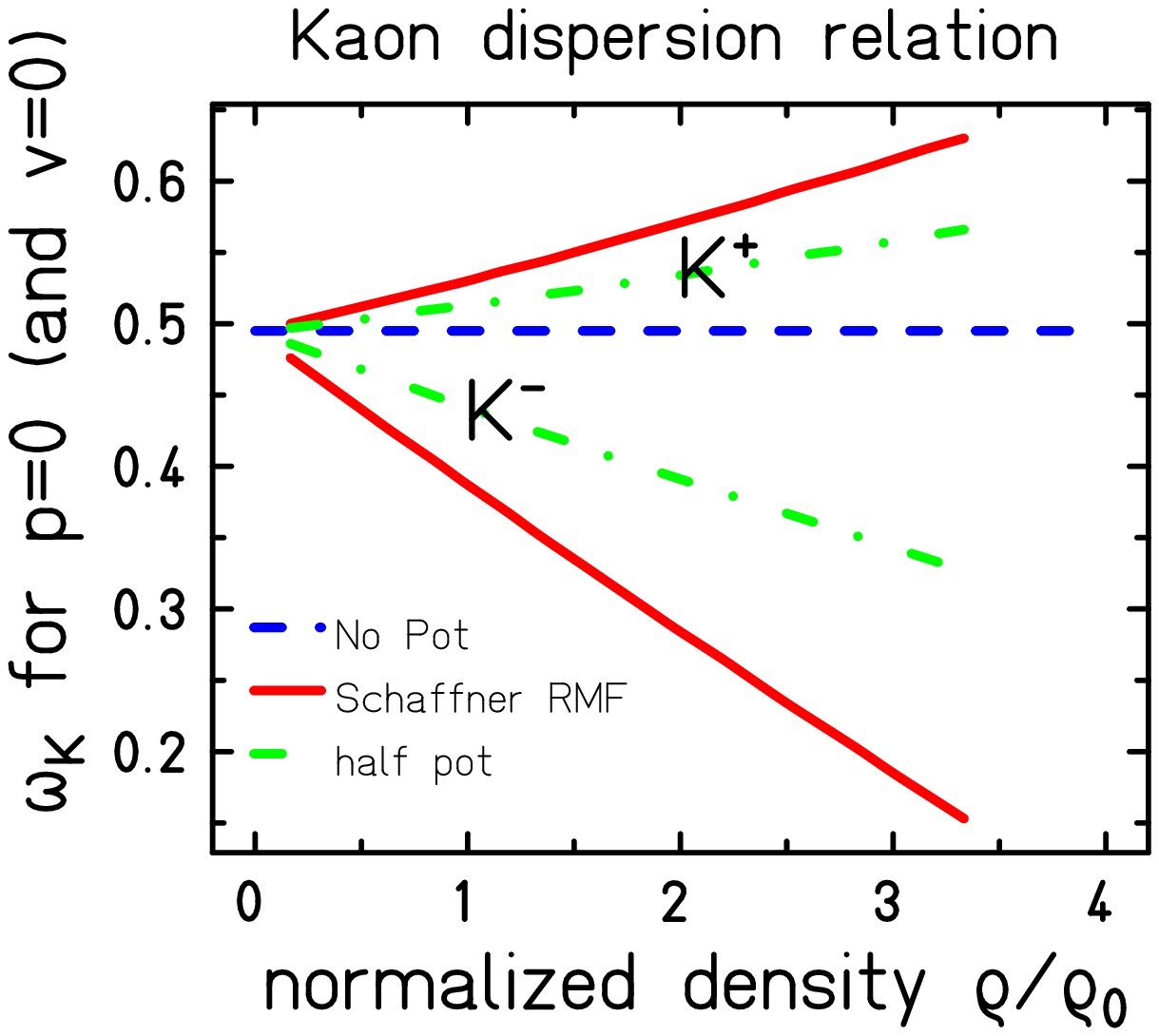,width=0.4\textwidth} &
\epsfig{file=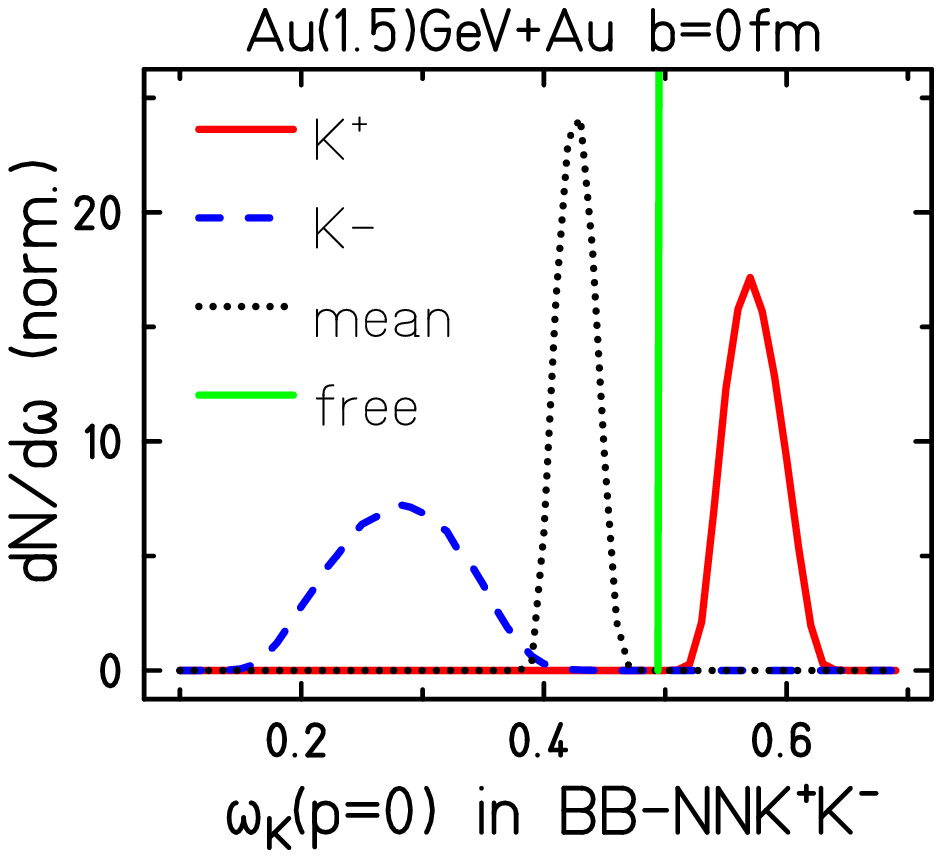,width=0.4\textwidth} \\
\end{tabular}
 \caption{The used parametrization of the optical potential and its influence
 on the kaon production in $BB\to NN K^+K^-$ reactions in Au(1.5 AGeV)+Au 
 collisions}
\Label{opt-pot-1}
\end{figure}

We see that the potential (red full line) enhances the zero-momentum
energy $\omega_K$ (which corresponds to an `in-medium mass' ) 
for the $K^+$ and reduces it for the $K^-$ if the system
becomes more dense. Without potential (blue dashed line) this energy corresponds
exactly to the free kaon mass. Thus, the production of a $K^+$ will require
more energy in a dense system. On the contrary, the production of a $K^-$ will
require less energy. The curves reach the value of the free kaon mass if the
density becomes small.

A calculation with reduced parameters (half pot, green dash-dotted line)
will yield less significant changes and simulates a weaker strength 
of the optical potential.  

The r.h.s of fig.\ \ref{opt-pot-1} shows the application of the RMF optical 
potential to a calculation of a Au+Au head-on (b=0fm) collision at 1.5 AGeV.
All baryonic collisions having sufficient energy for a $BB \to  NN K^+K^-$
reaction are analyzed concerning the density where the collision takes place.
The zero-energy of the $K^+$ (red full line) enhances by about 90 MeV, while
the zero-energy of the $K^-$ (blue dashed line) is reduced by about 200 MeV.
The average energy of both particles (black dotted line) is reduced by about
50-60 MeV. Thus, the threshold of this reaction will be reduced in the average
by about 100 MeV.

\begin{figure}[hbt]
\begin{tabular}{cc}
\epsfig{file=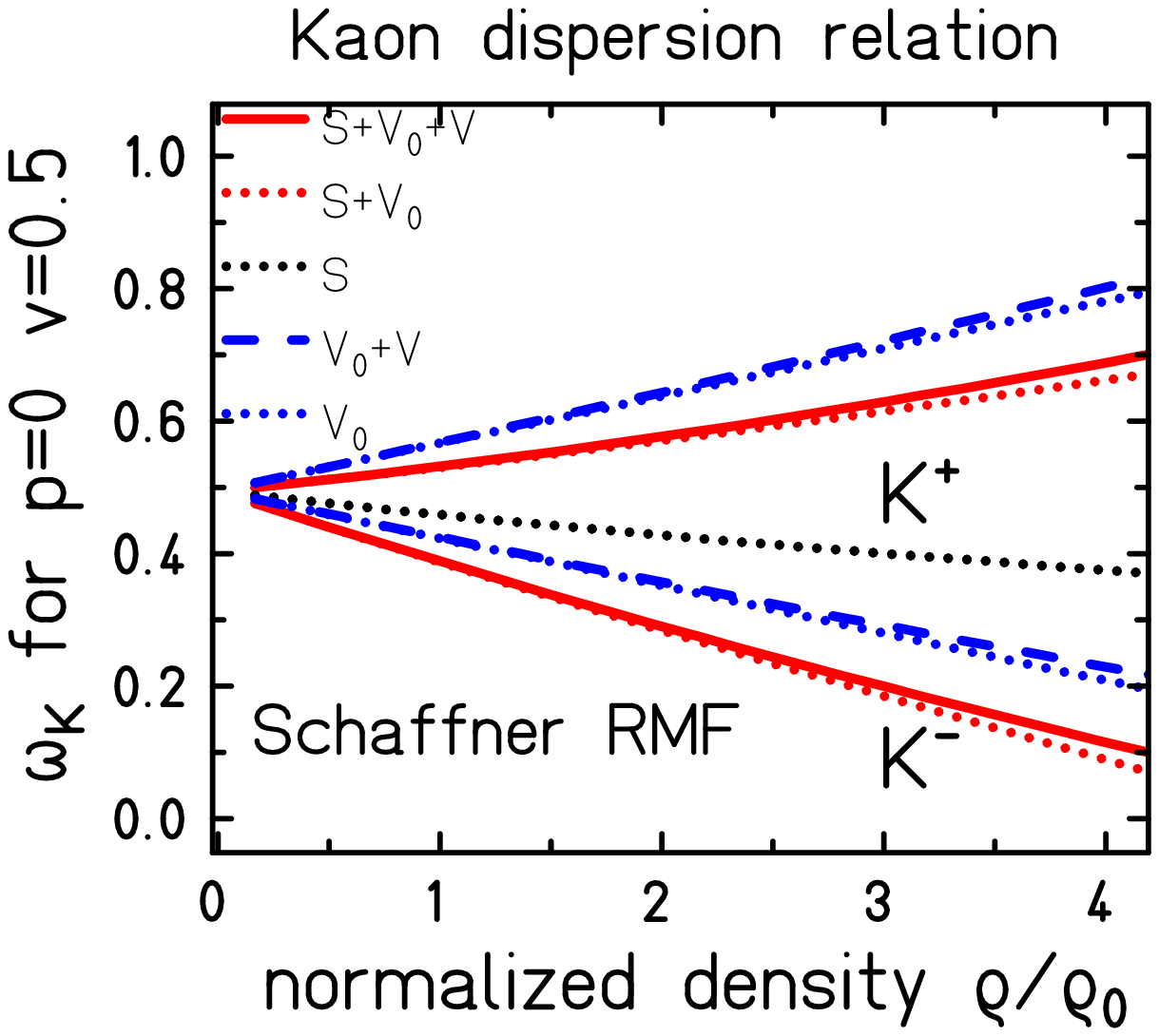,width=0.4\textwidth} &
\epsfig{file=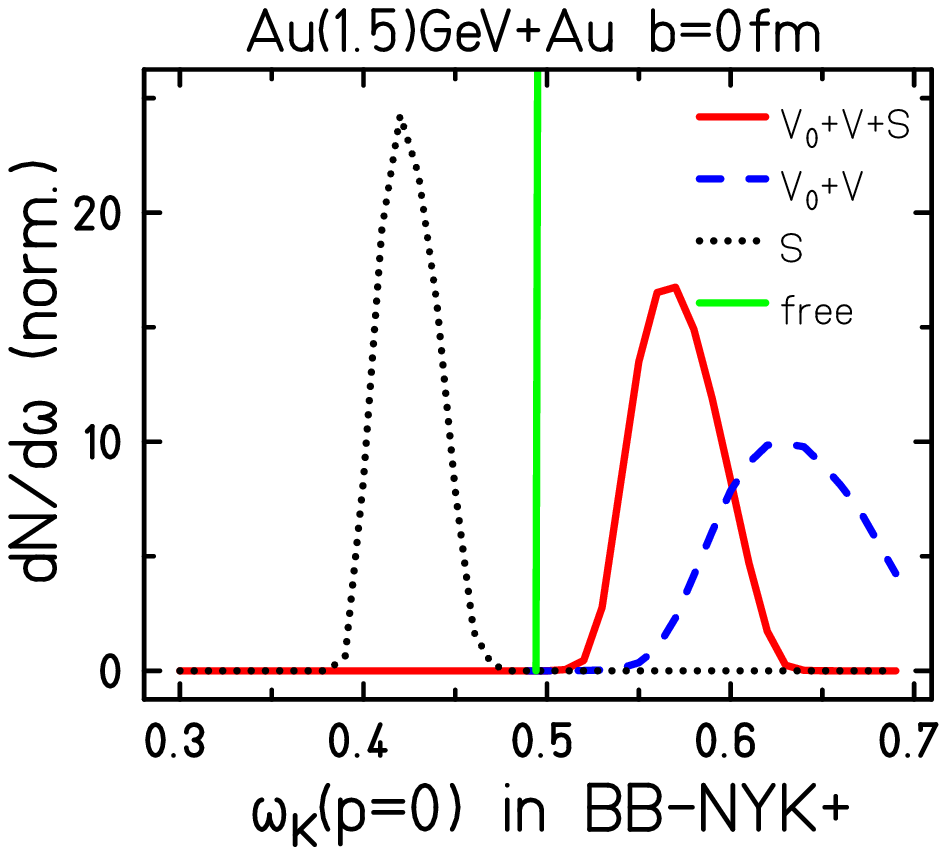,width=0.4\textwidth} \\
\end{tabular}
 \caption{The contribution of scalar and vector part 
 to the optical potential  and its influence
 on the kaon production in $BB\to NYK^+$ reactions in Au(1.5 AGeV)+Au 
 collisions}
\Label{opt-pot-2}
\end{figure}
The used optical potential is a combination of scalar and vector
potentials. 
The scalar part of the potential couples in the same way to $K^+$ and $K^-$ while
the vector part couples to both particles with an inverted sign. Thus, the
difference between  $K^+$ and $K^-$ in fig.\ \ref{opt-pot-1} is due to the vector
part, while the lowering of the average energy of   $K^+$ and $K^-$  (r.h.s.)
is due to the scalar part. Fig.\ \ref{opt-pot-2}  shows on the l.h.s.\ the variations
which occur, when instead of the full potential (red full line) only parts
of the potential are used. If we only employ the scalar part (black dotted line)
both $K^+$ and $K^-$ will show a lowering of the energy, which will be identical
for both $K^+$ and $K^-$. The lowering of the $K^-$ is weaker than in the full
combination of scalar and vector part.
If we only use the vector parts (blue lines) the difference
of  $K^+$ and $K^-$ will remain but the energies will be quite higher for both
$K^+$ and $K^-$. A reduction of the vector potential to the temporal part only
$g_v\Sigma_v^0$ will slightly lower the energy. This effect depends on the
relative velocity $v$ of the nuclear medium.

On the r.h.s.\ of fig.\ \ref{opt-pot-2} the effect of the different parts on 
the $K^+$ production is shown. For a Au+Au head-on collision at 1.5 AGeV all
baryon-baryon collisions with sufficient energy to produce a kaon (the lowest
threshold is the threshold of the $BB\to NYK^+$ reaction) have been analyzed.
A use of the scalar potential (black dotted line ) lowers the kaon energy and 
thus the threshold by about 70 MeV, while the vector potential only (blue
dashed line) yields a strong enhancement. The full potential (red full line)
shows an average enhancement of about 70 MeV. This value is slightly lower than
the average value obtained from fig. \ref{opt-pot-1} since the threshold for the reactions
regarded here is lower than in \figref{opt-pot-1}. 
For those reactions with lower thresholds the mean density of the collisions
is lower.

\begin{figure}[hbt]
\begin{tabular}{cc}
\epsfig{file=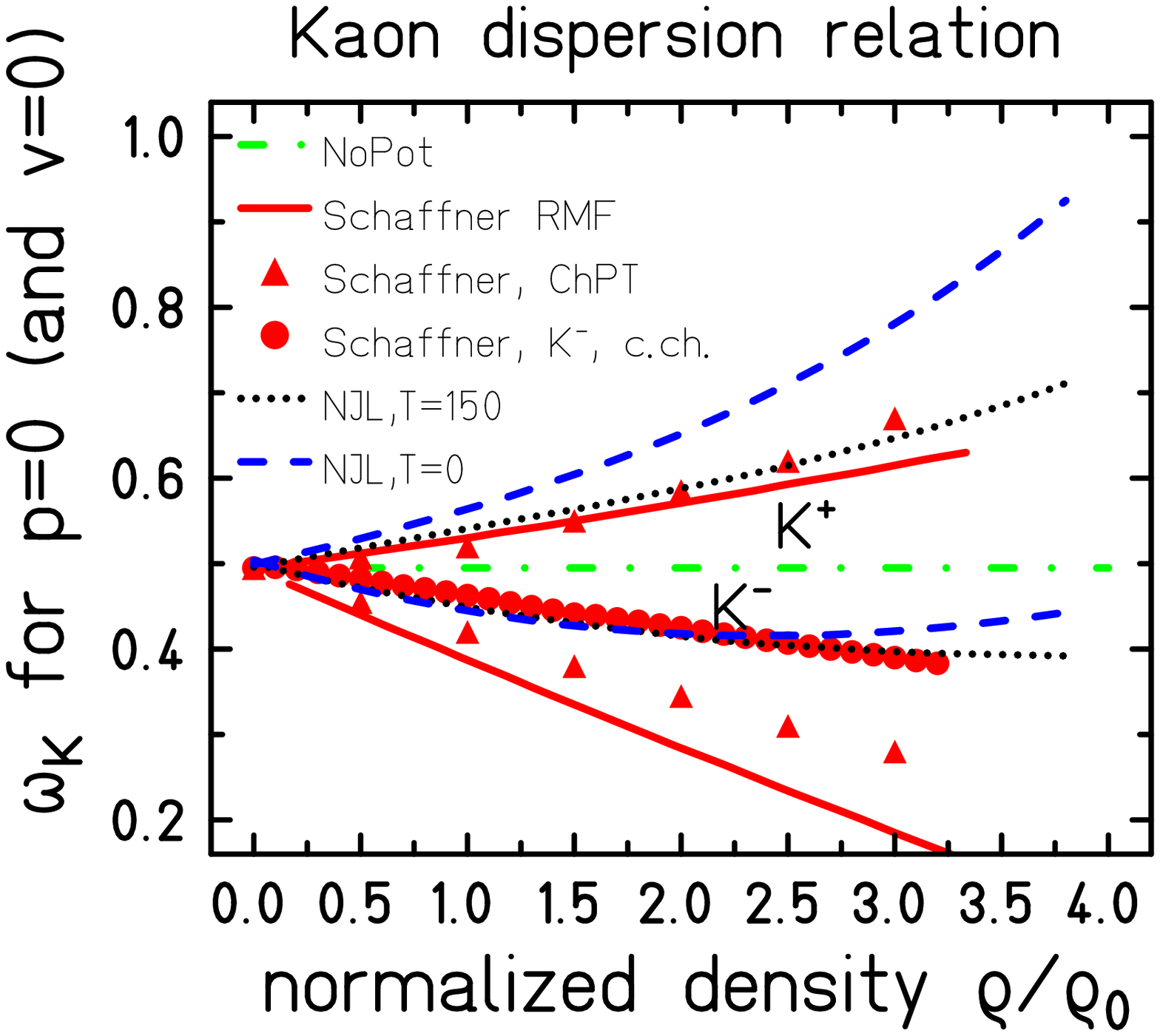,width=0.4\textwidth} &
\epsfig{file=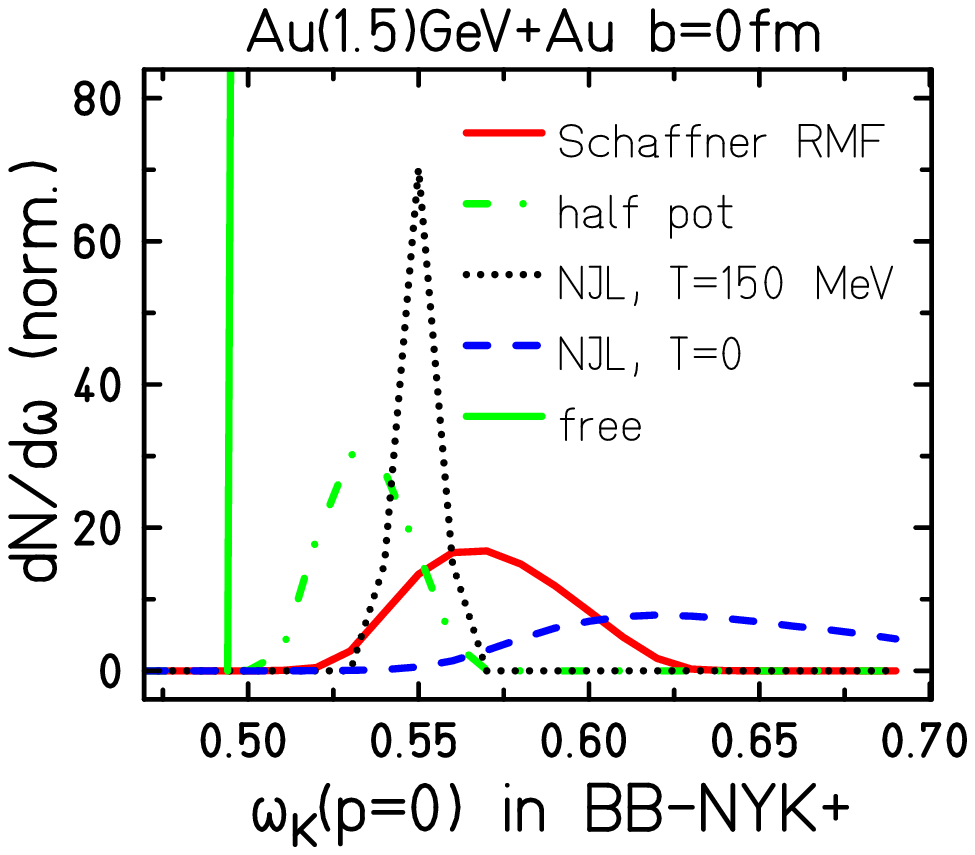,width=0.4\textwidth} \\
\end{tabular}
 \caption{ Comparison of the used parametrization of the optical
 potential to other parametrization and their influence
 on the kaon production in $BB\to NYK^+$ reactions in Au(1.5 AGeV)+Au 
 collisions}
\Label{opt-pot-3}
\end{figure}

Finally, fig. \ref{opt-pot-3}  compares the used optical potential 
(Schaffner RMF,
red full line) to other calculations like Chiral Perturbation Theory (ChPT, red
triangles) or calculation resulting from the Nambu-Jona-Lasinio model (NJL)
using different assumptions for the temperature ($T=150$MeV, black dotted and
$T=0$, blue dashed line). 
Except for the NJL, $T=0$ calculation the values for the $K^+$ are quite similar.
However, the values for the $K^-$ differ visibly. 
In a Au+Au collision the different parametrization will yield different
changes of the threshold (r.h.s.). Therefore, some observables might be different
when using different potential parametrization. We will come to this point
later on.

\subsection{Dynamics of heavy ion collisions}

In order to understand the dynamics of kaon production we want first to
sketch rapidly the dynamics of the whole heavy ion collision in which the
production and propagation of strange particles is embedded.

\begin{figure}[htb]
\epsfig{file=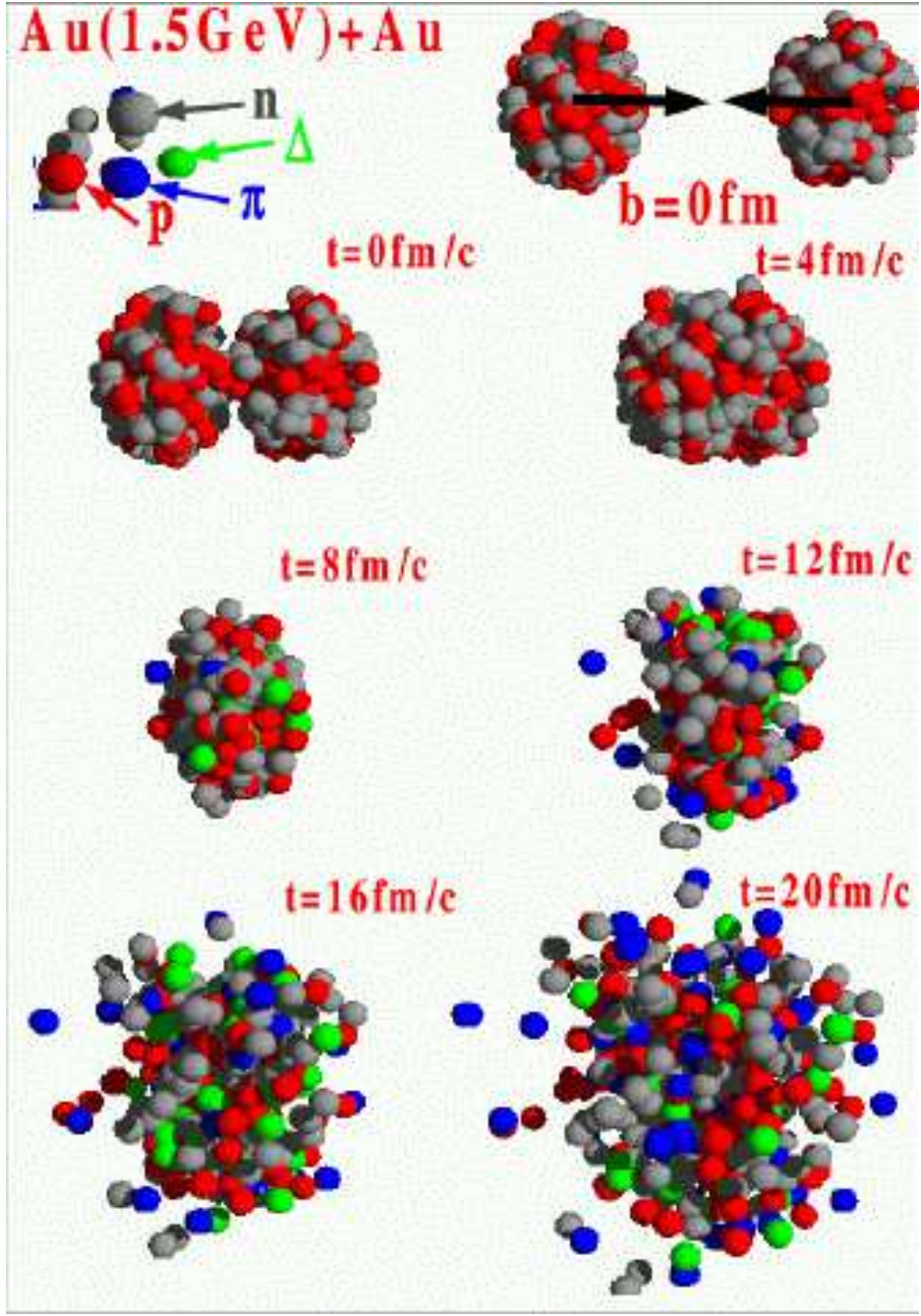,width=0.75\textwidth}
 \caption{Time evolution of a Au(1.5AGeV)+Au collision at 
 $b=0$ fm seen in coordinate space}
\Label{hicpic}
\end{figure}

Fig. \ref{hicpic} shows some `photos' of a heavy ion reaction of Au+Au at
1.5 AGeV incident (lab) energy at an impact parameter of $b=0$fm. 
The scene is seen in the centre-of-mass frame. Therefore, the projectile
and target approach each other in a symmetric way and both show a 
Lorentz contraction of their longitudinal size. Protons are shown as
red balls while neutrons are shown as gray balls.
The time of the whole reaction is very short (less than $10^{-22}$ seconds).
Therefore, it is commonly used to measure the time in units of fm/c.
\begin{equation}
1 \mbox{fm/c} = \frac{10^{-15} \mbox{m} }{3 \cdot 10^{8} \mbox{m/s} }
\approx 3 \cdot 10^{-24} \mbox{s} 
\end{equation}

At $t=0$ fm/c projectile
and target have first contact. First nucleon-nucleon collisions take place
and first resonances are built up in the centre of the reaction. 
The nuclear matter starts to become stopped by inertial confinement and the
density is increasing rapidly. 

At about $t=4$ fm/c the production of resonances is strongly rising. 
However, they are mostly produced in the dense centre and not
`visible' from outward.

At about $t=8$ fm/c the system is reaching maximum density in the centre of
the reaction. Nevertheless, the density in the peripheral region of the reaction
is rather low. First resonances (deltas, green balls) reach the surface.
Some of them decay into pions (blue balls).

At about $t=12$ fm/c the maximum number of deltas is reached. The system is still
highly compressed. Most of the nucleons have collided and do no more carry
the initial momentum. The spectra of the nucleons are highly non-thermal.

At about $t=16$ fm/c the pions have overtaken the dominance over the Deltas.
The system is in expansion. The central density falls down rapidly.

At about $t=20$ fm/c the system is dominated by the expansion.
The number of high energetic collisions drops strongly. There are
still some deltas who will decay into nucleons and pions. 
The momentum distributions of the nucleons start to approach their 
final values.

\begin{figure}[hbt]
\begin{tabular}{cc}
\epsfig{file=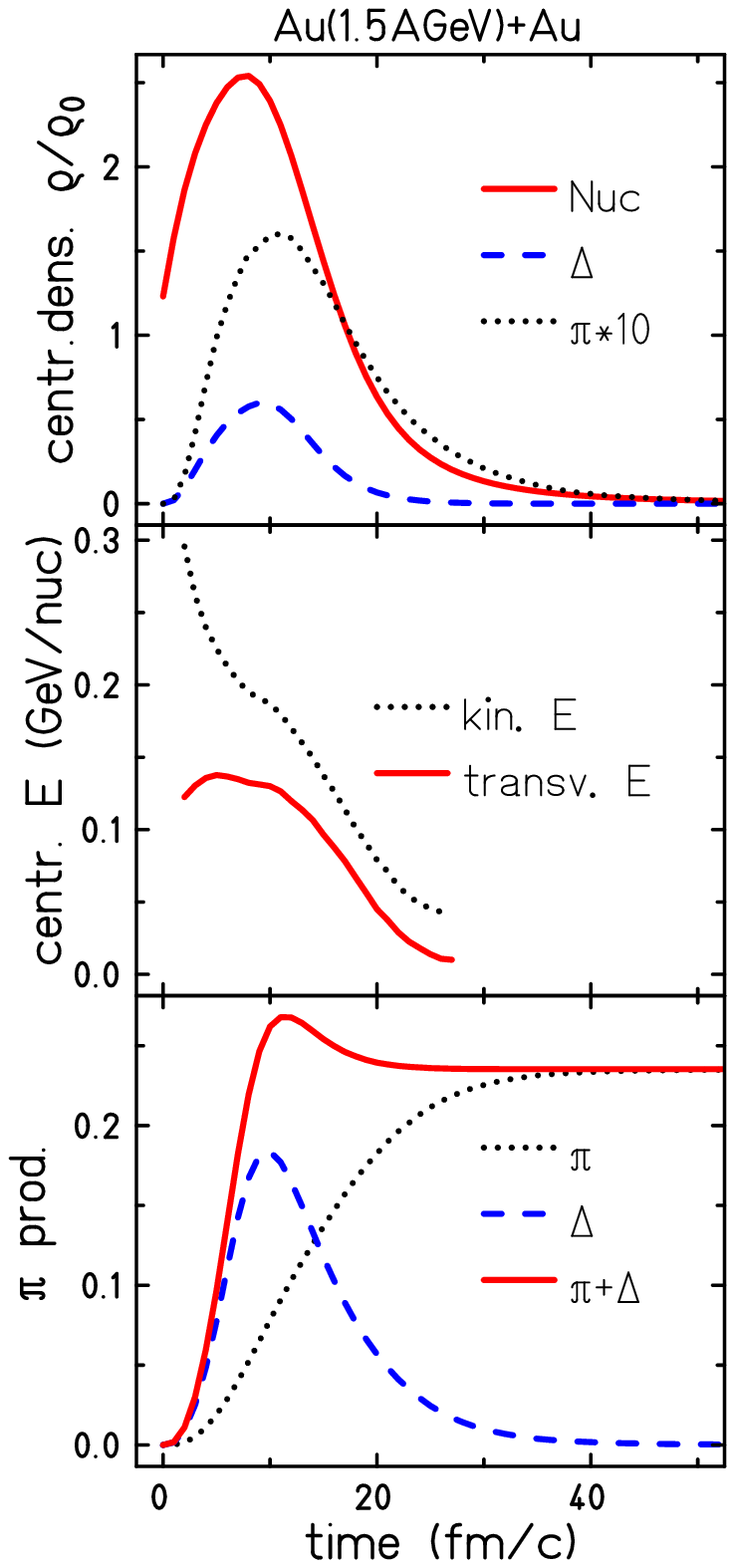,width=0.3\textwidth} &
\epsfig{file=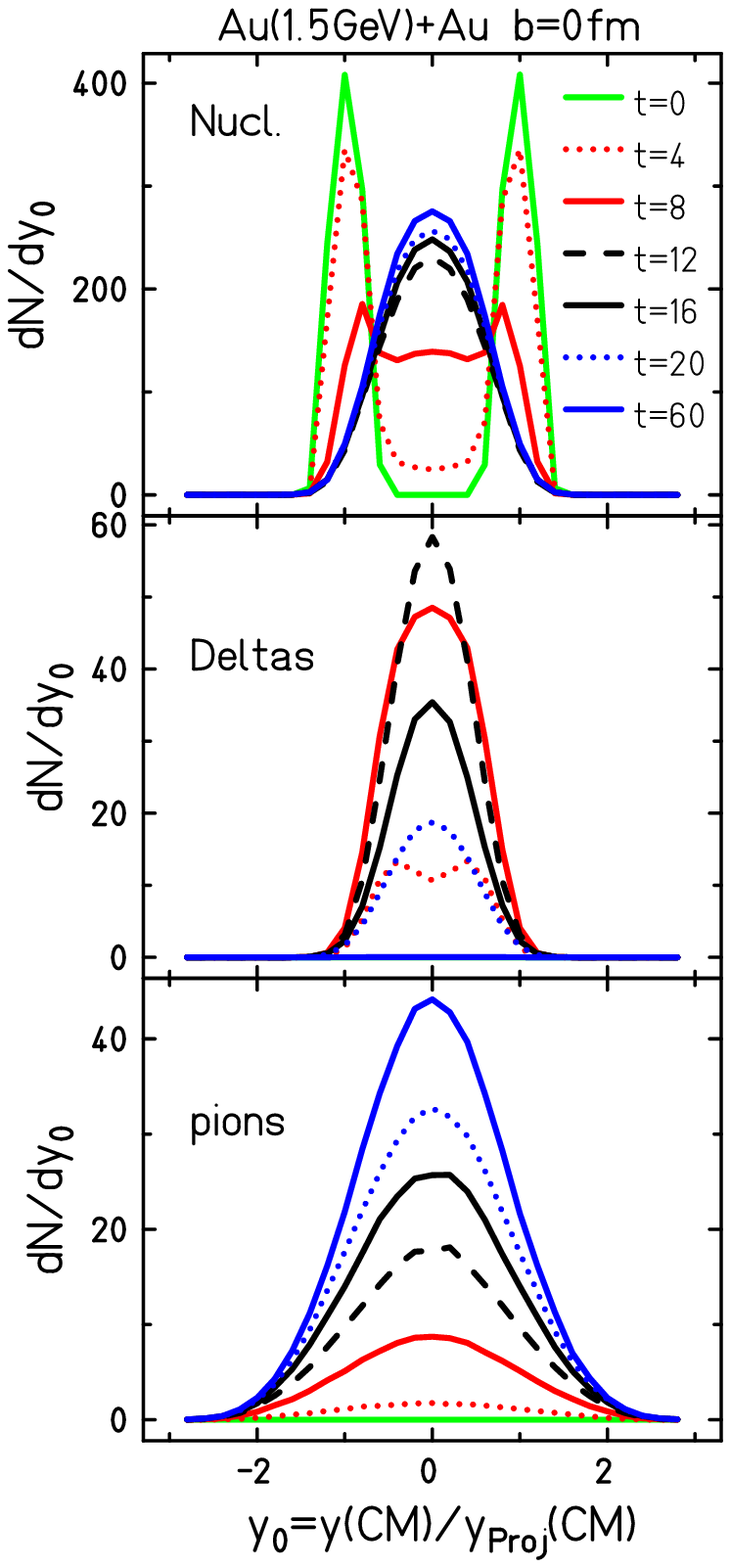,width=0.32\textwidth} \\
\end{tabular}
 \caption[Time evolution of density, energy, pion number and 
 rapidity distributions]{Time evolution of a Au(1.5AGeV)+Au collision at b=0fm.
 Left: Evolution of central density, central energy and pion number
 as function of time. Right: Evolution of the normalized rapidity
 distributions }
\Label{time-evol-dns}
\end{figure}

Fig. \ref{time-evol-dns} shows on the left hand side the time evolution
of the density in the central region (top), the mean total and transverse energy
in the central region (mid) and the number of deltas and pions (bottom).
The nucleon density (top, red full line) grows up rapidly, reaches a maximum
of about three times ground state density 
at about 8 fm/c and falls down afterward. The delta density (black dotted line)
starts later, reaches a maximum  of nearly half ground state density 
at about 9 fm/c and falls down rapidly. 
The pion density (multiplied by 10, blue dashed line) starts even later and
reaches  a maximum  of around one sixth ground state density at about 10 fm/c.

In the central region the kinetic energy (middle, black dotted line)
drops down rapidly. The incident energy
in the centre-of-mass frame is due to the big longitudinal initial momentum.
This energy is partly eaten up by the creation of resonances (deltas). 
Another part of the longitudinal momentum is redirected by nuclear collisions into
the transverse direction. Therefore, the transverse energy (red full line)
is built up rapidly in the
central region. In the final expansion phase the fast (high energy) particles
are leaving the central region quite rapidly, leaving the slower particles
behind. The energies are falling down. Finally, the particles are leaving
the central region.

The fast dropping of the energy in the mid-part of the lhs of fig. \ref{time-evol-dns} 
is accompanied by a fast increase of the number of deltas (bottom, blue 
dashed line). As already stated, the resonances eat up a big amount of the 
energy. This energy will be released later on in form of pions (black dotted line). 
Their number is continuously increasing to reach the final number at about 
30-40 fm/c. Pions are strongly interacting with nucleons. Therefore, pions
decaying in the dense medium have a high chance to be reabsorbed and to refeed
the number of deltas. This effect explains the slow increase of the pion number
when the density is high. The deltas themselves can be reabsorbed by the nuclear
matter. This effect can be seen when looking at the total number of deltas and
pions (full red line). It shows a maximum at about 10-12 fm/c and decreases
slightly afterward. After about 20 fm/c it stays roughly constant. Now 
only the contribution of deltas and pions to the total number is changing.

The right hand side of fig. \ref{time-evol-dns}
shows the time evolution of the rapidity distribution of nucleons (top), 
deltas (middle) and pions (bottom). The rapidity has been scaled to the 
projectile rapidity in the centre-of-mass. Thus, 1 corresponds to projectile
rapidity, -1 corresponds to target rapidity.

At $t=0$ fm/c (\gfl) projectile and target show their incident momenta as peaks at
projectile and target rapidity. The broadening of the peaks is due to the
Fermi momentum of the nuclei. Deltas and pions do not exist at this time.

At $t=4$ fm/c (red dotted line) first nucleons have been stopped to mid-rapidity.
First deltas have been produced. The stopped nucleons collide with incoming
nucleons of projectile and target causing a slight double peak of the delta
rapidity distribution at about the middle of cm-rapidity (0) and projectile
(1) resp. target rapidity (-1). There are nearly no pions.

At $t=8$ fm/c (\rfl) when maximum density is reached, the nuclei have still not been
completely stopped. There are still remnants of the peaks at projectile and
target rapidities. The number of deltas has strongly increased. Its distribution   
is now peaked around mid-rapidity. The rather narrow width is due to the high
mass of the delta which eats up a big part of the energy available in the
first collisions. After the decay of the deltas a big amount of momentum is
given to the light pions. Therefore, the rapidity distribution of them becomes
quite broad.

At $t=12$ fm/c (black dashed line) the nucleon rapidity is peaked at mid-rapidity.
The delta distribution shows its maximal values while the pion distribution
rise up continuously.

At $t=16$ fm/c (full black line)  and $t=20$ fm/c (blue dotted line) the 
nucleon rapidity distribution is slightly growing up in the centre. This is
due to the feeding by the decay of the deltas whose rapidity distribution
is continuously falling down. For kinematic reason the nucleon only gets
few energy from the decay while most of the energy is given to the light
pion whose broad rapidity distribution is still rising in number.

At $t=60$ fm/c (full blue line) the reaction is in the final state. There are
no more deltas. The system is expanding and the particles will direct outward
and finally touch the detectors.    
 
\subsection{Comparison to experiment}
In the discussion of the kaon dynamics we will later on find a lot of cross talk
between the nucleons and the kaons. Therefore we want first to assure that the
dynamics described above is comparable to experiment. This will allow us as well
to study some ingredients which may have some influence on the kaon production.
 
\begin{figure}[hbt]
\begin{tabular}{cc}
\epsfig{file=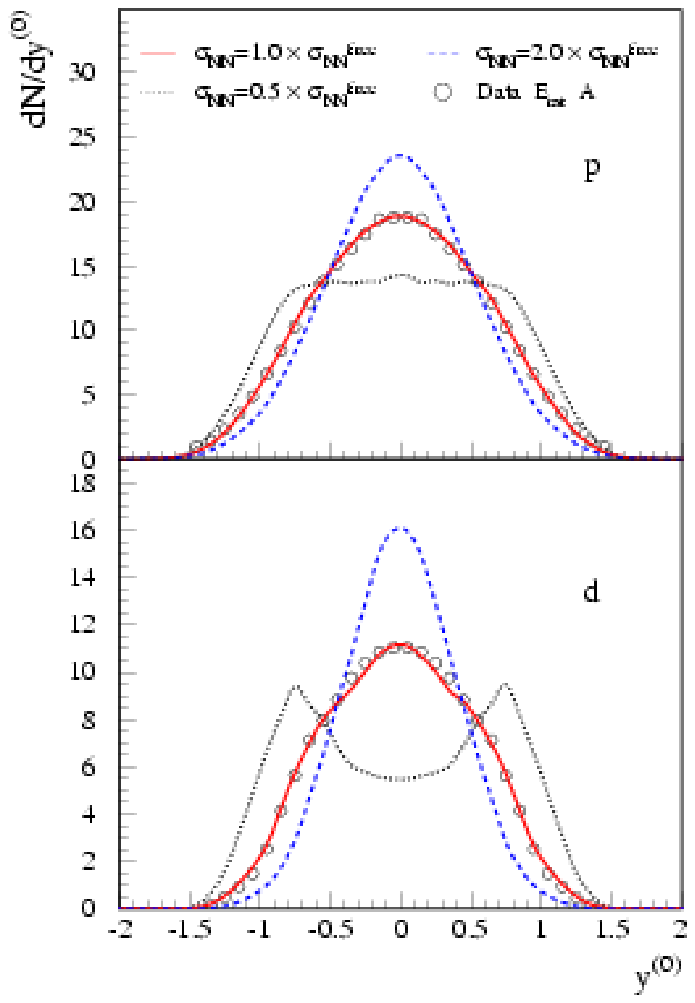,width=0.4\textwidth} &
\epsfig{file=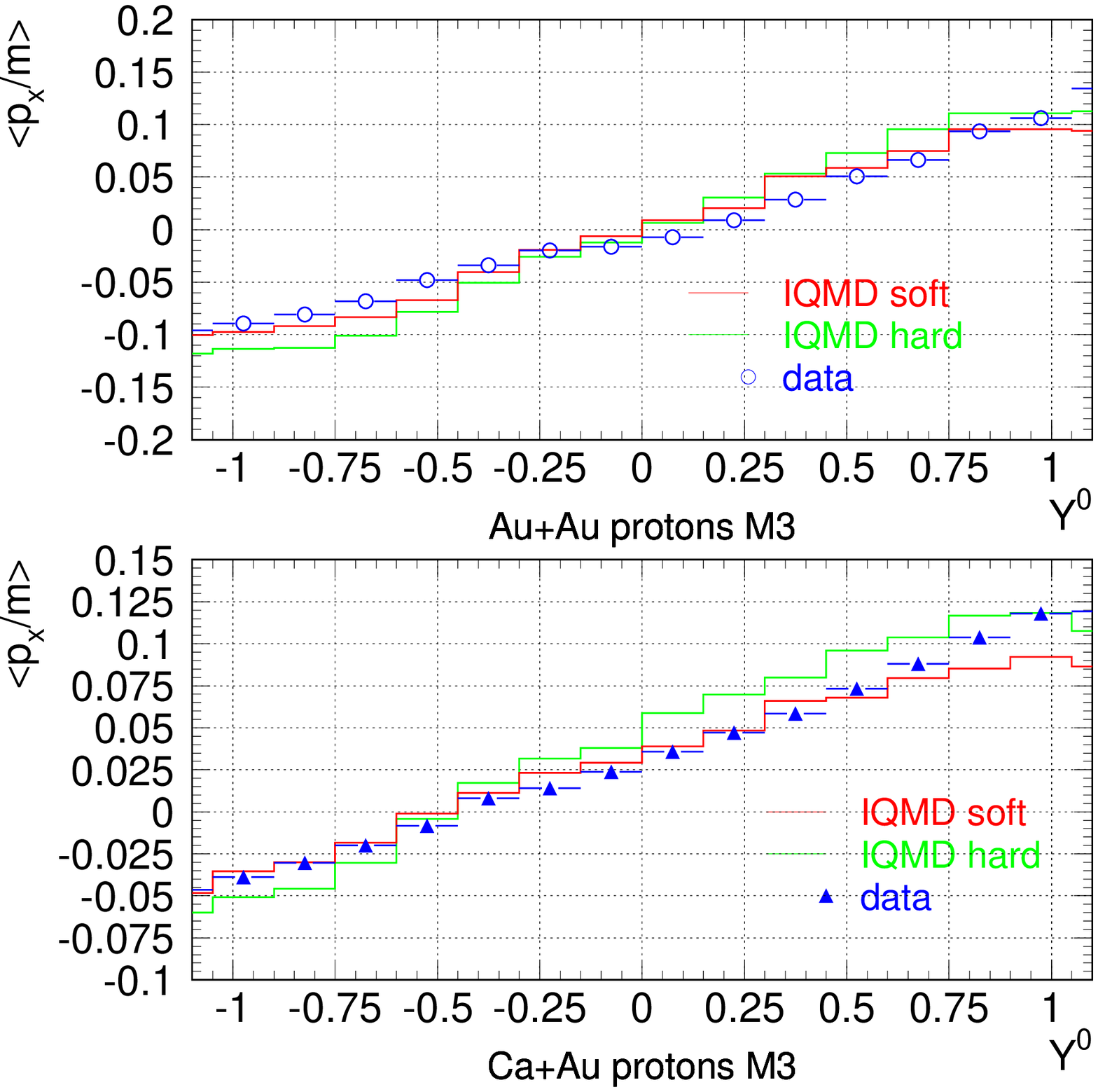,width=0.5\textwidth} \\
\end{tabular}
 \caption{Comparison of IQMD calculations using different nucleon-nucleon 
 cross sections to Ru+Ru data of FOPI (left) and comparison of Ca+Au and
 Au+Au flow data (right)}
\Label{nuc-compa}
\end{figure}

On the \lhsref{nuc-compa} we see a comparison of IQMD results with FOPI data
performed by the FOPI collaboration \cite{hong} on the system Ru+Ru at 400 AMeV
incident energy. Here the rapidity distribution of protons (top) and
deuterons (bottom) are studied. IQMD uses as cross section the free 
nucleon-nucleon cross sections $\sigma_{NN}^{free}$ (full line). 
The effect of the surrounding
medium is taken into account by requiring the validity of the Pauli principle
in the final state. In dense matter the cross section is effectively reduced.
There is the possibility to apply additional factors the cross section as it is
done on \lhsref{nuc-compa}. A reduction of the cross section to half of the value
($0.5\times\sigma_{NN}^{free}$, dotted line) yields less stopping and a broader
rapidity distribution while the doubling of the cross section 
($1.0\times\sigma_{NN}^{free}$, dashed line) yields a smaller rapidity
distribution. The data (circles) support an unscaled free cross section with
Pauli blocking.

The \rhsref{nuc-compa} presents the comparison of transverse flow data
of Au+Au and Ca+Au at 1.5 AGeV incident energy to IQMD results 
also performed by the FOPI collaboration \cite{hartmann}. Here the influence
of the nuclear equation of state is studied. The nuclear equation of state
describes the repulsion of the nuclear matter against compression.
A hard equation of state has a stronger repulsion and thus yields stronger
transverse flow \cite{st86} while a soft equation of state causes less repulsion
and less flow. The comparison seems to prefer the soft equation of state,
an equation of state which will also be supported later on in the discussion
of the influence of the equation of state to the kaon production. 
In general it can be stated that IQMD seems to describe the nucleon dynamics
reasonably well.
 
\subsection{Comparison of pion spectra}
The dynamics of the pions is strongly coupled to the dynamics of the
nucleons. The pions are produced in inelastic collision by the production 
of delta resonances. The delta decays into a nucleon and a pion. The pion
has a high cross section of being absorbed by a nucleon. This absorption 
majorly creates a delta which may again decay into pion and nucleon.
However, it may also be possible that the delta is reabsorbed in an 
inelastic collision with a nucleon.
Thus, long chains of delta-nucleon-pion interactions of these types are
possible \cite{hart}:
\begin{equation}
\begin{array}{rcccccccl}
NN &\to & N \Delta &  & & &   &  & \\ 
 & & N \Delta &\to & NN & & & & \\
  & & \Delta &\to & N \pi & &  & &\\
& &  & & N\pi &\to & \Delta & & \\
&  & & & &  & N \Delta &\to & NN \\ 
 & & & & & & \Delta & \to & N \pi \\
 & & & & & & & &  N \pi \to \ldots \\
\end{array}
\end{equation} 

One major ingredient in this dynamics is the delta decay width 
far off the pole. There are parametrizations of the
lifetimes of deltas which differ especially at low delta masses.
Some examples of them are given on the \lhsref{pi-lifetime}. 
\begin{figure}[hbt]
\begin{tabular}{cc}
\epsfig{file=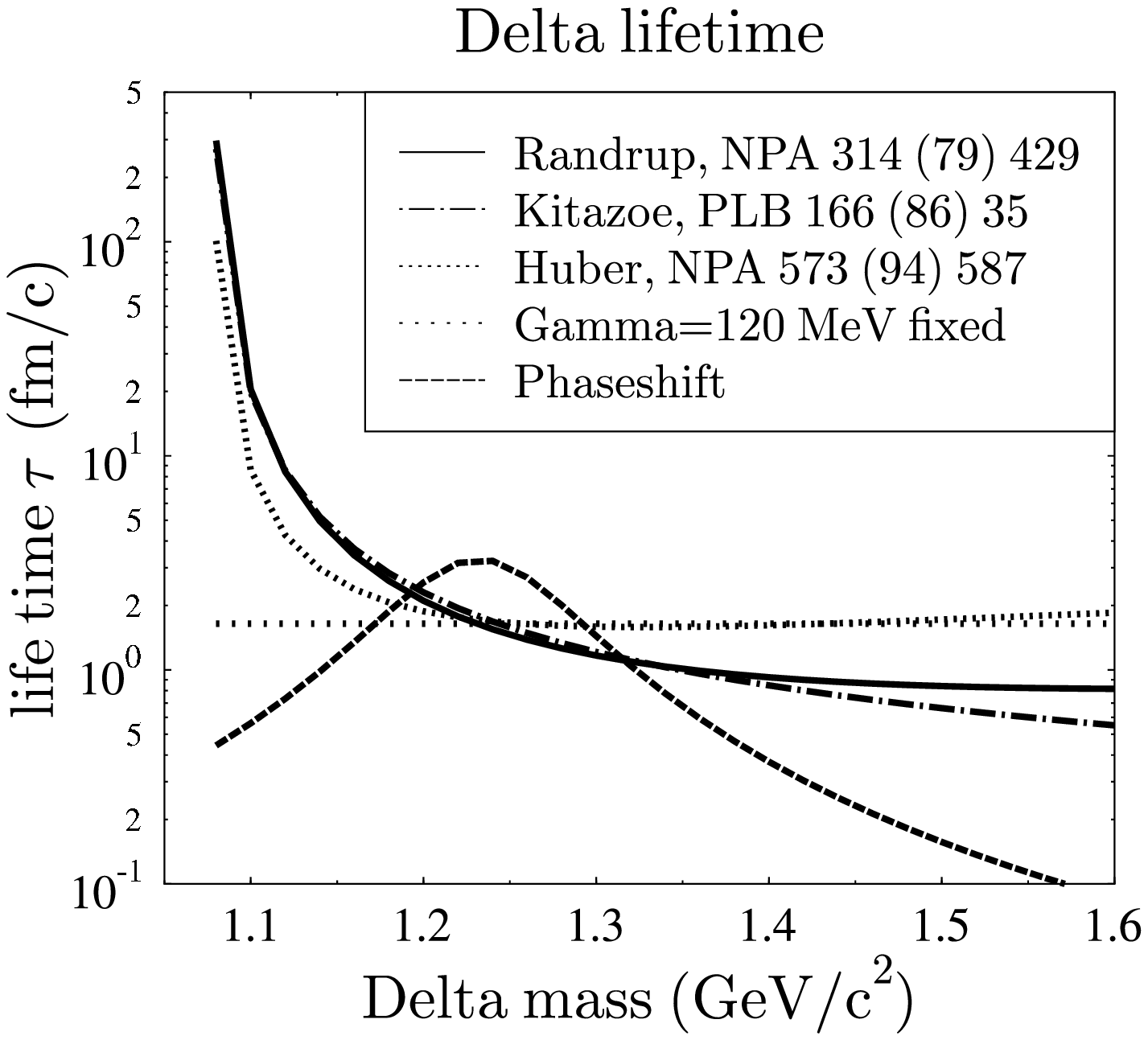,width=0.45\textwidth} &
\epsfig{file=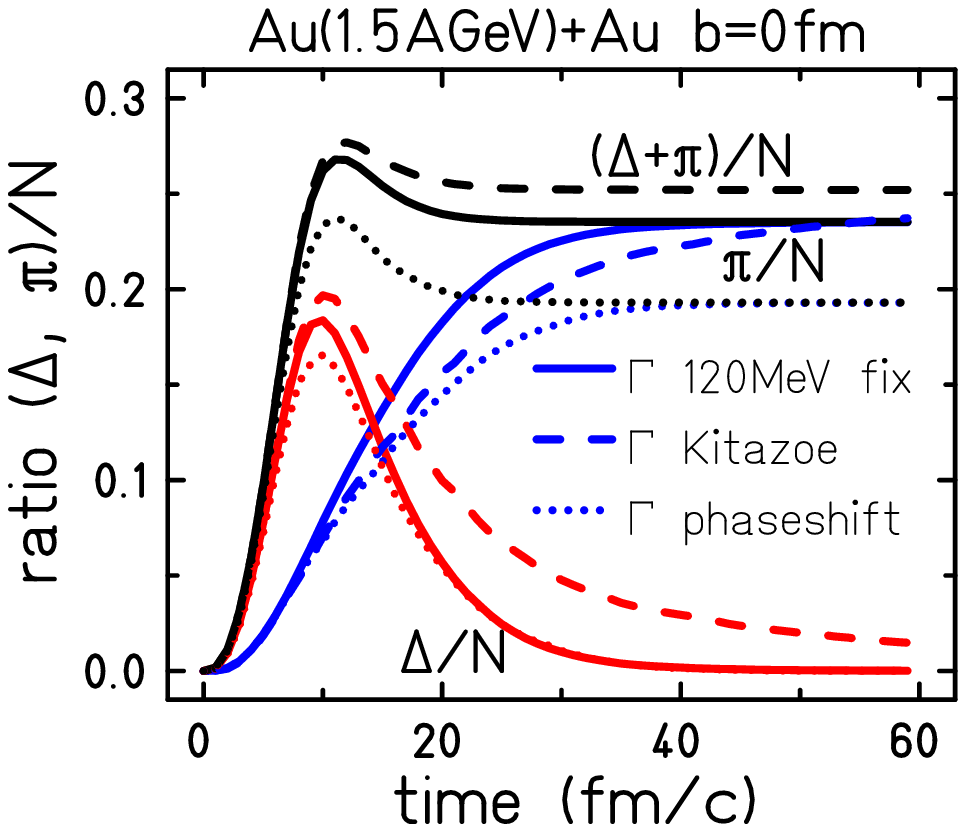,width=0.45\textwidth} \\
\end{tabular}
 \caption{Different parametrizations of the delta lifetime (left)
 and their influence on the delta and pion time evolution (right)}
\Label{pi-lifetime}
\end{figure}
The effect of these parametrizations is shown on the \rhsref{pi-lifetime}
where the yield of deltas (red curves), pions (blue curves) and of their
sum (black curves) normalized by the number of nucleons is shown as a function
of time. The Kitazoe parametrization (dash-dotted line on the \lhs, 
dashed lines on the \rhs) produces deltas which live for longer times and
freeze out quite late. The phase-shift parametrization (dashed line on the \lhs,
dotted line on the \rhs) very short-living deltas far off the resonance and
longer living deltas around the resonance. In the interplay of 
production and absorption they finally produce less pions. This interplay 
effects especially the low-momentum pions, as we can see in 
\figref{pi-compa} where we compare pion spectra of the KaoS collaboration
\cite{Sturm} with IQMD data.

\begin{figure}[hbt]
\begin{tabular}{cc}
\epsfig{file=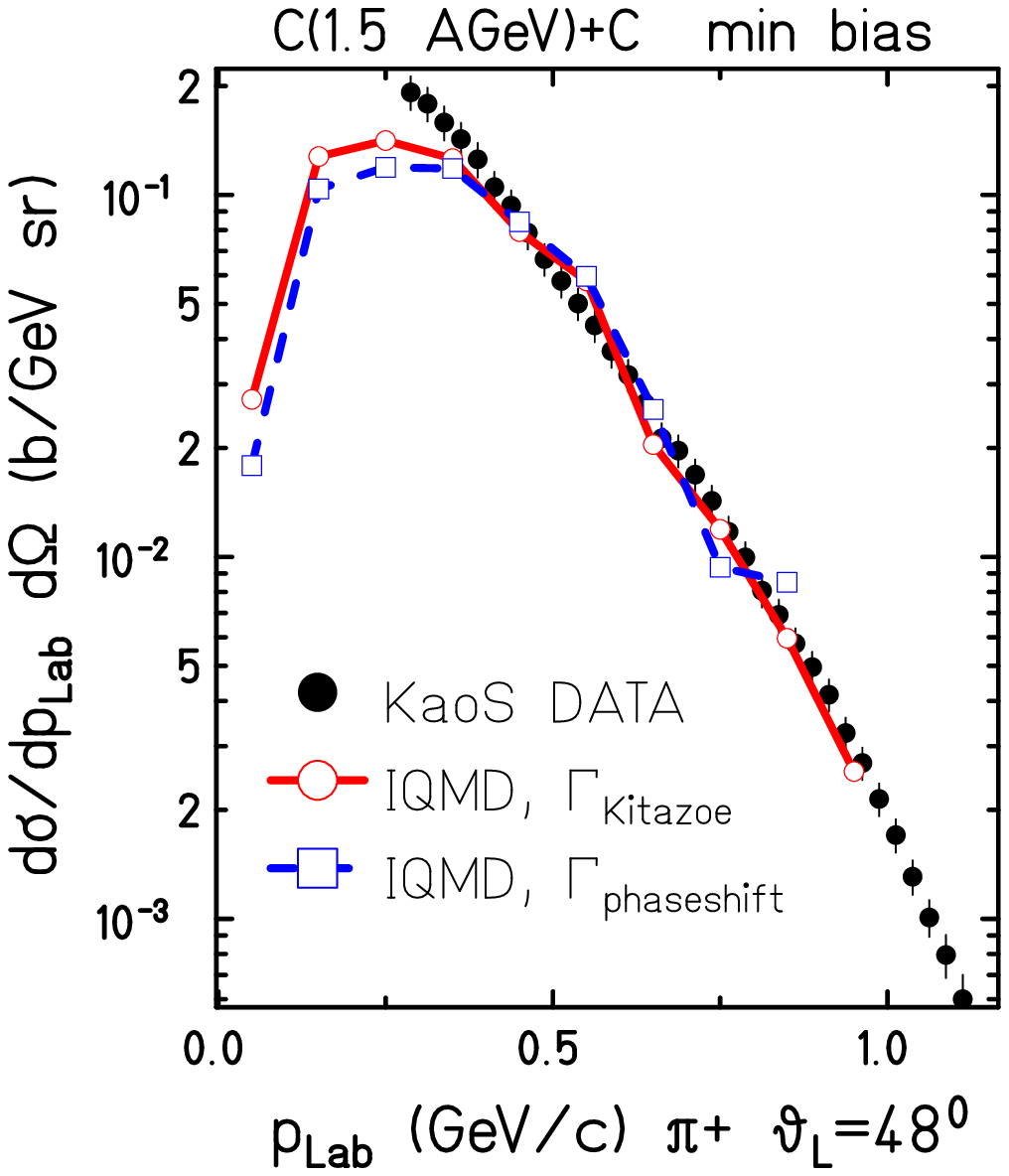,width=0.45\textwidth} &
\epsfig{file=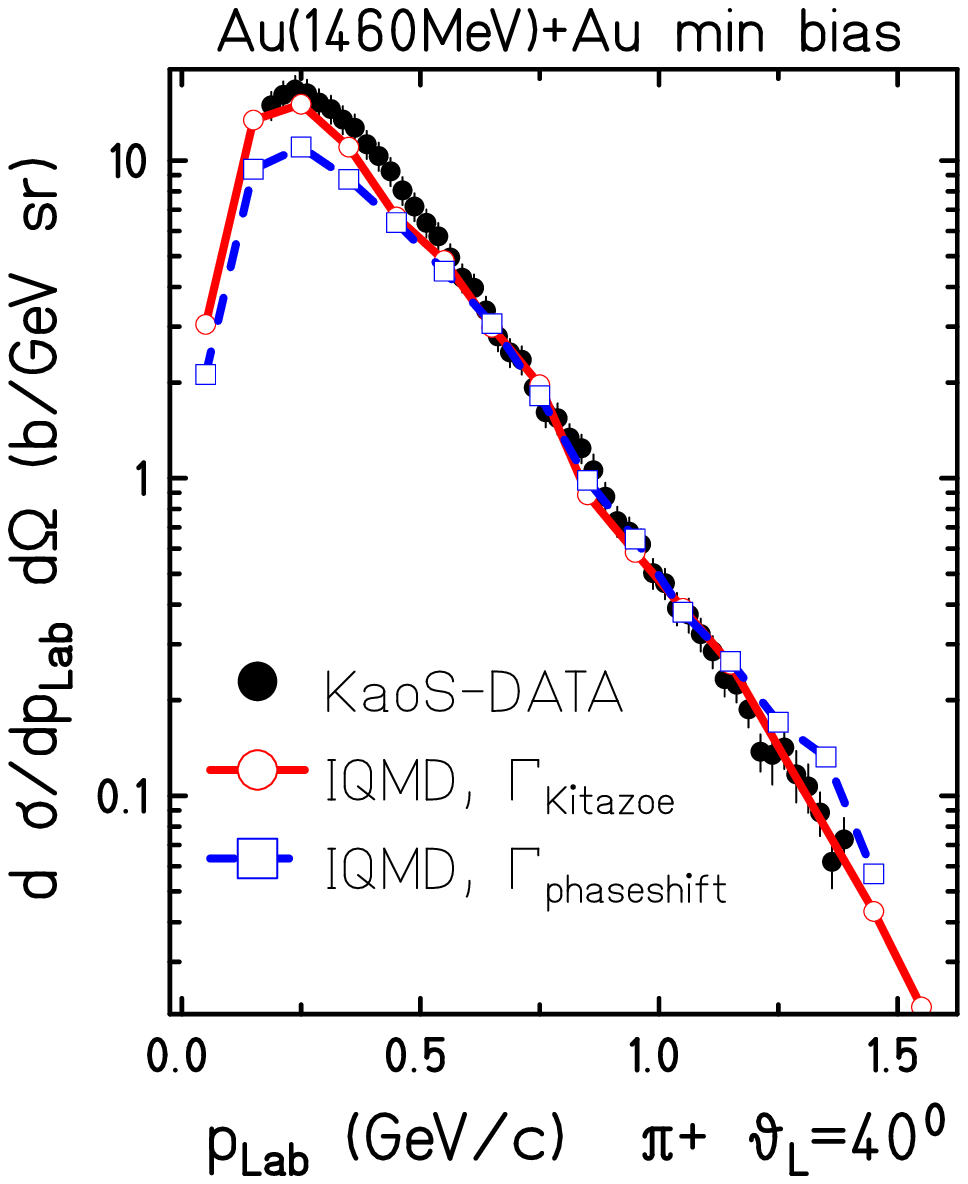,width=0.45\textwidth} \\
\end{tabular}
 \caption{Comparison of IQMD pion spectra using different 
 delta lifetimes with KaoS data for C+C (left) and Au+Au (right) 
 at 1.5 AGeV incident energy.}
\Label{pi-compa}
\end{figure}
We see that at these energies the data support rather the Kitazoe parametrization.
However, at lower incident energies the yields are better described by the
phase-shift parametrization. Furthermore it should be noted that both
parametrizations agree in the high energy part.
We see that also the dynamics of the pions seems to be described 
sufficiently.

\section{Production of $K^+$}

\subsection{The elementary production}
\begin{figure}[hbt]
\begin{tabular}{cc}
\epsfig{file=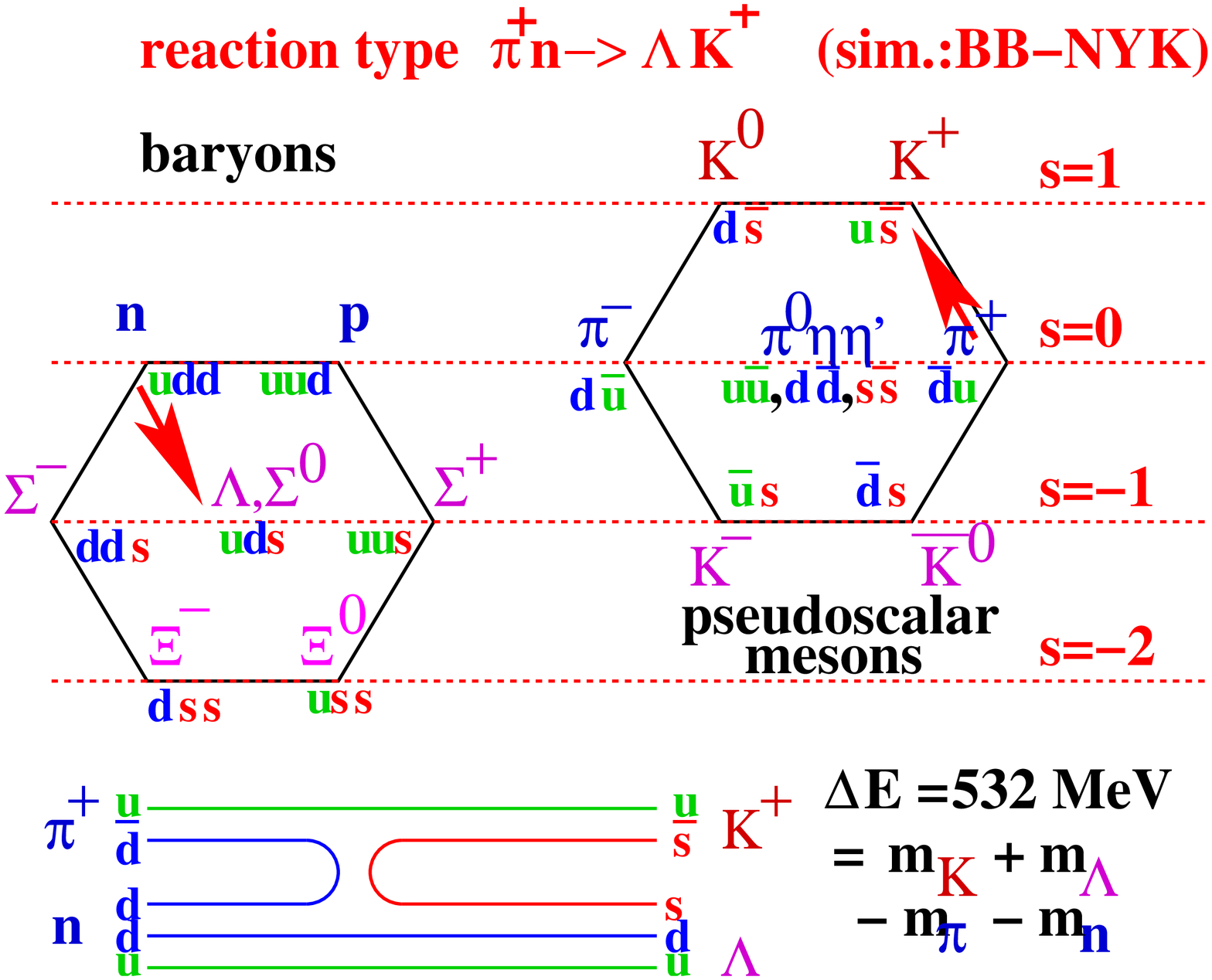,width=0.45\textwidth} &
\epsfig{file=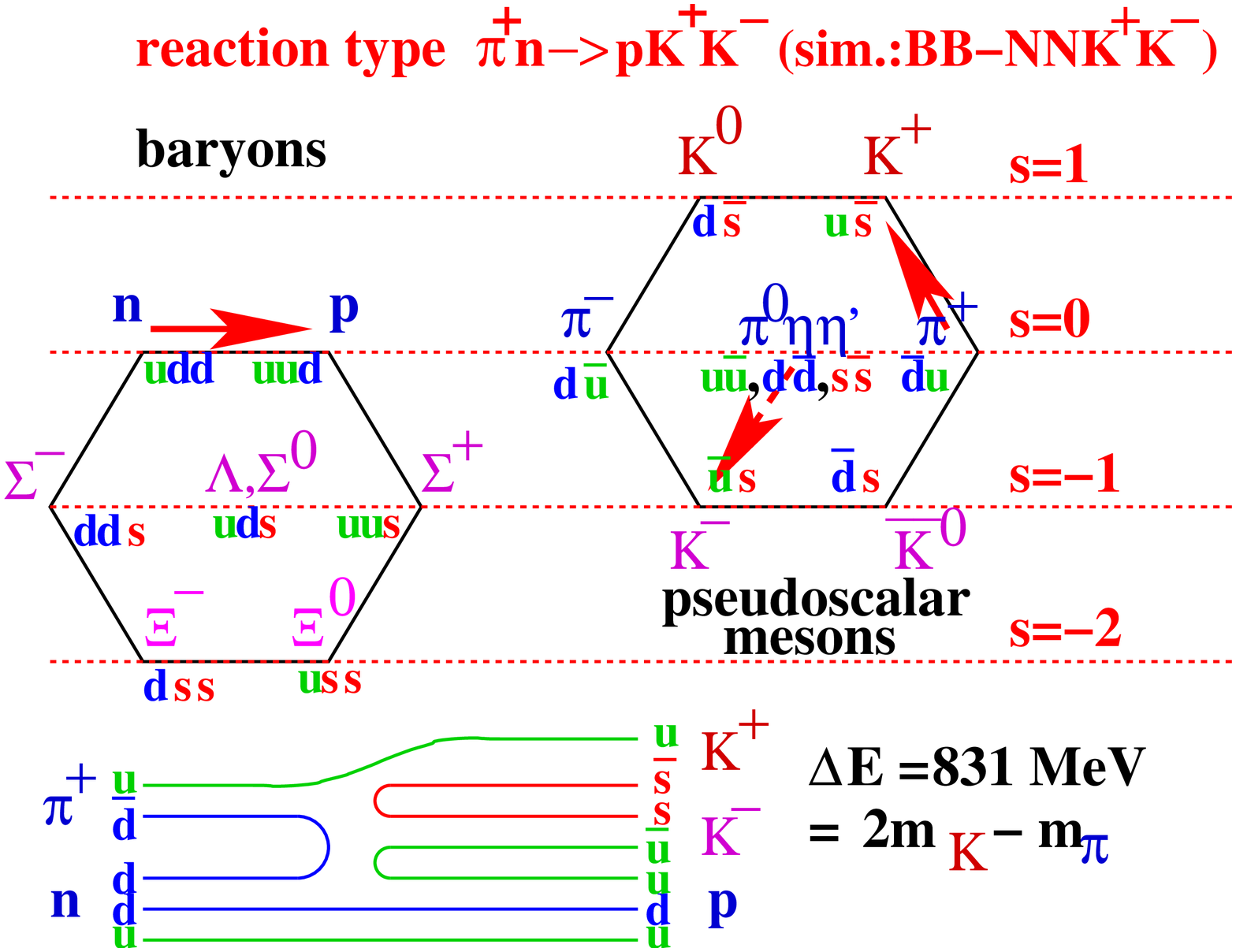,width=0.45\textwidth} \\
\end{tabular}
 \caption{Description of the major elementary channels of $K+$ production
 in a $\pi^+n$ collision}
\Label{kp-prod-multiplett}
\end{figure}
The elementary production of $K^+$ is governed by the conservation of 
strangeness.
The initial net strangeness is zero, thus strange quarks can only be 
produced together with strange anti-quarks. In the multiplett scheme 
shown in fig. \ref{kp-prod-multiplett} this means that the sum of the 
red arrows should be zero. 
The most economic way to do this is to create an $s$-quark which remains in 
a baryon (and thus transforms the nucleon into a hyperon) together with a
$\bar{s}$-quark which joins the kaon. An example of such a process is given
on the l.h.s. of  fig. \ref{kp-prod-multiplett}, where the reaction
$\pi n \to \Lambda K^+$ is described. In this reaction a $d \bar{d}$-pair
annihilates to form a $s\bar{s}$ pair. This reaction only needs an energy
of 532 MeV available in the centre-of-mass while a reaction of producing
a $K^+ K^-$ pair (r.h.s.) and leaving the nucleon a nucleon needs much
more energy. When regarding the quark diagrams on bottom of  
fig. \ref{kp-prod-multiplett} we see that the latter process creates an
additional  $u \bar{u}$ pair and should thus be suppressed by OZI rules.

The same argument holds when regarding baryon-baryon reactions. The
channel $B_1B_2 \to B_3 Y K^+$ has lower thresholds than the channel 
$B_1B_2 \to B_3B_4 K^+K^-$. 
Here $B$ may be a nucleon $N$ or a delta $\Delta$ and $Y$ may be a $\Lambda$ or 
a $\Sigma$ 
and we end up to a large amount of reactions which need to be described.
The list of all $K^+$ production reactions  parametrized in IQMD is shown
in table \ref{list-k-channels}. 

\begin{table}[hbt]
\begin{tabular}{cccc}
$ N N \to N \Lambda K^+$ &
$ N N \to N \Sigma K^+$ &
$ N N \to \Delta \Lambda K^+$ &
$ N N \to \Delta \Sigma K^+$ \\
$ N \Delta \to N \Lambda K^+$ &
$ N \Delta \to N \Sigma K^+$ &
$ N \Delta \to \Delta \Lambda K^+$ &
$ N \Delta \to \Delta \Sigma K^+$ \\
$ \Delta \Delta \to N \Lambda K^+$ &
$ \Delta \Delta \to N \Sigma K^+$ &
$ \Delta \Delta \to \Delta \Lambda K^+$ &
$ \Delta \Delta \to \Delta \Sigma K^+$ \\
$ \pi N \to \Lambda K^+$ & 
$ \pi N \to \Sigma K^+$ & 
$ \pi \Delta \to \Lambda K^+$ &
$ \pi \Delta \to \Sigma K^+$ \\
$ N N \to N N K^+ K^-$ &
$ N \Delta  \to N N K^+ K^-$ &
$ N \Delta  \to N \Delta K^+ K^-$ &
$ \Delta \Delta  \to N N K^+ K^-$ \\
$ \Delta \Delta  \to \Delta \Delta K^+ K^-$ &
$ \pi N \to N K^+ K^-$ &
$ \pi \Delta \to N K^+ K^-$ & \\

\end{tabular}
\caption{List of the $K^+$ producing reactions parametrized in IQMD}
\label{list-k-channels} 
\end{table}

In these channels different isospin combinations for $N, \Delta, \pi$
are possible which yields a further subdivision of these channels. 
Note that we only use the $\Delta (1232)$ which is the dominant resonance
channel in this energy domain. 

\begin{figure}[hbt]
\begin{tabular}{cc}
\epsfig{file=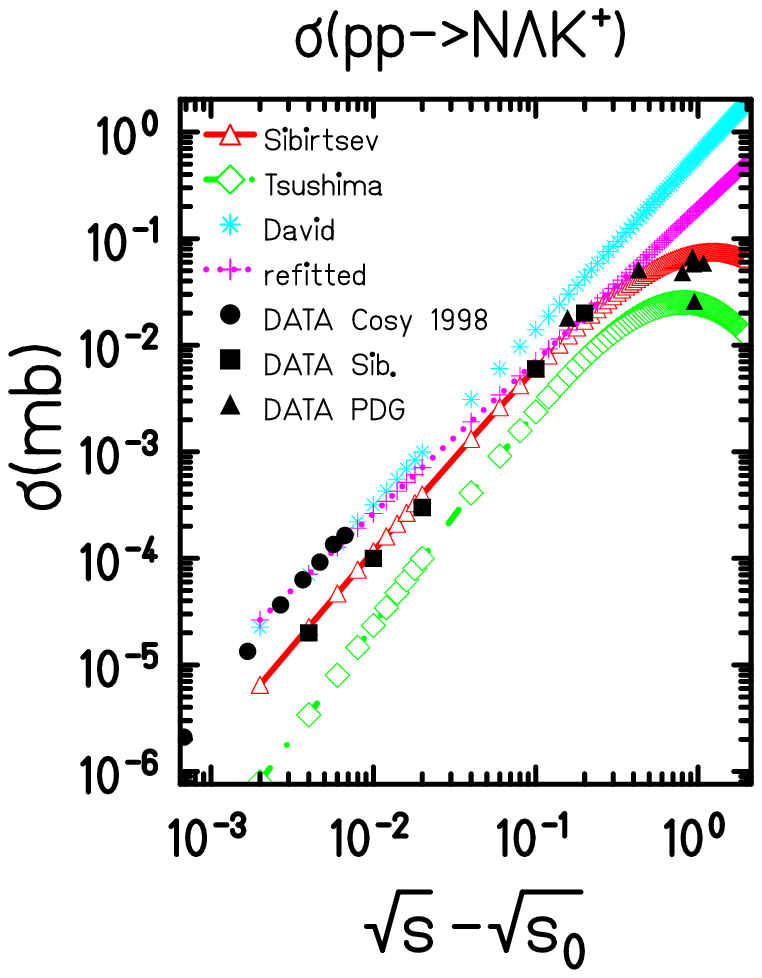,width=0.4\textwidth} &
\epsfig{file=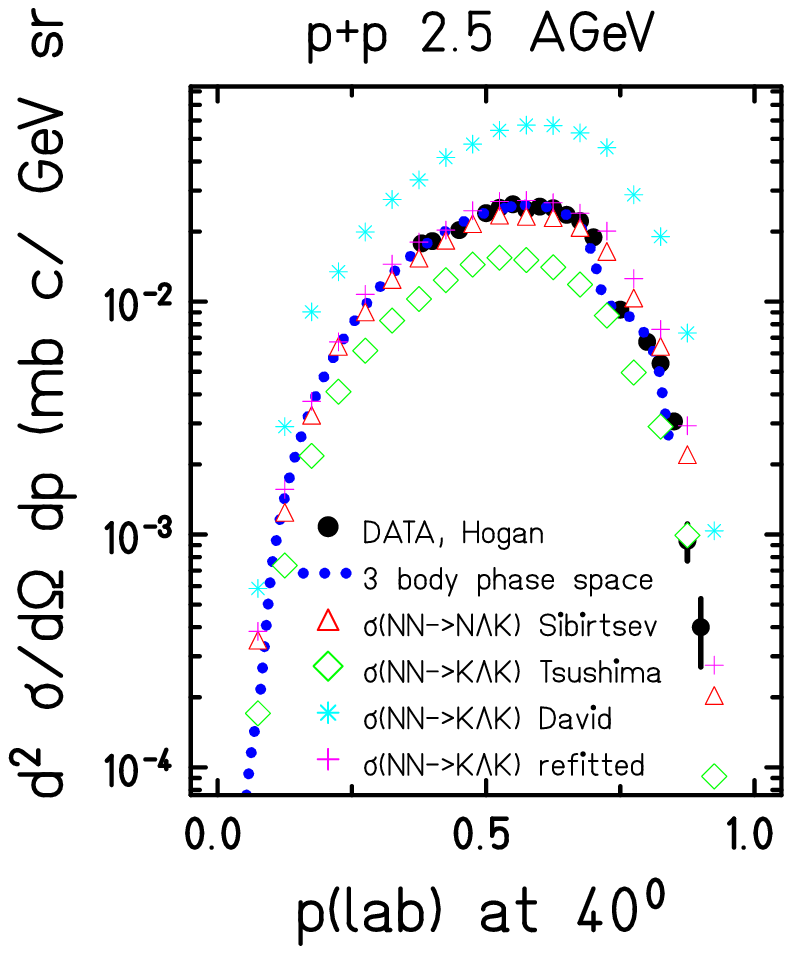,width=0.4\textwidth} \\
\end{tabular}
 \caption{Parametrizations of the $ pp  \to p \Lambda K^+$ channels (left)
 and comparison of IQMD with pp-data}
\Label{kp-prod-channels}
\end{figure}

Most of these channels are not accessible experimentally.
Even for the channel $ N N \to N \Lambda K^+$
only the isospin combination $ p p \to p \Lambda K^+$
is known by experiment. At energies close to the threshold still a lot
of uncertainties are remaining. Fig. \ref{kp-prod-channels} shows on the l.h.s.
several parametrizations of this channel used by our simulations.
Our standard parametrization is that of Sibirtsev et al. \cite{sibirtsev}
(\rfl{} with triangles) fitted to data cited in his publications. Other possibilities are the
parametrization by Tsushima et al \cite{tsu} (\gml{} with diamonds) 
and a parametrization of David et al \cite{david} (cyan line with stars)
fitting recent COSY data \cite{cosy98}. This fit has later on be modified
including data points at higher energies (magenta line with crosses)
and reads:
\begin{equation}
\sigma (pp \to p \Lambda K^+ ) =  191 \mu b \cdot \left ( \sqrt{s}-\sqrt{s_{\rm thres}}\right )^{1.43}
  \label{refit}
\end{equation}
This parametrization is of course only valid for energies not higher
than about 3 GeV, which is however a limit of IQMD.

Other channels have to be extrapolated by
isospin-considerations. When calculating in the One Boson Exchange model
the choice of the exchange particle (pion or kaon) changes the factor 
to extrapolate from pp to pn from 1 to $5/2$. In IQMD the latter factor of
5/2 is used which causes an isospin averaged cross section of 
\begin{equation}
\sigma (NN \to NYK^+)= \frac{1}{4}\bigl ( \sigma _{pp} + 2\sigma_{pn}
+ \sigma_{nn} \bigr ) = \frac{1+5+0}{4}\sigma_{pp}=\frac{3}{2}\sigma (pp \to NYK^+)
\end{equation}

For the channels including $\Delta$ no knowledge from experiment exists.
Here exists some freedom in assuming the cross sections which raises 
uncertainties in the interpretation of heavy ion data. We will come to this
point later on.
For our parametrization results of calculations performed by 
Sibirtsev \cite{sibirtsev}  are used for the channel $\sigma (NN \to N\Lambda K^+)$ 
while for all other channels the parametrizations of Tsushima
et al. \cite{tsu} are used. 

The right hand side of fig.\ref{kp-prod-channels} compares results of experimental
pp-data of Hogan et al \cite{hogan} (black bullets) 
with IQMD calculations (red triangles) using the Sibirtsev and Tsushima cross 
sections. The agreement shows that in the known sector
of cross sections our parametrizations work well. 
A replacement of the $\sigma (NN \to N\Lambda K^+)$ cross section by the 
corresponding cross section of Tsushima \cite{tsu}  (green diamonds) would 
yield a too small yield which explains the choice taken for IQMD.
The parametrization of David \cite{david} (cyan stars) yields too much kaons 
since its cross section is too high at high energies. The refitted parametrization
(magenta crosses, see eq. \ref{refit}) shows again a spectrum quite comparable to that using
the Sibirtsev parametrization.

Another important point to be discussed in this figure is the dynamics of
the collision itself. Since there are three particles in the outgoing channel,
there is some freedom in determining the dynamics of the particles within the
constraints of the conservation laws. One possibility is to create an 
intermediate resonance which decays into two particles.
\begin{equation}
NN \to R N \qquad R \to Y K^+ \qquad R = N^*(1650), N^*(1710) \mbox{etc}
\end{equation}
This procedure is e.g. done in UrQMD \cite{urqmd}. In IQMD the two colliding particles
form one intermediate resonant two particle state which decays directly into
three particles by a three-body decay.
\begin{equation}
NN \to R \qquad R \to N Y K^+ \qquad M(R)=\sqrt{s}
\end{equation}
The experimental data at these energies support the dynamics of a three-body
phase space model (blue dotted line) as it is implemented in IQMD. 
However, at higher energies the phase space model describes the data less well.

\subsection{Subthreshold production}
Subthreshold production is the production of particles at an incident energy
at which a production in a free p+p collision is no more permitted due to 
energy-momentum conservation. For the production of kaons the production threshold is
constrained by the conservation of strangeness. The lowest threshold for a
reaction producing a kaon is given by the channel $p+p \to p+\Lambda+K^+$
\begin{equation}
\sqrt{s}= \sqrt{\sum P_\mu P^\mu} \ge m_p+m_\Lambda+m_{K^+}=2.55 {\rm GeV}\qquad P=p_1+p_2
\end{equation}
where $p_1$ and $p_2$ are the four-momenta of the colliding particles.
This implies a minimum incident energy of around 1.58 AGeV of the proton projectile
on the proton target.
In heavy ion collisions kaons can also be produced at lower incident energies.
There are majorly two effects which allow this subthreshold production
\begin{enumerate}
\item Due to the Fermi momentum of the nucleons in  projectile and target
two colliding particles may have a higher energy than given by the incident
energy. This may allow to reduce the threshold incident energy up to around
1 AGeV.
\item Due to rescattering of the nucleons some nucleons may accumulate 
energy such that finally sufficient energy is given for the production of
kaons. Resonances may serve as energy storage since they transform kinetic
energies into mass.  Two nucleons of projectile and target may collide and
create a delta which due to its higher mass will be quite slow. Another
fast projectile nucleon may collide with this delta and produce a kaon.
Or, if another delta is formed by two other  nucleons of projectile and target
and create a delta and these two delta collide, there is no much need for high
additional energy. The sum of the two masses (the peak of the delta mass
distribution is at around 1.232 GeV) is already nearby the threshold.
With this mechanism  one may create kaons even at a few hundred MeV of incident
energy. However the probability will be quite small and depend strongly on the
system size.

\end{enumerate}
After fixing the cross section parametrizations to pp-data
we now can use our models for the simulation of  heavy ion collisions.
Let us first start with some analysis of p+A.

\begin{figure}[hbt]
\begin{tabular}{cc}
\epsfig{file=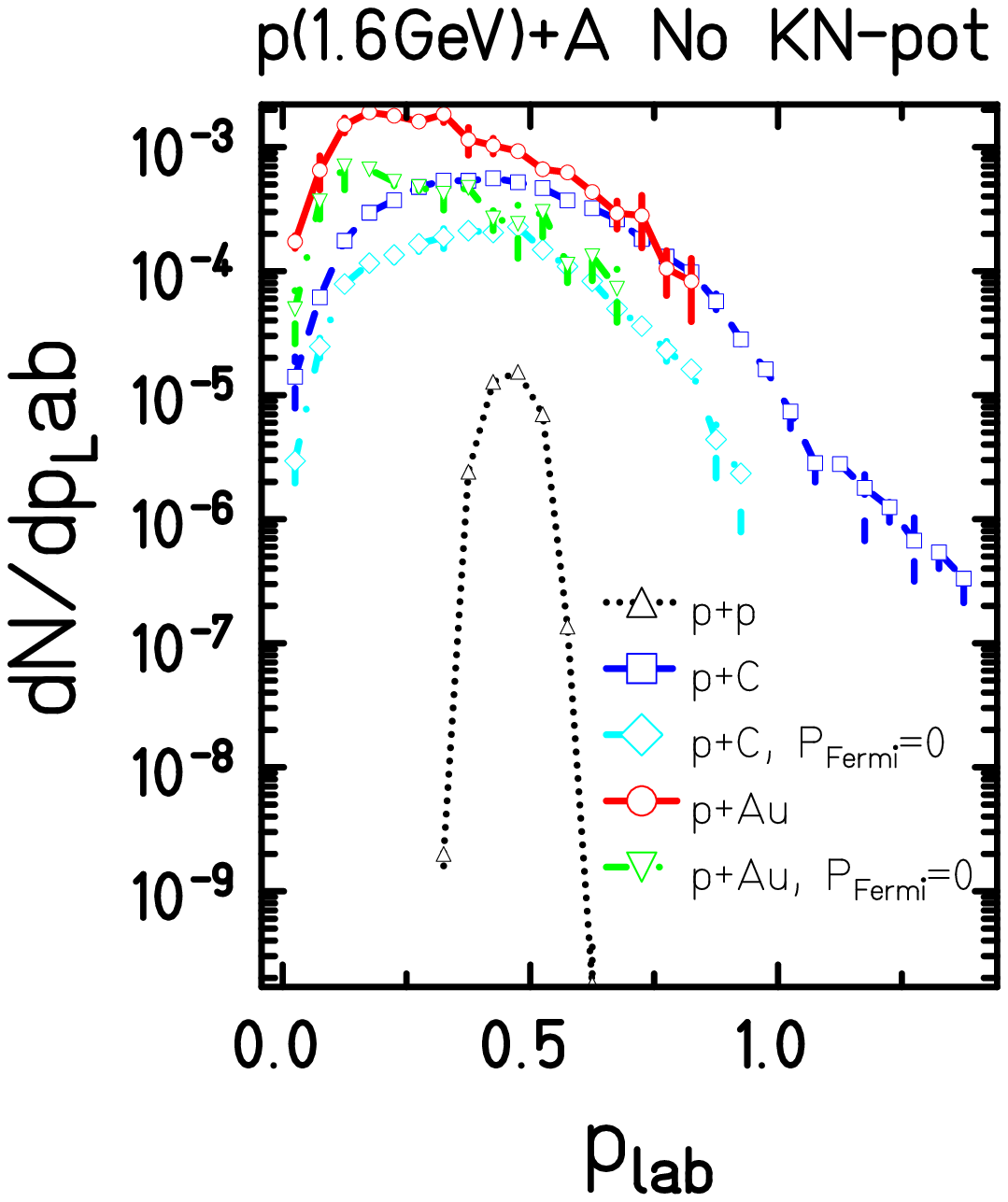,width=0.4\textwidth} &
\epsfig{file=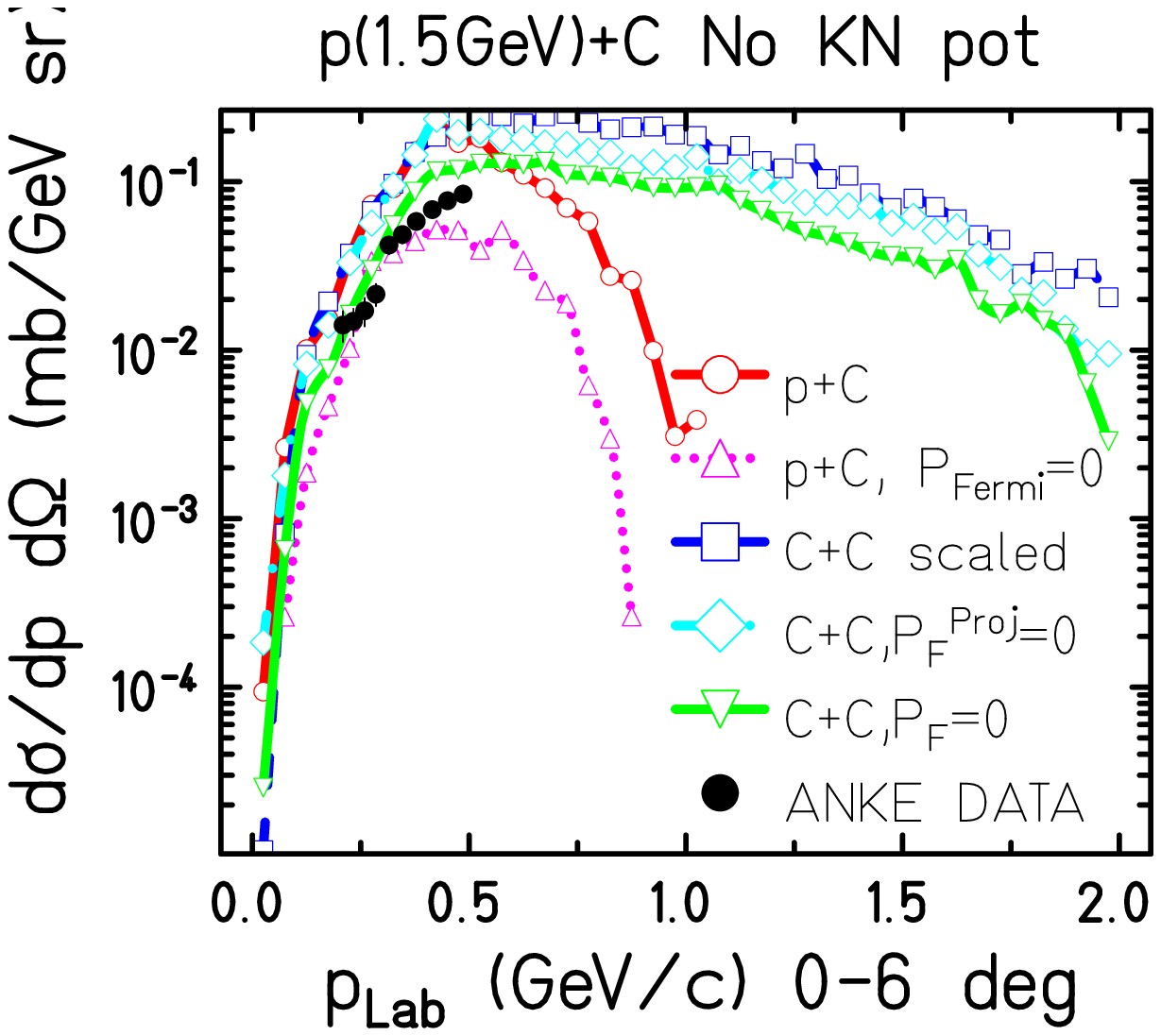,width=0.4\textwidth} \\
\end{tabular}
 \caption{Comparison of p+p, p+C and p+Au
 momentum spectra at 1.6 GeV (left) 
 and of p+C and C+C spectra at 1.5 AGeV (right)}
\Label{pa-nopot}
\end{figure}

The \lhsref{pa-nopot} shows the lab momentum distribution of  p+A collisions
at 1.6 GeV incident energy, i.e. slightly above the threshold.
In a p+p case (\bpl) there is a quite narrow distribution. Only little energy
is available to the kaon, which is produced in the centre-of-mass of the
reaction. Thus it carries only a momentum slightly different to the
centre-of-mass momentum. 
In the p+C collision (\bdl) there is a much wider distribution.
This is only in part due to the
Fermi momentum in the carbon target where the projectile may collide with a particle
with a Fermi momentum in the opposite direction of its momentum and thus
enhance the available energy of the kaon. This can be seen when comparing the
p+p curve (\bpl) and the p+C curve (with Fermi momentum, \bdl) with a calculation
without Fermi momentum (p+C, $\rm P_{Fermi}=0$, cyan curve with diamonds).
The major part of the difference between p+p and p+C stems from the opening
of additional channels due to multi-step processes: $N_1N_2 \to N\Delta, 
\quad N_3 \Delta \to NYK$ ($N\Delta$ channel) and  $N_1N_2 \to N\Delta,
\quad \Delta \to N \pi \quad \pi N_3 \to YK$ ($N\pi$ channel) which in the 
case of p+C contribute to more than 80 \% to the total kaon number.
A further rise is gained by the Fermi momentum, which allow for enhancing the
available energy by selecting a partner which opposite Fermi momentum.
Since the cross section increases strongly nearby the threshold (see 
\figref{kp-prod-channels}) it allows for a strong enhancement of the kaon
production especially for the $NN$ channel which regains the dominance
in a p+C collision if the Fermi momentum is on.  
All effects contribute to the effect that the total production probability
is strongly enhanced with respect to the p+p case. Finally if the first
collision takes the elastic channel (which at high energies is strongly
forward-backward peaked) there is a further chance in producing a kaon
in another collision of the projectile nucleon with a target nucleon.
These effects  explain the huge increase of the kaon multiplicity when going
from p+p to p+C. When going from p+C (\bdl) to p+Au (\rfl) there is an
additional increase at low laboratory momenta. This is probably due to second
chance effects. If the projectile looses only little energy in its first
collision(s) it may still produce a kaon in a secondary collision if the
Fermi momentum of the target nucleon supports this. For the Au case there is much more
chance to find such a particle. The kaon is no more produced in the p+p
centre-of-mass but in the centre-of-mass of the slowed down projectile with
a target nucleon. This explains the shift toward lower lab momenta which
can be seen as well in calculations with (\rfl) and without (\gml) Fermi
momentum.

The \rhsref{pa-nopot} compares calculations at 1.5 AGeV incoming energy, i.e.
already below the threshold. A p+p collision cannot produce a kaon any more.
The forward angle laboratory momentum distribution of IQMD events (\rfl) is
compared to data of the COSY-ANKE collaboration \cite{anke}.
We see a qualitative agreement which supports the existence of the effects 
stated above. A p+C calculation without Fermi momentum (magenta dotted line
with upward triangles) shows again a strongly reduced kaon yield which again
demonstrates the importance of multi-step processes. 
Furthermore a minimum bias reaction of C+C (\bdl) has been
scaled by the different reaction cross sections and the average number
of participant projectile nucleons. This allows to compare a C+C collision to
a $A_{part}\times$p+C collision, an idea which corresponds to the Glauber-model.
We see a similar behavior at low lab momenta but a much wider range of 
available laboratory momenta. This is less due to the additional
Fermi momentum in the projectile but majorly there is now the possibility
of accumulating the energy of several collisions of projectile nucleons
and target nucleons (for instance in forming two deltas which collide).
We see that calculations where the Fermi momentum of the carbon has been
removed for the projectile only (cyan curve with diamonds) or for both
nuclei (green curve with downward triangles) show a similar
enhancement at high energies.
This already demonstrates the importance of the full nucleon dynamics,
multi-collision effects, resonance production and so on.
We will therefore address this subject more in detail.


\subsection{The major production channels}
At first, we want to address the question, which channel is 
important for heavy ion collisions.
\Figref{nk-b} shows the kaon yield in a collision of Au+Au at 1.5 AGeV
incident energy as a function of the impact parameter(\bpl). We see a strong
centrality dependence, where the production of kaons is strongest supported
in central collisions. A decomposition of the kaon yield into the
producing channels shows that the $NN$ channel (\bdl) only contributes quite
weakly while the addition of the $N\Delta$ channel already describes
the major part of the yield. 

\begin{figure}[hbt]
\begin{tabular}{cc}
\epsfig{file=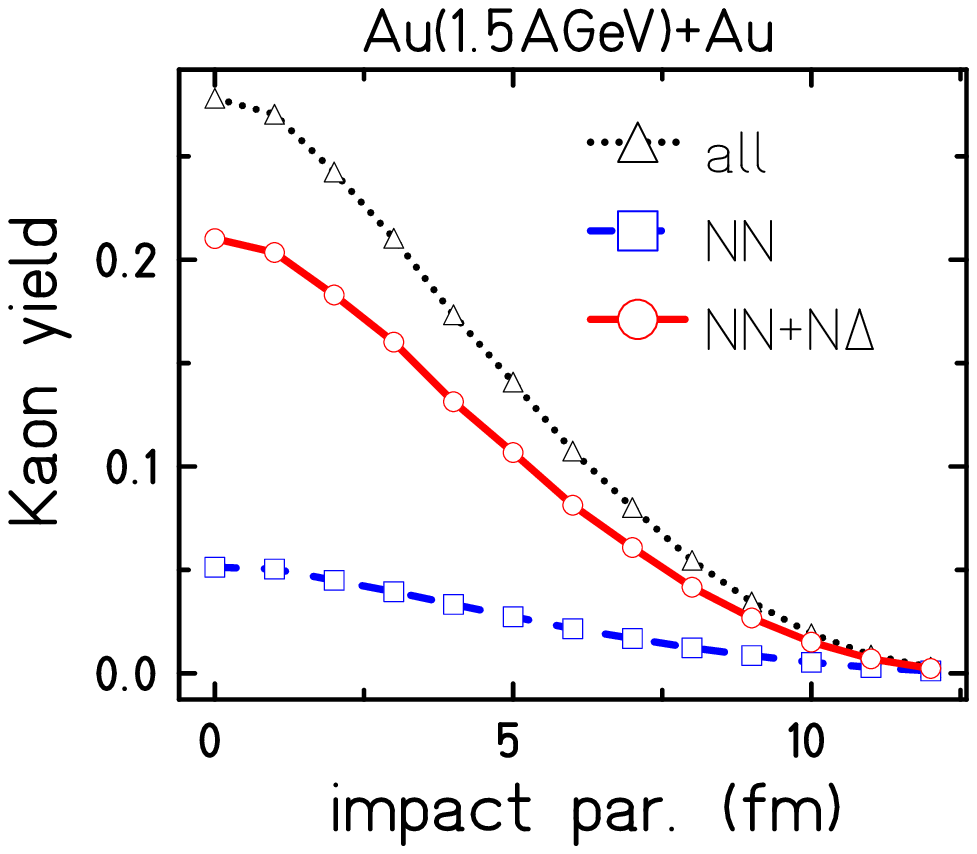,width=0.4\textwidth} &
\epsfig{file=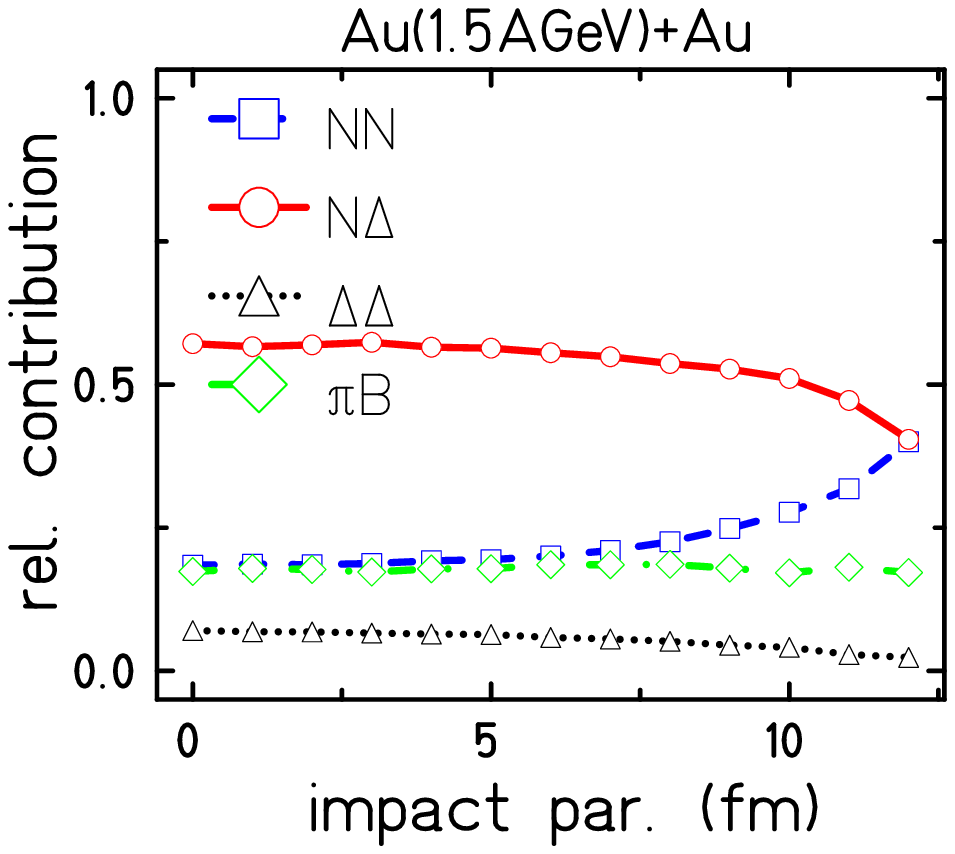,width=0.4\textwidth} \\
\end{tabular}
 \caption{The absolute kaon yield (left) and the relative contribution
 (right) of different production channels as function of the impact
 parameter in a Au+Au collision at 1.5 AGeV incident energy.}
\Label{nk-b}
\end{figure}
The \rhsref{nk-b} shows the relative contribution of the different channels
implying nucleon-nucleon collisions ($NN$, blue dashed line), 
delta-nucleon collisions ($N\Delta$, red full line), delta-delta
collisions ($\Delta\Delta$, \bpl) and pion-baryon collisions ($\pi B$,
\gml ). The $N\Delta$ channel dominates the production over a large impact
parameter range. Only in very peripheral collisions the rather weak
$NN$ channel runs up to become as important as the $N\Delta$ channel.

Fig. \ref{ncps-dist} shows the relative contribution of the
different channels  ($NN$ - blue dashed line, 
$N\Delta$ - red dashed line, $\Delta\Delta$ - \bpl, $\pi B$ -
\gml ) as function of energy and system size.. 
At the left hand side the excitation function of Au+Au is shown,
while on the left hand side the dependence on the system size for
b=0 collisions at 1.5 AGeV incident energy is studied.

 \begin{figure}[hbt]
\begin{tabular}{cc}
\epsfig{file=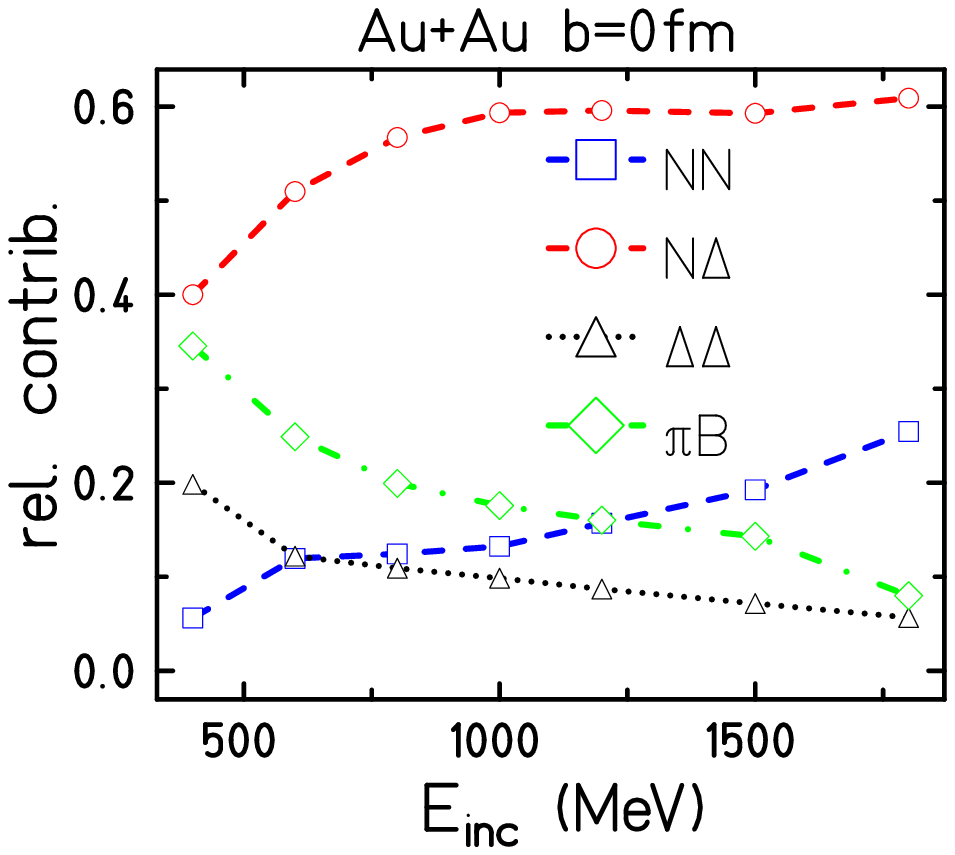,width=0.4\textwidth} &
\epsfig{file=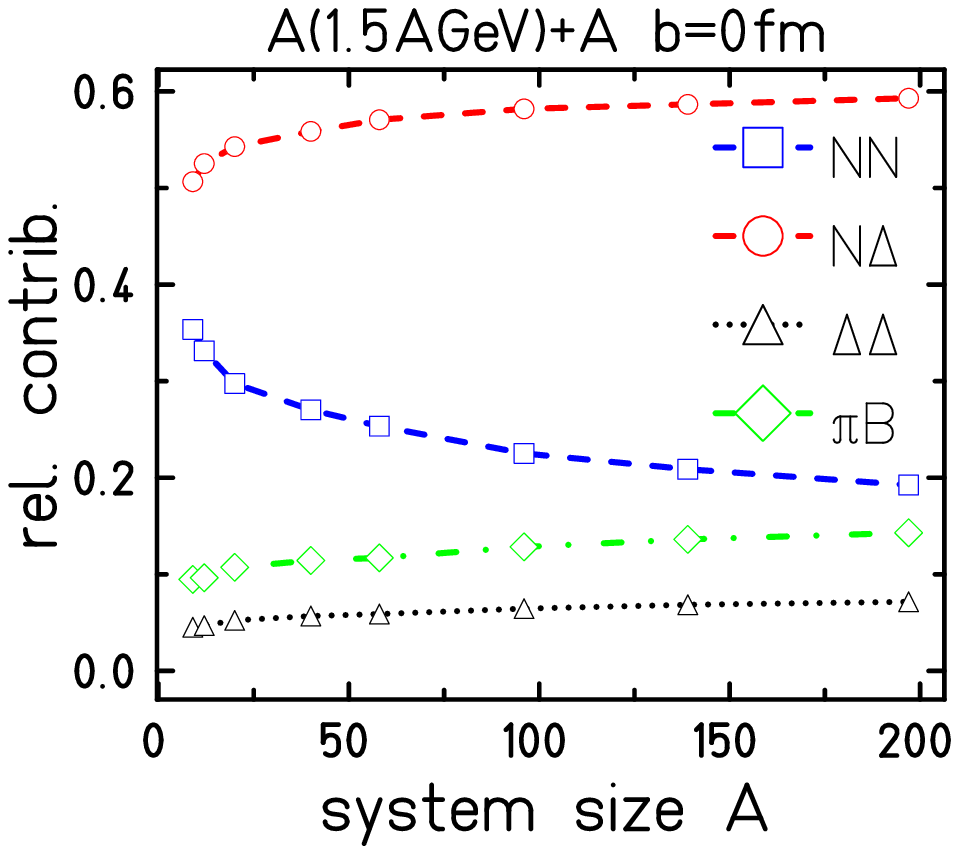,width=0.4\textwidth} \\
\end{tabular}
 \caption{Contribution of different production channels
 to the kaon yield as function of incident energy (left) 
 and system size (right)}
\Label{ncps-dist}
\end{figure}

We see the contribution of $NN$ is rather small especially at low
energies. Note that only the highest energy point of 1.8 AGeV has enough
energy to produce a kaon directly. The same reaction done with p+p would
not produce any kaon at incident energies below 1.6 GeV. In order to get
the energies necessary for a kaon production, the system has to cumulate
energy. This might be done by the creation of resonances (like the $\Delta$)
or pions. Therefore, the $N \Delta$ channel has the strongest contribution.
The $\Delta \Delta$ channel gains even more energy but the possibility that
two deltas  collide is rather small. But if the system goes down to very
low incident energies, the need for energy cumulation is that high that even 
these rare channels start to play an important role. The same is true for the
$\pi B$ channel where a high energy pion has to find a $\Delta$ or a high
energy nucleon for producing a kaon.

Let us now look at the system size dependence (r.h.s.) of the channel
contributions. We are at rather high energies (1.5 AGeV, slightly below 
the threshold) where already the Fermi momentum of the nucleons may help
the nucleons to produce kaons in first collisions. Therefore, the contribution
of the rare combinations $\Delta \Delta$ (\bpl ) and $\pi B$ (\gml ) is 
rather small. The $N\Delta$ (red dashed line) channel dominates, but the
$NN$ channel contributes quite well especially for small systems. For these
systems we have a larger contribution of the surfaces, where we could expect
first $NN$ collisions. This effect is in agreement with the impact parameter
dependence seen in \figref{nk-b} where the $NN$ channel becomes important
in very peripheral collisions where only the surfaces of the nuclei touch.

\subsection{The collision history}

Next we want to discuss the aspect of cumulating energy in analyzing
the collision number of the parents. 
\Figref{nuc-col} shows on the \lhs{} the relative fraction of nucleons
as a function of the number of collisions the particles had during the reaction.
We see that for a light C+C system (\bdl) the maximum is reached for about 
2 collisions while for the heavy Au+Au system (\rfl) most of the particles had
about 6 collisions. 

\begin{figure}[htb]
\begin{tabular}{cc}
\epsfig{file=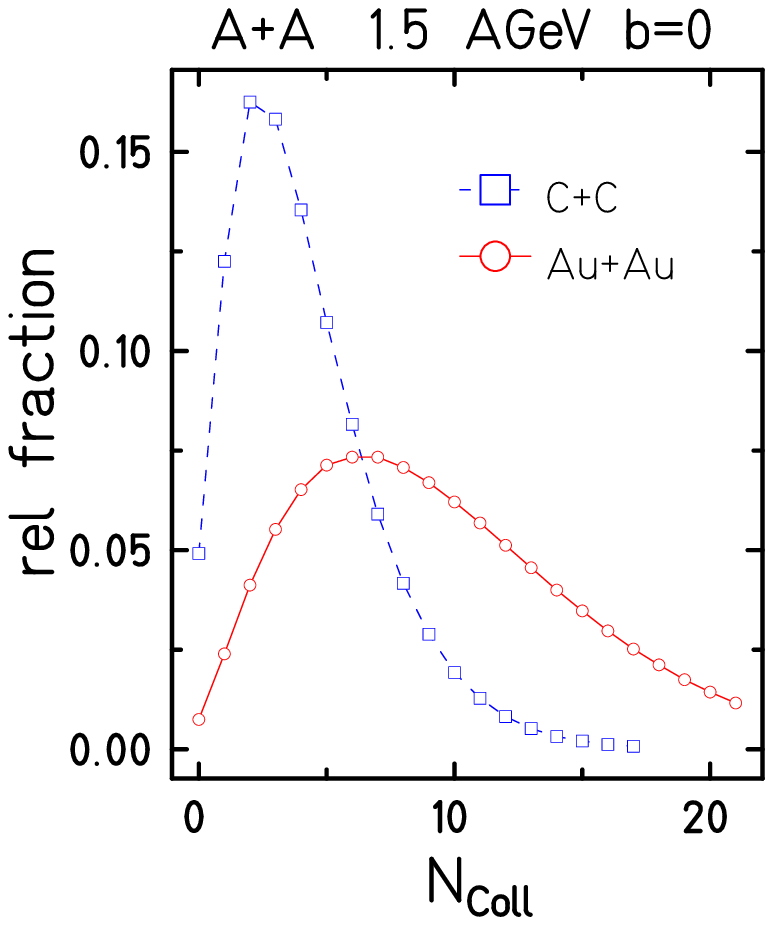,width=0.4\textwidth} &
\epsfig{file=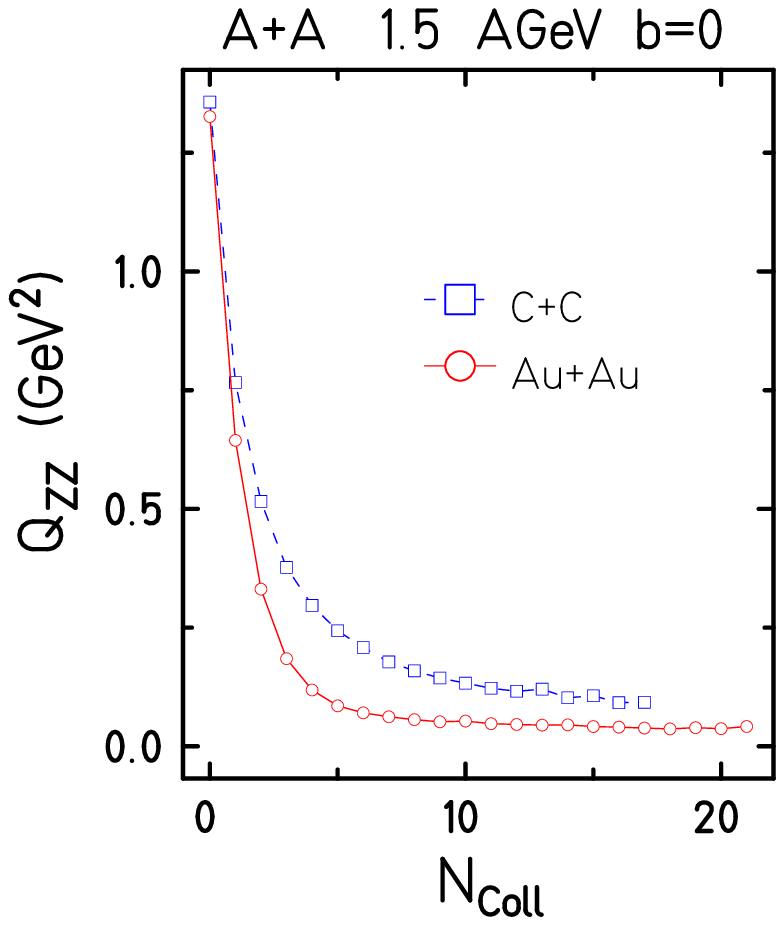,width=0.4\textwidth} \\
\end{tabular}
 \caption{Fraction of nucleons having a given collision number (left) and
 sphericity $Q_{ZZ}$ as function of the collision number (right)
  for Au+Au and C+C}
\Label{nuc-col}
\end{figure}
The \rhsref{nuc-col} shows the sphericity value $Q_{ZZ}$ as a function of the
collision number. $Q_{ZZ}$ is defined by
\begin{equation}
Q_{ZZ}= 3 p_Z^2 -p^2 = 2 p_Z^2 -p_T^2 = 2 p_Z^2 -p_X^2 -p_Y^2
\end{equation}
$Q_{ZZ}>0$ defines a prolate momentum ellipsoid and $Q_{ZZ}<0$ an oblate one.
An isotropic distribution yields $Q_{ZZ}=0$.
We see that the system approaches isotropy with increasing collision number
and that at least 3-6 collisions are necessary to come nearby equilibration.

We now analyze for each kaon the sum of the collisions each partner had
up to the production of the kaon.
If the kaon is produced in one of
the first collisions, the sum of the collision numbers of the parents
should be 2 (first collision for each partner). 
If the kaon is produced in an `equilibrated medium' than
this number should be much higher. As each particle needs more than 3 collisions
for equilibrating, therefore this number should be  at least 6 or higher. 

\begin{figure}[ht]
\begin{tabular}{cc}
\epsfig{file=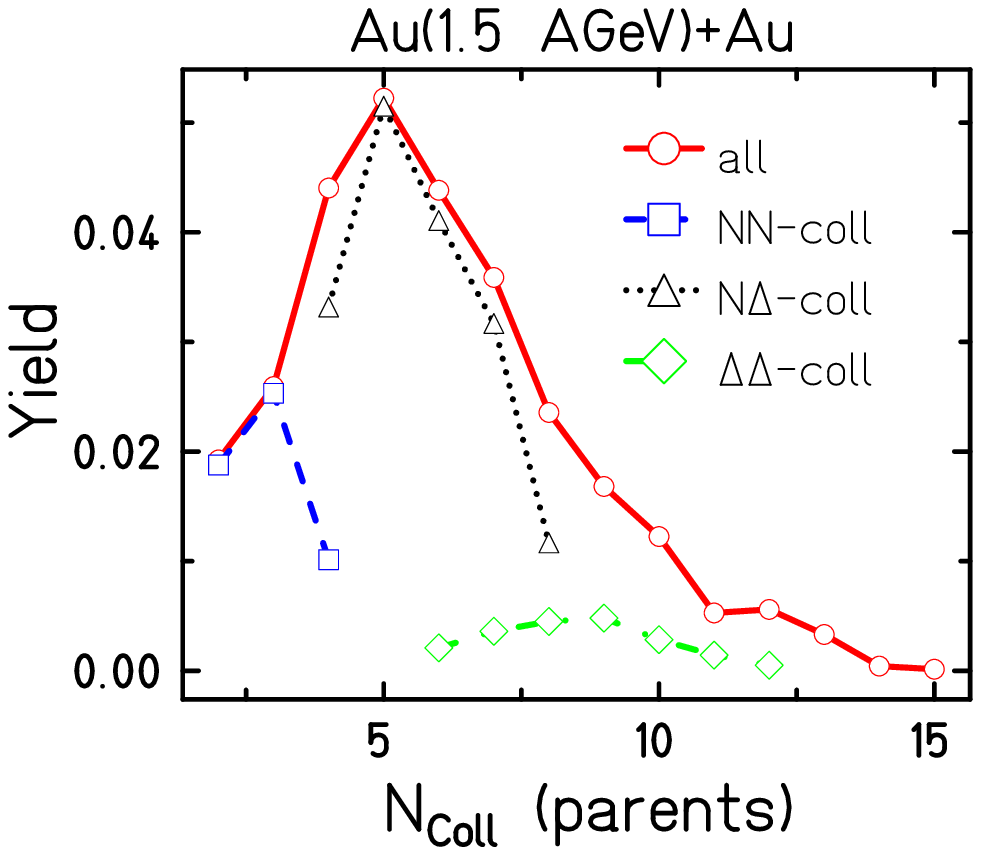,width=0.4\textwidth} &
\epsfig{file=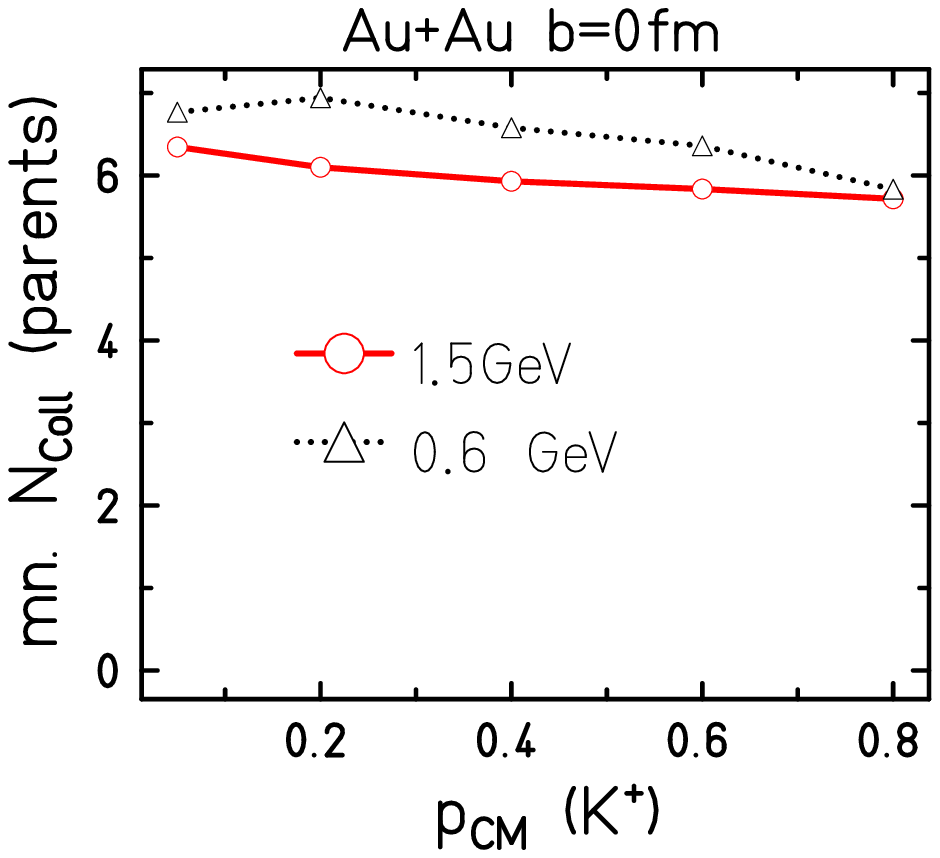,width=0.4\textwidth} \\
\end{tabular}
 \caption{Distribution of the number of collisions the parents had
 when producing of the kaon selected according to different
 production channels (left) and to the momentum of the kaon (right)}
\Label{ncps-select}
\end{figure}

The \lhsref{ncps-select} shows the distribution of the number of collisions the parents had
 when producing of the kaon selected according to different
 production channels.
 In the $NN$-channel (\bdl) the parents did undergo only a few collisions.
 The very first collisions ($N_{Coll}=2$) contribute strongly and the other
 contributions are majorly the first collision of an incoming particle with 
 the stopped matter in the centre. This is easy to understand since the 
incident energy is weakly below the threshold energy. Thus, already the Fermi
momentum of the particles is sufficient to overcome the threshold.

In the $N\Delta$-channel (\bpl) the maximum of the distribution is at $N_{Coll}=5$,
i.e. for collisions where at least one particle is nearby equilibration.
In the $\Delta\Delta$-channel (\gml) the numbers are even higher.

The mean value of the distribution is around 6 as we can easily see on the
\rhsref{ncps-select}, where we describe the mean value of the sum of the parent
collision numbers as function of the centre-of-mass momentum of the outcoming
kaon. The function is flat, which means the the production of higher energy
kaons does not need more collisions of the parents. With about 6 collisions
at least one partner has become equilibrated and more collisions do not
equilibrate more. The only questions is now to find a particle in the tail 
of the energy spectrum which has sufficient energy for producing a kaon.
This finding is supported by the observation that the number of parental 
collisions does not change very much with the incident energy. A reaction
far below threshold (0.6 GeV,\bpl) gives only slightly higher numbers than
a reaction nearby the threshold (1.5 GeV, \rfl).
Moreover there is no significant system size dependence. A C+C collision
yields at 1.5 GeV about the same values than a Ni+Ni or Au+Au collision.


\subsection{Production time and density}
Let us now look at the question when the kaons are produced. The
\lhsref{time-rho-prod} shows the time profile of the produced kaons selected according to their
production channels. 
\begin{figure}[htb]
\begin{tabular}{cc}
\epsfig{file=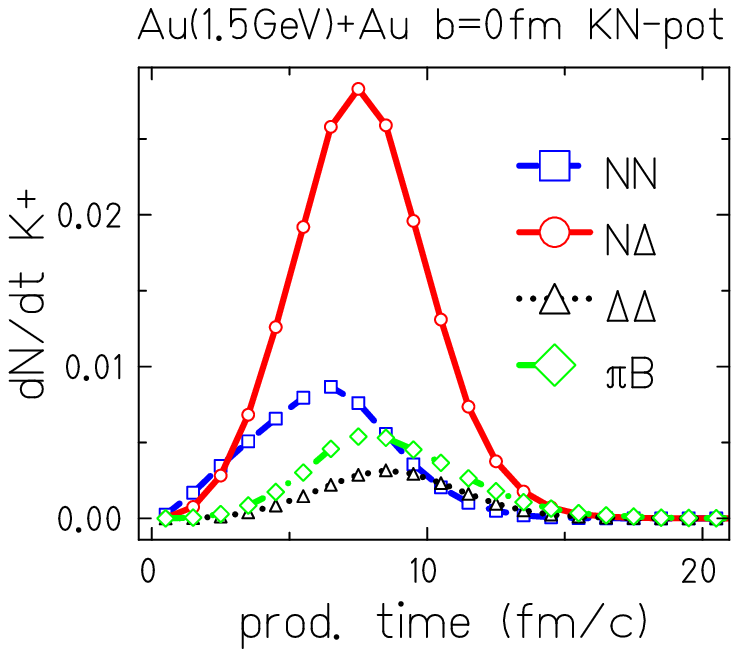,width=0.4\textwidth} &
\epsfig{file=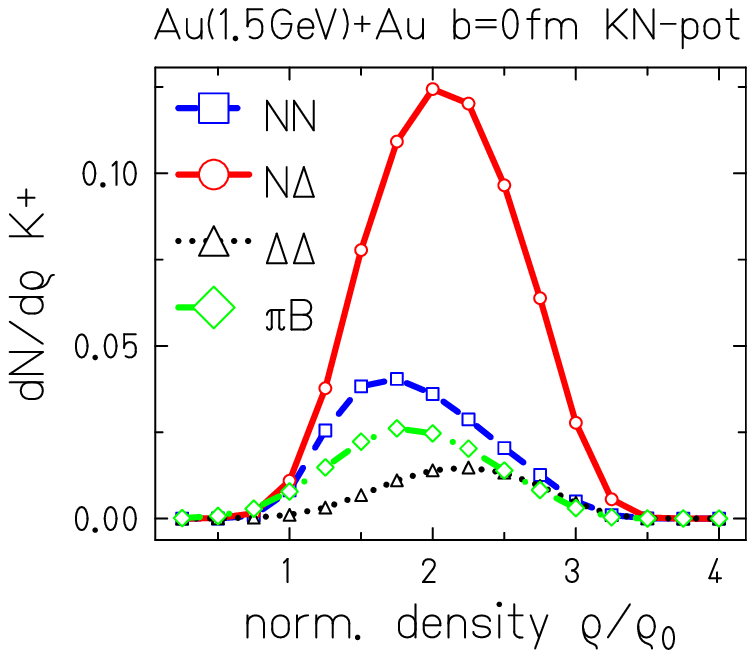,width=0.4\textwidth} \\
\end{tabular}
 \caption{Time and density profile of kaon production}
\Label{time-rho-prod}
\end{figure}
We see that the $NN$ channel (\bdl) starts first, shows a maximum
at about 6 fm/c and falls down rapidly afterward. This corresponds to the
assumption that this channel is majorly fed by the first collisions at the
beginning of the reaction. The $N\Delta$ channel (\rfl) starts a little bit later to
show a maximum at about 8 fm/c (which corresponds to the time of maximum
compression. The $\Delta\Delta$ channel (\bpl) starts even later. This 
corresponds to the observation of the \lhsref{ncps-select} that the 
$\Delta\Delta$-channel shows the biggest number of parental collisions 
for the production. This creation of highly stopped matter, of course,
needs some time, before this channel becomes active.
The $\pi B$ channel (\gml) also starts late. Here we have to keep in mind that in
the beginning the pions are rapidly reabsorbed into deltas and come out quite
late. Therefore, the pion number increases slowly which allows the pion channel
only to start later.

The \rhsref{time-rho-prod} shows the density profile of the different channels.
The early  $NN$ channel (\bdl) reaches quite moderate densities which corresponds
to the previous picture that this channel is happening early in the non-equilibrated
matter which is still in compression. The $N\Delta$ channel (\rfl) reaches higher
densities. The distribution peaks at about two times normal nuclear matter.
The maximum of the distribution of the $\Delta\Delta$ channel (\bpl) is even a 
little bit higher. This is in alignment to the high number of parental collisions
(which can be reached best in the high density zone) and to a production time
in the range of highest compression.
The $\pi B$ channel (\gml) finally prefers less high densities. In the high
density zone pions are rapidly absorbed to deltas and thus more pions are
available at lower densities.   


\subsection{Where are the kaons produced?}
Let us now address the question, where the kaons are produced.
\Figref{r-prod} shows the radial density-profile of the kaon production with
respect to the centre of the reaction. 
 
\begin{figure}[ht]
\begin{tabular}{cc}
\epsfig{file=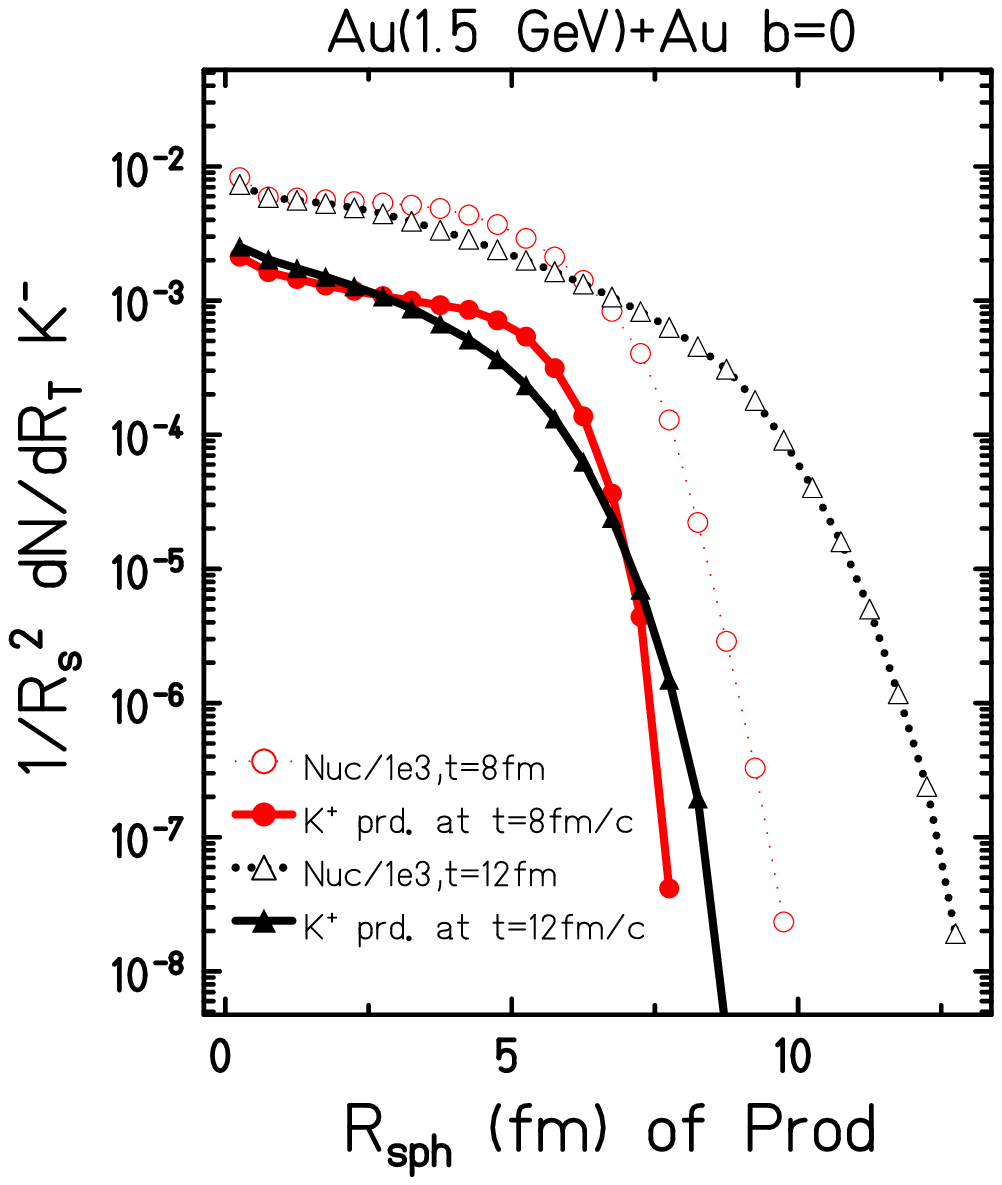,width=0.4\textwidth} &
\epsfig{file=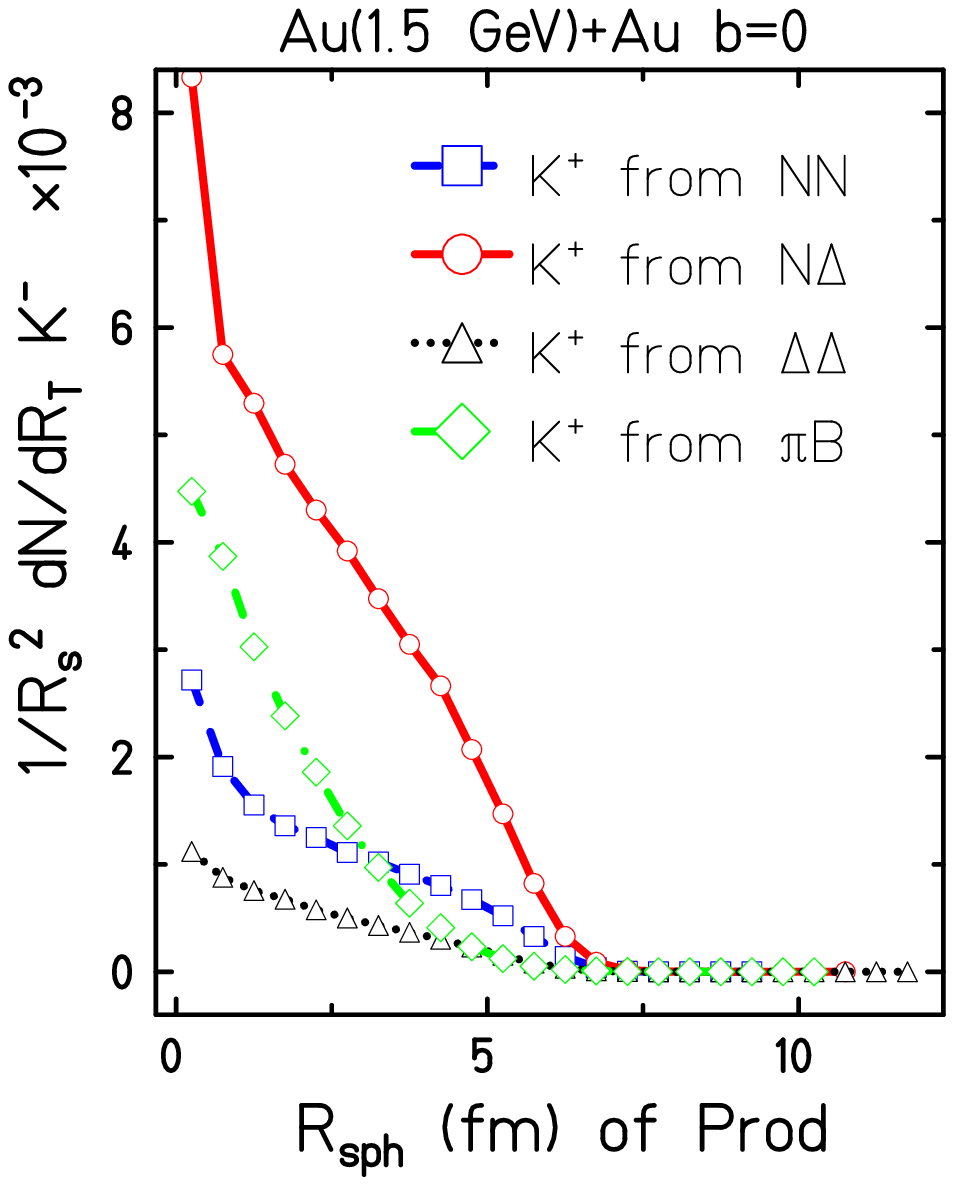,width=0.4\textwidth} \\
\end{tabular}
 \caption{Radial distribution of kaon production selected according to
 production time and to channel decomposition}
\Label{r-prod}
\end{figure}
We see on the \lhs\ a comparison of the nuclear radial distribution 
(dotted curves with open symbols)
at $t=8$ fm/c (max. compression, red circles) and $t=12$ fm/c (end of kaon
production, black triangles) with the distribution of the production points
of kaons produced at the same time windows (full curves with full symbols).
The nuclear distributions are broader than the kaon production distributions
and the differences become stronger at the surfaces. From this we can  conclude
that the kaons are really produced in the central part of the reactions.

The \rhsref{r-prod} shows the time integrated radial distribution of the
kaon production on a linear scale selected according to different channels.
We see that for all channels there is a peaking of the distribution at $R=0$,
i.e. in the centre of the reaction.


\subsection{Is there sensitivity to the nuclear eos?}
We can conclude from the previous analysis that the kaon
production at the discussed energies is strongly related to
a high compressed region where energy may be cumulated in multi-step
processes. We may therefore rise the question whether the kaon production
might be a good signal for determining the nuclear equation of state.

The nuclear equation of state describes the property of nuclear matter
to be compressed. A hard equation of state (having a high compressibility) 
gives a strong repulsion to the compression. More compressional energy is
needed to compress the system to a given density.
A soft equation of state (with less compressibility) resists less to the
external compression and allows for higher densities for a given compressional
energy. 
In a heavy ion collision thus a higher maximum compression can be reached 
when employing a soft equation of state (\rfl) than when using a hard one (\bdl), as it
can be seen on the \lhsref{eos-dns}\ where we compare the time evolution of the
central density for both equations of state. 
  
\begin{figure}[ht]
\begin{tabular}{cc}
\epsfig{file=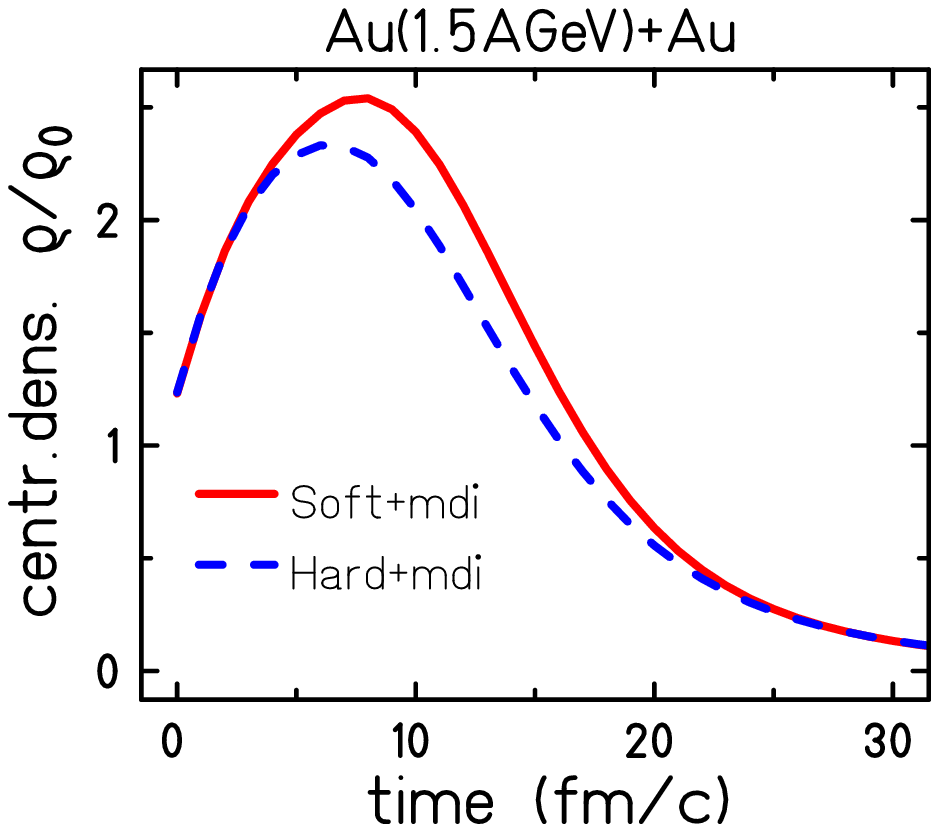,width=0.4\textwidth} &
\epsfig{file=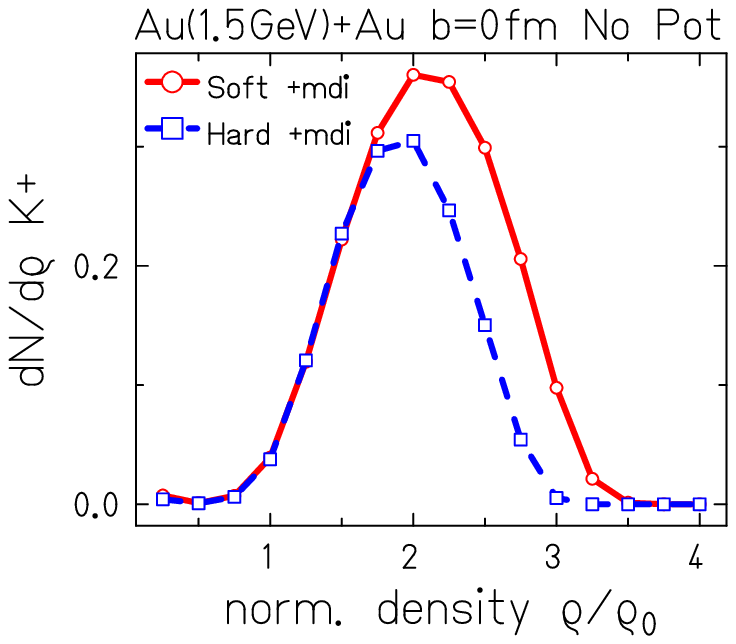,width=0.4\textwidth} \\
\end{tabular}
 \caption{Time evolution of the central density in a calculation with
 hard and soft eos (left) and density distribution of
 the kaon production within a calculation without potentials (right)}
\Label{eos-dns}
\end{figure}

A higher density means a smaller mean free path and thus a better possibility
to gain the energy necessary for kaon production via multi-step processes.
Therefore, the kaon production should be enhanced for a soft eos \cite{aik}.
The \rhsref{eos-dns} shows that indeed the number of kaons is enhanced when
a soft equation of state is employed (\rfl) and that the additional kaons just come
from the regions with higher densities.


\begin{figure}[ht]
\begin{tabular}{cc}
\epsfig{file=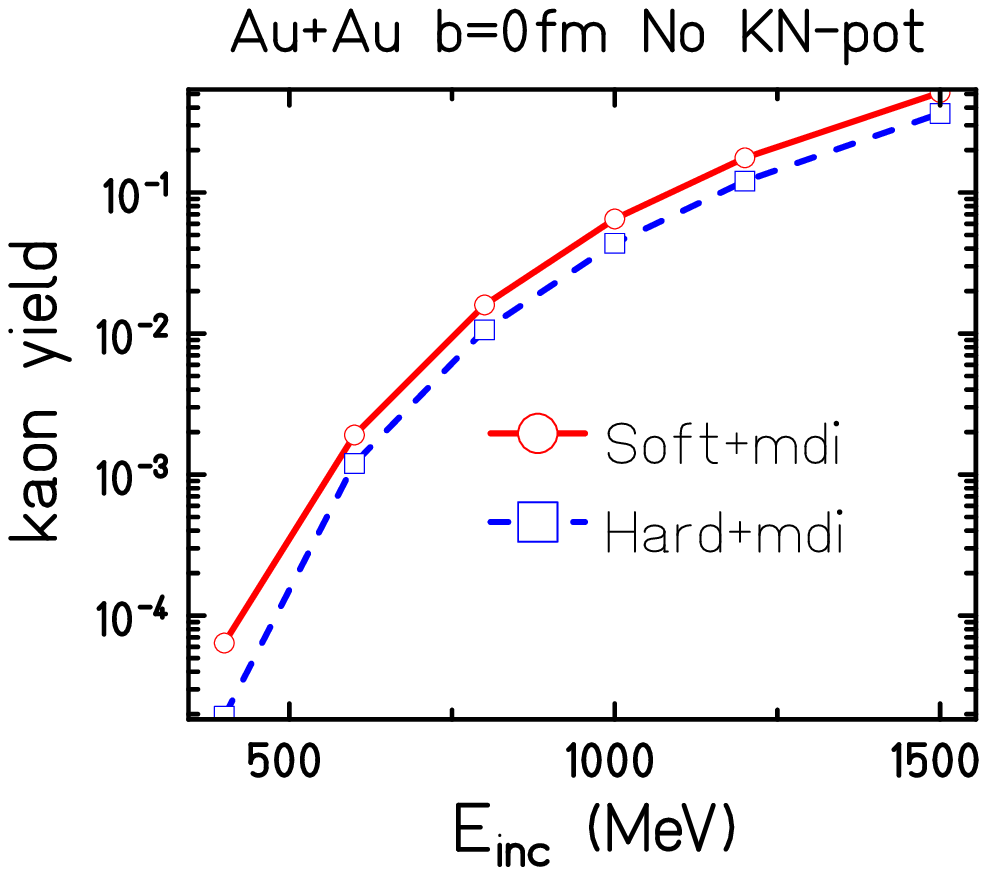,width=0.4\textwidth} &
\epsfig{file=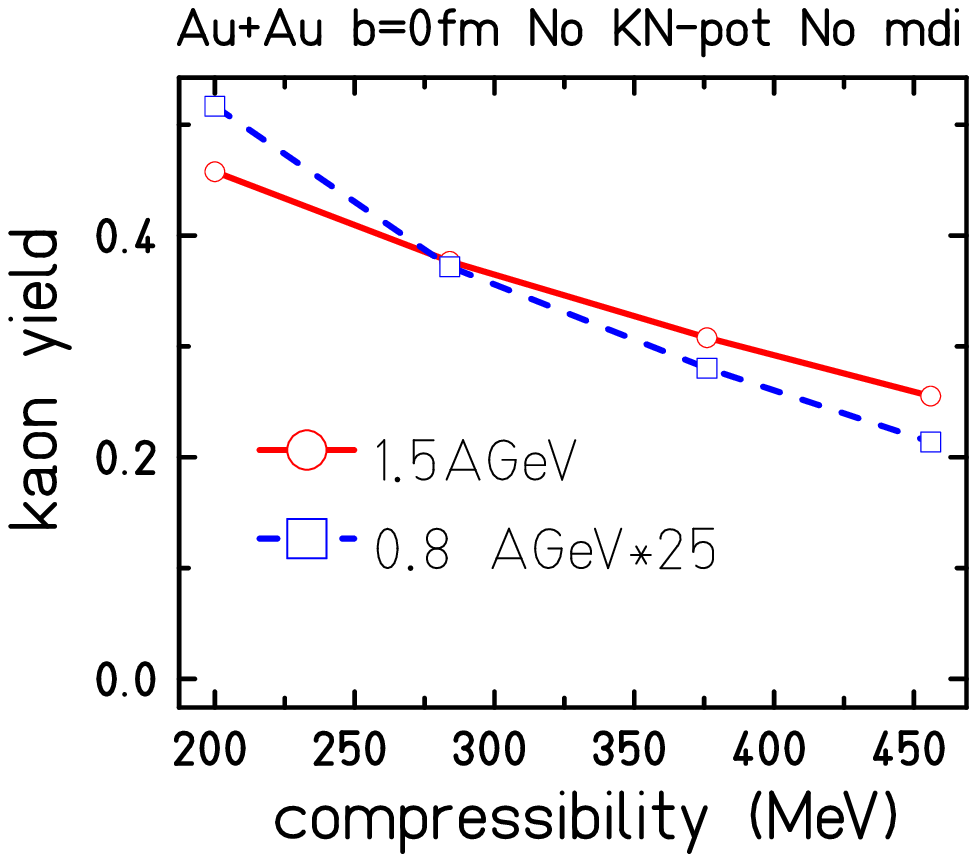,width=0.4\textwidth} \\
\end{tabular}
 \caption{Excitation function of kaon yield for a calculation 
 without KN-potentials (left) and kaon yield as function of
 compressibility}
\Label{e-dep-nopot}
\end{figure}

Therefore, the kaon production shows a sensitivity to the equation of state
as we can see it on the \lhsref{e-dep-nopot}\  where we compare the excitation function
of kaon production with a hard (\bdl) and a soft (\rfl) eos.
The \rhs\  shows the dependence of the kaon production on the compressibility
of the equation of state. A softer eos (small compressibility) yields a higher
kaon number than a harder one. This effect is even stronger when we reduce the
incident energy from 1.5 AGeV  (\rfl) to 0.8 AGeV (\bdl). At lower energies
there is more sensitivity to the available density since we are more  dependent
on the production of kaons in multi-step processes.  

However, the argument is not as simple as shown. We will first have to attack
some questions on the nuclear medium and uncertainties of the cross section
before we can revisit the question of the nuclear equation of state.

\section{Kaons in the medium}
When the kaons are in the nuclear medium of a heavy ion collision
several effects will have to be taken into account.
\begin{enumerate}
\item The kaons may rescatter. This may influence dynamical observables
as we will see later on.
\item The repulsive optical potential of the $K^+N$ interaction may deviate
the trajectories of the kaons.
\item The repulsive optical potential of the $K^+N$ interaction may penalize
the production of kaons and change the threshold effectively since the kaons
are produced at higher effective masses.
\end{enumerate}

\subsection{Rescattering of kaons}
Kaons rescatter with nucleons with a cross section of about 13 mb at lower
relative momenta. Since the nuclear matter is highly compressed in the region
of kaon production there is a high chance of kaon rescattering. 
\begin{figure}
\epsfig{file=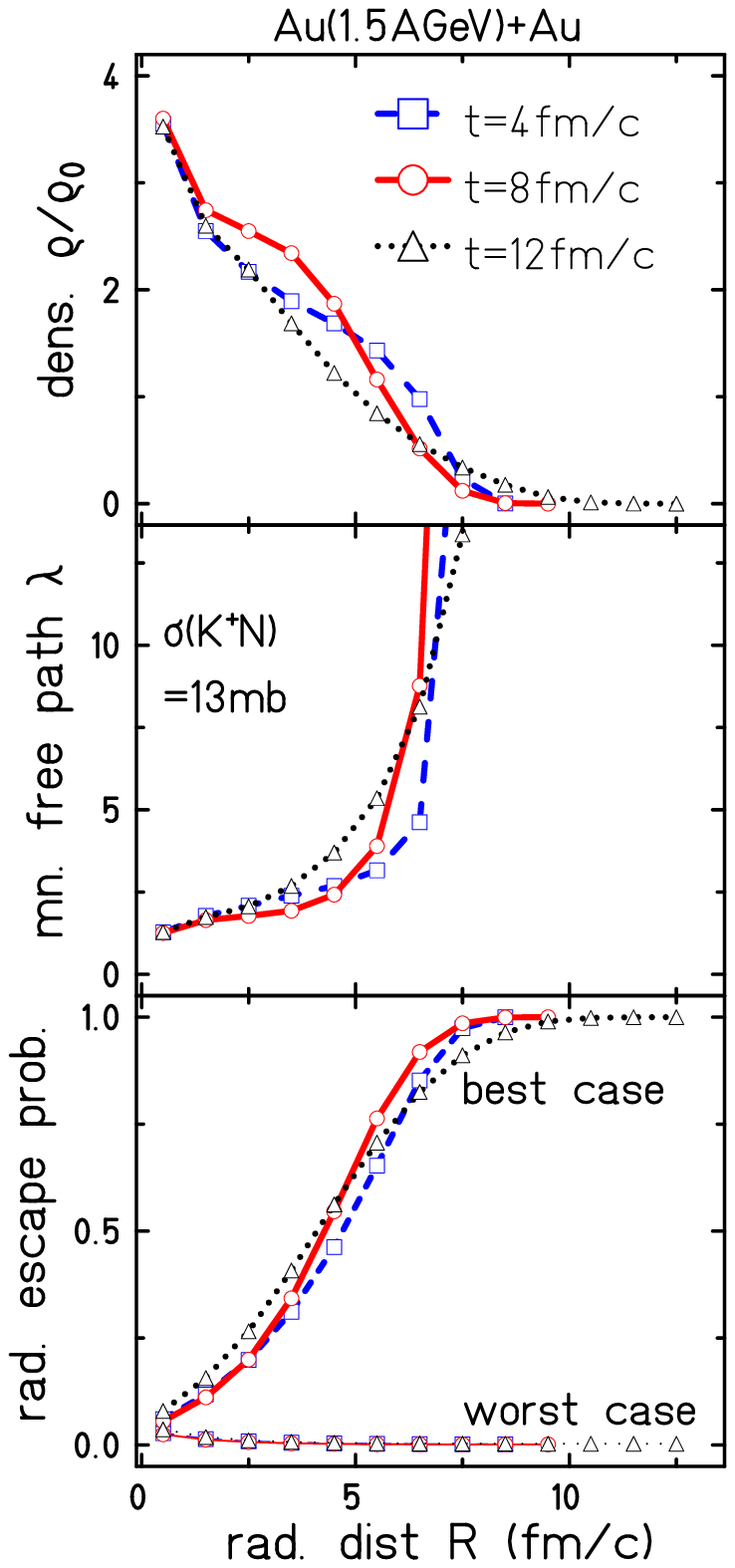,width=0.3\textwidth}
\epsfig{file=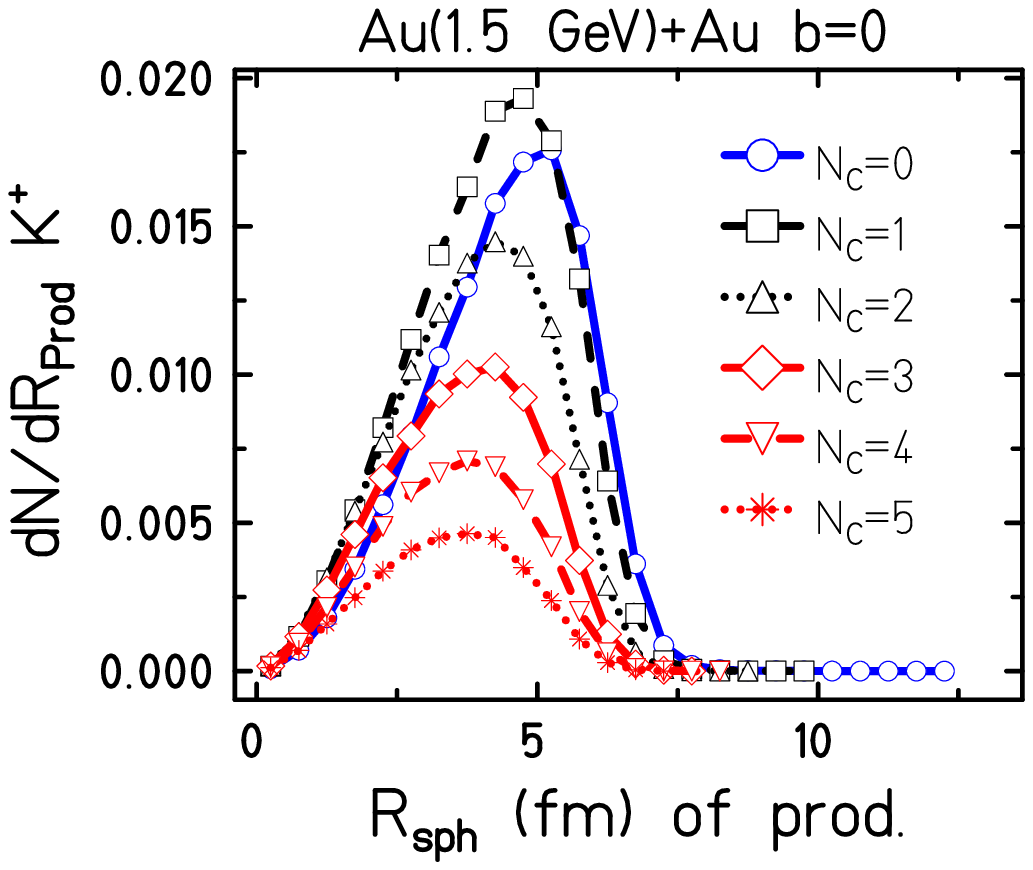,width=0.5\textwidth}
\caption{Radial densities of nucleons and the resulting escape probabilities
of $K^+$ (left) and the radial production profile of  $K^+$ selected according
to the number of collisions (right)
}
\Label{KP-rescattering}
\end{figure}

The \lhsref{KP-rescattering} shows on top the radial density distribution of the
nucleons at time steps of $t=4$ fm/c (\bdl), $t=8$ fm/c (\rfl) and $t=12$ fm/c (\bpl).
The times of 4 and 12 fm/c correspond to the time window when most of the
kaons are produced, while 8 fm/c corresponds to the maximum compression and the
maximum kaon production. From this density we can derive by assuming a cross
section of 13 mb a mean free path as depicted in the mid of the  \lhsref{KP-rescattering}.
An integration over the mean free path finally gives an escape probability
of leaving the system without a collision (bottom). 
This probability of course depends on the exact way the kaon takes. We therefore
indicate the best case (a direct radial escape) and a worst case (an escape in
the opposite direction). We should note that most of the kaons are produced
in a three body decay and therefore the kaons take a high momentum 
with an isotropic distribution in the centre-of-mass frame of the collision.
The collision frame itself has a rather small velocity for maximum kinematic
use of the energy of the colliding particles. Therefore, the velocity of the
sources plays no important role and the kaon will choose its direction 
randomly. 
The \rhsref{KP-rescattering} shows the radial distribution of the kaons selected
according to their collision number. Note that we plot the radial distribution
while in \figref{r-prod} the radial density was plotted, which takes into account
the volumes of the radial cells.
We see, that at the production radii the probability of escape is rather
reduced. We see further that the multicolliders really stem from rather
small radii.


Therefore, we may assume that the kaons really stemming from highest densities
have the highest chance to collide.

\begin{figure}
\epsfig{file=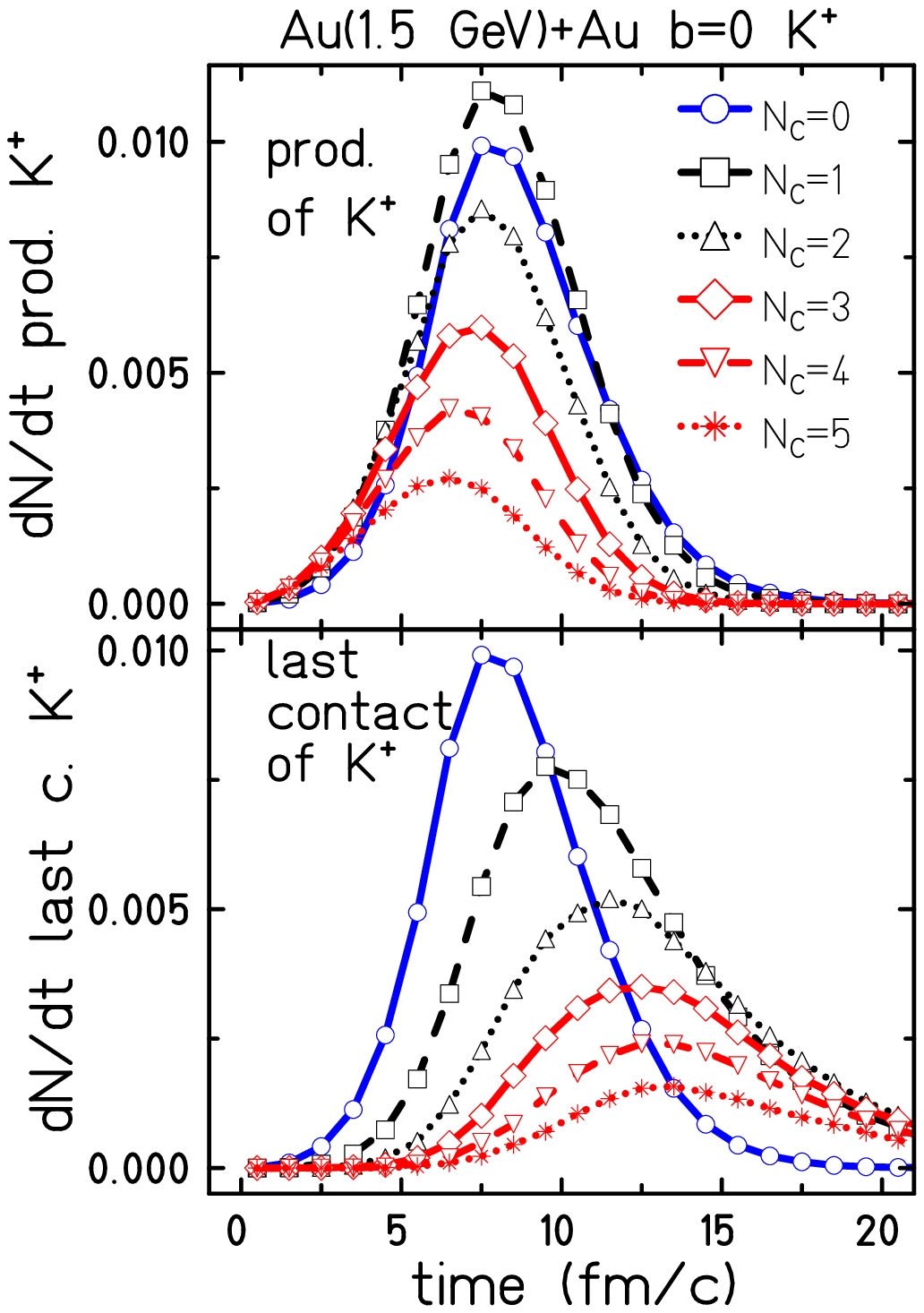,width=0.4\textwidth}
\epsfig{file=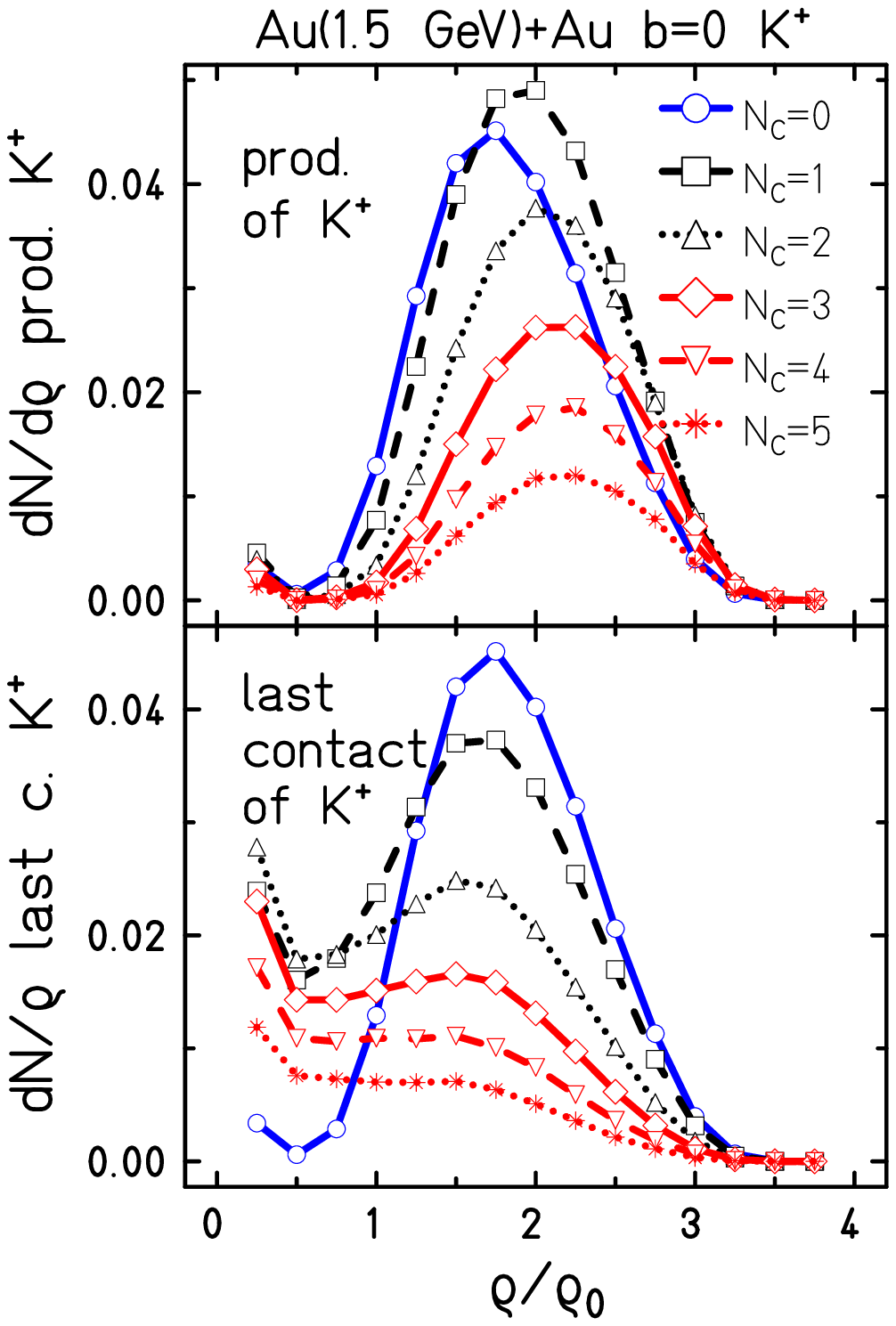,width=0.4\textwidth}
\caption{Time and density profiles for production and last contact
selected according to collision number
}
\Label{dns-tim-resc}
\end{figure}

\Figref{dns-tim-resc} shows again the profiles of production times (left)
and density (right) selected according to the number of collisions.
Multicolliding kaons (red curves) are produced very early and at very high density
but their last collision contact is quite late and happened at rather low 
density. Therefore, the idea to look for the dynamics of kaons in order
to learn something on the high density region will be constrained by the
effects of rescattering. 

\begin{figure}
\epsfig{file=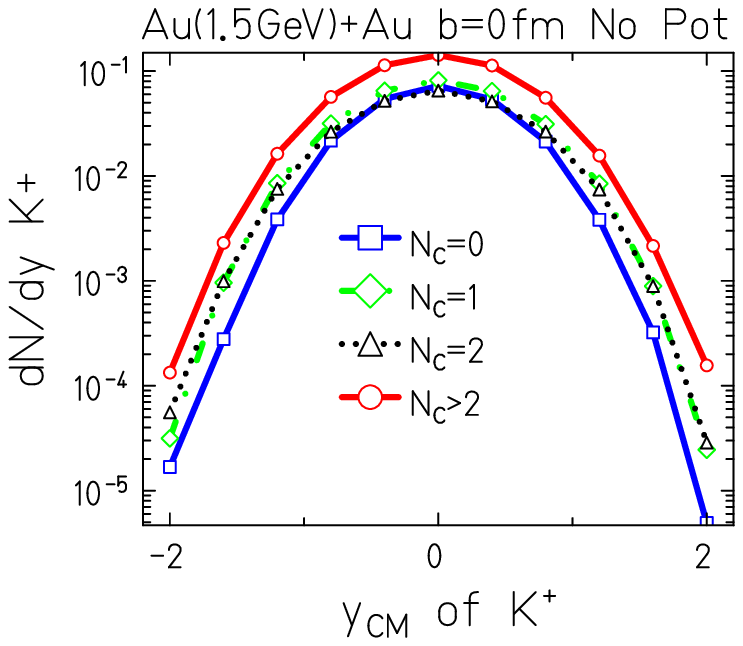,width=0.4\textwidth}
\epsfig{file=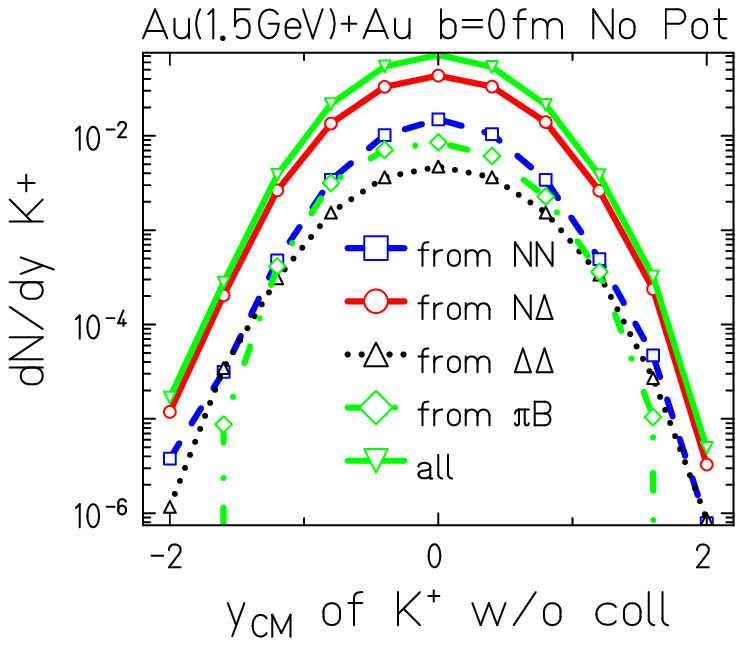,width=0.4\textwidth}
\caption{Rapidity distribution of $K^+$  selected according to
their collision number (left) and of $K^+$ 
without collisions selected
according to their production channel (right) 
}
\Label{dndy-resc}
\end{figure}
The effect of rescattering to dynamical observables is illustrated in 
\figref{dndy-resc} where we show the rapidity distribution of kaons 
selected according to their collision numbers. 
We see on the \lhs\ that kaons without collisions (full blue line) show
a smaller distribution than kaons which did undergo collisions.
Therefore the rapidity distribution of kaons is broader when rescattering 
is active.
For comparison we see on the \rhsref{dndy-resc} the comparison of the effect
of the incident channels on the rapidity distribution for particles without
collisions. Kaons stemming from  high-density $\Delta\Delta$ collisions (\bpl) 
show a broader distribution than kaons stemming from $NN$ (\bdl) but this
effect will be overruled by the effect of rescattering. 

Further discussion of the effects of rescattering on dynamical observables
can be found later on.


\subsection{Influence of the optical potential}
The optical potential of the kaon in the nuclear medium is repulsive for
$K^+$ and enhances its effective energy in the medium. Thus, the production of
a kaon in the medium needs more energy than in the free case and enhances
the threshold of its production. Since we are at energies below the elementary
threshold and since we already need to cumulate energy for getting a kaon,
this up-shift of the threshold yields a strong reduction of the kaon yield. 

\begin{figure}
\epsfig{file=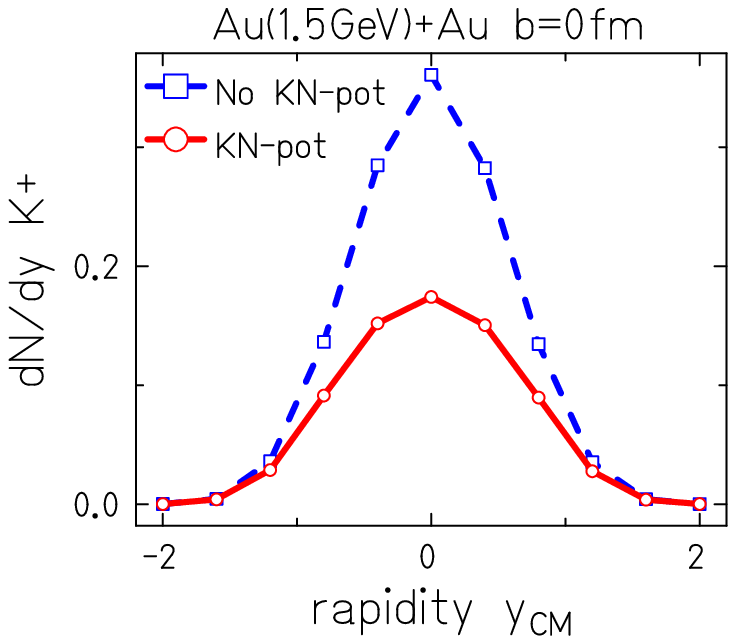,width=0.4\textwidth}
\epsfig{file=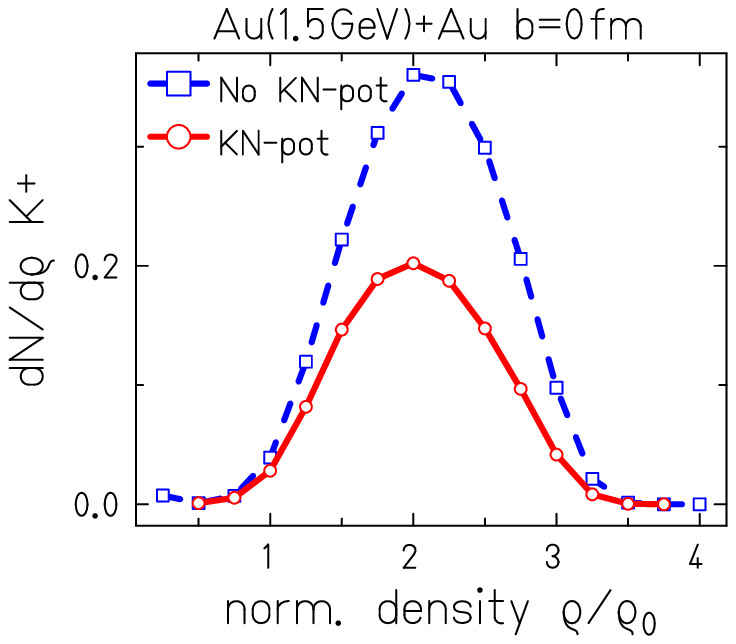,width=0.4\textwidth}
\caption{ Rapidity distribution and density profile of kaon production
with and without kaon potential. 
}
\Label{dns-pot-nopot}
\end{figure}

\Figref{dns-pot-nopot} shows on the \lhs\ the effect of the optical potential on the rapidity
distribution of kaons at Au(1.5AGeV)+Au b=0. A calculation with optical
potential (KN-pot, \rfl) yields much less kaons than a calculation without
optical potential (\bdl). Especially at mid-rapidity the discrepancy is most
prominent. Here the kaons with least energy can be found. They are produced
by collisions slightly above the production threshold, thus only few energy 
is available for the kinematics. Here a shift of the threshold inhibits their
production. Kaons with high absolute values of the rapidity have rather high
energies. In their production the shift of the threshold reduced the remaining
energy for the kinematics. However, the repulsive potential gives back this
energy when the kaon will leave the medium in the late expansion phase.
We will see later on similar effects for the spectra.

The \rhsref{dns-pot-nopot} shows the density profile of the kaon production in
both calculations. We see that the penalty of the potential acts especially at
high densities since here the shifts of the threshold are the strongest.

This penalty also effects the excitation function of the kaons as we can see
from the \lhsref{pot-yields}. At low incident energies the production of kaons
via multi-step processes is more important than close to the threshold. 
Remember that the effect of the eos was also stronger at lower energies due to 
the same argument. Therefore, the penalization of reactions at high densities
shows stronger effects which causes a strong influence of the calculation
with (\rfl) and without (\bdl) an optical potential  
The effect of the energy loss can also be seen when comparing the calculation 
with optical potential (and full Fermi momentum, \rfl) with a calculation
with optical potential where the Fermi momentum has been suppressed (\bpl).
The lack of Fermi momentum gives an additional lack of energy which shows
strongest effects at low energies where the multi-step processes very important
and where a multiple lack of energy reduces the yield drastically.

\begin{figure}
\epsfig{file=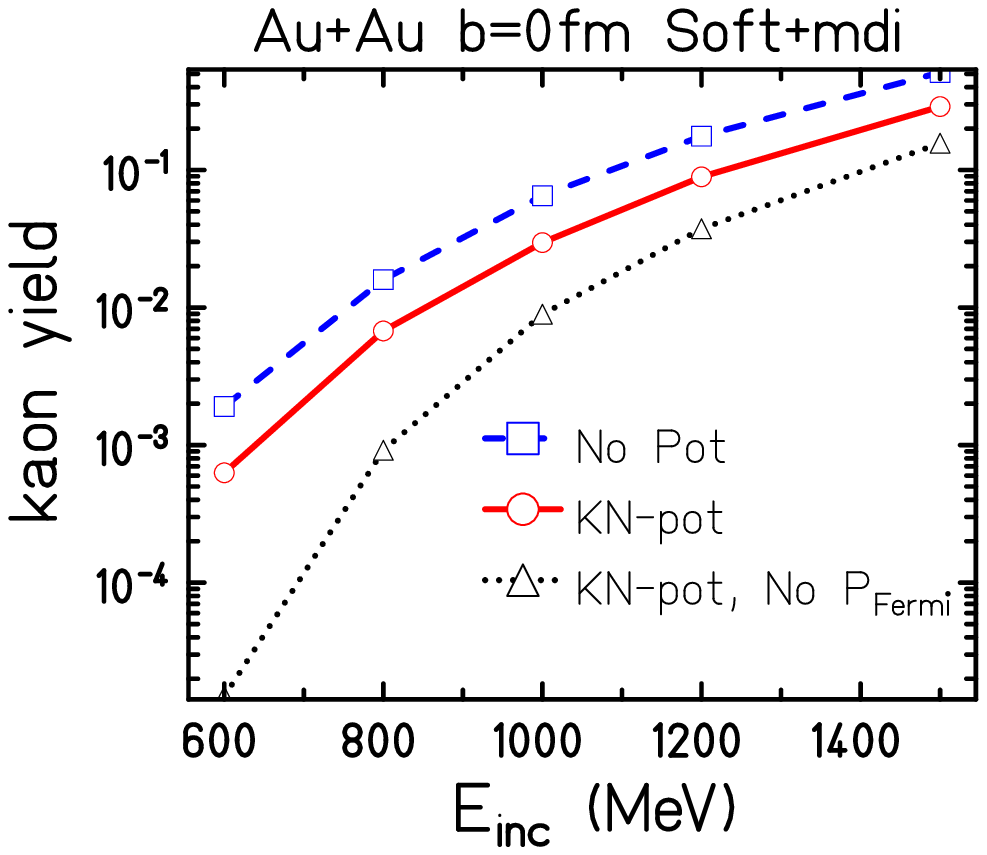,width=0.4\textwidth}
\epsfig{file=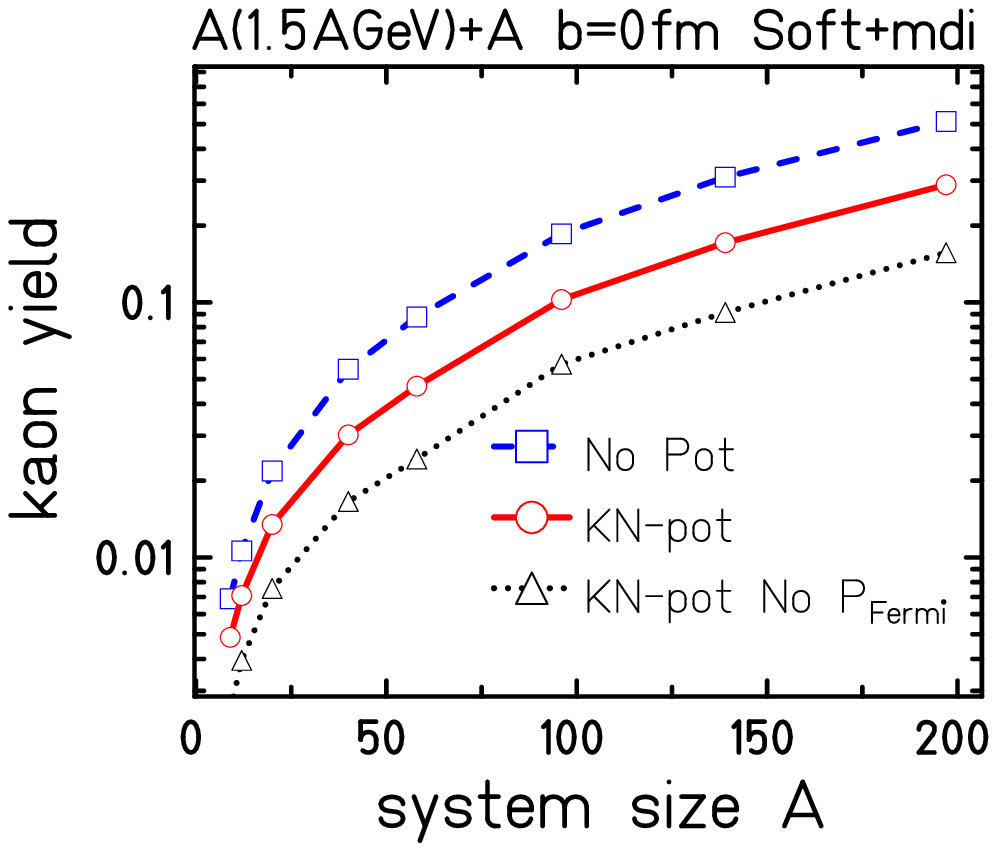,width=0.4\textwidth}
\caption{Excitation function (left) and system size dependence (right) 
 of the kaon yield with and without KN-potential and with KN-potential
 but no Fermi momentum. 
}
\Label{pot-yields}
\end{figure}

The system size dependence of kaon production shown in \rhsref{pot-yields} 
shows only weak effects for small systems but strong effects for large
systems. In small systems the compression is weaker and less high densities
are reached. Therefore, the penalty of the potential is weaker. 
Also the effect of the missing Fermi momentum is smaller.

This interplay of density and penalty can also be demonstrated nicely when
analyzing the effects  of the scalar and vector part of the optical potential
as it is shown in \figref{dns-scalar-vector}.
\begin{figure}
\epsfig{file=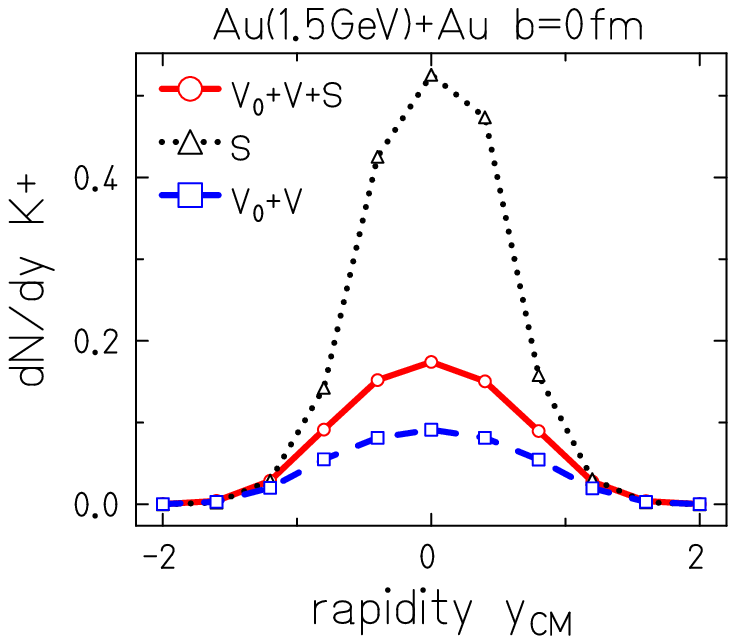,width=0.4\textwidth}
\epsfig{file=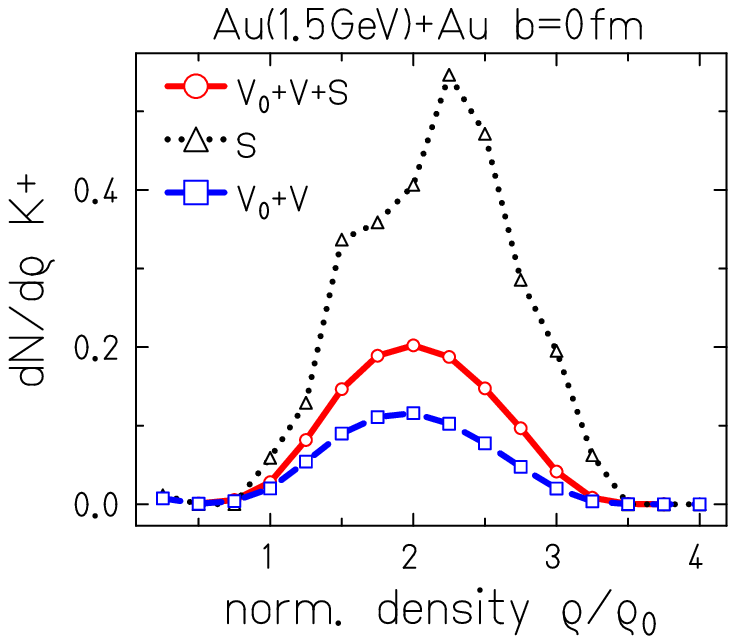,width=0.4\textwidth}
\caption{ Rapidity distribution and density profile of kaon production
and the contribution of scalar and vector part. 
}
\Label{dns-scalar-vector}
\end{figure}
The scalar potential (\bpl) is attractive (see \figref{opt-pot-2}). It enhances the kaon yield
especially at mid-rapidity. Since the penalty at high densities is changed
into a gain, it strongly enhances the kaon yield. The gain is stemming
dominantly from the high densities. The vector potential is very repulsive.
A calculation using only the vector potential (\bdl) reduces the kaon number
particularly at mid-rapidity and reduces the production especially at high
densities.


The effect of the strength of the optical potential can finally be seen
in \figref{dns-RMF-NJL} where we compare different parametrizations of the
optical potential. A parametrization with less strength
(half pot, \gml, see \figref{opt-pot-1}) yields higher kaon numbers than
our standard optical potential (Schaffner RMF, \rfl), a parametrization
with a comparable strength (NJL, T=150, \bpl,see \figref{opt-pot-3})
a comparable kaon distribution and a calculation with a higher strength 
of the potential (NJL, T=0, \bdl,see \figref{opt-pot-3}) a smaller kaon number.
\begin{figure}
\epsfig{file=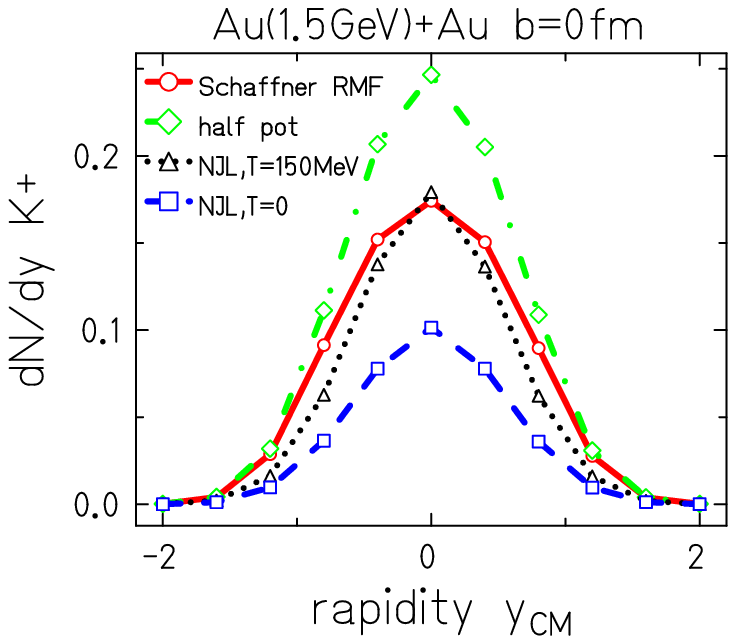,width=0.4\textwidth}
\epsfig{file=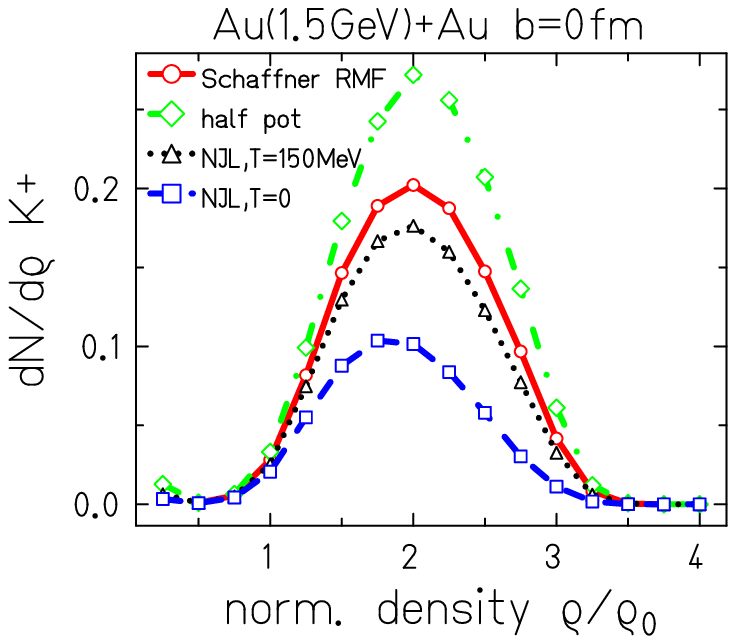,width=0.4\textwidth}
\caption{ Rapidity distribution and density profile of kaon production
for different parametrizations of the optical potential. 
}
\Label{dns-RMF-NJL}
\end{figure}

Again, the effect is most significant at mid-rapidity and effects mostly the kaons
stemming from high densities.
Different parametrizations of the optical potential may thus yield different
kaon yields.

The effect of the optical potential on spectra, temperatures and angular
distributions will be discussed later.

\section{Can we derive the potential from kaon yields?}
As we have seen the optical potential shifts the up the production
threshold and thus reduces the kaon yield.
A strong repulsive potential yields therefore a strong reduction of the
kaon yields while a weak repulsion yields a weak reduction of the yield.
We will now address the question whether this effect might be used to determine
the kaon optical potential.

\subsection{Comparison to p+A}
Let us start with p+A data which have already been shown to be quite
sensitive to the kaon production mechanisms.

\begin{figure}[hbt]
\epsfig{file=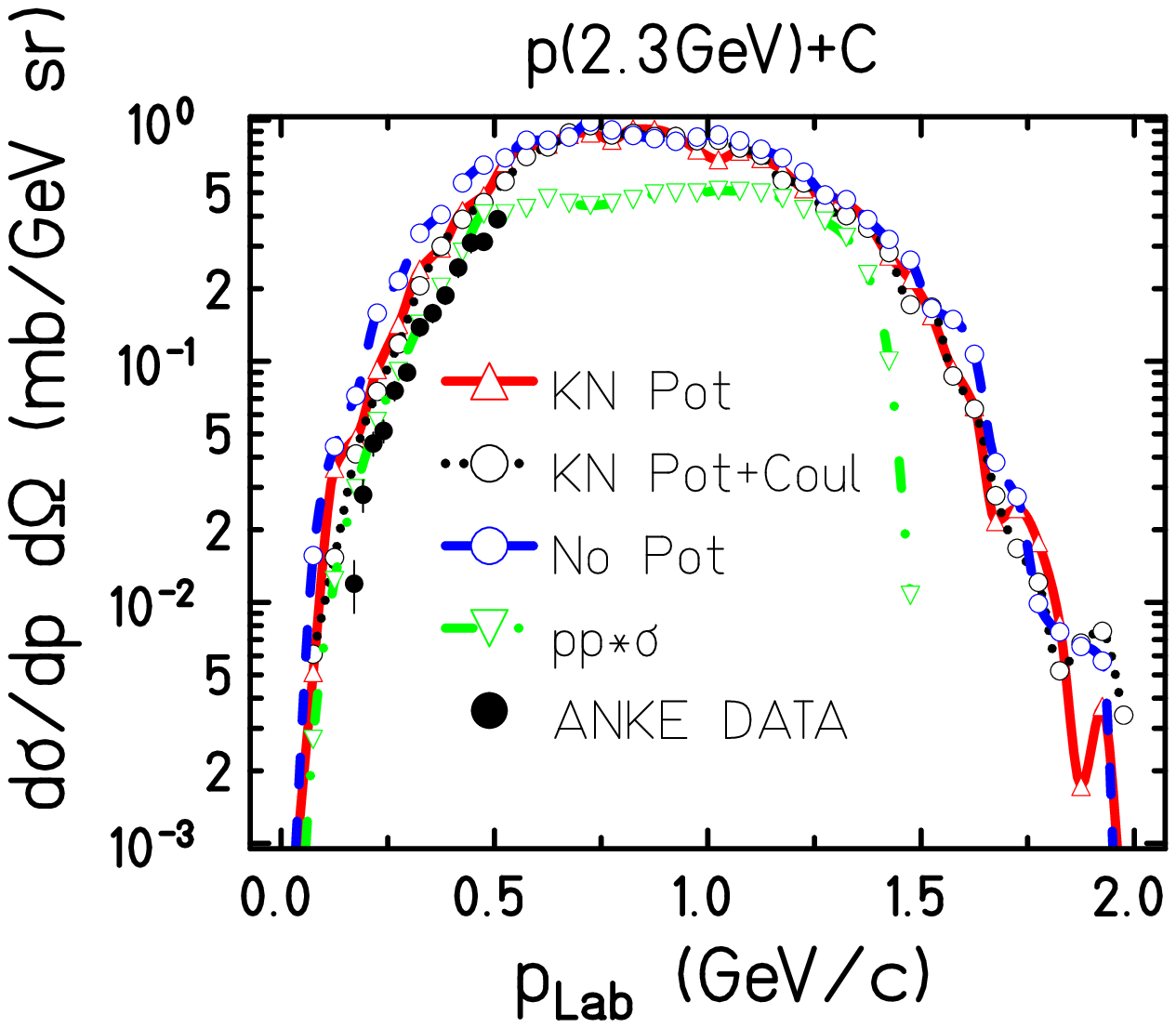,width=0.4\textwidth}
\epsfig{file=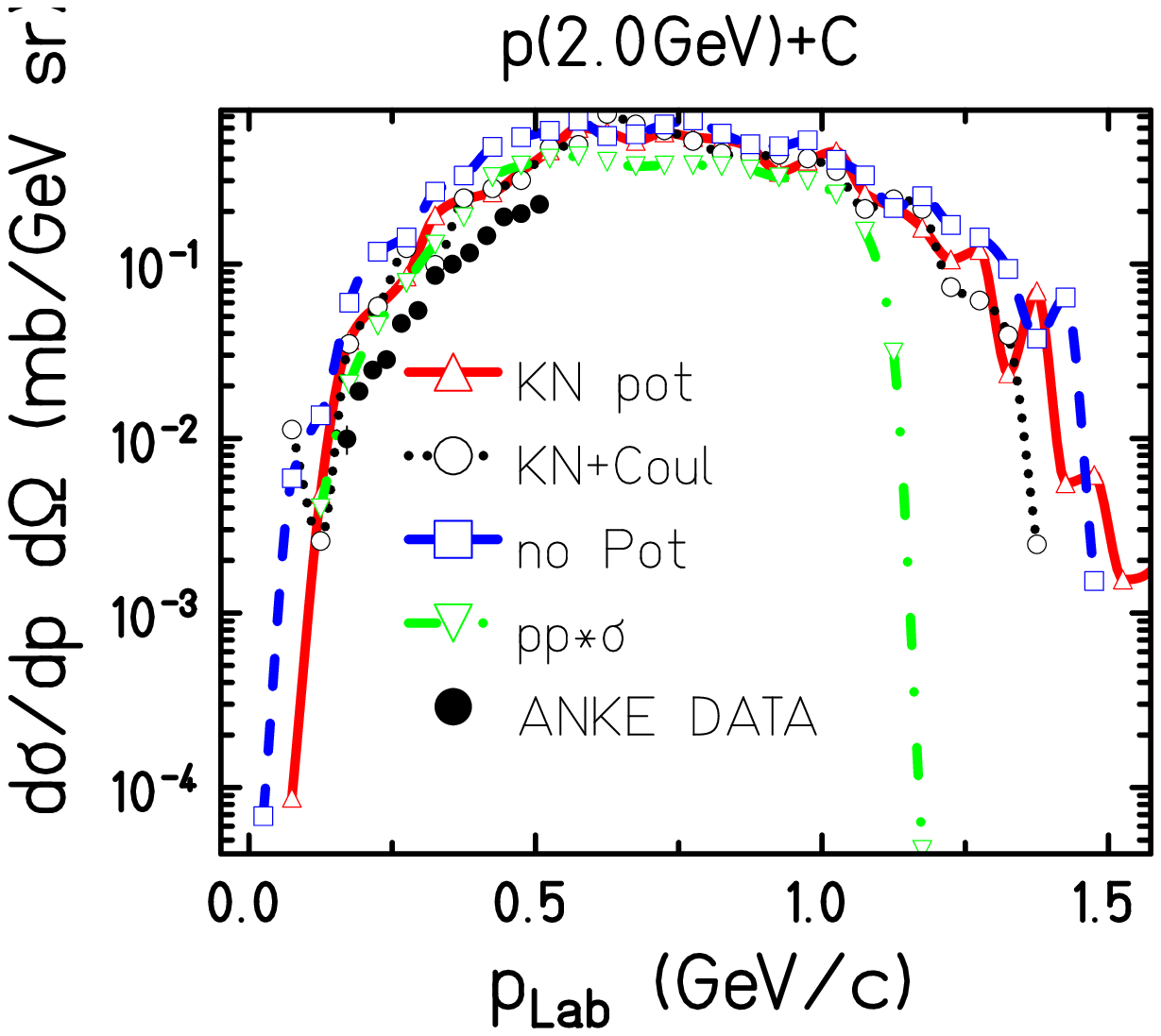,width=0.4\textwidth}
\caption{Comparison of p+C data at 2.3 and 2.0 GeV  
}
\Label{pC-high}
\end{figure}

In \figref{pC-high} a comparison of COSY-ANKE p+C data (bullets) \cite{anke} 
and IQMD calculations with (\rfl{} with triangles) and without KN potential
(\bdl{} with squares) for energies of 2.3 (left) and 2.0 GeV (right)
is shown. We see
that the calculation without potential yields slightly higher results than the
calculations with potential. The additional inclusion of Coulomb forces (\bpl{} 
with circles )
does not yield a significant change since the target is quite light.
The data support rather the calculations with potential.
A comparison of p+C with a scaled p+p reactions (\gml) shows again the importance
of the Fermi momentum to describe the high momenta. Unfortunately no experimental
data are available at these high momentum values. 

\begin{figure}[hbt]
\epsfig{file=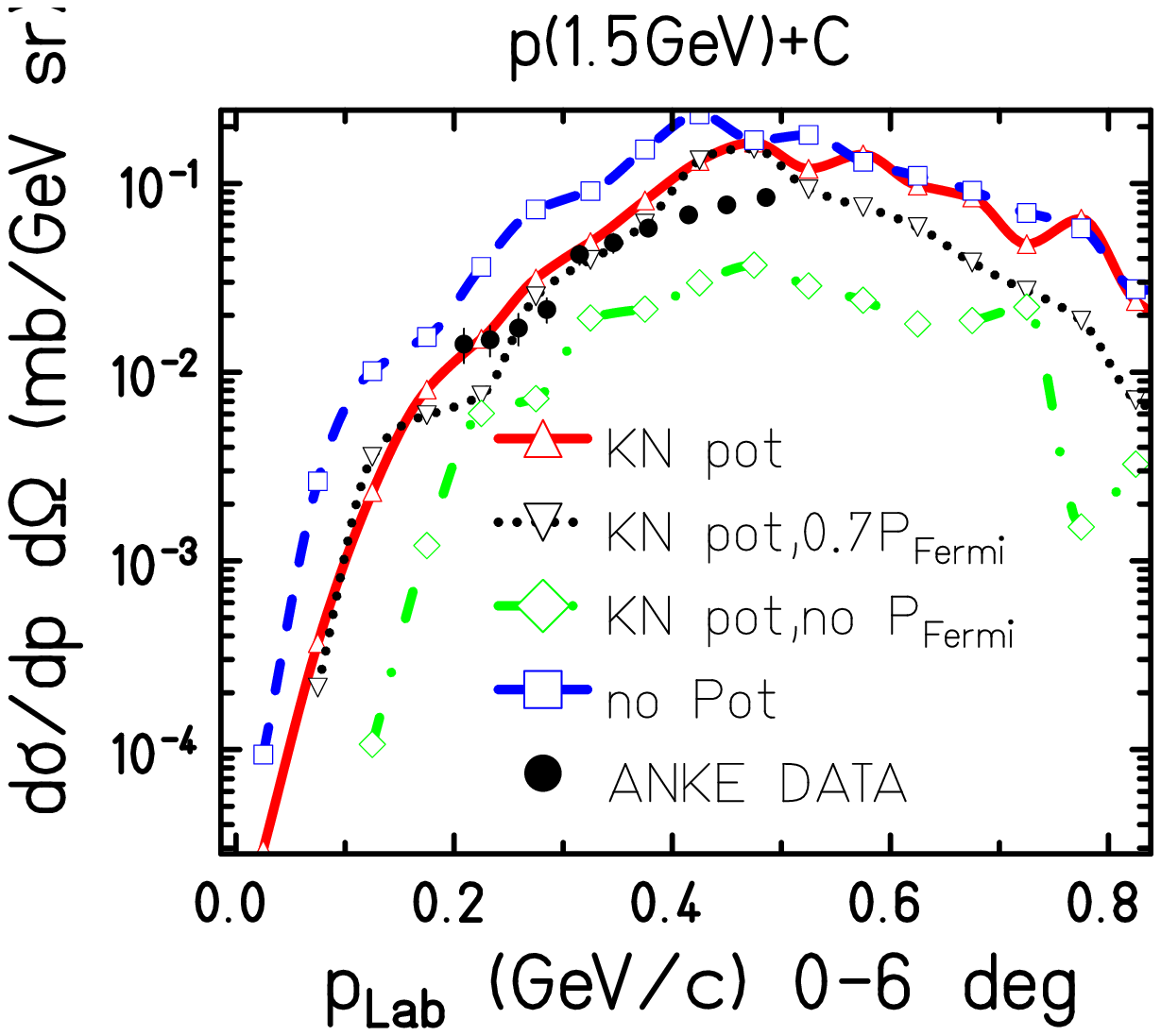,width=0.4\textwidth}
\epsfig{file=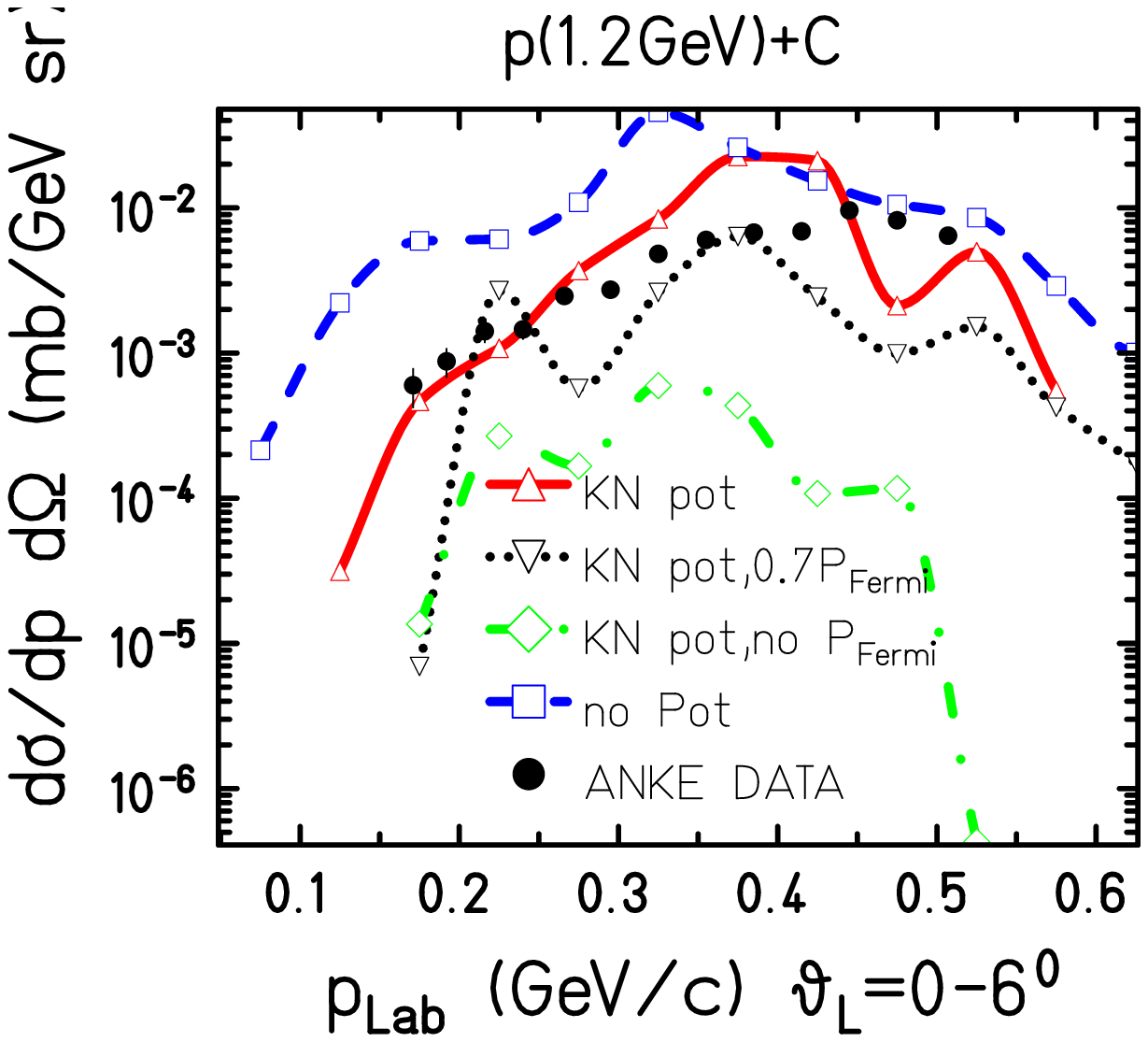,width=0.4\textwidth}
\caption{Comparison of p+C data at 1.5 and 1.2 GeV  
}
\Label{pC-low}
\end{figure}
However, the difference between calculations with and without potentials
is quite small. This is due to the high incident energy. The higher the energy 
above the threshold the less important becomes the penalty of the optical
potential caused by the threshold shift. 
Therefore \figref{pC-low} presents calculations below the threshold.  Here the
importance of the potential becomes stronger when going further down in energy
as we can see when comparing p+C at 1.5 GeV (left) and 1.2 GeV (right).
The data seem to comply better with the calculation with an optical
potential (\rfl) than with a calculation without potential (\bdl). However,
the result far below subthreshold depend strongly on the description of the
energy available in the nucleus. If we use a  calculation with potential
and reduce the Fermi momentum from its full value (\rfl) to only 70 \%
(\bpl) we also reduce the kaon production significantly. A complete
suppression of the Fermi momentum (\gml)  shifts the results visibly beyond
the experimental data.

\begin{figure}[hbt]
\epsfig{file=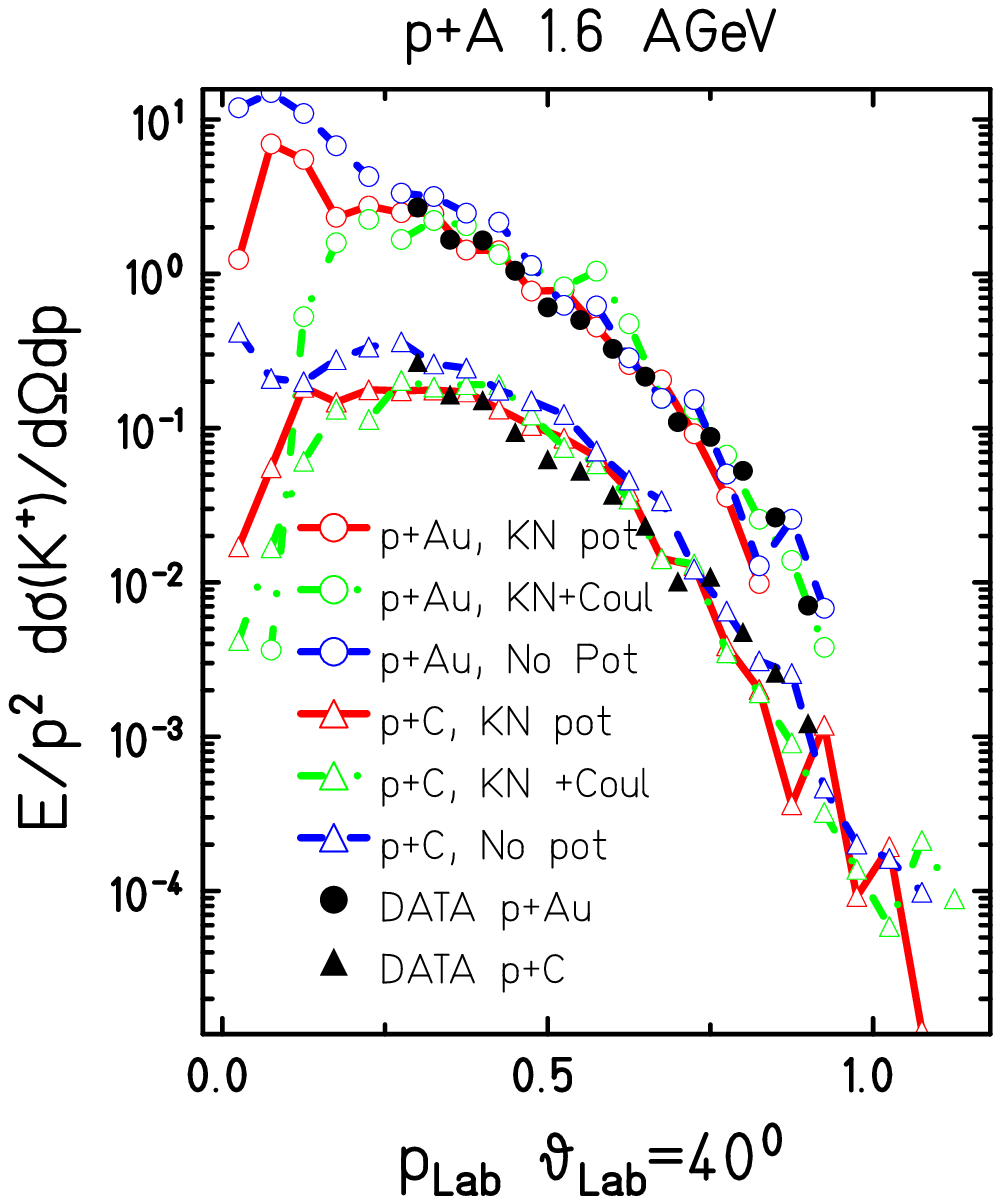,width=0.4\textwidth}
\epsfig{file=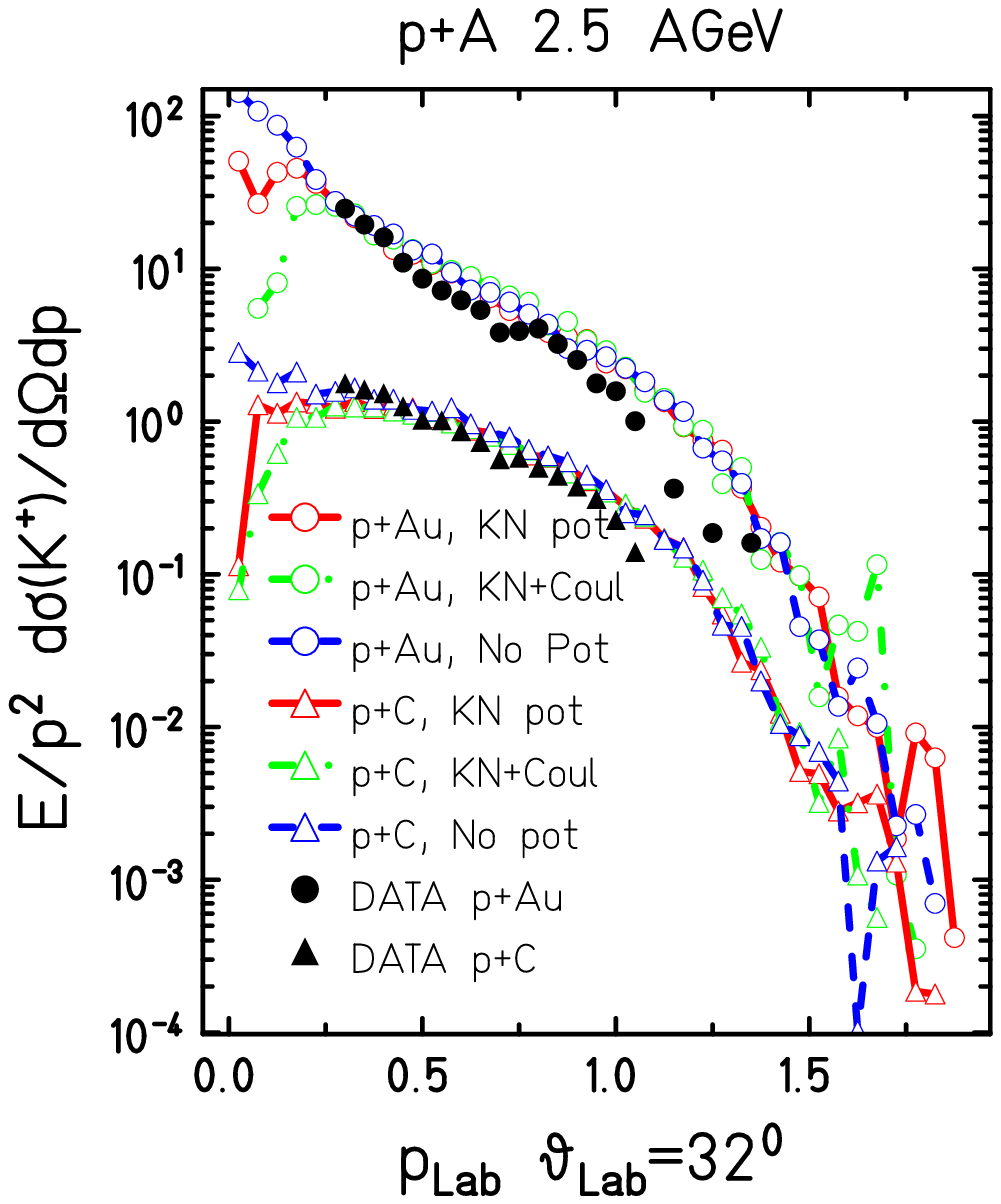,width=0.4\textwidth}
\caption{Comparison of p+C and p+Au KaoS data at 1.6 and 2.5 GeV 
}
\Label{pA-kaos}
\end{figure}

The ANKE data are measured at very small laboratory angles. 
As an effect, comparison these spectra might be influenced by rescattering.
A rescattered kaon may easily leave that small detector angle.
Therefore it is interesting to compare kaons also at other laboratory angles.
In \figref{pA-kaos} we see recent spectra of the KaoS collaboration \cite{scheinast}
taken at laboratory angles of $40^0$ (1.6 GeV, left) and $32^0$ (2.5 GeV, right).
We find a visible influence of the optical potential. However, a conclusion
on the potential is quite difficult. The inclusion of Coulomb forces
(\gml{}) only changes the spectra at very low momenta where no experimental
data are available. We see that the Coulomb forces show a stronger effect for
the heavy Au system than for the light C system due to the larger charge of
the nucleus.

\subsection{Influence of rescattering in p+A}
Let us now analyze the influence of the rescattering on the kaon spectra.
As indicated rescattering may change the direction of the outcoming kaon
and also change its energy.
\Figref{pA-resc} shows how the rescattering effects the spectra. The full
lines show spectra with normal rescattering, the dashed lines calculations
without rescattering and the dotted lines calculations with a doubled value
of the rescattering cross section.

\begin{figure}[hbt]
\epsfig{file=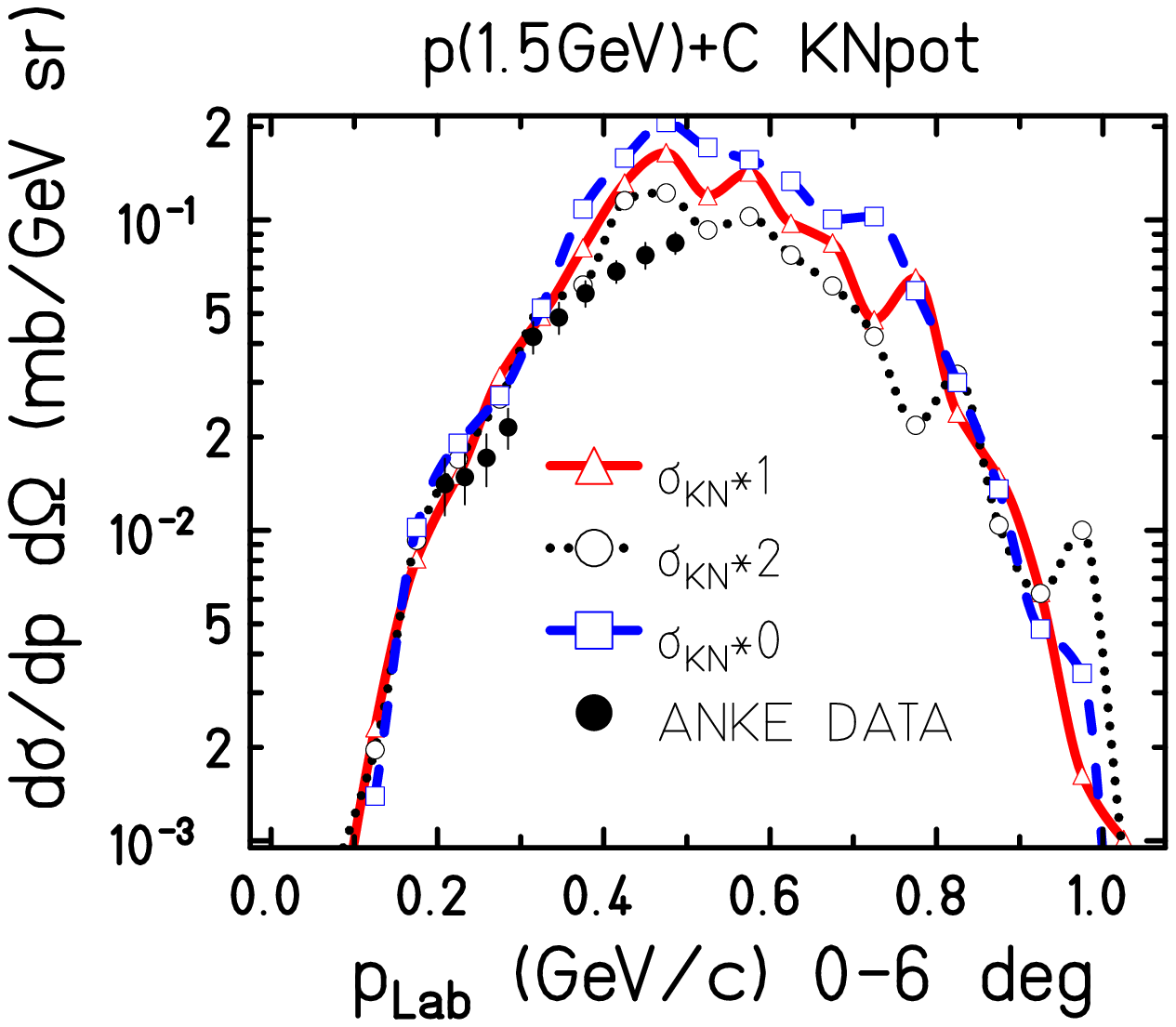,width=0.4\textwidth}
\epsfig{file=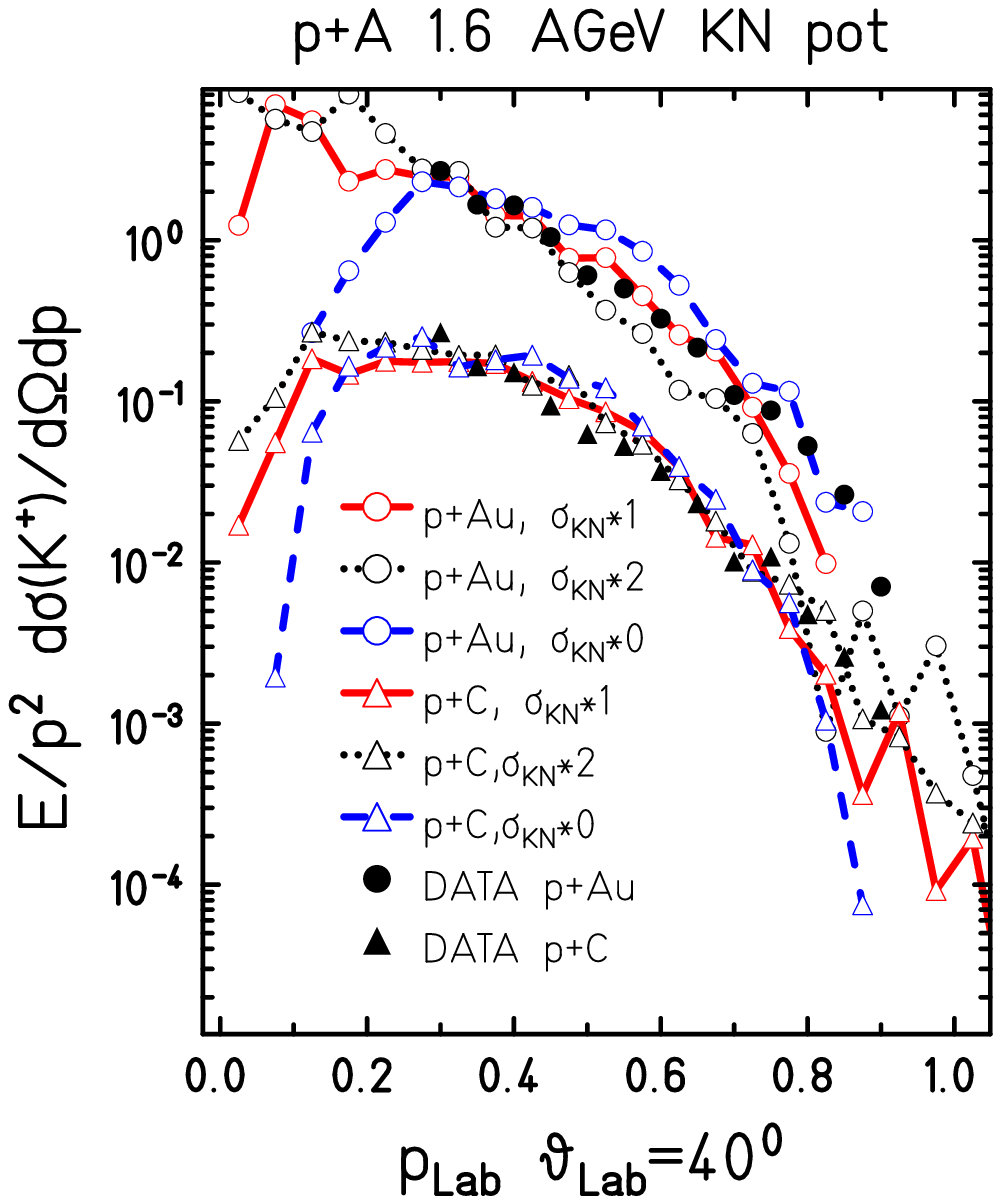,width=0.4\textwidth}
\caption{ Comparison of p+A spectra with normal rescattering, 
disabled rescattering and rescattering with the double cross section. 
}
\Label{pA-resc}
\end{figure}

We see for the spectra at small angles (\lhs) an enhancement of the absolute
yield when suppressing rescattering and a diminution of the yield when enhancing
the rescattering. A rescattered kaon leaves that detector angle. 
For the spectra at higher angles (\rhs) the low momentum part increases with
rescattering and decreases when turning off the rescattering. The effect is
stronger for a heavy system (p+Au) which allows for more rescattering partners
than for a light system.

\subsection{Influence of delta lifetime and of the nucleon-nucleon
cross section on the kaon production in p+A}
Let us now shortly discuss the effect of the delta lifetime
on the spectra in p+A. Since at low incident energies the 
$NN$ channel still plays an important role and no high density
region is built up there should be no strong influence of the lifetime
of the delta on the spectra. This can be seen in \figref{pA-gamma}.
\begin{figure}[hbt]
\epsfig{file=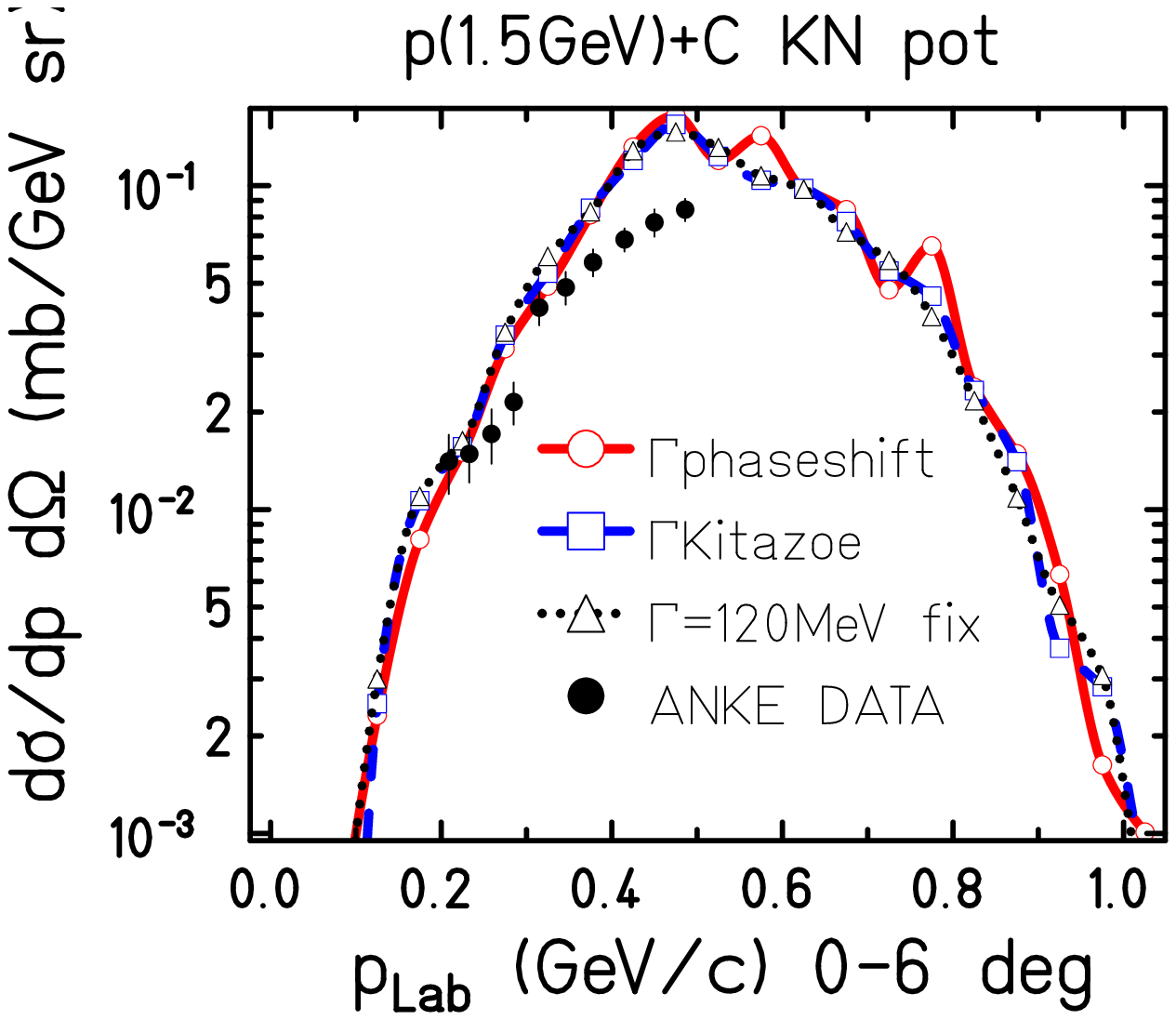,width=0.4\textwidth}
\epsfig{file=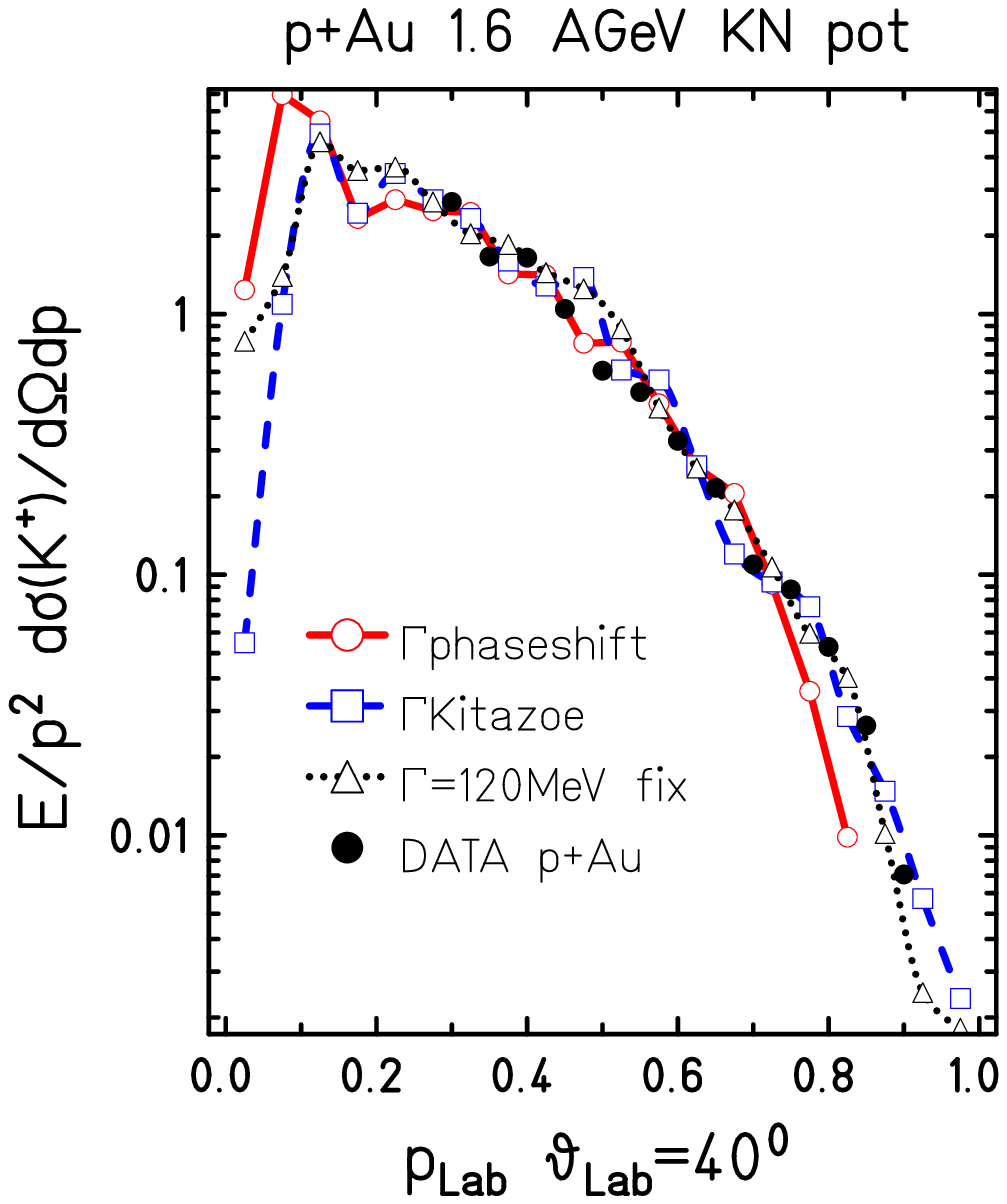,width=0.4\textwidth}
\caption{ Comparison of p+A spectra with different parametrizations
of the delta lifetime 
}
\Label{pA-gamma}
\end{figure}

However this delta lifetime may play a role when discussing A+A. We will
soon come back to this point. We should also note that a strong reduction 
of the Fermi momentum enhances the contribution of the $N\Delta$ and $N\pi$
channels and may enhance the significance of the delta lifetime for this case.

The nucleon-nucleon cross section influences the dynamics of the
nucleons and the production of resonances. However, like in the discussion
of the delta lifetime there is no big effect on the p+A results if we
use an unchanged production cross section of the kaons.
This can be seen in \figref{pA-sigbb}.

\begin{figure}[hbt]
\epsfig{file=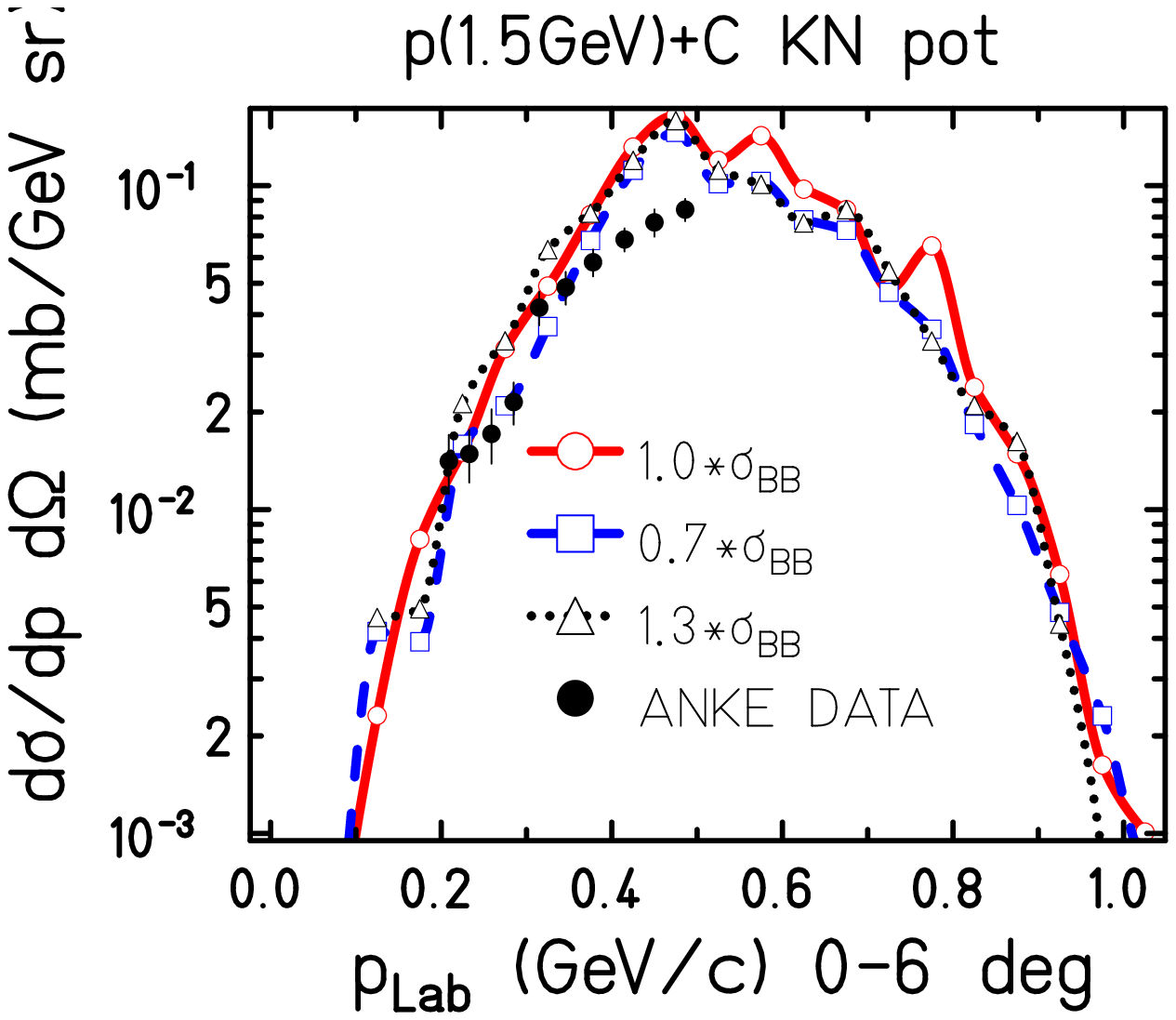,width=0.4\textwidth}
\epsfig{file=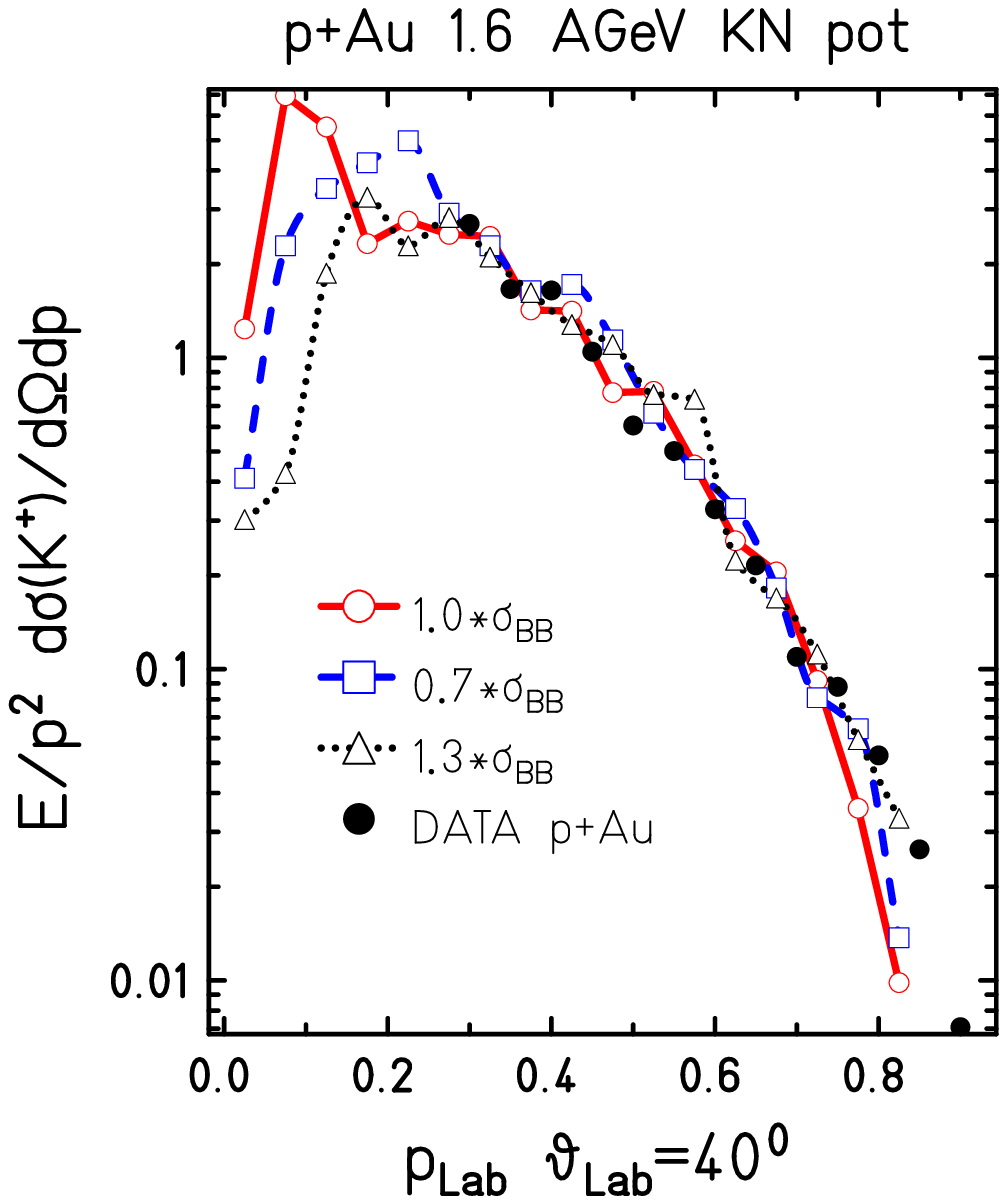,width=0.4\textwidth}
\caption{ Comparison of p+A spectra with different factors
to the nucleon cross section. 
}
\Label{pA-sigbb}
\end{figure}

\subsection{Uncertainties of unknown  production cross sections}

As is was shown in \figref{ncps-dist} the major contribution to kaon production
is the $N\Delta$ channel. Unfortunately, this channel is not accessible
experimentally. Thus, cross section parametrizations of this channel have some
relative freedom relying on different assumptions.

\begin{figure}[hbt]
\epsfig{file=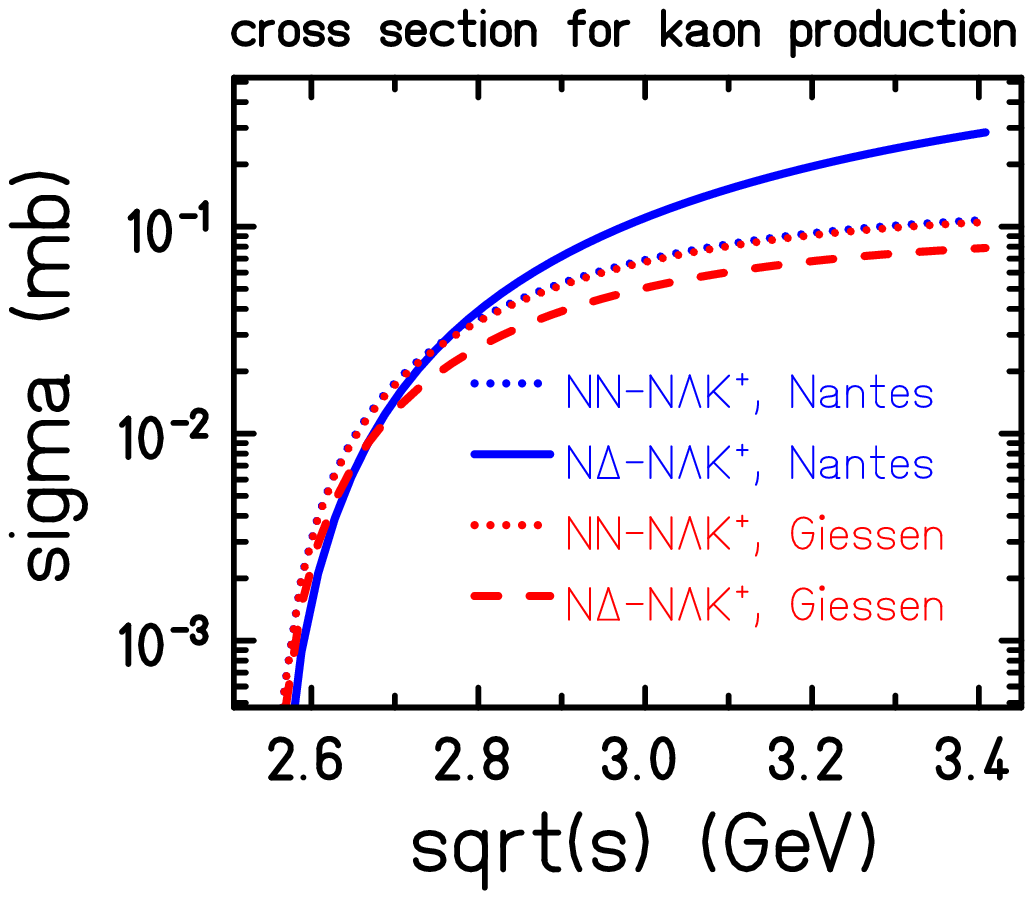,width=0.4\textwidth}
\epsfig{file=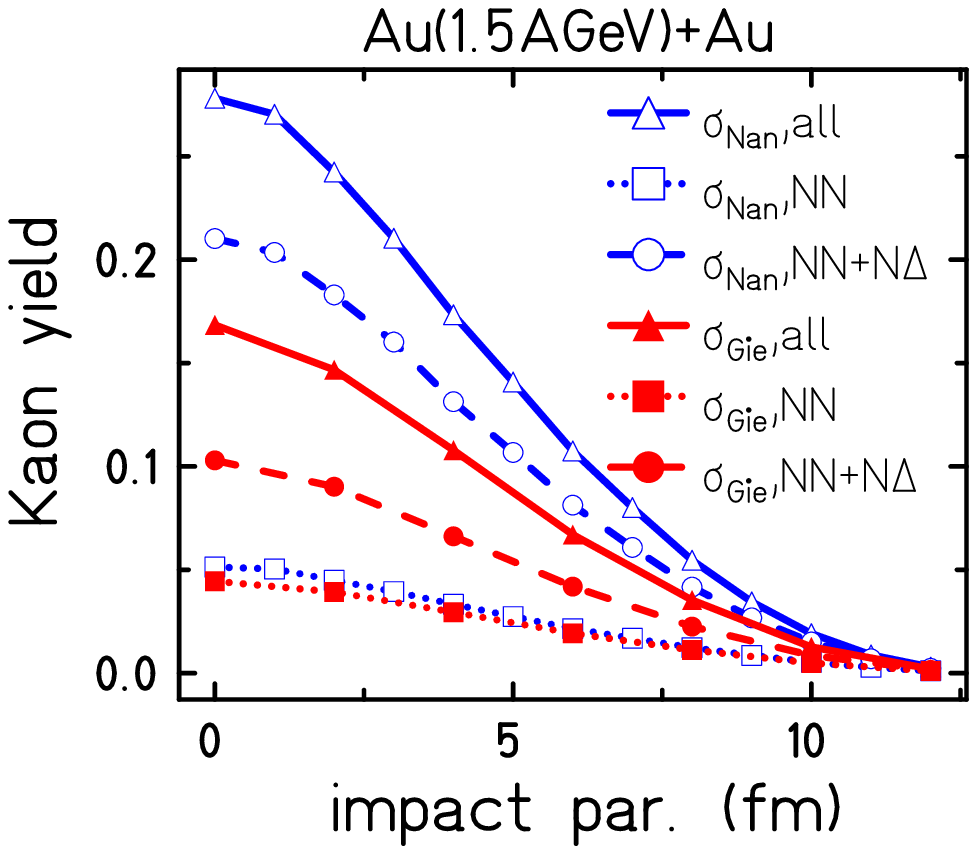,width=0.4\textwidth}
\caption{ Different parametrizations of the $N\Delta\to N\Lambda K$
cross sections and its influence to the kaon yield.  
}
\Label{xsections-ND}
\end{figure}
As an example the HSD-group in Giessen \cite{cassing} used a scaled
$NN$ production cross section for describing $N\Delta$ while
our calculations use the parametrizations of Tsushima et al. \cite{tsu}.
The \lhsref{xsections-ND} shows a comparison of the $NN$ and $N\Delta$
cross sections between our calculation (Nantes, blue curves) and the Giessen
group (red curves). While the  $NN$ cross sections (dotted curves) show
the same parametrizations, the $N\Delta$ cross sections (full blue line
for Nantes, red dashed line for Giessen) are strongly different.

This effects directly the kaon yield as it can be see on the 
\rhsref{xsections-ND} where we compare the impact parameter dependence of 
Au+Au. The calculation with the Nantes cross sections are given by the blue
curves with open symbols while the calculations using the Giessen cross section
are represented by red curves with filled symbols. 
The absolute yield of kaons is quite identical for kaons produced in $NN$ 
collisions (dotted line with squares), as it should be expected from the 
use of similar cross sections. However the dominating $N\Delta$ channel 
(dashed line with circles) yield a stronger enhancement when using the Nantes
cross section. This discrepancy cannot be counterbalanced by other channels.
Thus the total kaon yield is much higher for the Nantes cross sections than for
the Giessen cross sections.
It should be noted that in the mean time the Giessen group has changed its cross
section parametrization and implemented the Tsushima cross section in a similar
way than the Nantes group. Nevertheless we will keep the names "Nantes" and
"Giessen" in the following pages in order to study the effect of different
parametrizations. 


\subsection{Influence of the uncertainties on p+A spectra}

\begin{figure}[hbt]
\epsfig{file=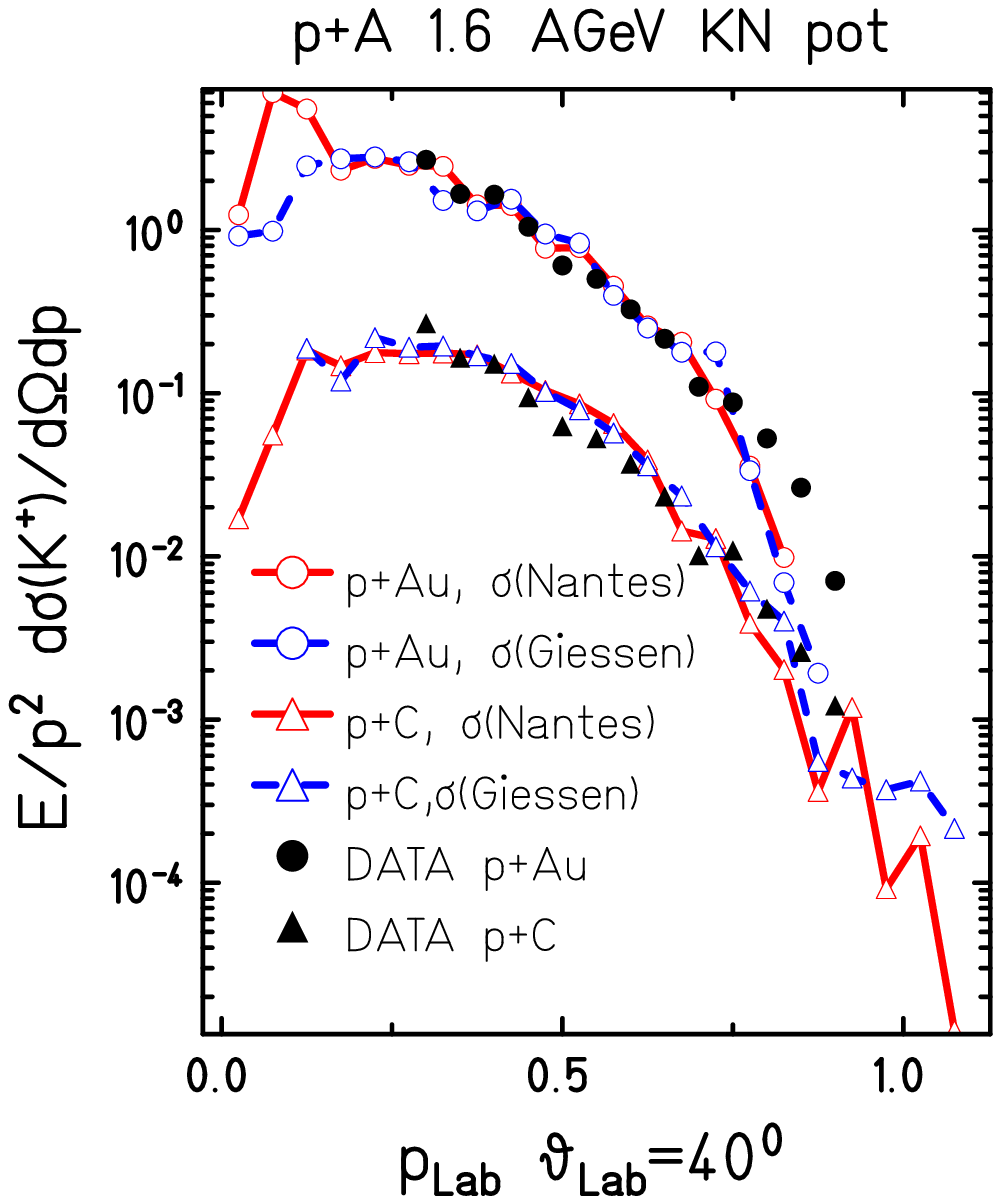,width=0.4\textwidth}
\epsfig{file=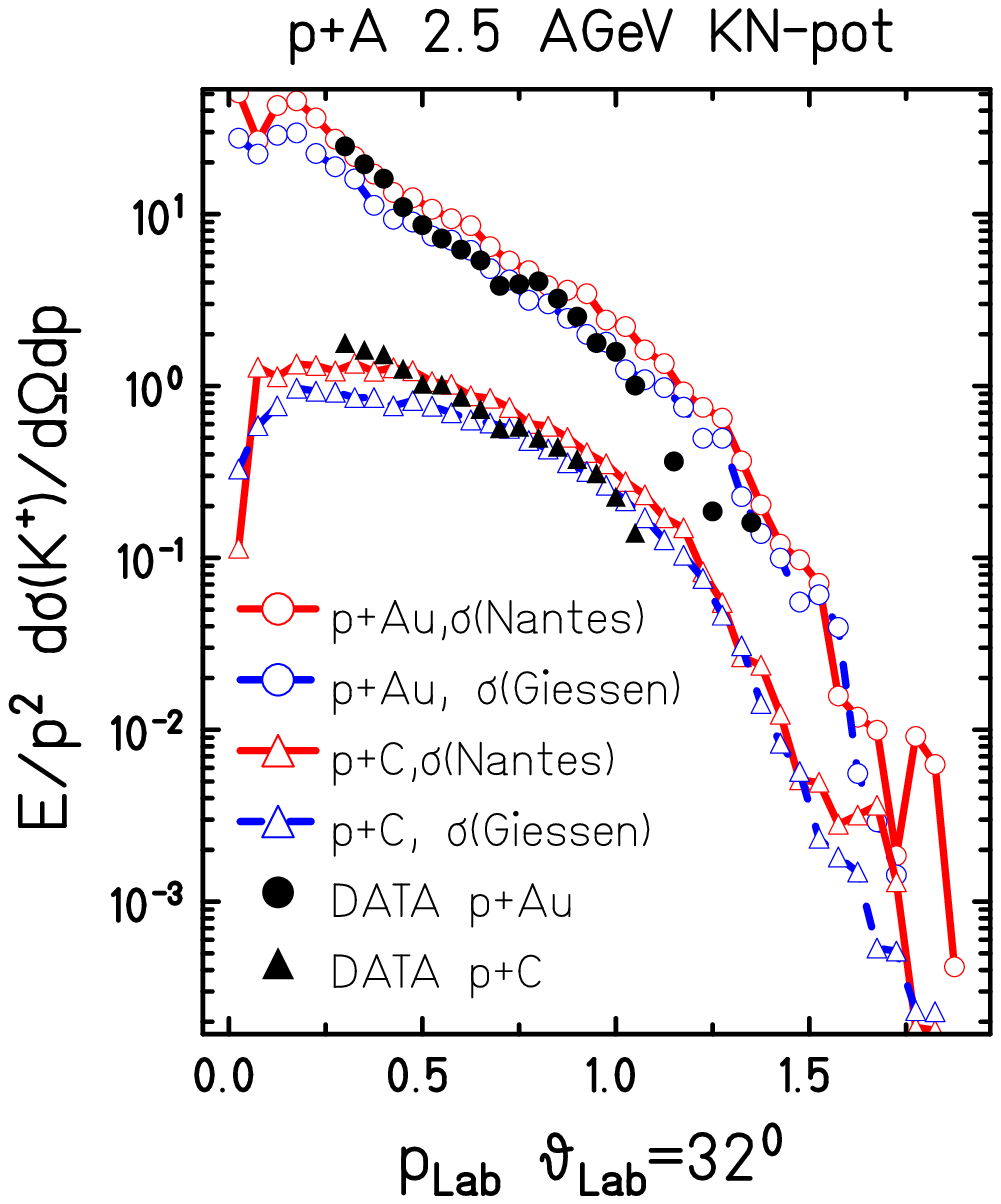,width=0.4\textwidth}
\caption{Influence of the different cross section parametrizations
 to  p+C and p+Au KaoS data at 1.6 and 2.5 GeV 
}
\Label{pA-xsect}
\end{figure}
Let us revisit the comparison of the p+A spectra. From
\figref{pA-xsect} we find a small influence of the cross sections
to the spectra when regarding low energies as we can also see
for forward angles on the \lhsref{pA-e}.
For this reaction the dominance of $NN$ channel yield on the other
hand a influence of the $NN\to N\Lambda K$ channel on the spectra. If we 
replace our parametrization (Sibirtsev, \cite{sibirtsev}) by that of 
Tsushima \cite{tsu} (\gml{} with diamonds), we find a visible 
lowering of the spectrum, which now is in better
agreement to the data.
However we should keep in mind that for other energies the Tsushima
cross section also shows deviation from the data (see \rhsref{kp-prod-channels}).
The parametrization of David \cite{david} (\bpl{} with triangles) overshoots
the spectra obtained with the Sibirtsev cross sections while the refitted
parametrization of David (cyan line with triangles) shows again a spectrum
comparable to that obtained with the Sibirtsev cross section.

However the effect is increasing with energy (see \rhsref{pA-xsect}
and \rhsref{pA-e}). 
At these higher energies
the effect of the optical potential is decreasing. Therefore
there might be a chance to see the potential at low energies.
However we have to remember the influence of the Fermi momentum.
A significant reduction of the Fermi momentum enhances the contribution
of the $N\Delta$ channel and thus the influence of its cross section
uncertainties.

\begin{figure}[hbt]
\epsfig{file=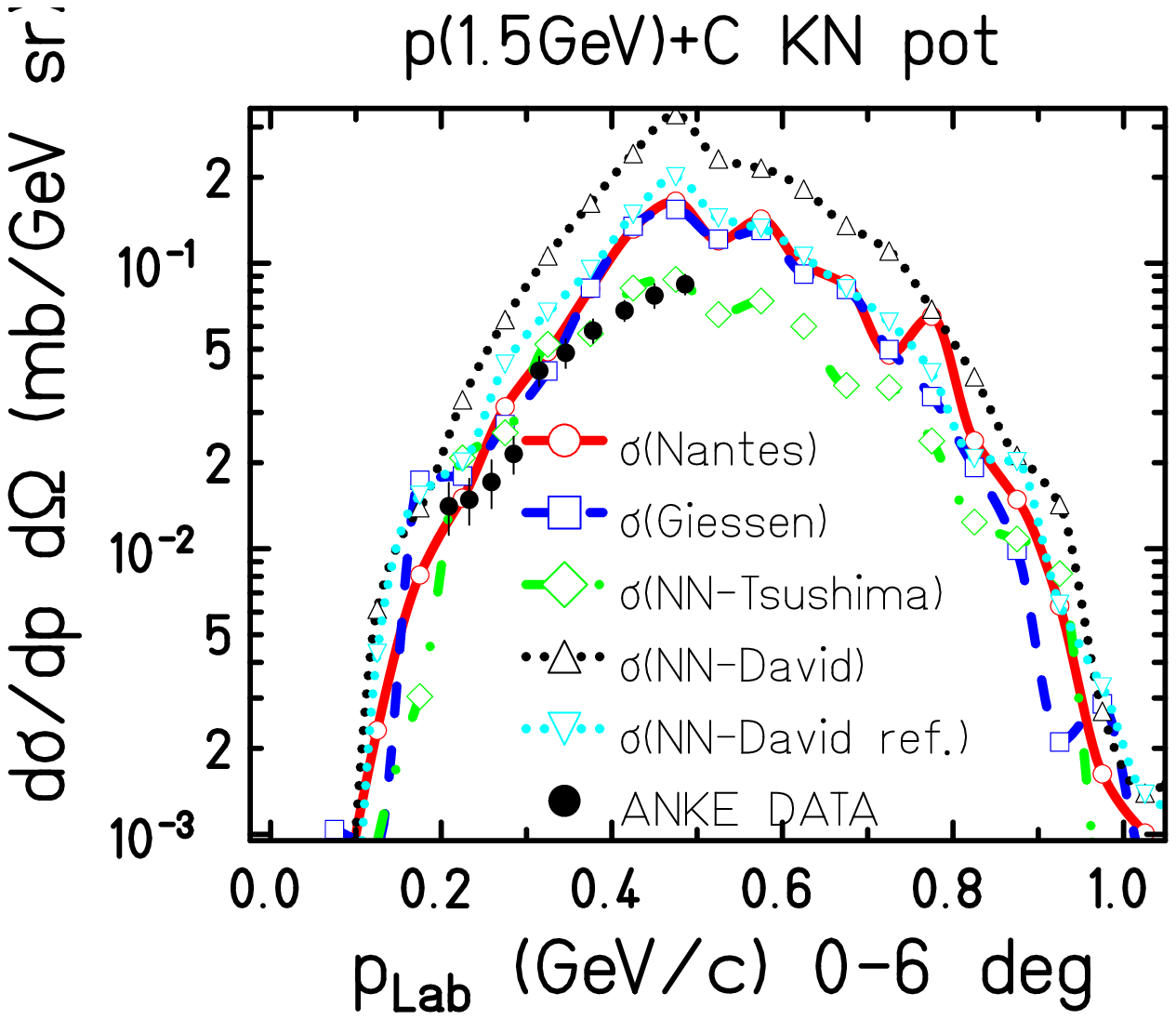,width=0.4\textwidth}
\epsfig{file=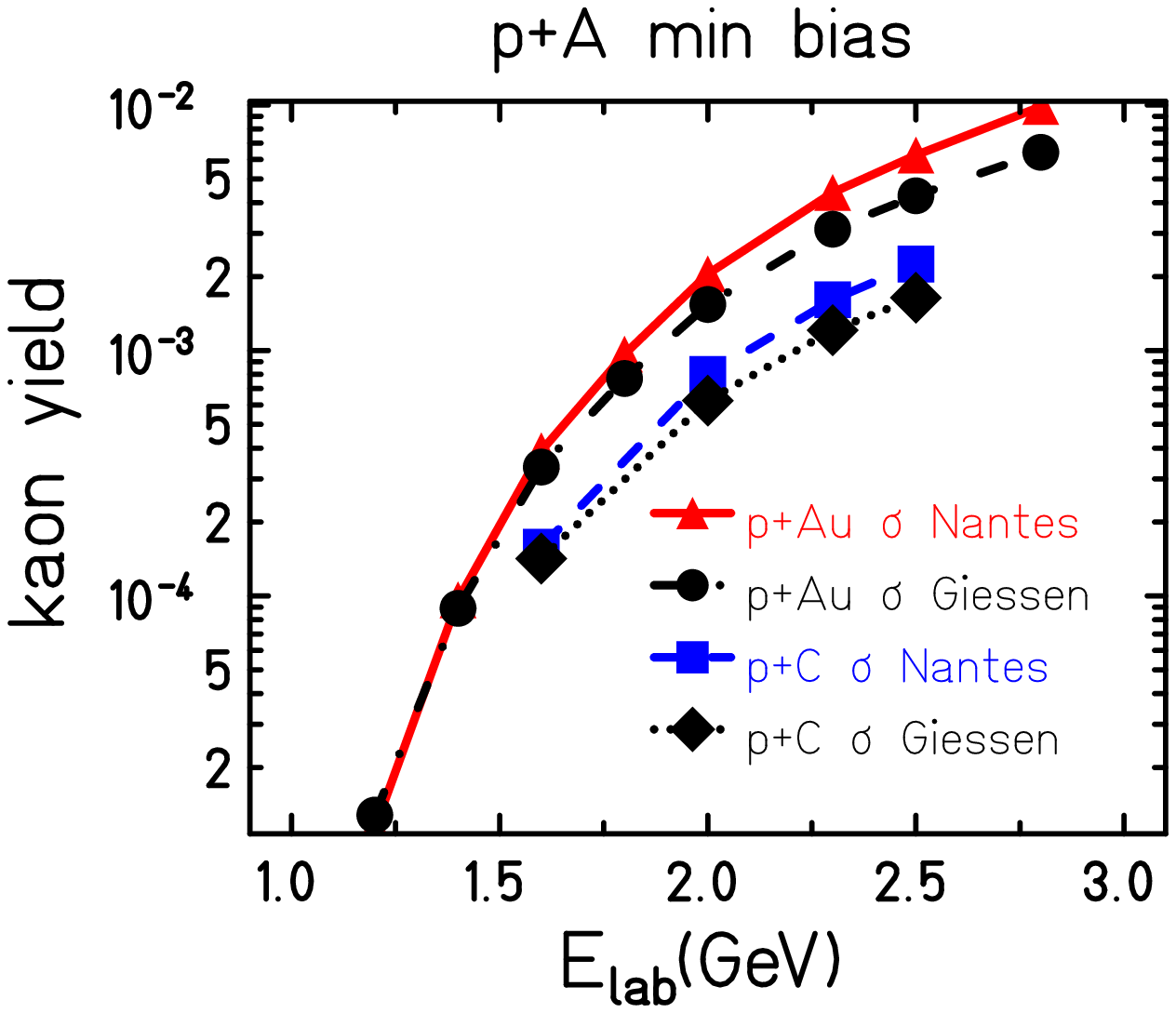,width=0.4\textwidth}
\caption{Influence of the different cross section parametrizations
 to  p+C ANKE data at 1.5 GeV (left) and excitation function of the
 kaon yield (right) 
}
\Label{pA-e}
\end{figure}
However we have to keep in mind that the rescattering
influences the spectra, especially at forward angles.

\subsection{Influence of the cross section uncertainties on A+A results}
As we have already seen in \figref{xsections-ND} 
there is a strong influence of the uncertainties of the cross section
in A+A collisions. This influence is most prominent where the contribution
of the $N\Delta$ channel is largest.

\begin{figure}
\epsfig{file=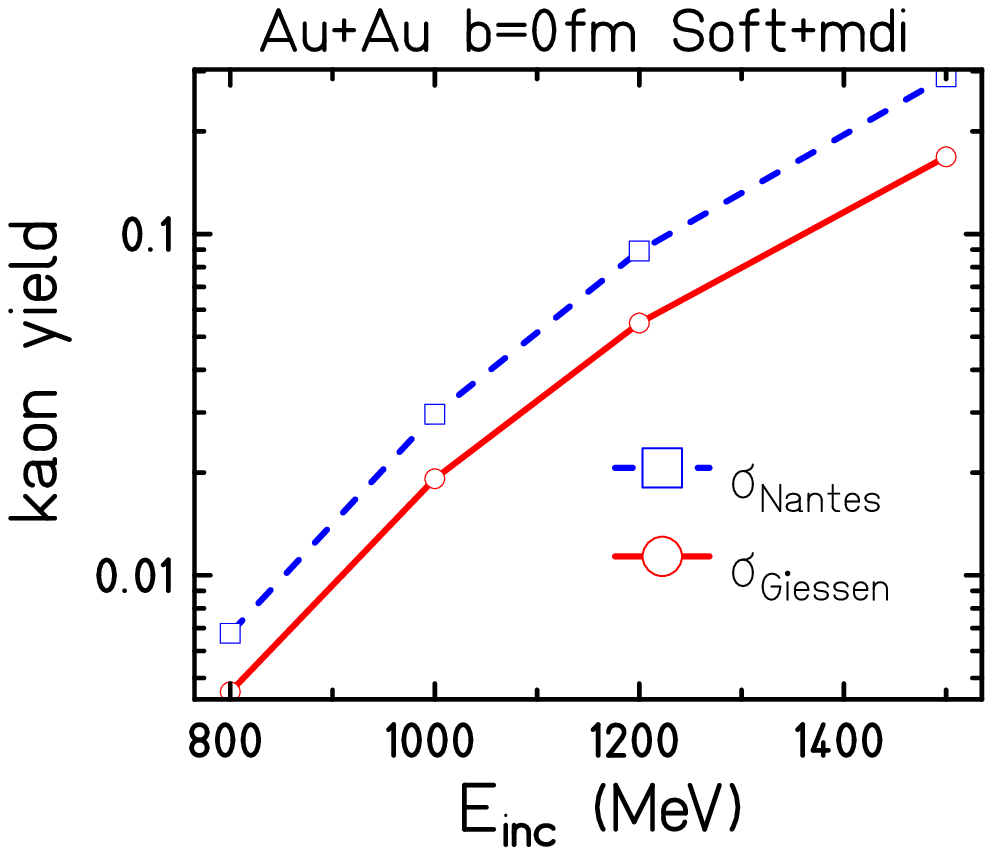,width=0.4\textwidth}
\epsfig{file=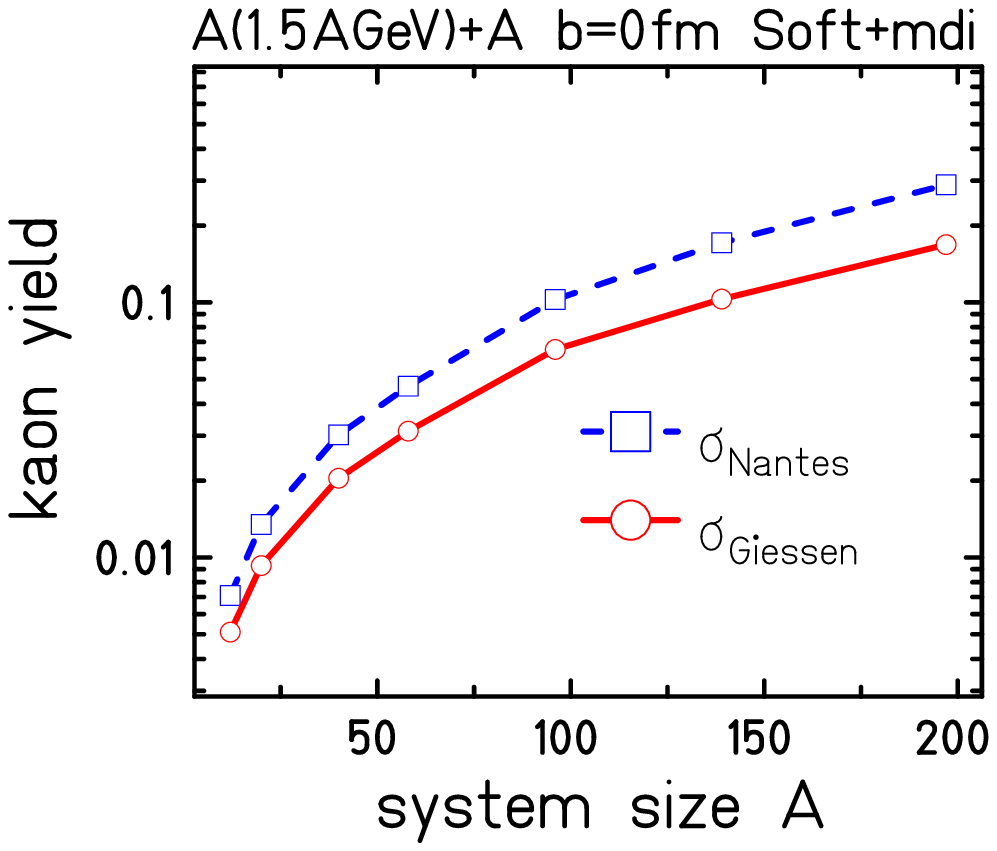,width=0.4\textwidth}
\caption{Excitation function (left) and system size dependence (right) 
 of the kaon yield with and without KN-potential. 
}
\Label{sig-yields}
\end{figure}
\Figref{sig-yields} shows the excitation function of Au+Au (\lhs) and
the system size dependence at 1.5 AGeV (\rhs) for both cross section
parametrizations. We see that all the time the calculations using the
Giessen cross sections (full lines) yield less kaon than the calculations
with the Nantes cross sections.
The difference is smaller for smaller systems which corresponds to a smaller
contribution of the $N\Delta$ channel. However the difference is also smaller
at lower energies although here the contribution of the $N\Delta$ channel
is higher. This effect is caused by the parametrizations themselves which
show a larger discrepancy for high $\sqrt{s}$. This high values of $\sqrt{s}$
are only available at high incident energies.

\begin{figure}[hbt]
\epsfig{file=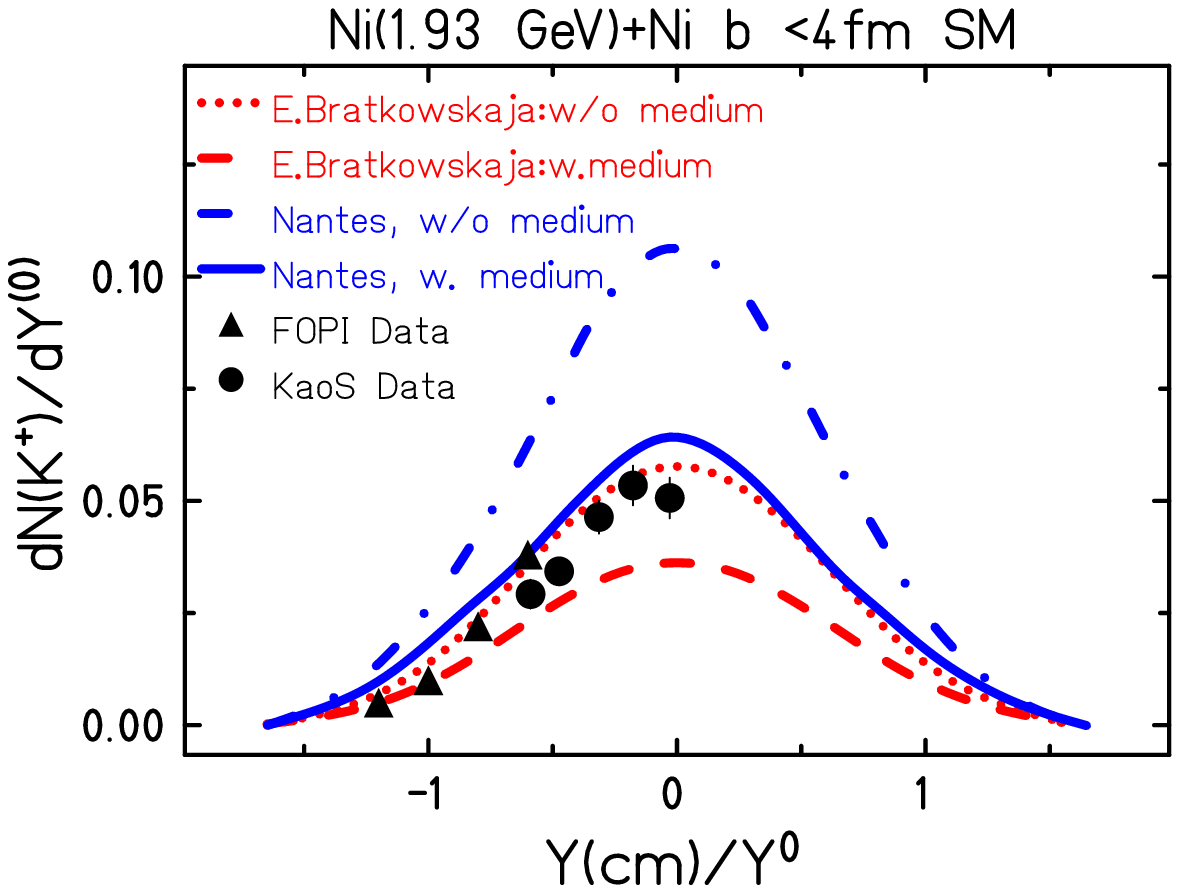,width=0.4\textwidth}
\epsfig{file=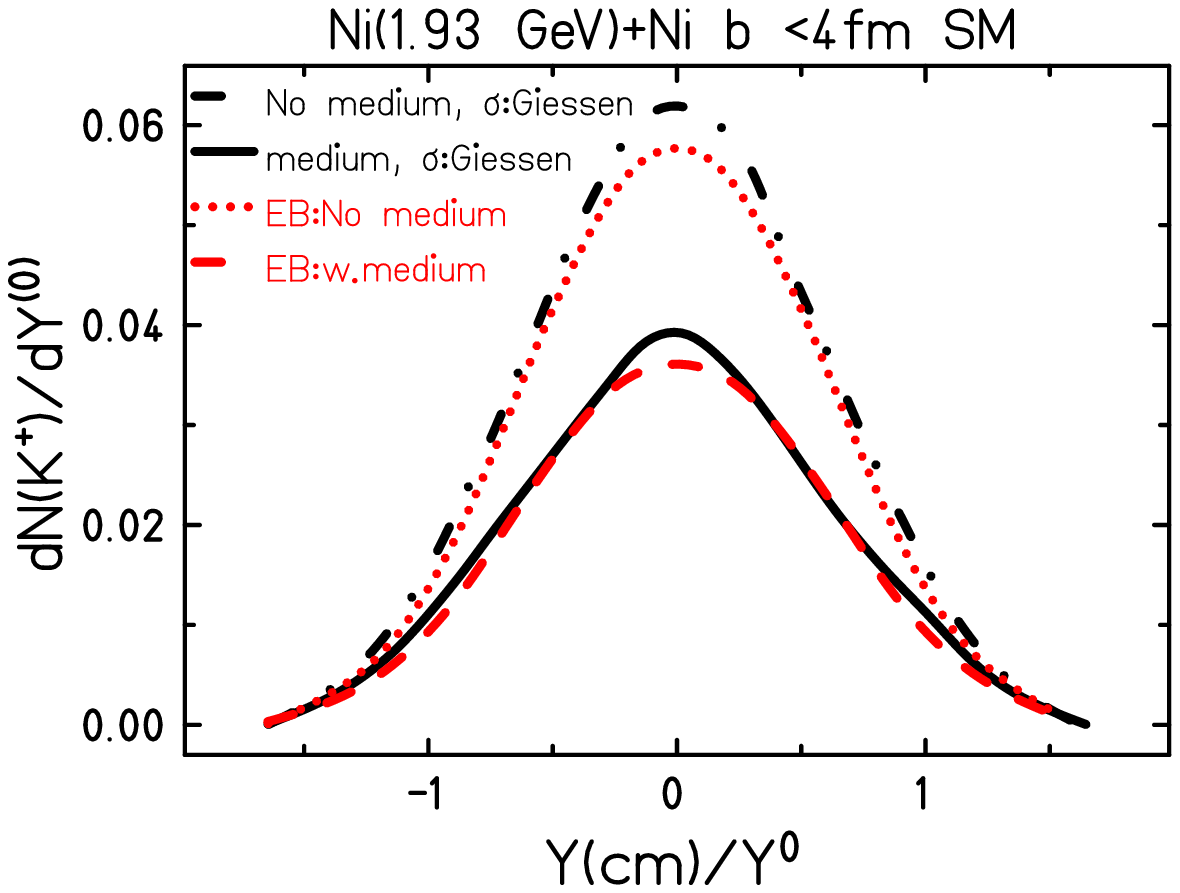,width=0.4\textwidth}
\caption{ Comparison of FOPI data to different microscopic calculations
and the contribution of the different cross section parametrizations.  
}
\Label{dndy-explain}
\end{figure}

\Figref{dndy-explain} shows on the \lhs\ a comparison \cite{sqm2001} 
of experimental data of
the FOPI \cite{rit} and KaoS collaboration \cite{menzel} with calculations 
from Giessen (RBUU) \cite{cassing}
and Nantes (IQMD) which puzzled the kaon community for a while.
While our calculations (blue lines) can reproduce the data by assuming 
a kaon optical potential (full blue line),
the Giessen results (red lines) could only explain the data when calculating 
without an optical potential (red dotted line). 
The \rhsref{dndy-explain} reveals the effect of the $N\Delta$
cross section to this puzzle. When using the same cross section parametrization
as the Giessen group, our IQMD calculations (black curves) reproduce nearly
the Giessen curves (red curves) as well with as without optical potential.

This example illustrates that the uncertainty on the cross section of inaccessible
channels still is an important constraint on the understanding of the
experimental results on kaon production.  
It should be repeated that the Giessen group has changed its cross section
parametrizations in the mean time and now also reproduces the Ni data
with an optical potential.

\subsection{Influence of delta lifetime and nucleon-nucleon cross sections
on the kaon production in A+A collisions}

Let us finally investigate the influence of the delta lifetime on 
the kaon yield in A+A collisions. Here the delta channel is much more important
as we have already seen in the discussion of the uncertainties of the
$N\Delta$ production cross section.

\begin{figure}
\epsfig{file=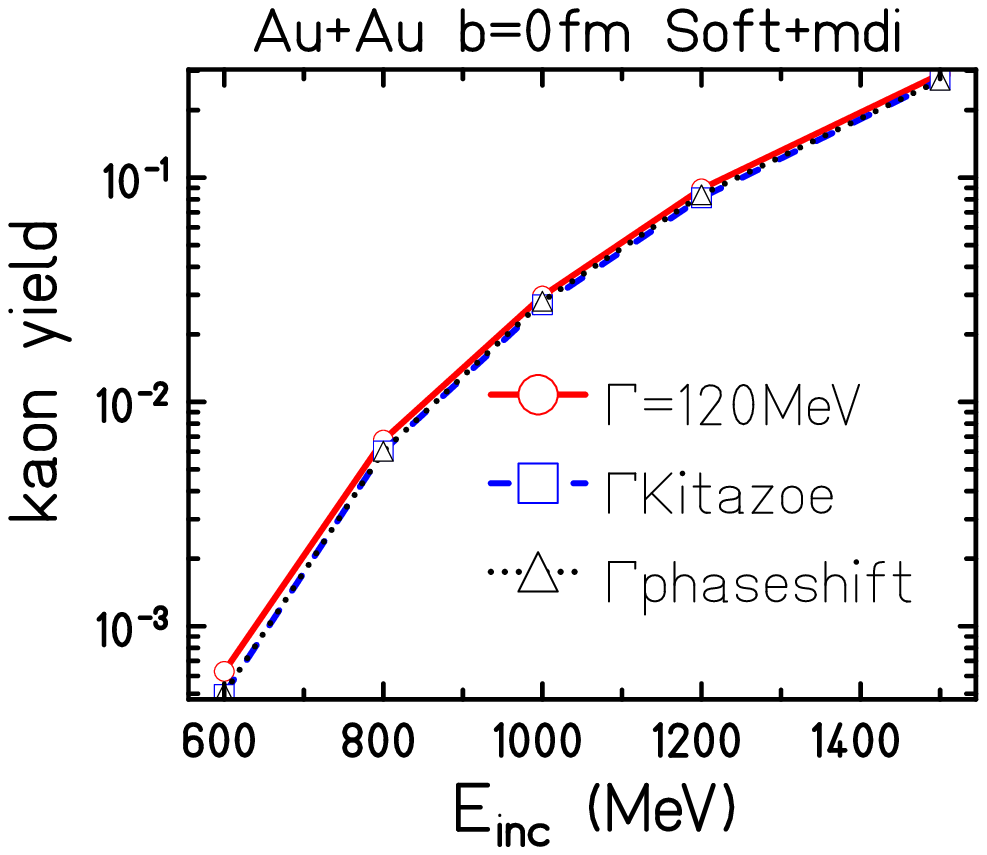,width=0.4\textwidth}
\epsfig{file=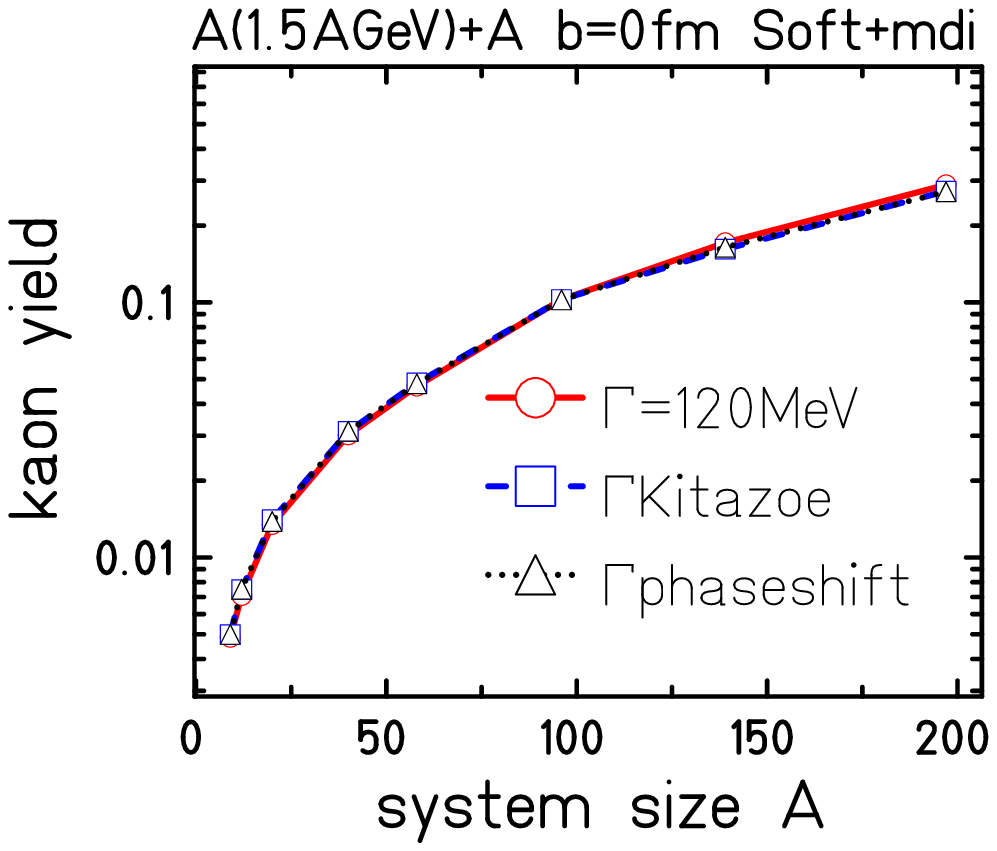,width=0.4\textwidth}
\caption{Excitation function (left) and system size dependence (right) 
 of the kaon yield with different delta lifetimes. 
}
\Label{nk-gamma-a}
\end{figure}

\Figref{nk-gamma-a} shows the excitation function (left) and system size dependence (right) 
 of the kaon yield with different delta lifetimes:
 a fixed lifetime of 120 MeV (\rfl), the Kitazoe parametrization (\bdl) and
 the phase-shift parametrization (\bpl). All parametrizations yield quite
 the same yields, the fixed width having slightly higher values. 
 This corresponds to the effect that Kitazoe and phase-shift parametrization do 
 not very much differ for high mass deltas, while the fixed value allows
 a longer lifetime of the high mass delta before it decays. High mass delta
 have a better chance for having sufficient energy for producing a kaon.

\begin{figure}
\epsfig{file=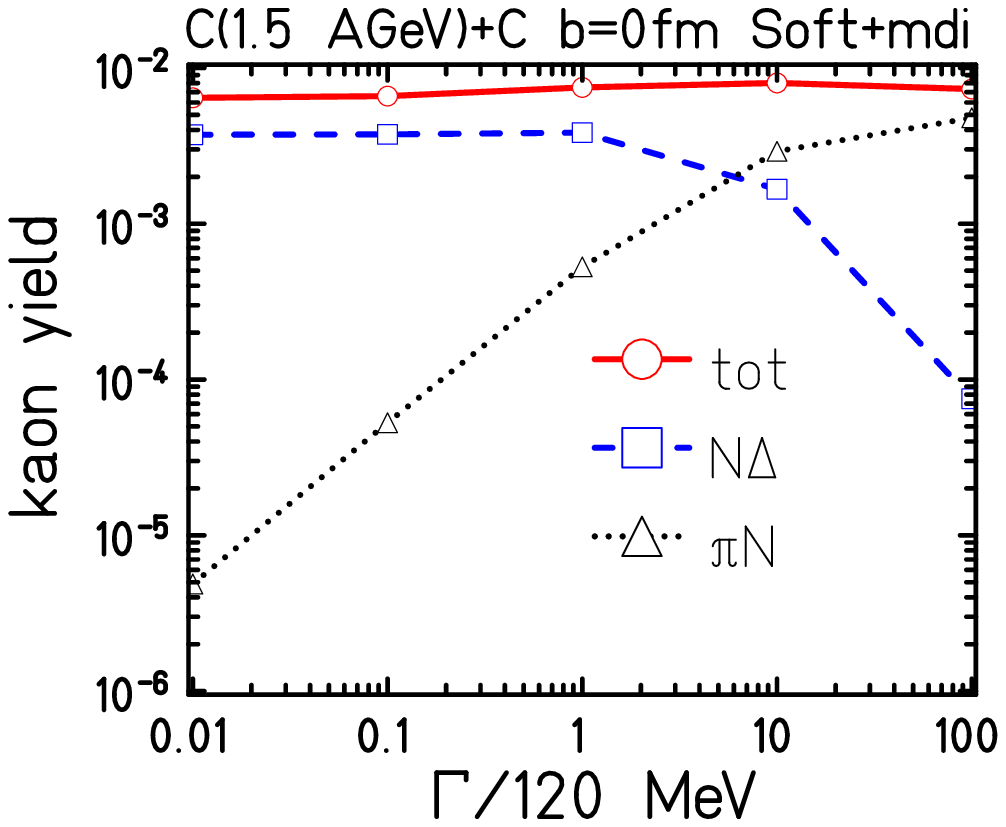,width=0.4\textwidth}
\epsfig{file=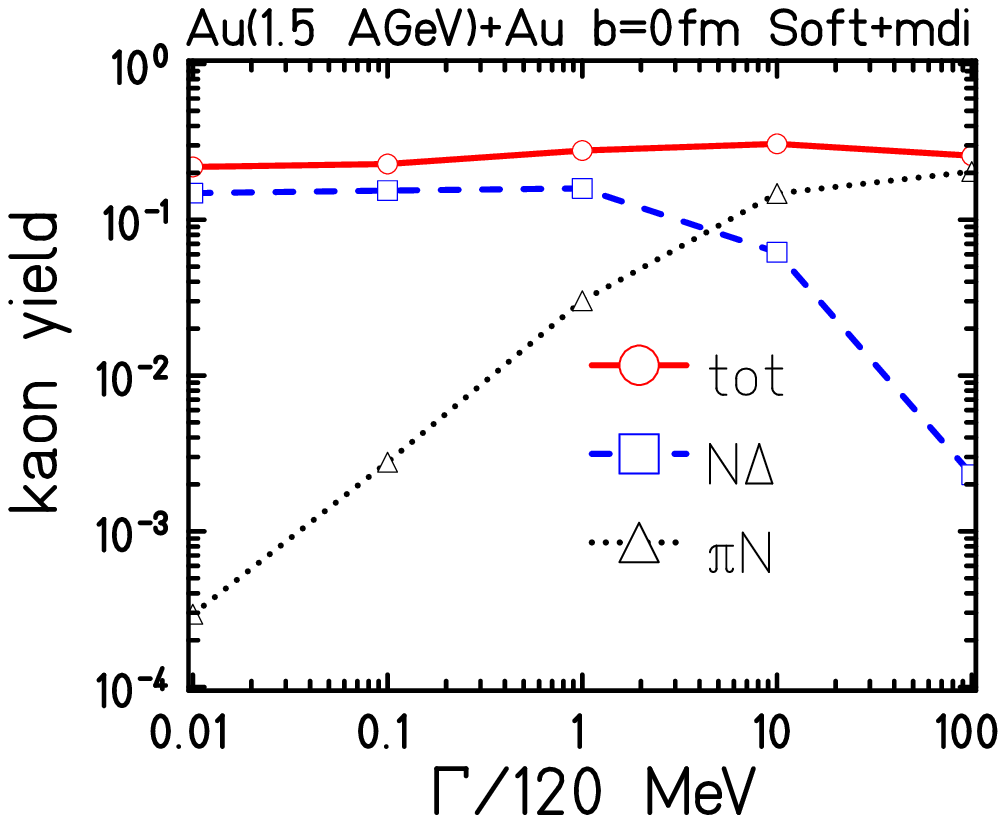,width=0.4\textwidth}
\caption{Dependence of the kaon yield on the delta lifetime for C+C (left)
and Au+Au (right) 
}
\Label{nk-gamma-f}
\end{figure}
A reason of the quite similar yields is the counterbalance of delta involved and
pion involved channels. \Figref{nk-gamma-f} shows the dependence of the kaon
yield (\rfl) on the (fixed) value of $\Gamma$ for C+C (left) and Au+Au (right).
Changing  $\Gamma$ to smaller values means enhancing the delta lifetime and thus
allowing the delta for a longer time to test collisions with nucleons. 
Enhancing  $\Gamma$ reduces the disponibility of the delta in the reaction zone
and thus reduces the kaon production via the $N\Delta$ channel (\bdl).
On the other hand, the pions are emitted earlier and thus can better produce
kaons via $\pi N$ collisions when the system is still dense. Both effects
compensate over a large scale in the delta lifetime.

\begin{figure}
\epsfig{file=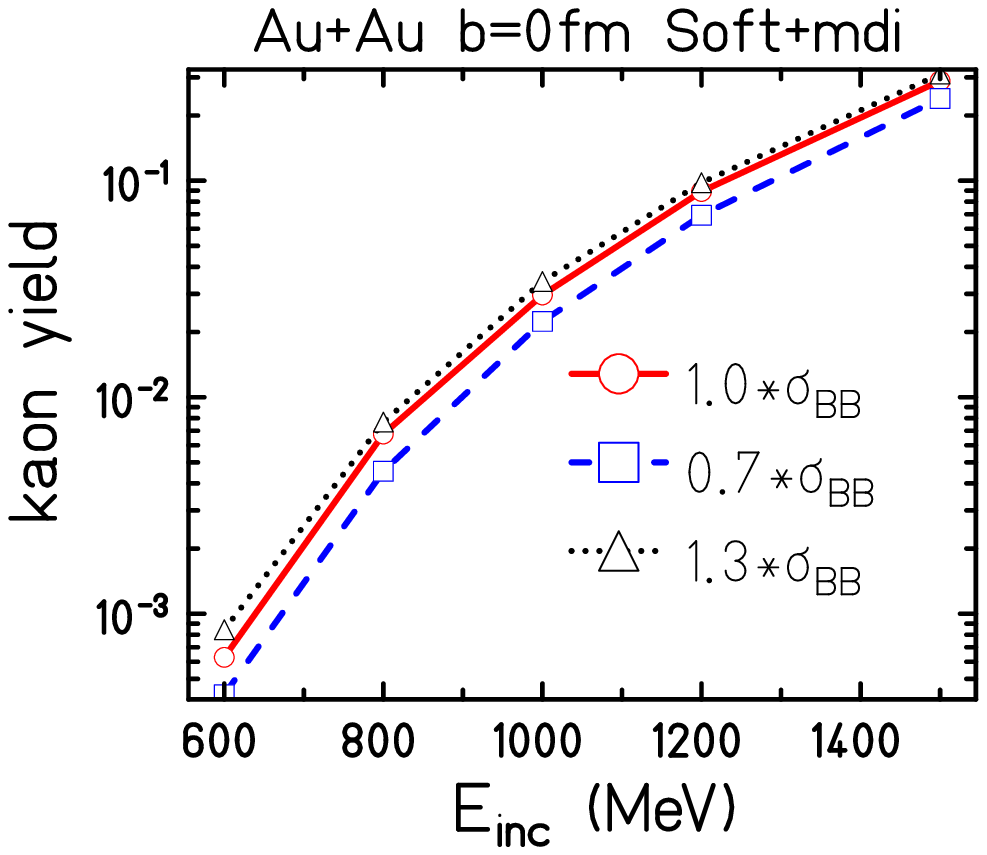,width=0.4\textwidth}
\epsfig{file=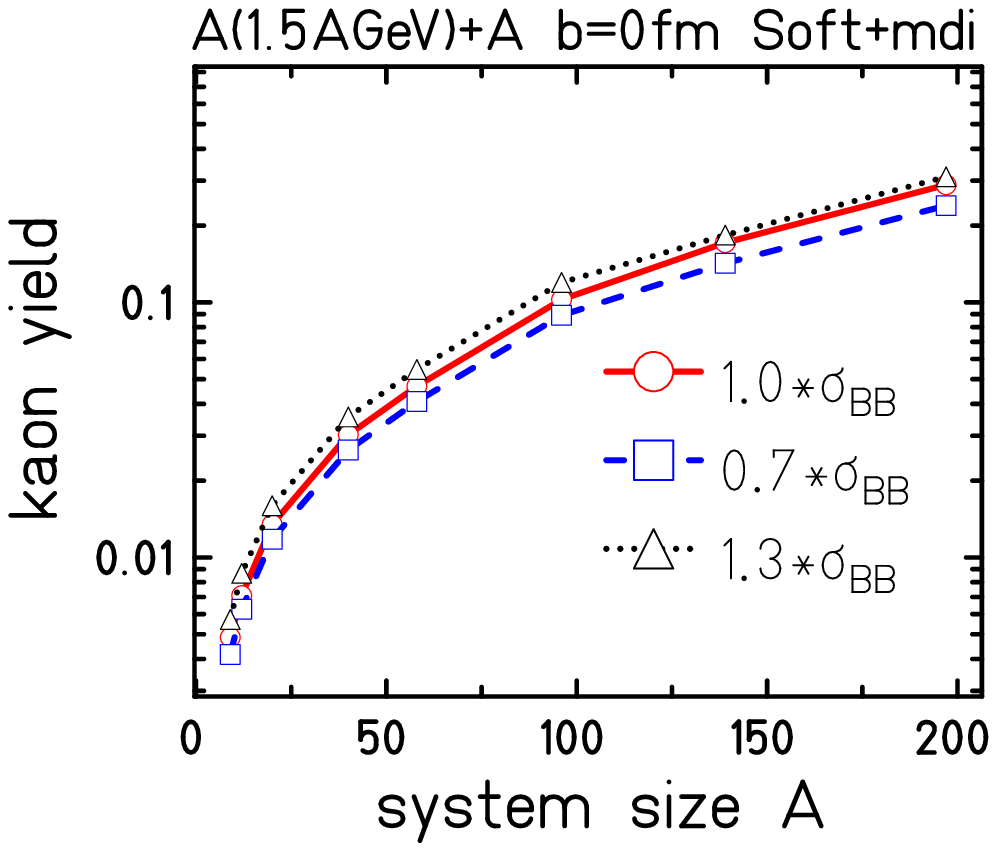,width=0.4\textwidth}
\caption{Excitation function (left) and system size dependence (right) 
 of the kaon yield with different factors to the nucleon-nucleon cross section. 
}
\Label{nk-sigmaBB}
\end{figure}
Let us finally study the influence of scaling the total 
nucleon-nucleon cross section but leaving the kaon production
cross section unchanged. \Figref{nk-sigmaBB}
shows the corresponding excitation function (left) and system size dependence
(right) for calculations with an unscaled cross section (\rfl), a cross section
reduced by a factor of 0.7 (\bdl) and a cross section enhanced by a factor of 
1.3 (\bpl). A reduced cross section (\bdl) yields less stopping and reduces the
number of particles equilibrated in the high density region. These particles
are the major producers of kaons. Therefore a reduction of the kaon number
is found. The opposite effect is seen for the enhancement of the nucleon-nucleon
cross section (\bpl) which enhances the kaon number. However, the effects are
still moderate since the nucleons have still a possibility to undergo a high
number of collisions. 
Finally it should be reminded that also  the properties of the
nucleus like a change in the Fermi momentum has a visible effect on the
kaon yield as it has already been shown in \figref{pot-yields}.

\section{Kaons and the nuclear eos}
As we have seen, there are two major problems on fixing the equation of state
by looking on the kaon multiplicities which are the influence of the
optical potential and the uncertainties of the cross sections.
We will soon discuss a method for resolving this problem but first do
some considerations on $K/A$ scaling.  

\subsection{K/A scaling}
As we have already seen, the number of kaons is depending on the size
of the participating system. In central collisions all particles of both 
nuclei are participating, in peripheral collision one has to describe
the number of participants by e.g. a geometrical model. 
Our calculations here are performed with b=0, thus the number of nucleons
$A$ in one nucleus is equal to the number of participating nucleons and half
the total participant number.  

If we scale the number of kaons by the system size $A$ and plot its dependence
on A in a double-logarithmic representation, a linear graph of a slope
$m$ would correspond to a scaling law of the type
\begin{equation}
N(K) \propto A^{1+m}
\end{equation}
For a flat curve $m=0$  the number of kaons is directly proportional to the 
total number of nucleons. We may assume that the kaons are produced in the
whole volume of the reactions. For negative $m=-1/3$ we may assume that the
kaons are only produced at the surface and scale with the number of nucleons
at the surface. For a positive number $m>0$ we may assume a collective 
production of the kaons, requiring a high density. This would be of interest
when searching for an effect of the nuclear equation of state, i.e. an observable
focusing on high densities.  

\begin{figure}[hbt]
\epsfig{file=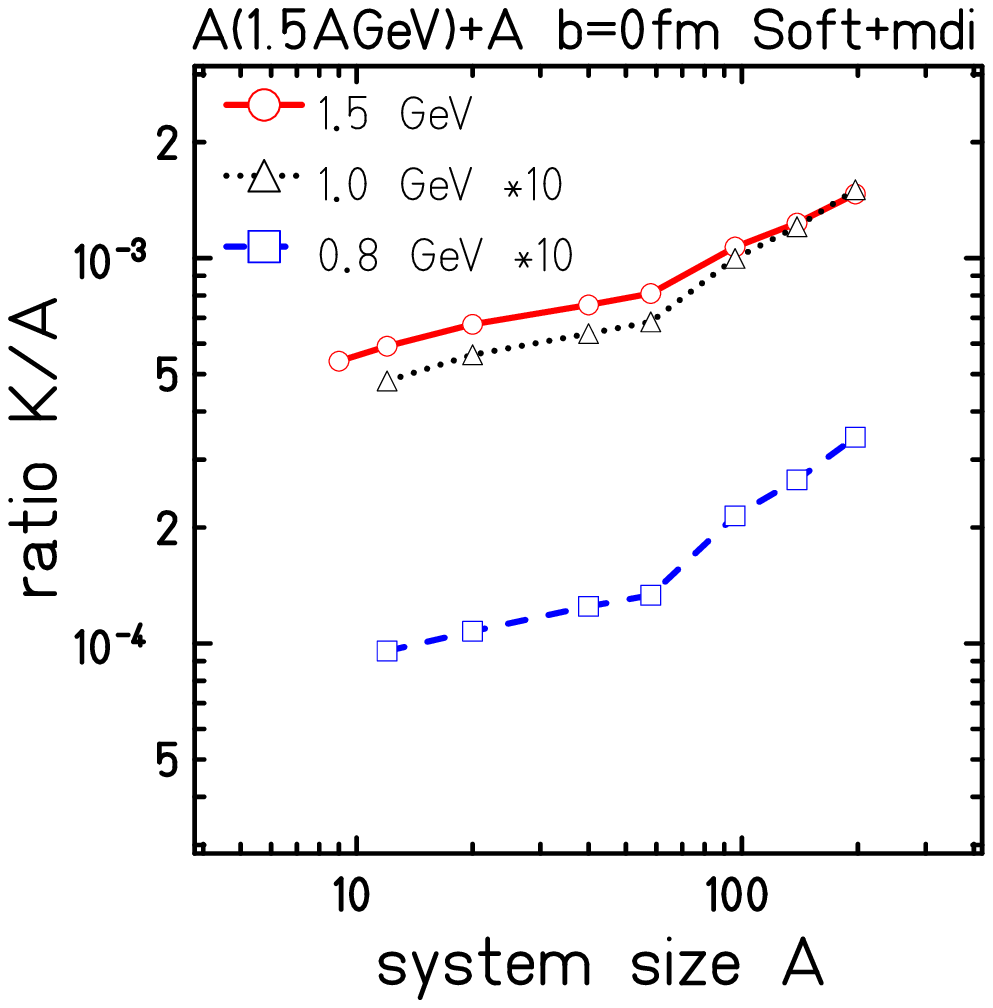,width=0.4\textwidth}
\epsfig{file=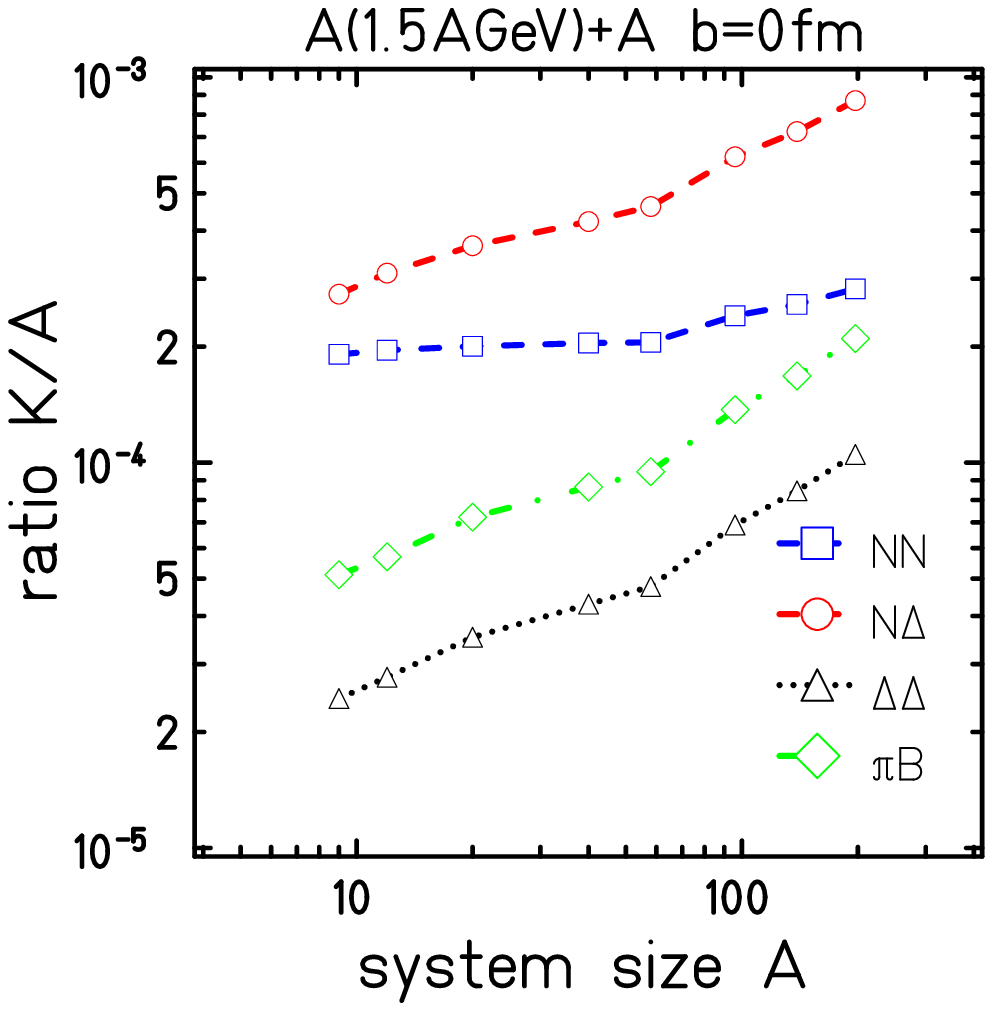,width=0.4\textwidth}
\caption{ System size dependence of the $K/A$ ratio in b=0 collisions
at different energies (left) and contributions of the different 
channels (right) in double-logarithmic representation.   
}
\Label{k-over-a}
\end{figure}

\Figref{k-over-a} shows the system size dependence of $K/A$ in a double
logarithmic representation. We see that the curves are not linear.
Thus, a direct $A^{1+m}$ scaling is not possible. Nevertheless, the curves
are continuously rising, giving significance for collective production.
The curve for 0.8 GeV at the \lhs\  (\bdl) increases stronger than the
curve for 1.5 GeV (\rfl), showing that the collectivity is more important
at low energies than at high energies. 
The \rhs\ shows the scaling for different channels at an energy of 1.5 AGeV.
The $NN$-channel (\bdl) is nearly flat. Here we are nearby the threshold so that
we could nearly produce a kaon in each collision of a projectile and a
target nucleon. The $N\Delta$ channel (\rfl) and $\Delta\Delta$ channel (\bpl)
show a stronger increase. This is in agreement with the previous findings
that these channels require many previous collisions and take place at high
densities. 

Let us now look on the centrality dependence of kaon production in collisions
at 1.5 AGeV. This analysis differs to the previous analysis performed at $b=0$fm 
where one could imagine that all nucleons were actively participating in the
collision. In less central collision we have to differentiate between
participating nucleons and spectators. This distinction is not unique and
depends on the criteria one may use to define a participant. Here we will
use the geometrical model which relates the impact parameter directly to 
a number of participants. The relation is shown on the \lhsref{apart}.

\begin{figure}[hbt]
\epsfig{file=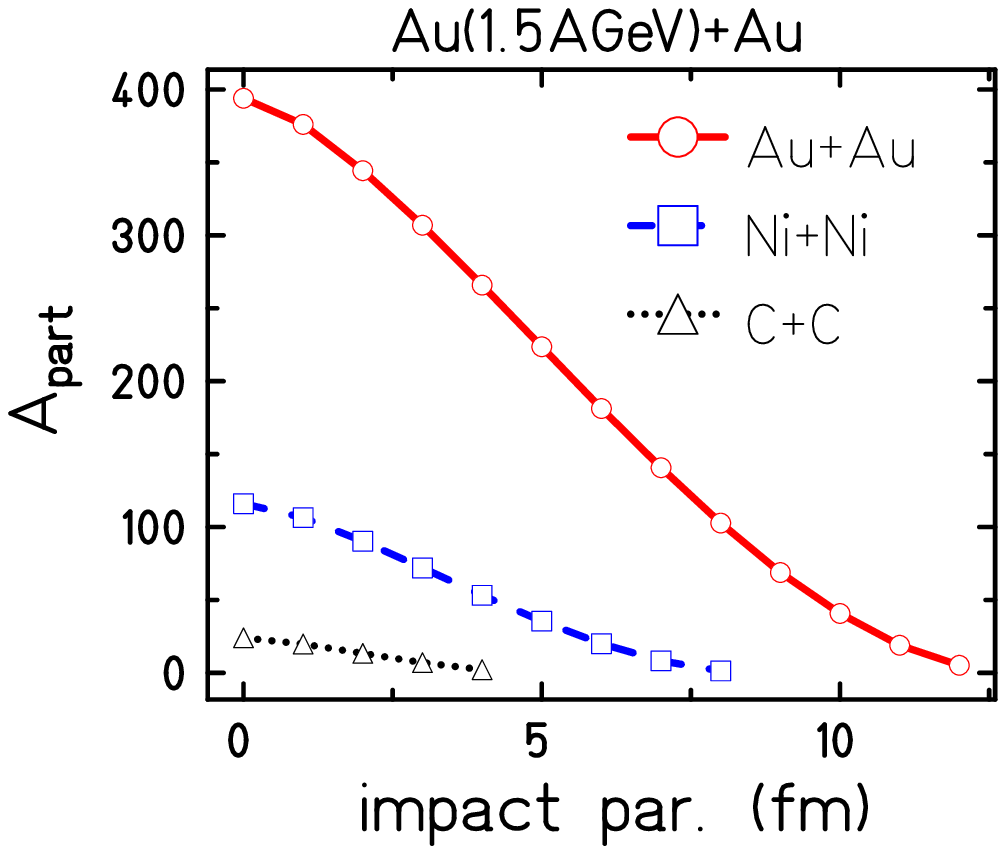,width=0.4\textwidth}
\epsfig{file=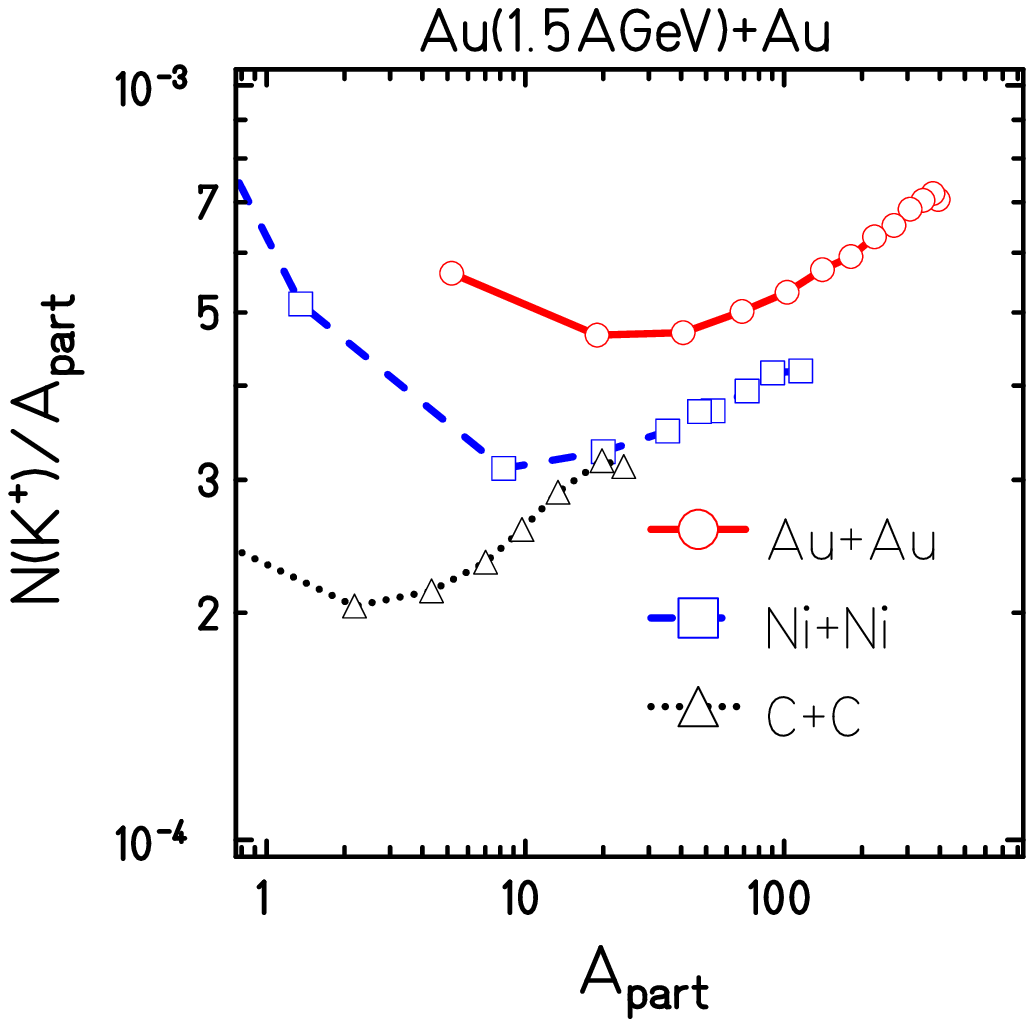,width=0.4\textwidth}
\caption{ Participant number as function of the impact parameter (left)
and kaon number per participant as function of the participant
number (right) for Au+Au, Ni+Ni and C+C 
}
\Label{apart}
\end{figure}

The \rhsref{apart} shows the dependence of the ratio $N(K^+)/A_{part}$ as a
function of $A_{part}$ for the systems Au+Au (\rfl), Ni+Ni (\bdl) and
C+C (\bpl) in double logarithmic representation. 
We see that all curves show a positive slope for central collisions
but a negative  one for very peripheral collisions. This effect might
be due to surface effects and problems in defining participants at
very central collisions. We will therefore skip the peripheral
collisions in the following.

We also find  that the slopes of C+C could be continued to Au+Au.
However, the curves of Ni+Ni ly  a little bit below which corresponds
to the edge structure seen on the \lhsref{k-over-a} where this system
already showed some different behavior.
We therefore plot the values of different systems into one graph a try 
to fit it with one global slope. In order to avoid problems with peripheral
collisions we require events with a participant number that is at least 
a quarter of the maximum participant number 
$A_{part}>0.25A_{part}^{max} =0.5A$. 

\begin{figure}[hbt]
\epsfig{file=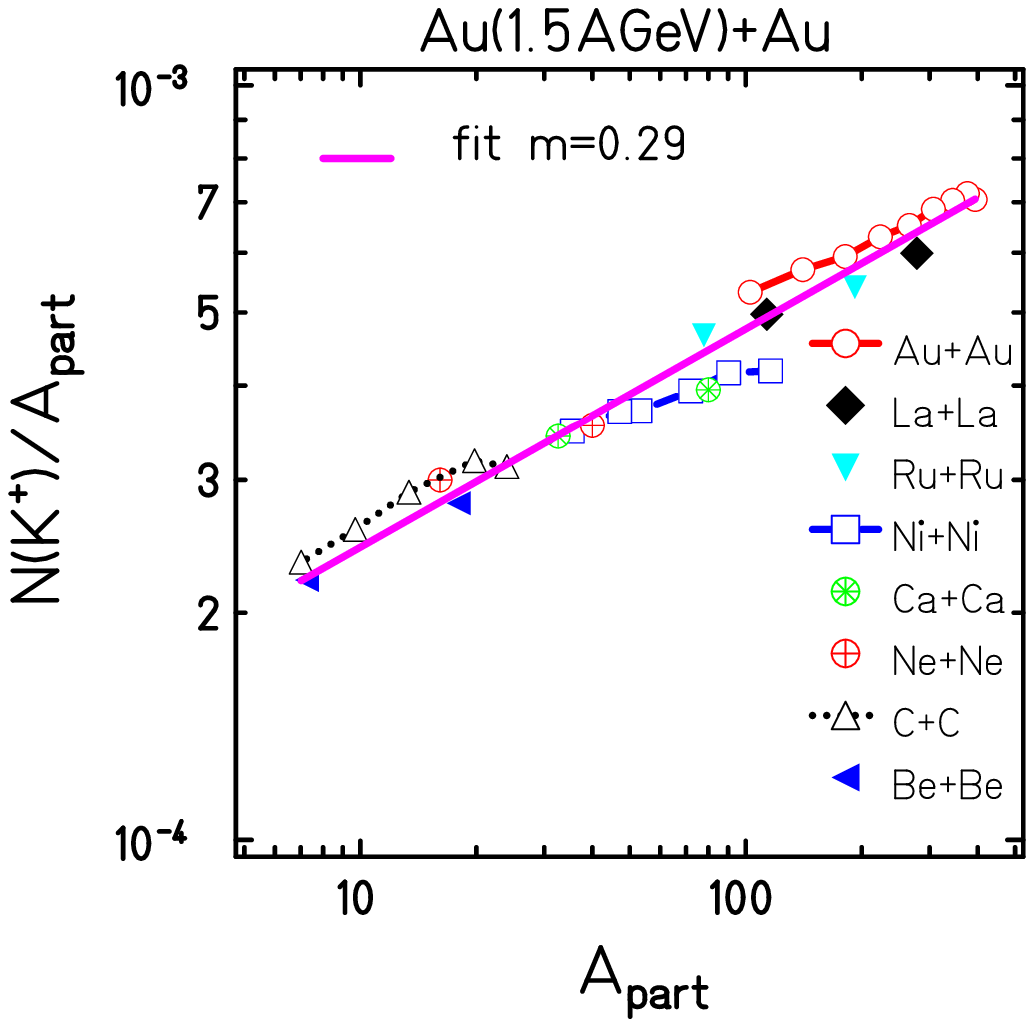,width=0.4\textwidth}
\epsfig{file=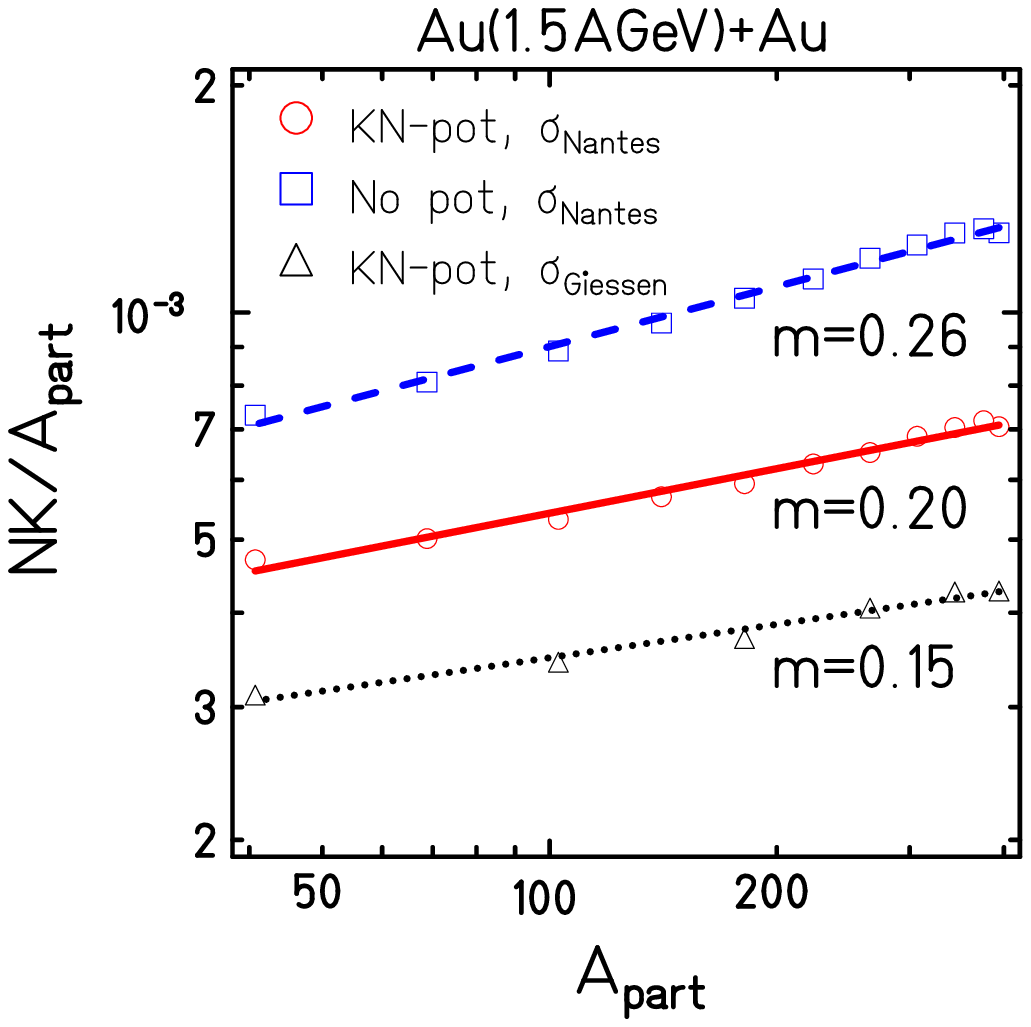,width=0.4\textwidth}
\caption{ Participant number dependence of the kaon yield for 
non-peripheral collisions of different systems (left) and influence
of potential and production cross section (right)   
}
\Label{k-a-p}
\end{figure}
\Figref{k-a-p} shows on the \lhs\  that all the kaon numbers of different
events could be roughly fitted by one function with
$N(K^+)\propto A^{1.29}$. This clearly signifies the existence of
collective effects for the kaon production.
  
This slope may depend on different ingredients as we can conclude from the
\rhsref{k-a-p} where we plot the centrality dependence in Au+Au for calculations
with different options for the KN optical potential and the production
cross section. If we use the Nantes cross section parametrization and
switch off the potential (\bdl) we enhance the slope parameter $m$ from
0.20 (with potential, \rfl) to 0.26. This corresponds to the effect that the
optical potential penalizes especially at high densities, which are most 
important for kaon production. When changing the cross section parametrization
to the Giessen type (\bpl) we reduce $m$ to 0.15. Here we reduce the
contribution of the $N\Delta$ channel which has a stronger slope than the
$NN$ channel as we have seen on the \rhsref{k-over-a}.

\begin{figure}[hbt]
\epsfig{file=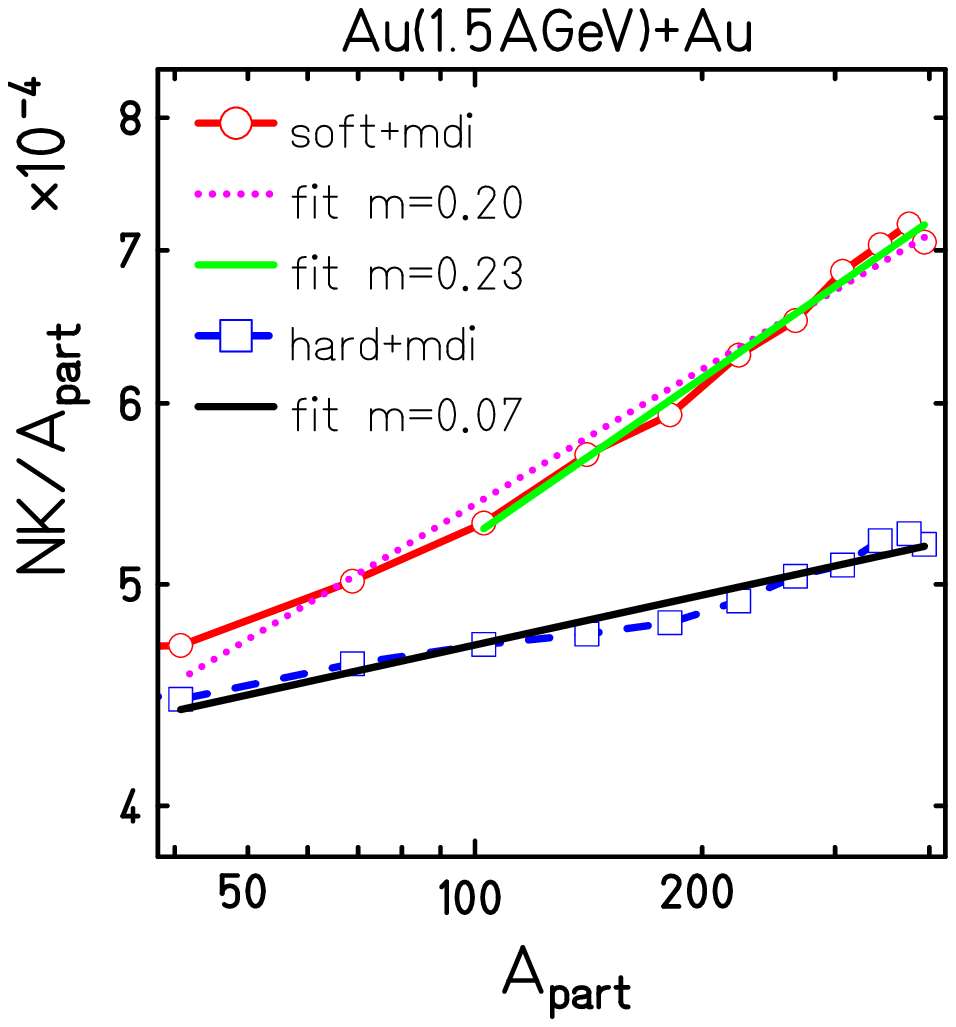,width=0.4\textwidth}
\epsfig{file=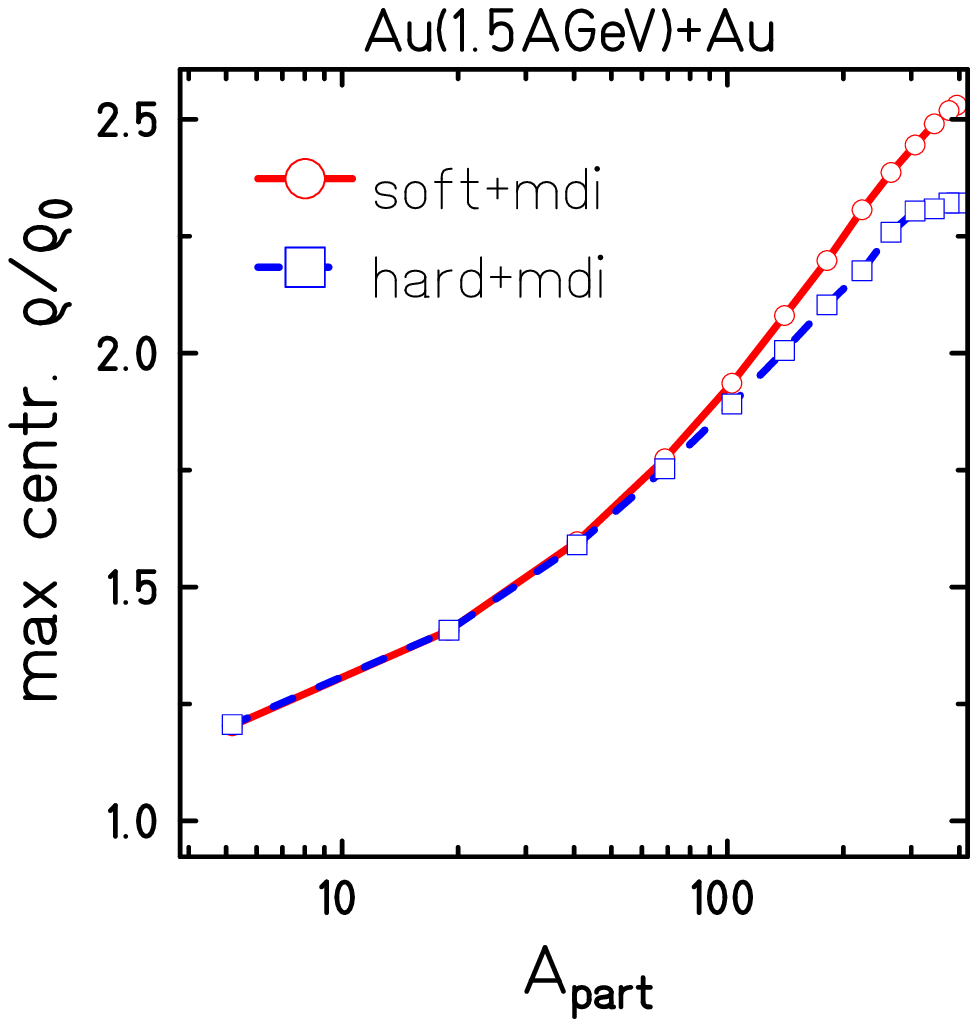,width=0.4\textwidth}
\caption{ Participant number dependence of the kaon yield  
(left) and of the maximum central density (right) 
for a hard and a soft equation of state
}
\Label{k-a-b}
\end{figure}
Finally the equation of state has also a strong effect on the slope parameter
$m$ as we can see in \figref{k-a-b} where we compare calculations with a hard
(\bdl) and a soft (\rfl) equation of state. The hard equation of state has
a small value of $m=0.07$ while the soft equation of state has a higher  
value $m=0.20$  which still increase to $m=0.23$ when we apply the condition
$A_{part}>0.25A_{part}^{max} =0.5A$. 
At small participant numbers the yields of hard and soft equation of state
become similar. For peripheral collisions both equations of state yield
about the same maximum densities while for central collisions the difference
of the maximum density increases which causes a stronger rise of the kaon number.

Therefore, an analysis of the kaon data toward the dependence of the nuclear 
equation of state seems to be interesting.


\subsection{The effect of the nuclear equation of state on the kaon yields}

Let us now look at the effect of the nuclear equation of state, when the
optical potential is included. We already saw that this optical potential
penalizes the kaon production especially at high densities and thus 
counterbalances (at least in part) the effect of the higher density reached
in a soft equation of state.

\begin{figure}
\epsfig{file=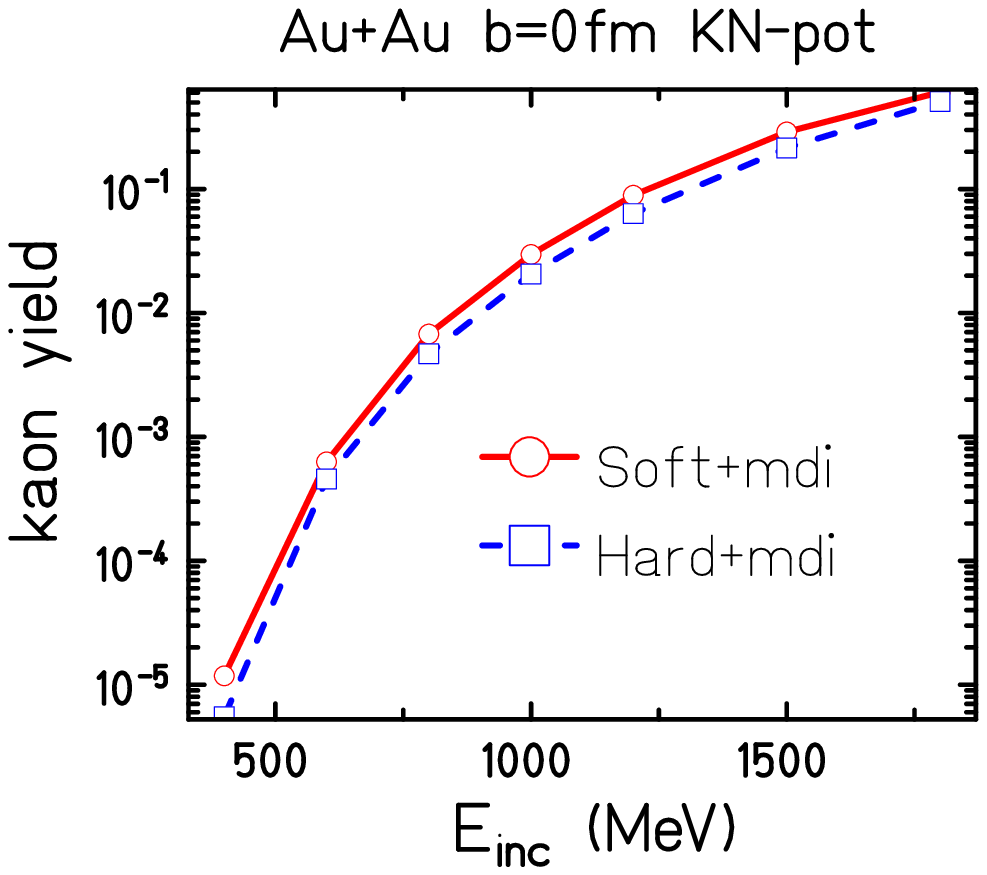,width=0.4\textwidth}
\epsfig{file=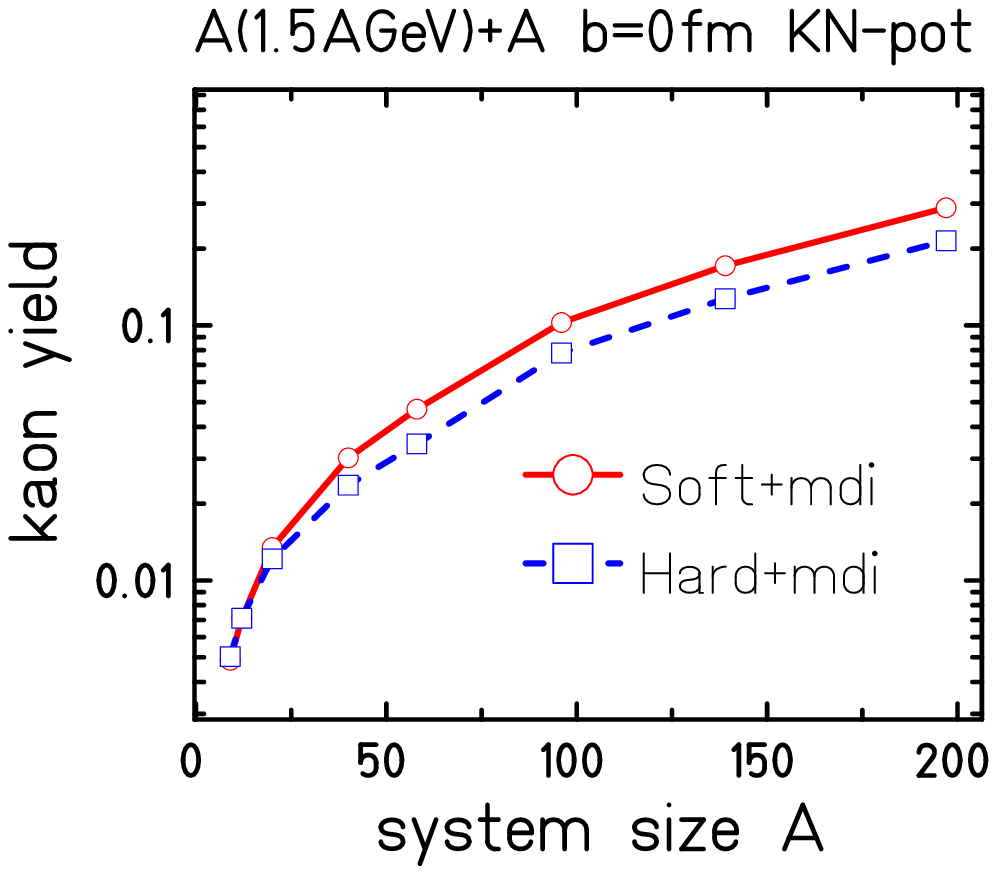,width=0.4\textwidth}
\caption{ Excitation function and system size dependence of the
kaon production with KN potential for a hard and a soft potential.  
}
\Label{k-eos}
\end{figure}

\Figref{k-eos} compares the effect of the nuclear equation of state 
and the kaon yield. We see a slight enhancement of the kaon number when a soft
equation of state is used (\rfl). This enhancement can be found at all energies
analyzed here but should vanish at energies far above the threshold.
The effect of the equation of state is strongest for big systems and
vanishes at small systems like C+C.

In order to explain this effect we analyze the influence of the
compressibility at lower incident energies where these effects have been
found to be especially sensitive. 

\begin{figure}
\epsfig{file=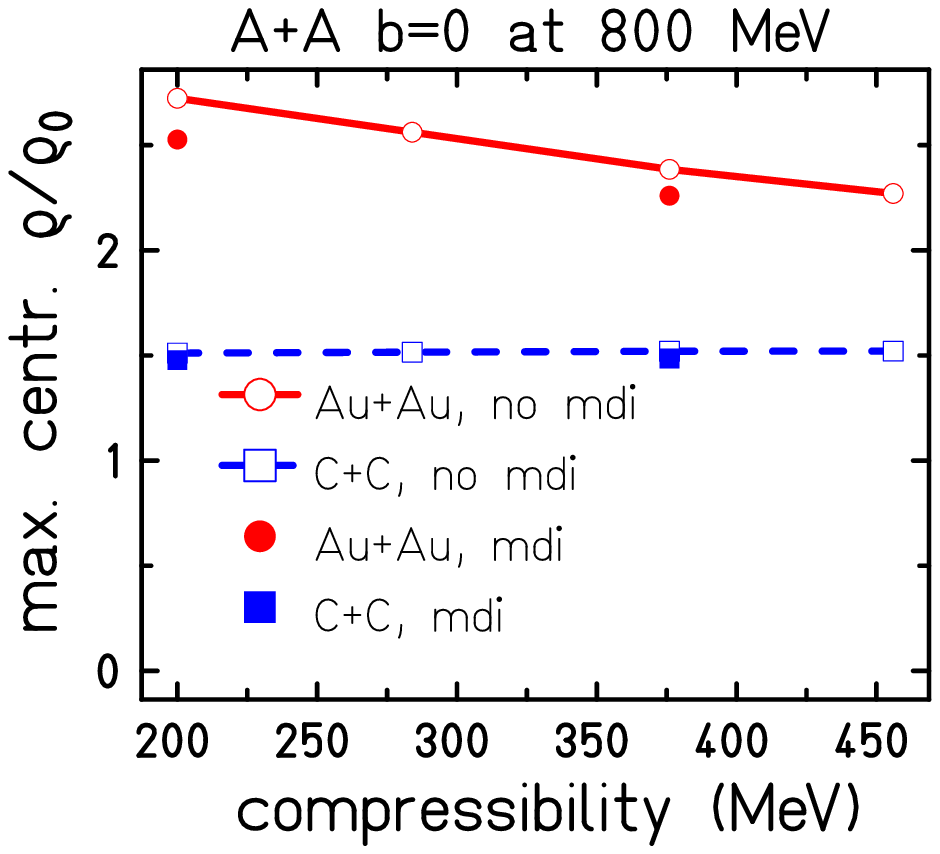,width=0.4\textwidth}
\epsfig{file=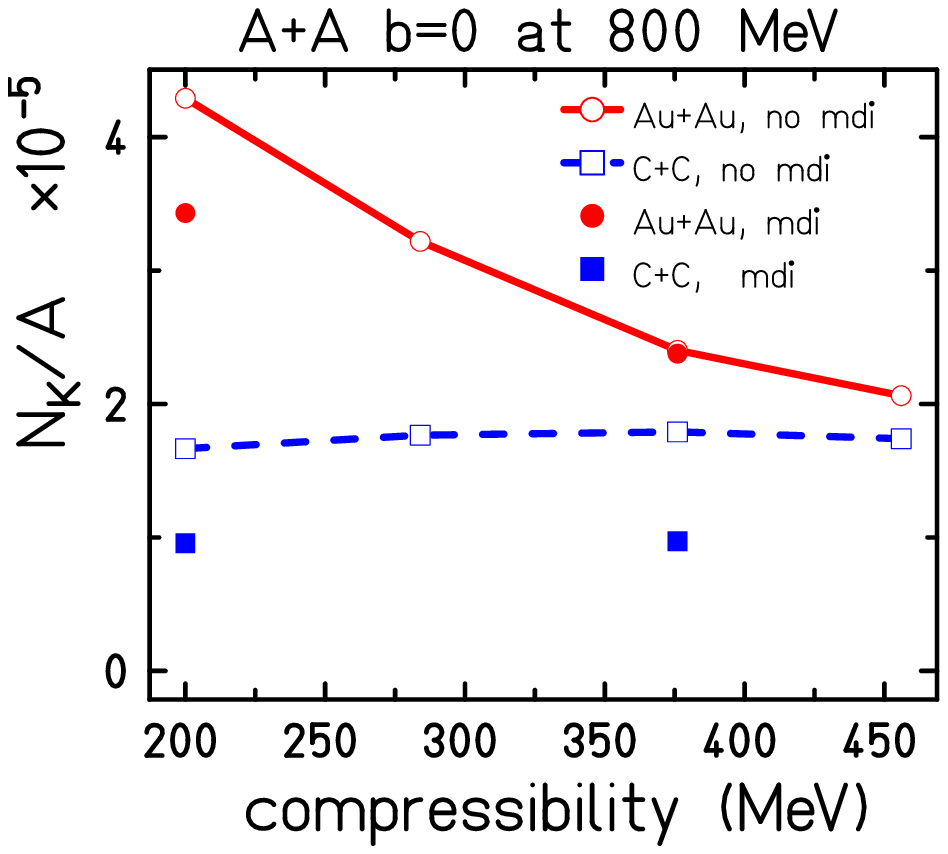,width=0.4\textwidth}
\caption{The maximum central density (left) and the kaon yield
normalized by the system size $N_K/A$ (right) as a function of the
compressibility for Au+Au and C+C reaction at 800 MeV incident energy  
}
\Label{au08-eos}
\end{figure}

The \lhsref{au08-eos} shows the maximum density reached in a central reaction
in calculations of Au+Au (\rfl) and C+C (\bdl) collisions at 800 MeV incident energy.
A smaller value of the compressibility corresponds to a softer equation of
state. For the heavy gold system we see a visible decrease of the density 
with increasing compressibility while for the light carbon system there
is no effect. Furthermore the absolute values of the maximum compression are
much smaller than the corresponding values of the gold system.
The carbon system is too small for allowing enough compression to built up a
high density region. Furthermore we see that the calculation without
momentum dependent interactions (open symbols with curves) yield higher
compression than the calculations including momentum dependence (full
symbols). Momentum dependent interactions yield additional repulsive forces
which add to the repulsion caused by the compressed matter.

The \rhsref{au08-eos} shows that the kaon yield (normalized by the system size) 
$N_K/A$ shows a similar behavior. For the heavy system the kaon yield
drops with increasing compressibility  (and decreasing maximum density)
while for the light system the yield stays constant.   
The effects is weaker than shown in \figref{eos-dns} since the kaon potential
which is applied for the calculations in \figref{au08-eos} penalizes the production
at higher densities and thus reduces the effect. Nevertheless a net effect
is remaining. It should also be stated that the application of momentum 
dependent forces (mdi, full symbols) reduces the kaon yield additionally.
This is due to the additional repulsion caused by the momentum dependent
potentials. The effect is weaker for the carbon system where less 
compression is obtained.

\begin{figure}
\epsfig{file=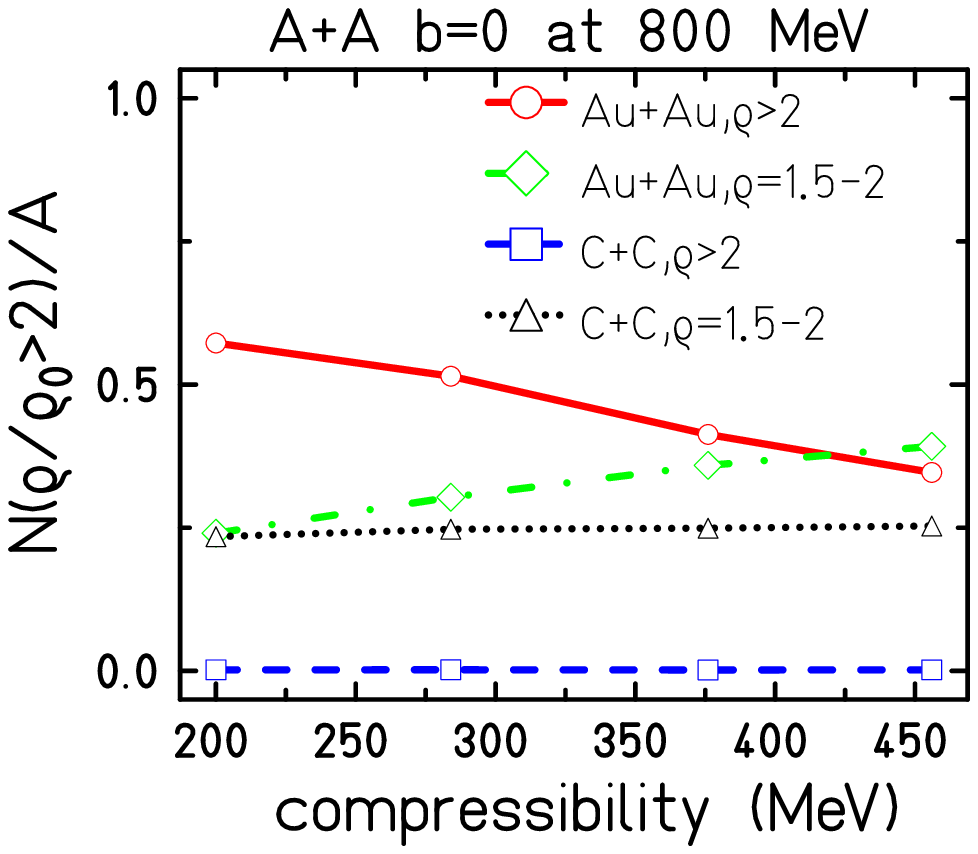,width=0.4\textwidth}
\epsfig{file=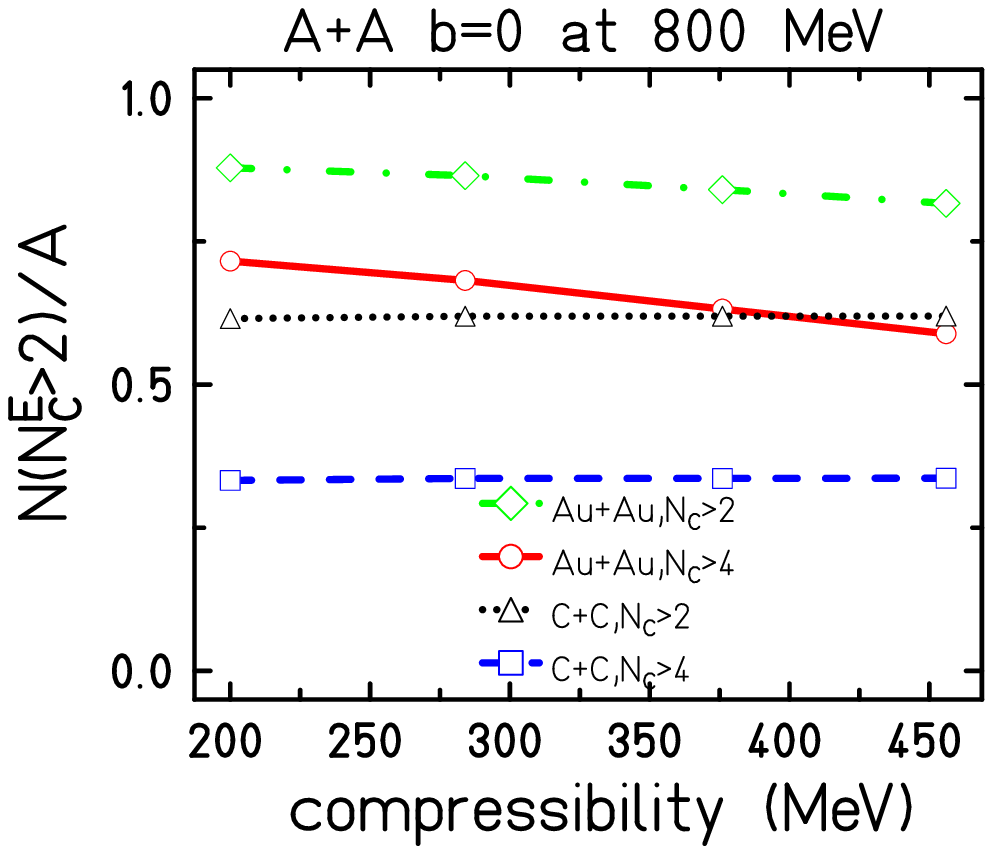,width=0.4\textwidth}
\caption{The fraction of particles having reached a certain maximum density
(left) and having reached a minimum number of collisions (right) as a function
of the compressibility for Au+Au and C+C.
}
\Label{au08-explain}
\end{figure}
In order to understand the effect of the compressibility let us look to the
fraction of particles which have touched maximum density of more than twice
times normal nuclear matter density. From the \lhsref{au08-explain} we conclude
that in heavy systems more particles enter a high density $\varrho/\varrho_0>2$
when using a low compressibility (\rfl) while for a high compressibility
more particles stay at intermediate densities ($\varrho/\varrho_0=1.5-2$, \gml).
For a light system there is no effect. In a high density region there is more
chance for having a collision. This is especially important for deltas
which have to find a collision partner before they decay. In the average they
only have 2-3 fm/c of time for having a $N\Delta$ collision, which as we have 
seen is the major channel at lower incident energies.

Starting from the \lhsref{au08-explain}   {}
it is not surprising to see on the \rhsref{au08-explain}  that the 
contribution of particles having more than 4 (\rfl{} for the Au case and
\bpl{} for the C case) or at least 3 collisions (\gml{} for Au and \bpl{} for C)
is decreasing with increasing compressibility for the Au case and staying
constant for the C case.
As we have already seen, kaons are predominantly produced at high densities 
by parents having undergone a large number of collisions. Therefore it is 
quite reasonable to see a strong dependence of the kaon yield on the reached
density.

The different effect of the compressibility to large and small
systems suggests to fix uncertainties of the production cross section on C+C
where the equation of state do not show an influence. Afterward one could
use the results of Au+Au for looking at the equation of state.
A similar ansatz would be to look directly at the ratios of kaon numbers
in Au+Au and C+C.


\subsection{Ratios of Au/C}

The KaoS collaboration has proposed to analyze  the ratio of the kaon numbers 
in Au+Au and C+C \cite{sturm,Sturm} in order to search for the nuclear equation
of state. The basic idea is
that all uncertainties on the optical potential and the cross sections 
should drop out that way. Furthermore experimental problems of 
detector efficiencies, acceptance cuts, total normalization etc. could be
minimized when measuring both systems with the same experimental setup. 

\Figref{ratio-sig} shows this ratio as a function of energy for a hard
and a soft equation of state.

\begin{figure}
\epsfig{file=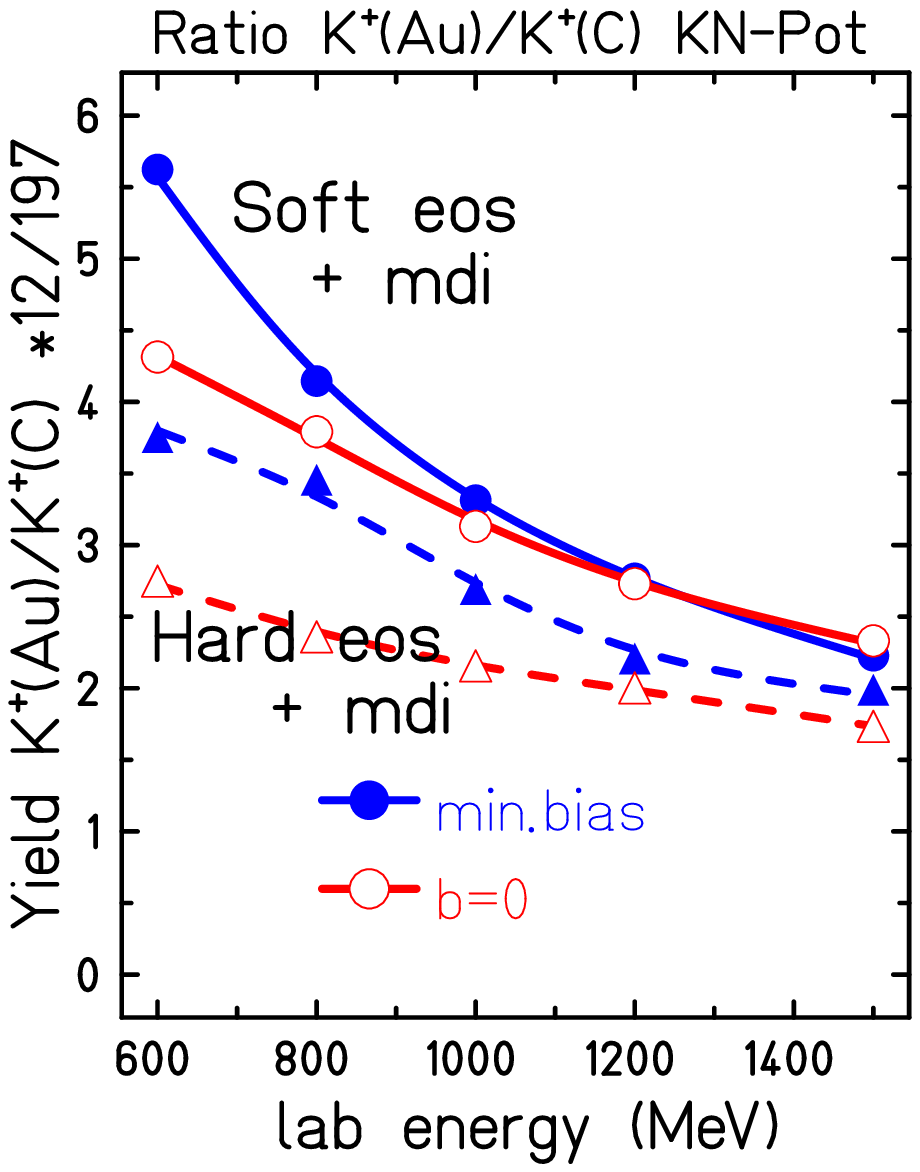,width=0.4\textwidth}
\epsfig{file=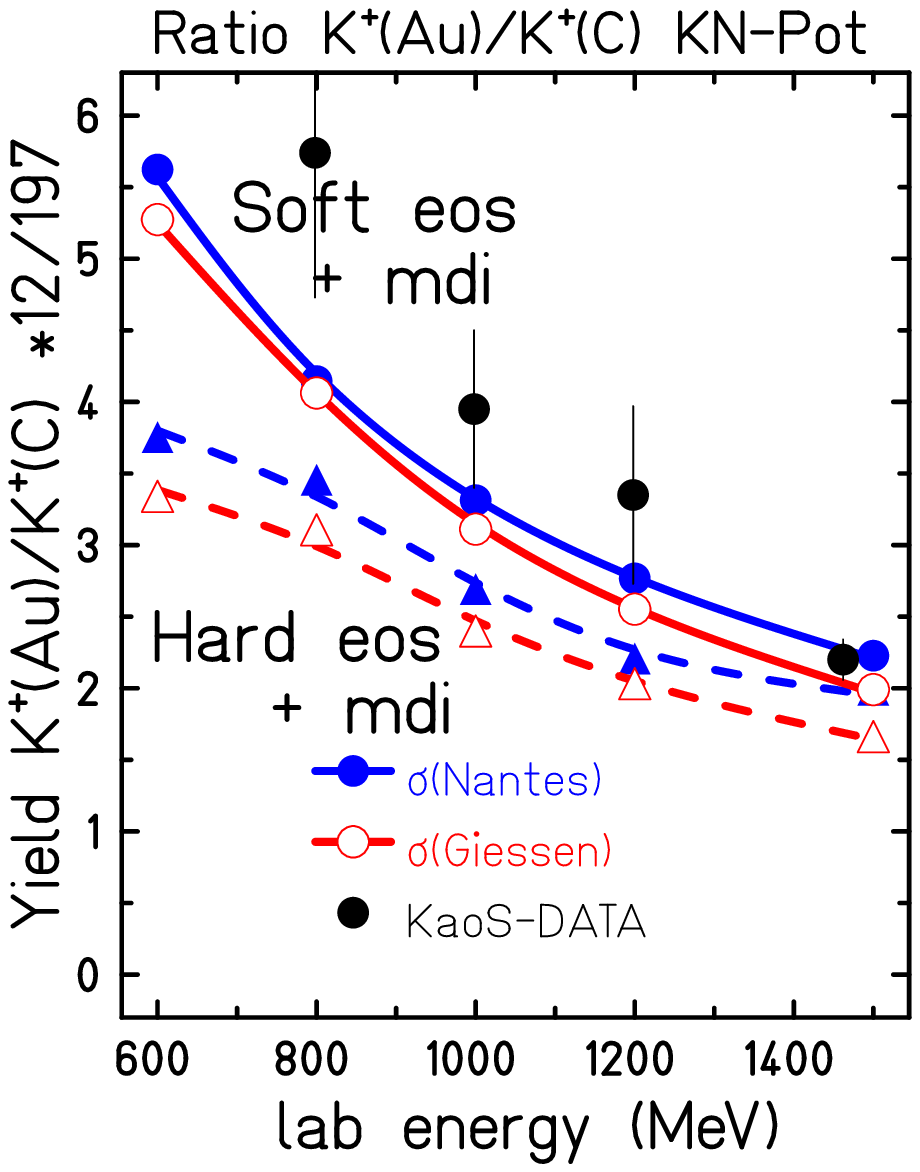,width=0.4\textwidth}
\caption{Kaon ratios Au/C for minimum bias and b=0 events
(left) and comparison to data using different cross
section parametrizations 
}
\Label{ratio-sig}
\end{figure}

A soft eos yielding higher kaon numbers shows higher values than a
hard eos. The values become larger for lower energies. The difference
between hard and soft eos also become more significant for lower
incident energies. This confirms the tendency which was already seen
in \figref{e-dep-nopot} where for a calculation without potential
the sensitivity to the eos was higher at 0.8 GeV than at 1.5 GeV.

Since an experiment may not measure $b=0$fm  we compare on the \lhsref{ratio-sig} 
 calculations with $b=0$fm (red curves) which correspond to the previous analysis
with calculation using the whole impact parameter range (min. bias, blue curves).
We see that both calculations show the same tendencies.

The \rhsref{ratio-sig} compares the minimum bias calculations with experimental
data (bullets) taken by the KaoS collaboration \cite{sturm,Sturm}. The data
are supporting the soft eos. In order to study possible incertitudes
of the results to the cross sections, we show two calculation sets,
one using our standard parametrization (Nantes, blue curves) and one 
using the parametrizations of the Giessen group (Giessen, red curves).

The uncertainties of the cross sections cancel in the ratio and the curves
of both parametrizations are quite similar. It should be noted that similar
results have also been obtained by independent groups \cite{fuchs}.


Next we want to study whether the optical potential could effect this
ratio.
\begin{figure}
\epsfig{file=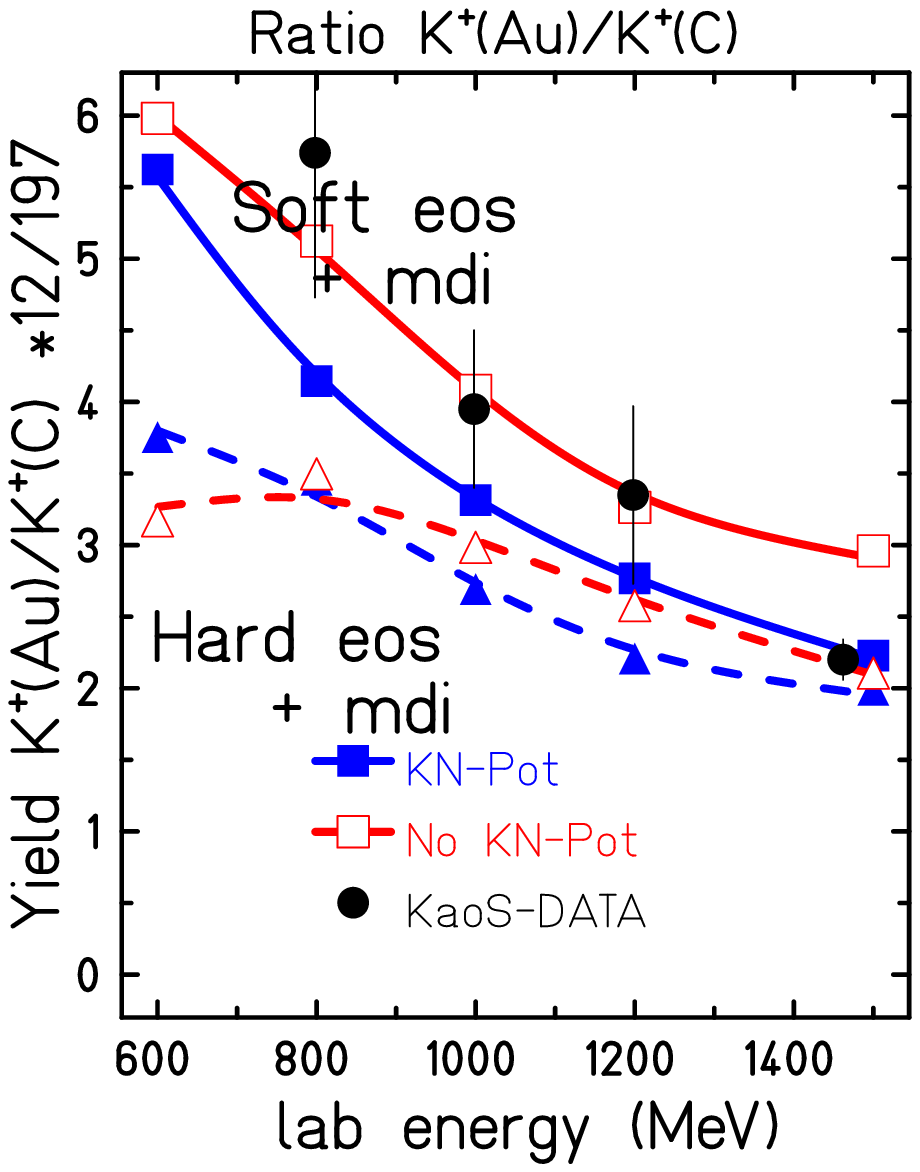,width=0.4\textwidth}
\epsfig{file=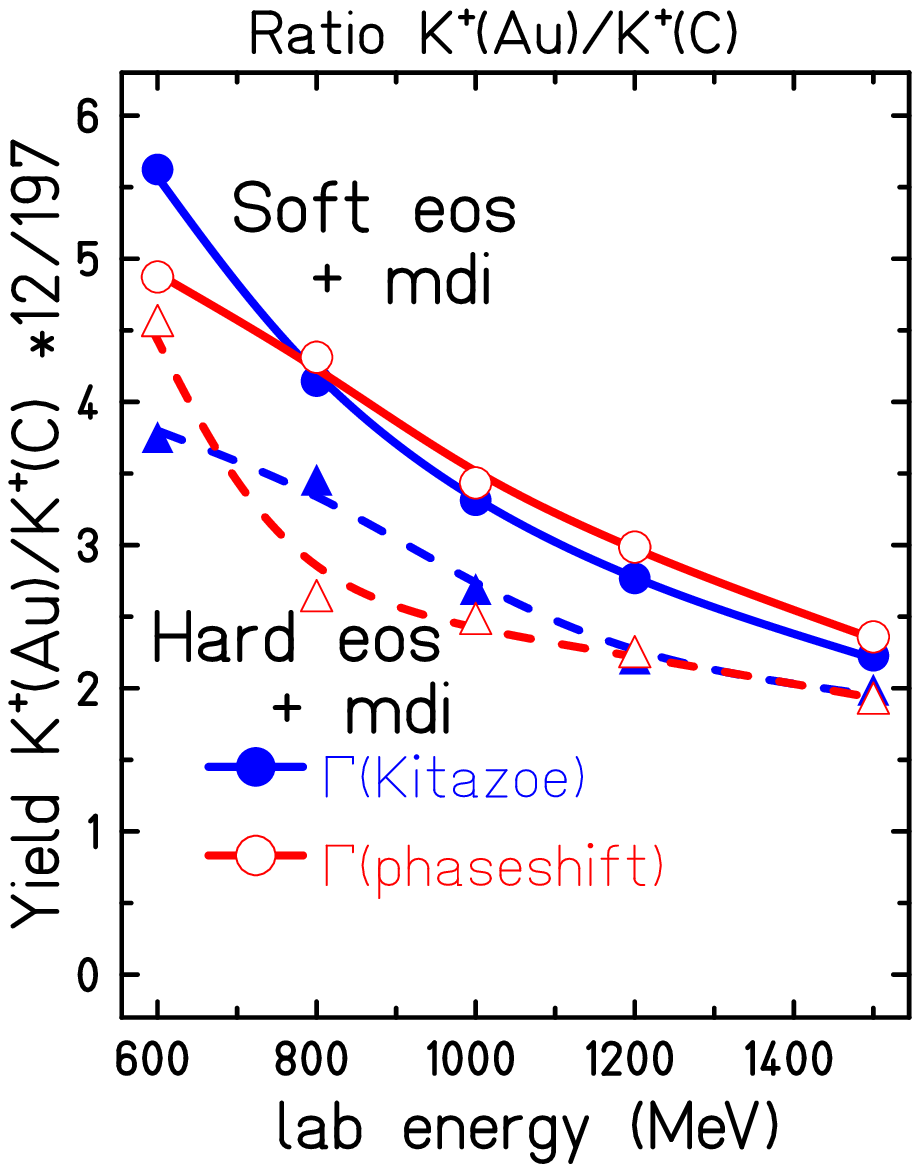,width=0.4\textwidth}
\caption{Kaon ratios Au/C calculated with and without 
KN potential compared to data (left) and influence of
different delta lifetimes (right) 
}
\Label{ratio-pot}
\end{figure}
The \lhsref{ratio-pot} compares calculations with (blue curves) and
without (red curves) optical potential with the KaoS data.
Again the general difference between hard and soft eos remains the
same. The calculations with soft eos without potential fit even better
to the data. 

In general this ratio seems to be quite robust. On the \rhsref{ratio-pot}
we changed some parametrization of the delta decay changing the lifetime 
of the delta without changing the ratio dramatically.


Even changes in the nucleon-nucleon cross sections as shown on the 
\lhsref{ratio-bbsig} do not harm this ratio. When we reduce the 
nucleon-nucleon cross section from normal values (blue curves) to 70 \% 
of this value (red curves) we may change the nucleon rapidity distributions,
transverse flow and pion numbers significantly \cite{hart} but not the
Au/C kaon ratio.
\begin{figure}
\epsfig{file=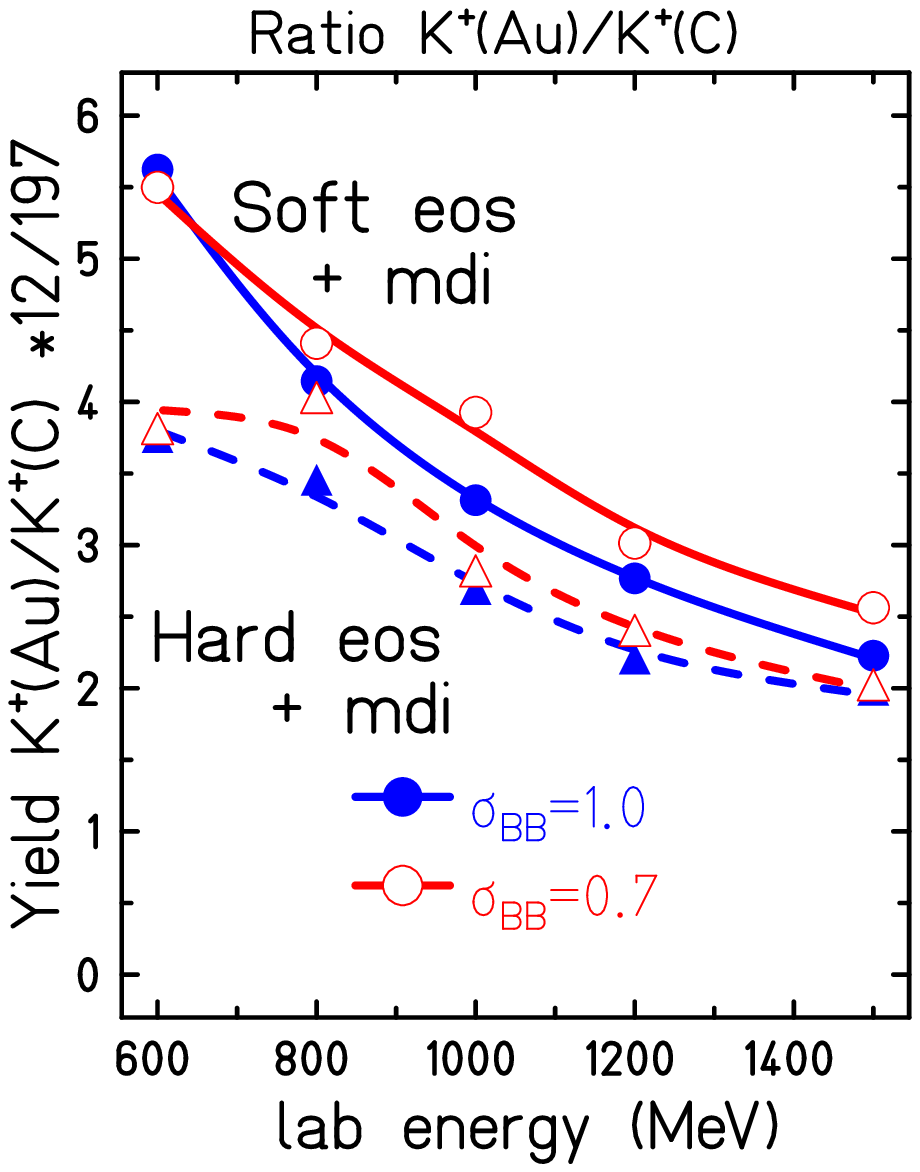,width=0.4\textwidth}
\epsfig{file=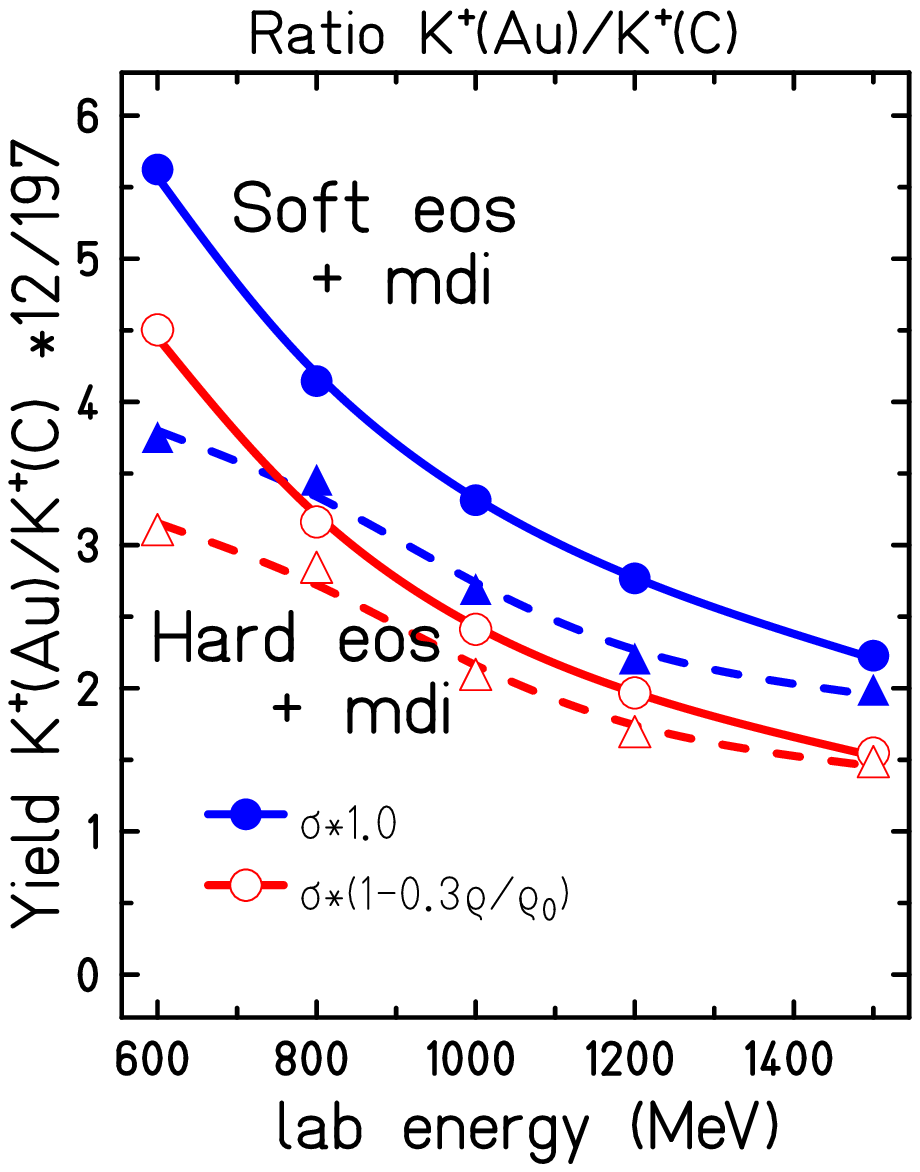,width=0.4\textwidth}
\caption{Kaon ratios Au/C for different cross section
parametrizations for BB collisions
}
\Label{ratio-bbsig}
\end{figure}

The only way to really harm this robust ratio is the use of density dependent
nucleon-nucleon cross sections (red curves) as shown on the \rhsref{ratio-bbsig}. 
However, these changes would yield drastic effects on all other observables.
Thus, we can state that the ratio of the kaon yields in Au/C is a very robust
observable supporting the soft equation of state.


\section{Dynamical observables and thermal properties}
After the discussion of the kaon yields we now want to describe
observables of the kaon dynamics and study the influence of the
optical potential and the rescattering. We shall keep in mind that
for the given distributions the optical potential influences the
absolute yield of the kaons and thus the absolute normalization.
If we state that the optical potential influences an observable more
or less significantly, the absolute yield will all the time remain
a difference.

\subsection{Radial expansion}
As we have seen in \figref{r-prod} the radial density profile 
of the kaon production is peaked in the centre at $R=0$.
The radial density of the kaons keeps this maximum during the
whole reaction as it can be seen on the \lhsref{au15-rr} where
we show snapshots of the radial density of the kaons taken at the
times $t=4$ fm/c (\bdl), $t=8$ fm/c (\rfl), $t=12$ fm/c (\bpl),
$t=16$ fm/c (\gml) and $t=20$ fm/c (blue dotted line). 

\begin{figure}
\epsfig{file=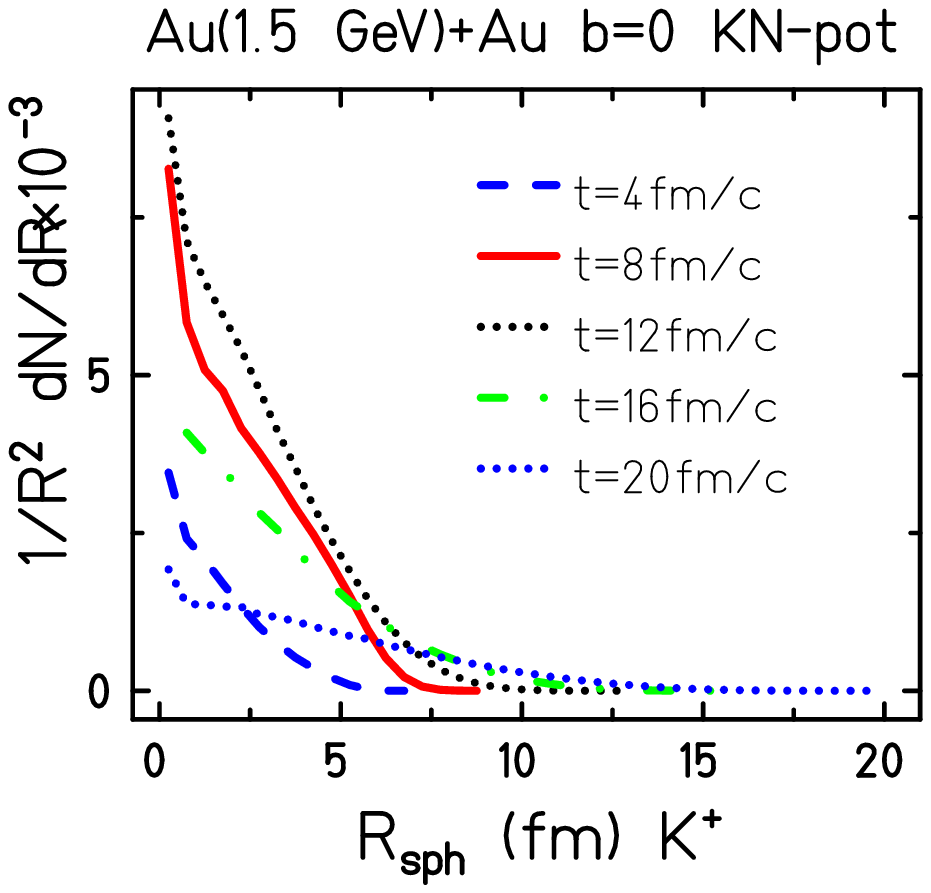,width=0.4\textwidth}
\epsfig{file=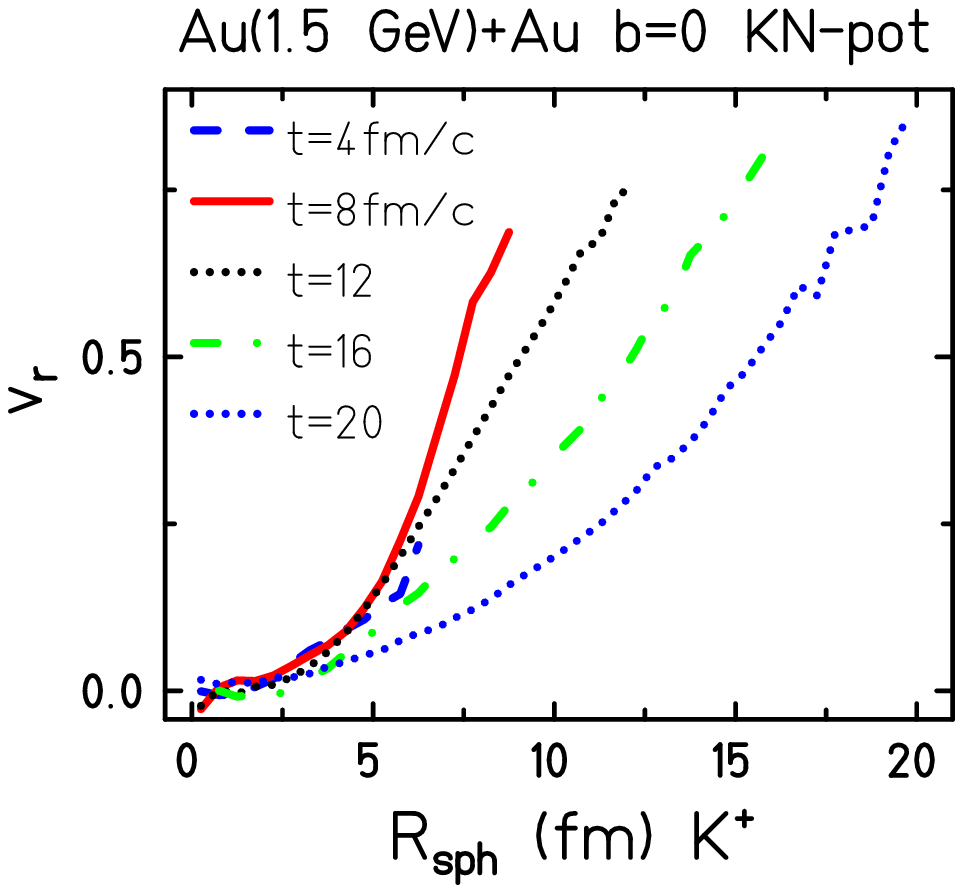,width=0.4\textwidth}
\caption{Time evolution of the radial density and the radial
velocity profile of $K^+$ in Au(1.5AGeV)+Au collisions at b=0fm. 
}
\Label{au15-rr}
\end{figure}

The \rhsref{au15-rr} shows the corresponding radial velocity components
$v_r= \vec{v}\cdot\vec{r}/r$ as a function of the radial distance $R$.
We see that the profiles are not linear, therefore a radial expansion
concept might have difficulties. Furthermore we see that the maxima of the
velocities increase with time. This may be due to an acceleration of the
kaons caused by the repulsive optical potential.

\subsection{Kaon spectra }
Let us now look at the kaon spectra. We present them in Lorentz invariant form
\begin{equation}
\mbox{xaxis}=E(cm) \qquad
\mbox{yaxis}=\frac{E}{p^2}\frac{dN}{dp \quad d\Omega} 
\label{spectra-pres}
\end{equation}
In this representation a thermal Boltzmann gas with the
distribution
\begin{equation}
dN \propto e^\frac{-E(cm)}{k_B T}\, d Lips \qquad
dLips = \frac{d^3p}{E} = \frac{p^2\, dp \, d\Omega }{E}
\label{Boltzmann}
\end{equation}
(where $dLips$ stands for a Lorentz invariant phase space element)
has a linear shape when presenting the y-axis in logarithmic scale
and the slope parameter corresponds to the inverse temperature.

\begin{figure}
\epsfig{file=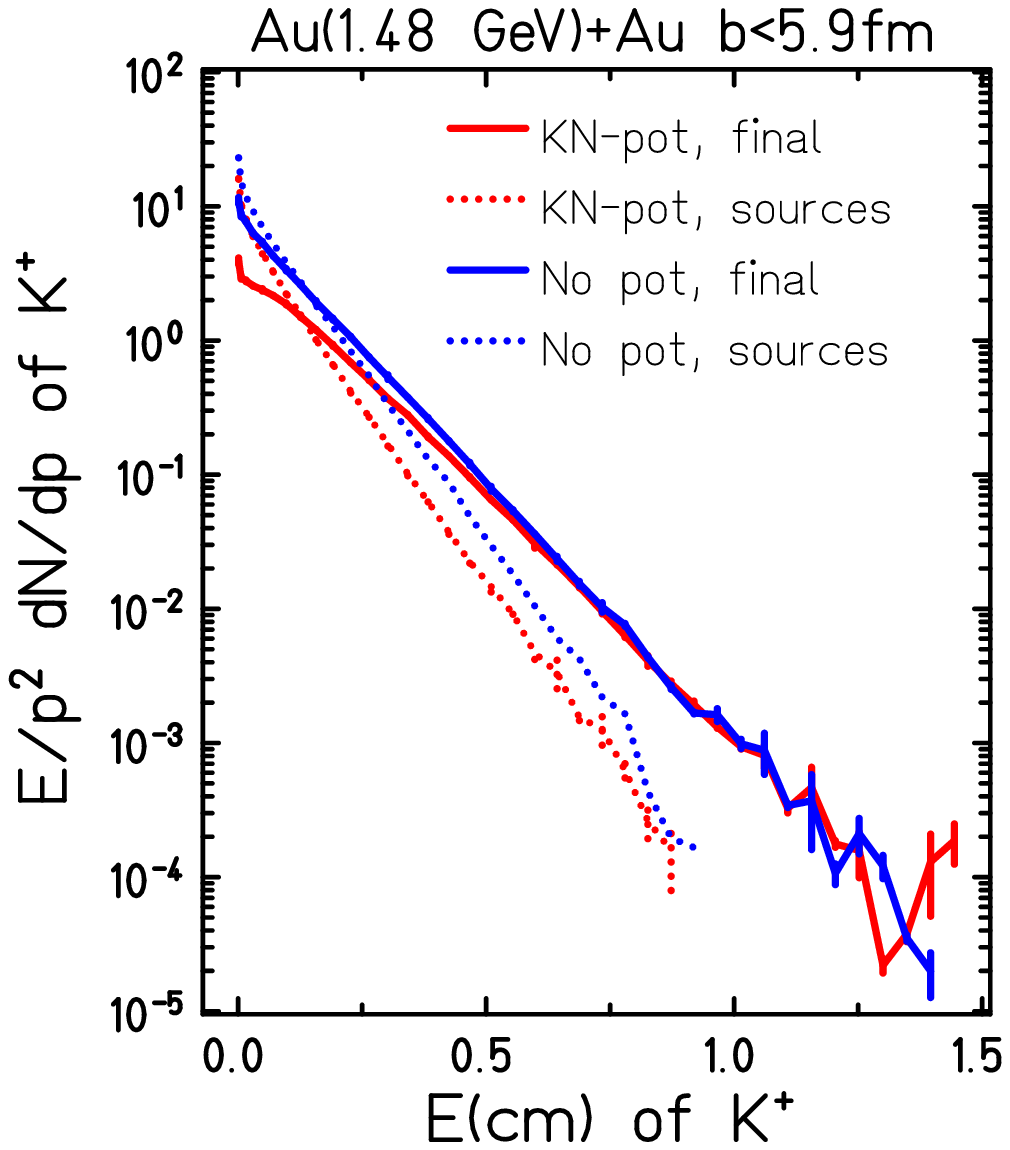,width=0.4\textwidth}
\epsfig{file=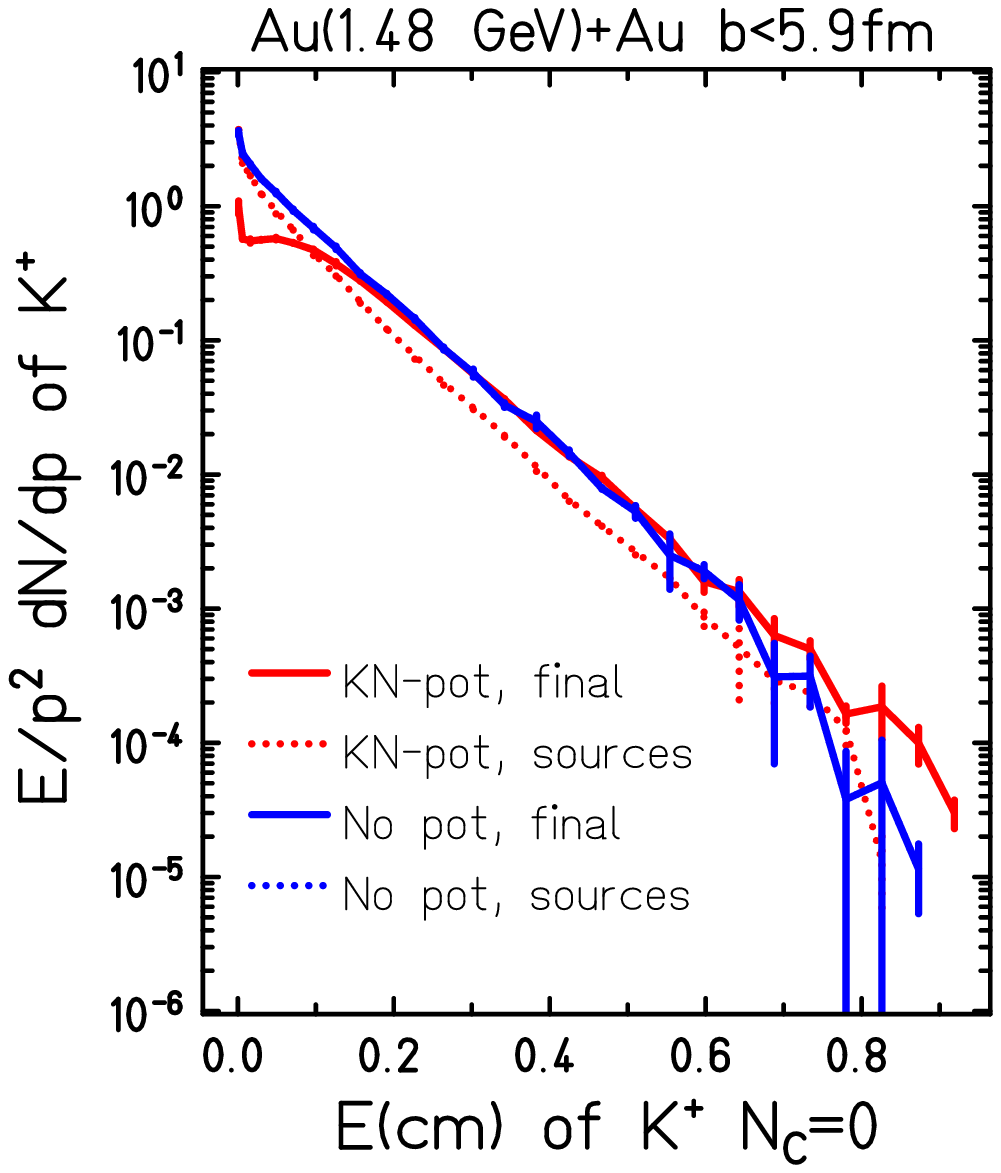,width=0.4\textwidth}
\caption{Comparison of initial and final spectra 
with and without potential.
}
\Label{au148-pep}
\end{figure}

\Figref{au148-pep} shows the temperature of semi-central
collisions in a Au(1.48AGeV)+Au collision for calculations
with (red curves) and without (blue curves) potential.
The solid lines show the spectra taken with the final momenta
of the kaons (after propagation and rescattering in the medium)
whereas the dotted lines show the momenta the kaon had directly
after their production (source spectra or initial spectra).
For both calculations (with and without potentials)
the initial spectra show similar slope at higher energies
and converge at very low energies.
The final spectra show higher temperatures than the initial spectra.
Calculations with and without potential show different values at low
energies and same values at high energies.
The difference of the initial spectra at high energies corresponds to
the shift of the threshold due to the potential penalty. At the final state
the repulsive potential pays back this energy when the kaon enters the
free mass. The regain of the energy loss in threshold can be nicely
demonstrated on the \rhsref{au148-pep} where initial and final spectra
are shown for those particles that did not collide. For the calculations
without potential initial and final spectra are the same. For the calculation
with an optical potential 
the difference between initial and final spectra is only due to the
repulsive optical potential. We see that the potential shifts the spectra
back to the values of the calculation without potential. Only close to the
threshold there is a lack remaining. This lack is caused by those reactions
which would have been above the threshold if the potential was not there
but which were forbidden by the threshold shift.

As we can see from the difference of the initial and final spectra
in calculation without the optical potential of the kaons, the rescattering
of the kaons plays an important role for understanding the final energy distribution.


\begin{figure}
\epsfig{file=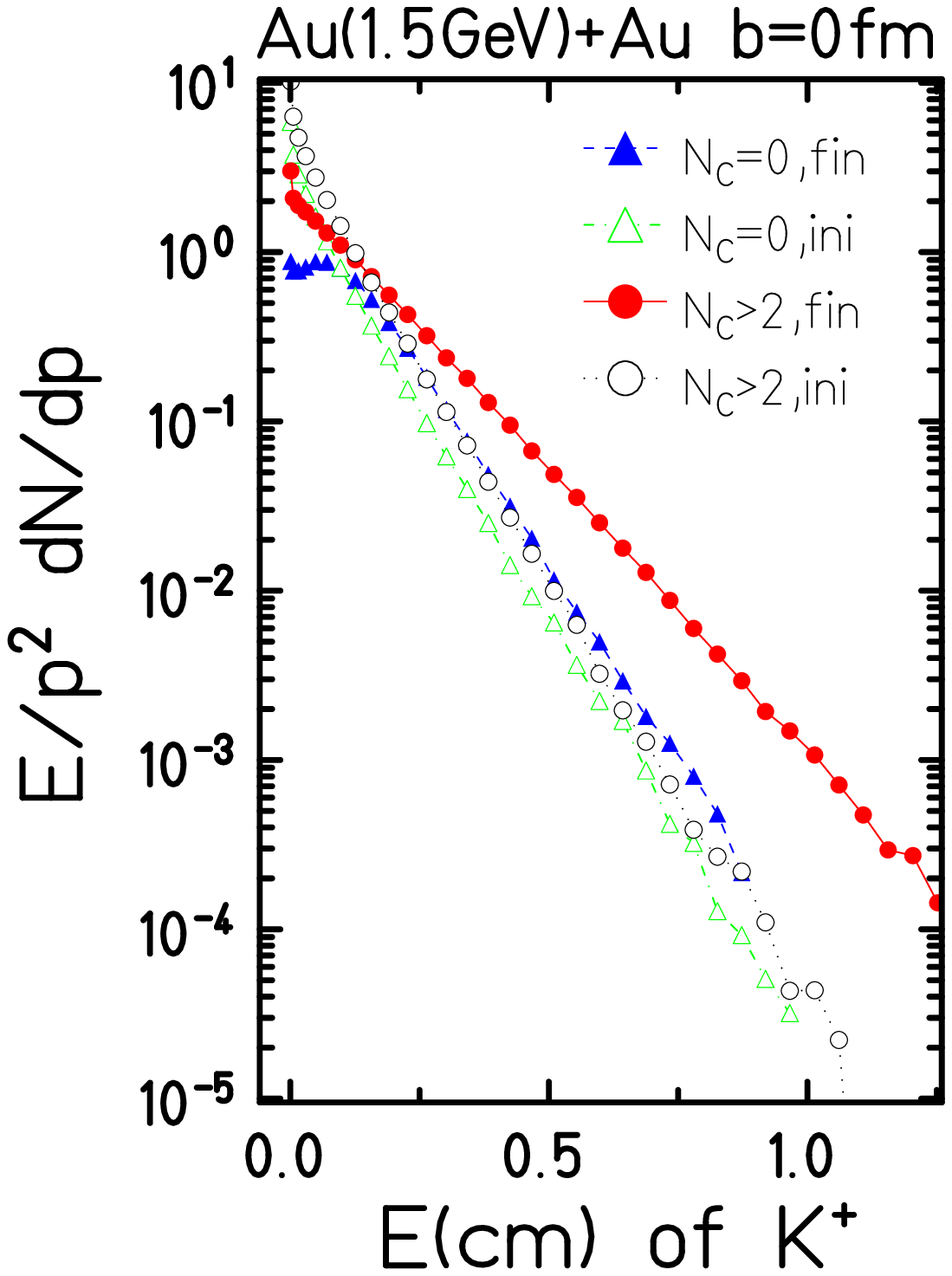,width=0.4\textwidth}
\epsfig{file=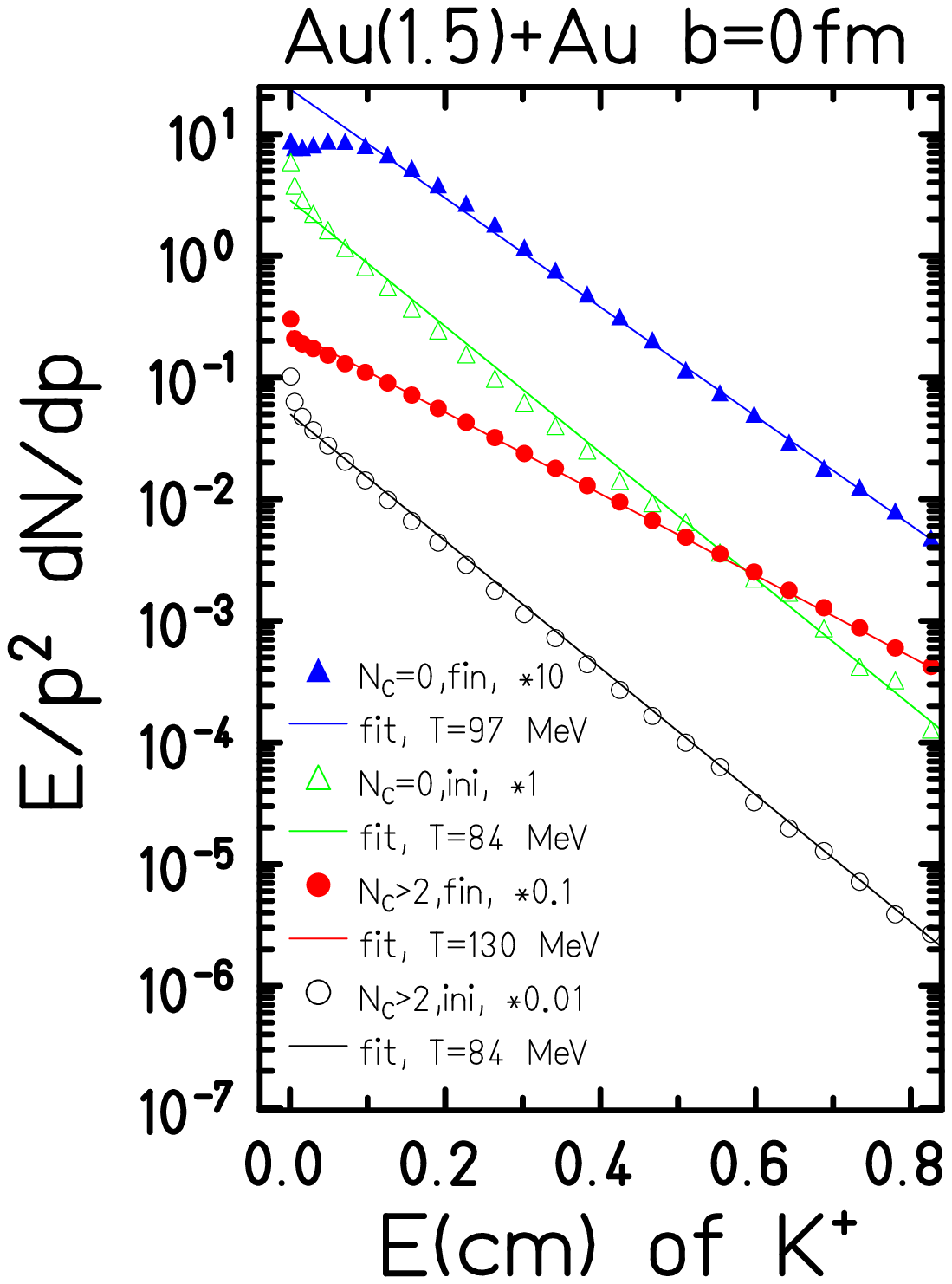,width=0.4\textwidth}
\caption{Comparison of the contribution from
kaons that collided frequently and those who did not collide
}
\Label{au15-pep}
\end{figure}

Therefore we study in \figref{au15-pep} those kaons which did not collide
(triangles) and those which did collide frequently (circles)
in separate analysis. We see on the \lhsref{au15-pep} 
that the initial spectra of the kaons without collisions (\gml\ with open
triangles) and of the multicolliders (\bpl\ with open circles) show
similar slopes. The multicolliders dominate in the absolute yield.
Thus, we can conclude that this selection to the collision numbers has
no bias. 
The final spectra are completely different.  
The \rhsref{au15-pep} shows the same spectra multiplied with additional
factors in order to disentangle the graphs. Each spectrum has been fitted by
a Boltzmann ansatz as seen in eq. \ref{Boltzmann}.
We see that the sources show a temperature of about 84 MeV. The potential
shifts the uncollided particles to a temperature of about 97 MeV.
The multicolliders who underwent more than two collisions (additional to the
potential repulsion) show final temperatures of about 130 MeV.
Therefore, we can conclude that the collisions are giving stronger effects than
the potentials. 

\begin{figure}
\epsfig{file=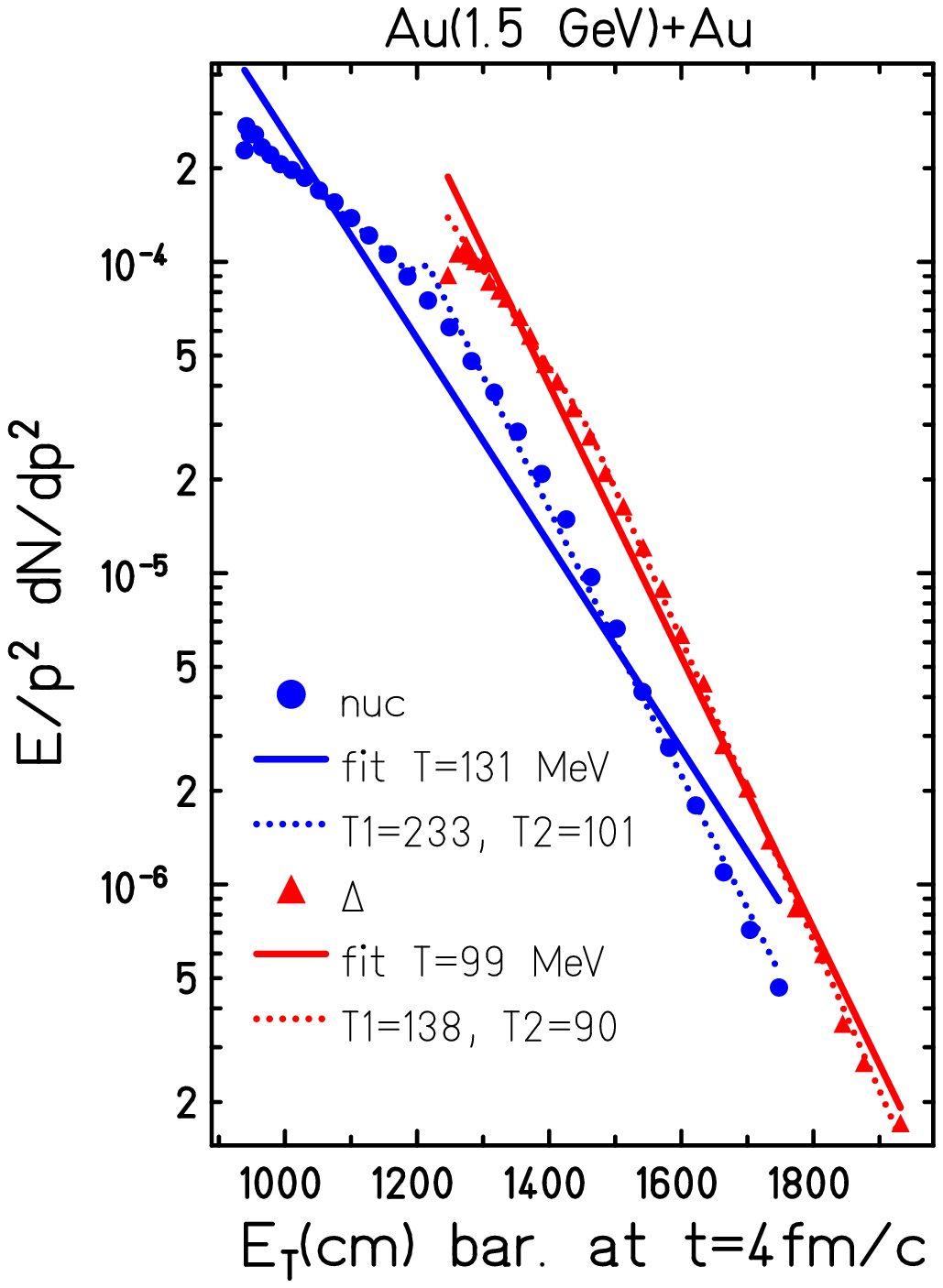,width=0.3\textwidth}
\epsfig{file=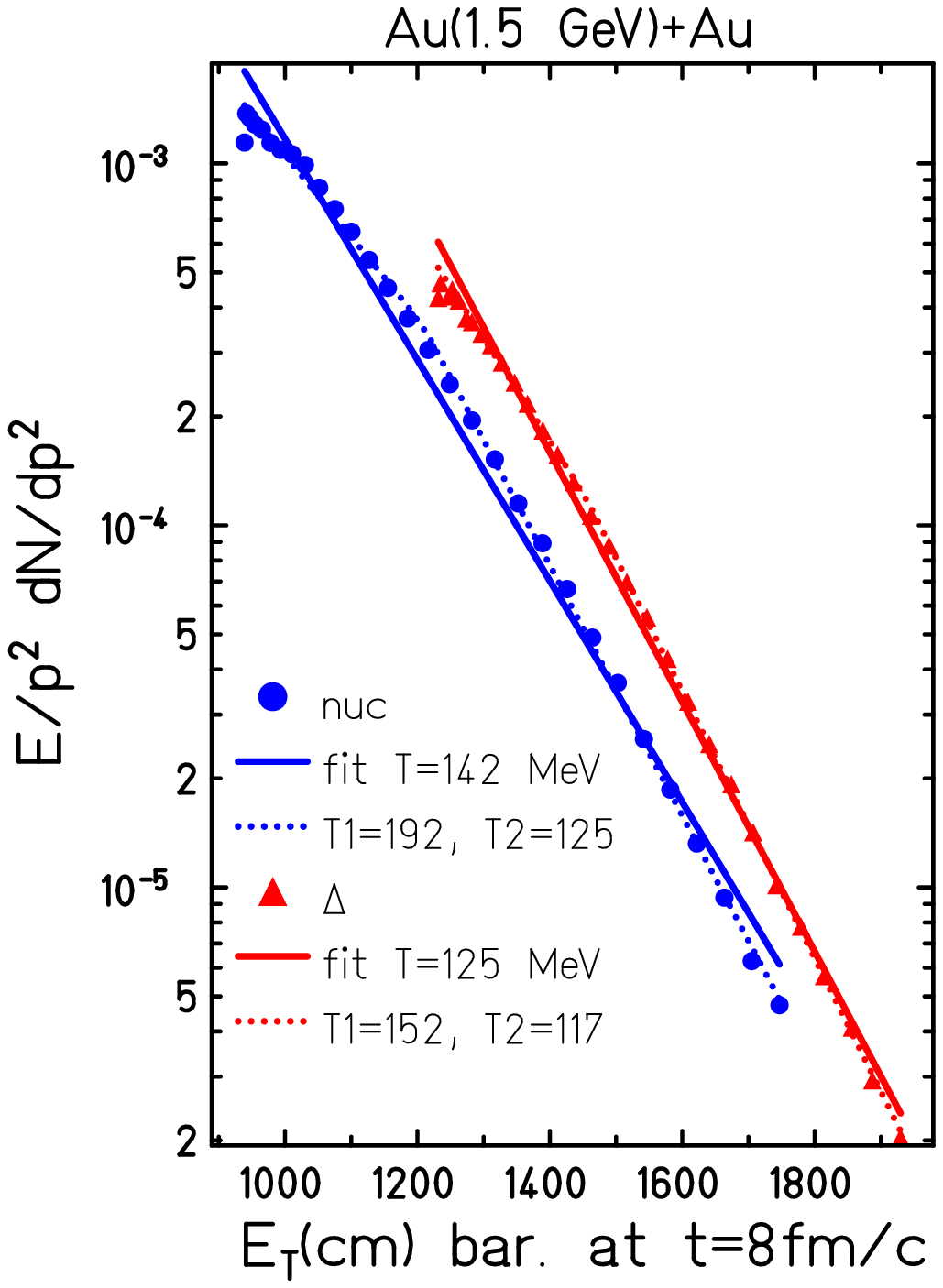,width=0.3\textwidth}
\epsfig{file=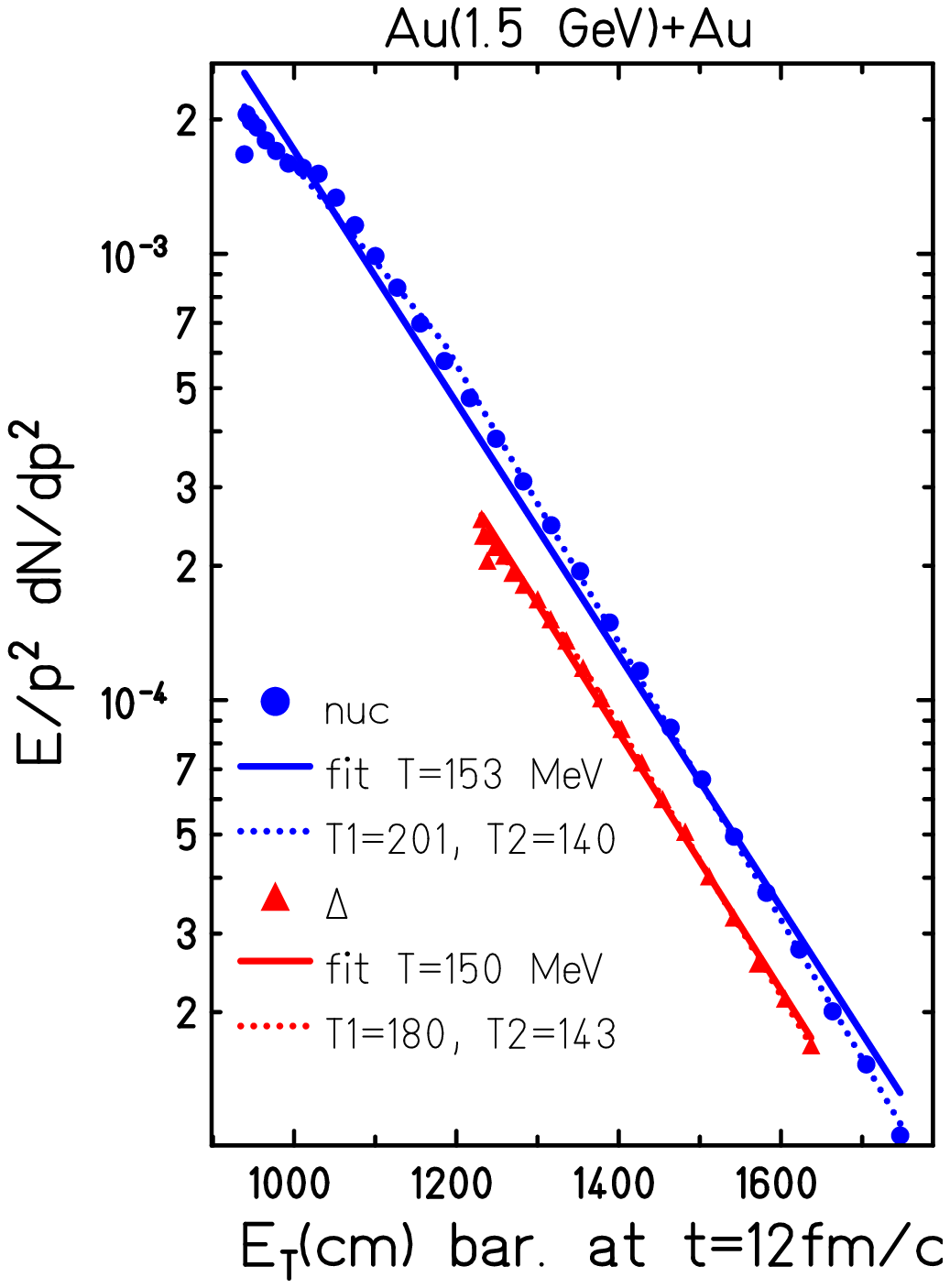,width=0.3\textwidth}
\caption{Time evolution of the nucleon and delta spectra
}
\Label{au15-nuc}
\end{figure}

A tentative explanation of this effect is shown in \figref{au15-nuc}.
We present the nucleon (blue curves) and delta (red curves) 
transverse spectra taken at the time steps $t=4$ fm/c
(left), $t=8$ fm/c (middle) and $t=12$ fm/c (right). The longitudinal spectra
might still be dominated by the projectile-target kinematics in this early
stage of the reaction.  
The chosen time steps correspond to the time window of kaon production.
We see that at early times the nucleons and deltas do not show a thermal
spectrum that could be described with one temperature only. 
They will reach this equilibrated spectrum  
at about 16-20 fm/c. The high energy components simulate lower temperatures
than the low energy components. The delta temperatures are lower than 
the nucleon temperatures. The production of a kaon takes place early and requires high energetic nucleons and (preferentially) deltas. 
Their temperatures are quite low at these early times. Therefore, 
the source temperature is quite low. Later on the kaons may collide with all
nucleons (where there are more nucleons with low energies) and thus see
the high temperature component in the collisions.  

\begin{figure}
\epsfig{file=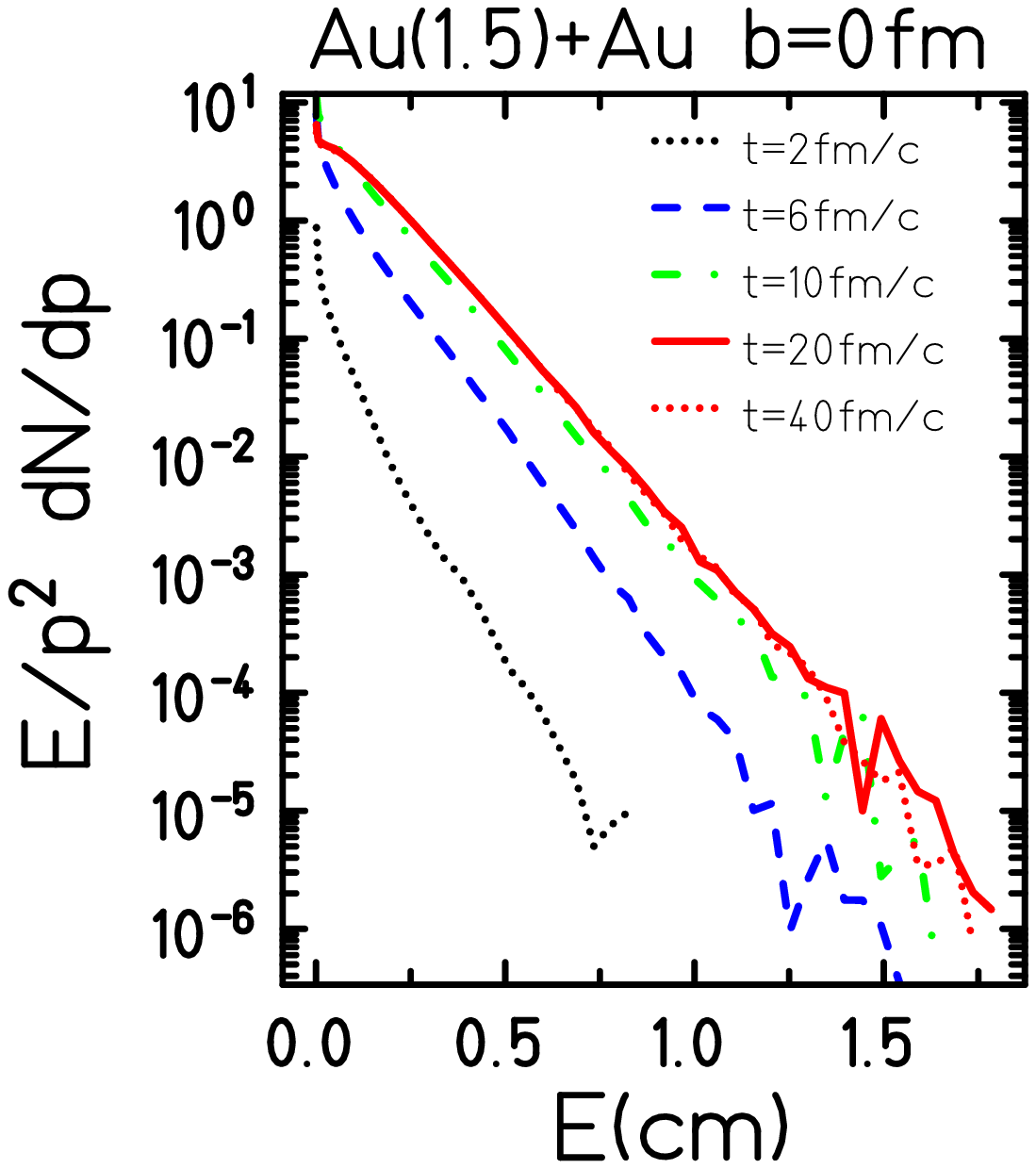,width=0.4\textwidth}
\epsfig{file=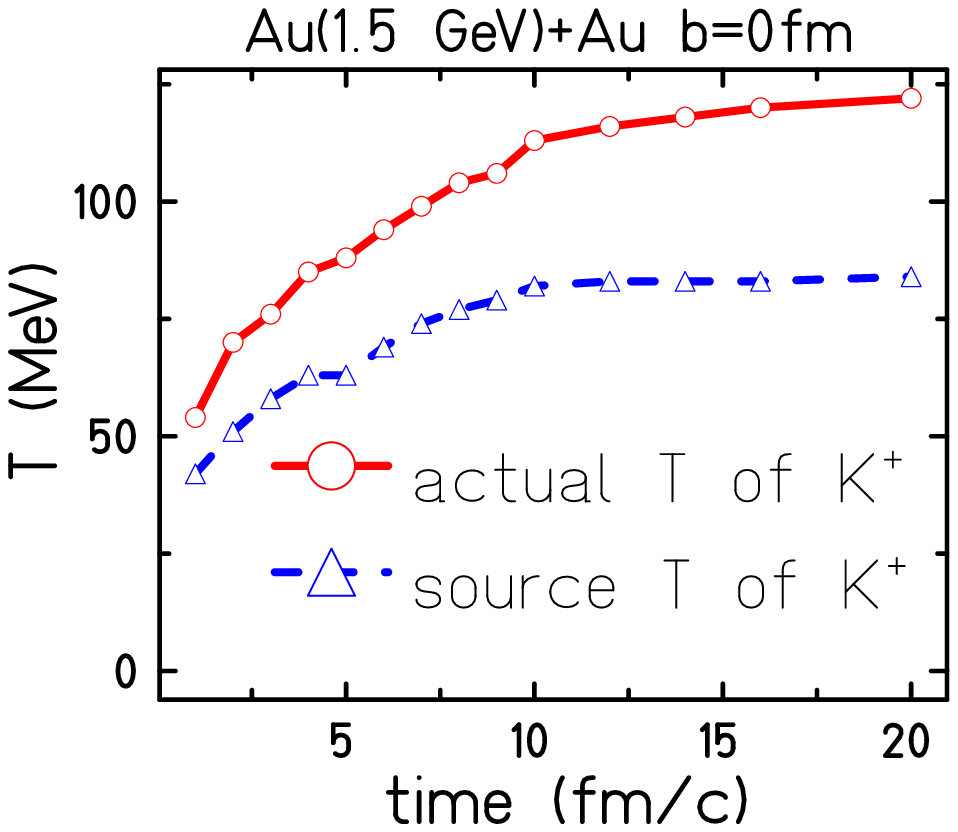,width=0.4\textwidth}
\caption{Time evolution of kaon spectra and temperatures
}
\Label{au15-pep-t}
\end{figure}
This assumption is supported by the analysis of the time evolution of the
spectra. The \lhsref{au15-pep-t} shows actual spectra at different time steps.
We see a rise of the temperature in time. The \rhsref{au15-pep-t} shows
the time evolution of the source and actual temperatures. We see that
both temperatures are rising in time. The source temperatures rise since up to
about 12 fm/c the high energy components show rising temperatures.
The rise of the actual temperatures is due to the collision in a system of nucleons
which is still heating up.

\begin{figure}
\epsfig{file=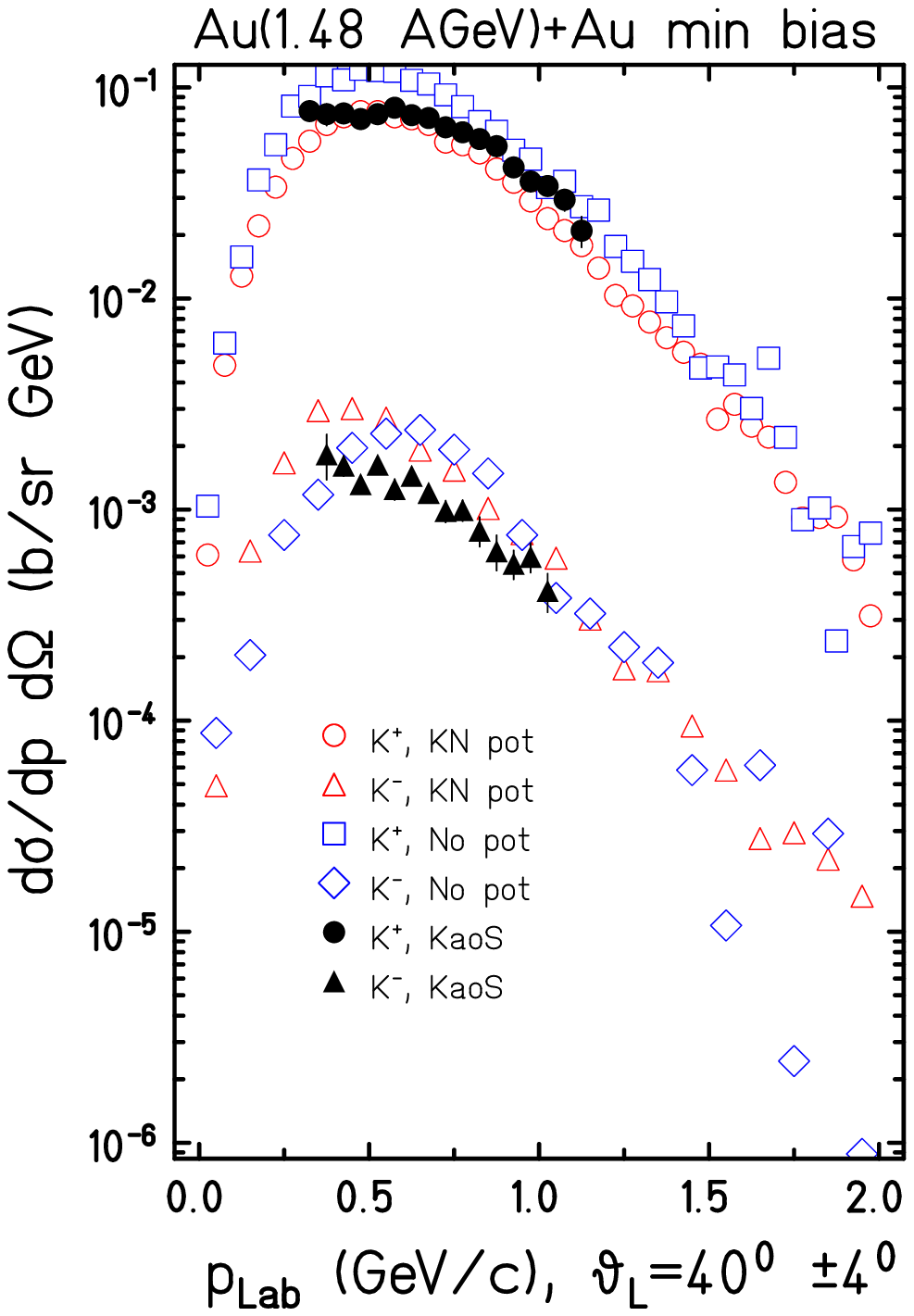,width=0.4\textwidth}
\epsfig{file=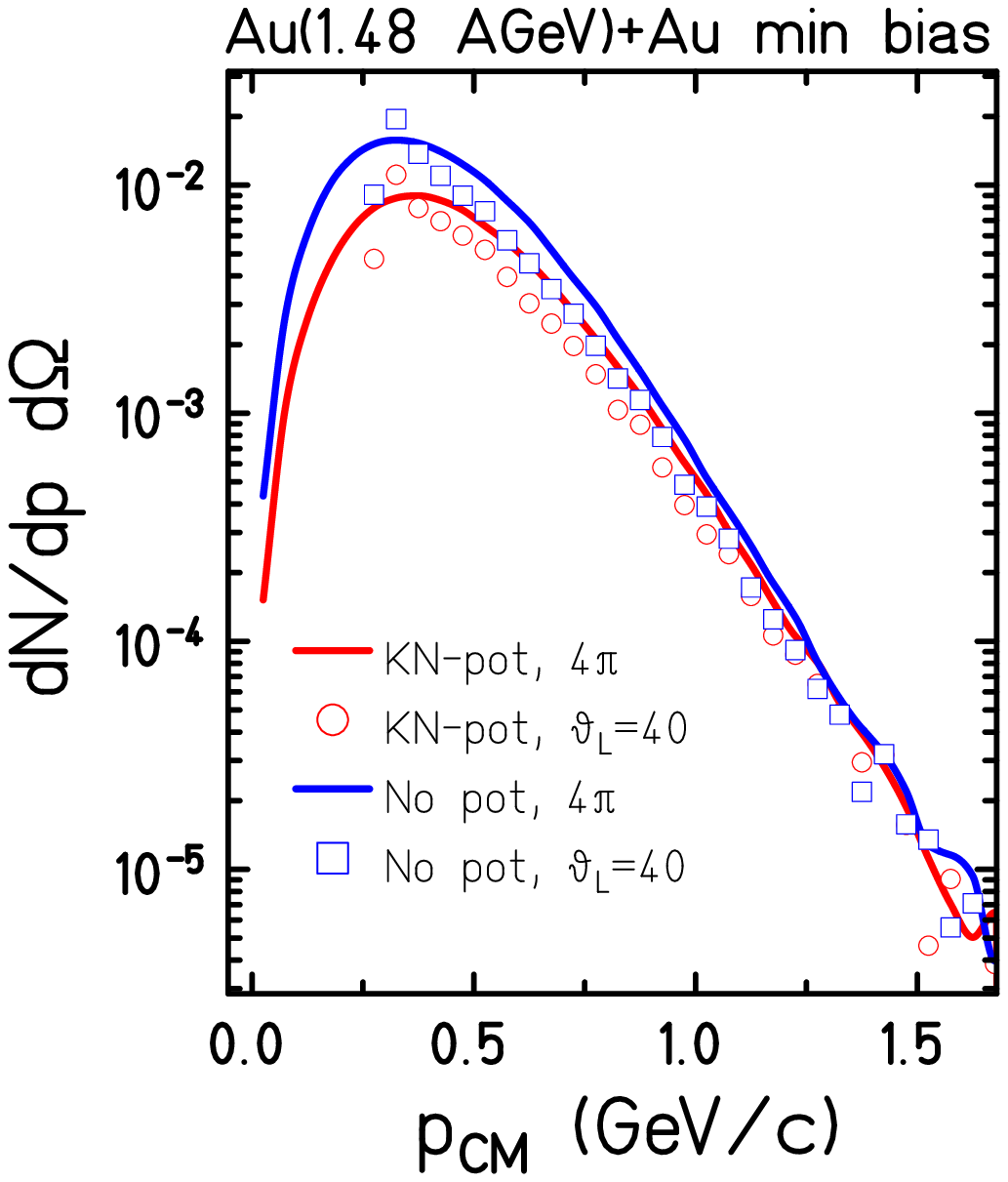,width=0.4\textwidth}
\caption{Comparison of lab momentum spectra and the effect of the potential
on the spectra in the cm frame}
\Label{au15-pep-compa}
\end{figure}

Let us now compare the spectra to experimental data.
The \lhsref{au15-pep-compa} compares the laboratory momenta of kaons
taken at $40^0$ degrees with experimental data of the KaoS-collaboration
\cite{foerster}. Blue symbols denote calculations without potential,
red symbols calculations with an optical kaon potential and the black
filled symbols the KaoS data. We see a preference for the calculation
with an optical potential but the differences are not very significant.
This is due to the effect that unfortunately the experiment cannot measure
the kaons up to lowest centre-of-mass momenta.
The \rhsref{au15-pep-compa} shows the corresponding centre-of-mass spectra
calculated with (red line and symbols) and without KN potential (blue line
and symbols). The lines correspond to spectra obtained from full $4\pi$
analysis while the symbols correspond to spectra only in the $\vartheta_{lab}=
40^0 \pm 4^0$ angle. We see that the latter spectra just end at the point when
the significance of the calculations becomes most prominent.

\subsection{Kaon temperatures}
Let us now compare the temperatures determined by experiment with calculations
of the IQMD model. The experiment divided their events in five centrality bins
selected by the participant number. IQMD divides its events into the same
centrality bins by direct selection of the impact parameter.
A further difference is that the experiment can only use an angular
segment determined by the detector position while IQMD can access the full
$4\pi$ event. Furthermore the experiment can only cover a range of detectable
energies. Very high and very low momenta are cut off.

\begin{figure}
\epsfig{file=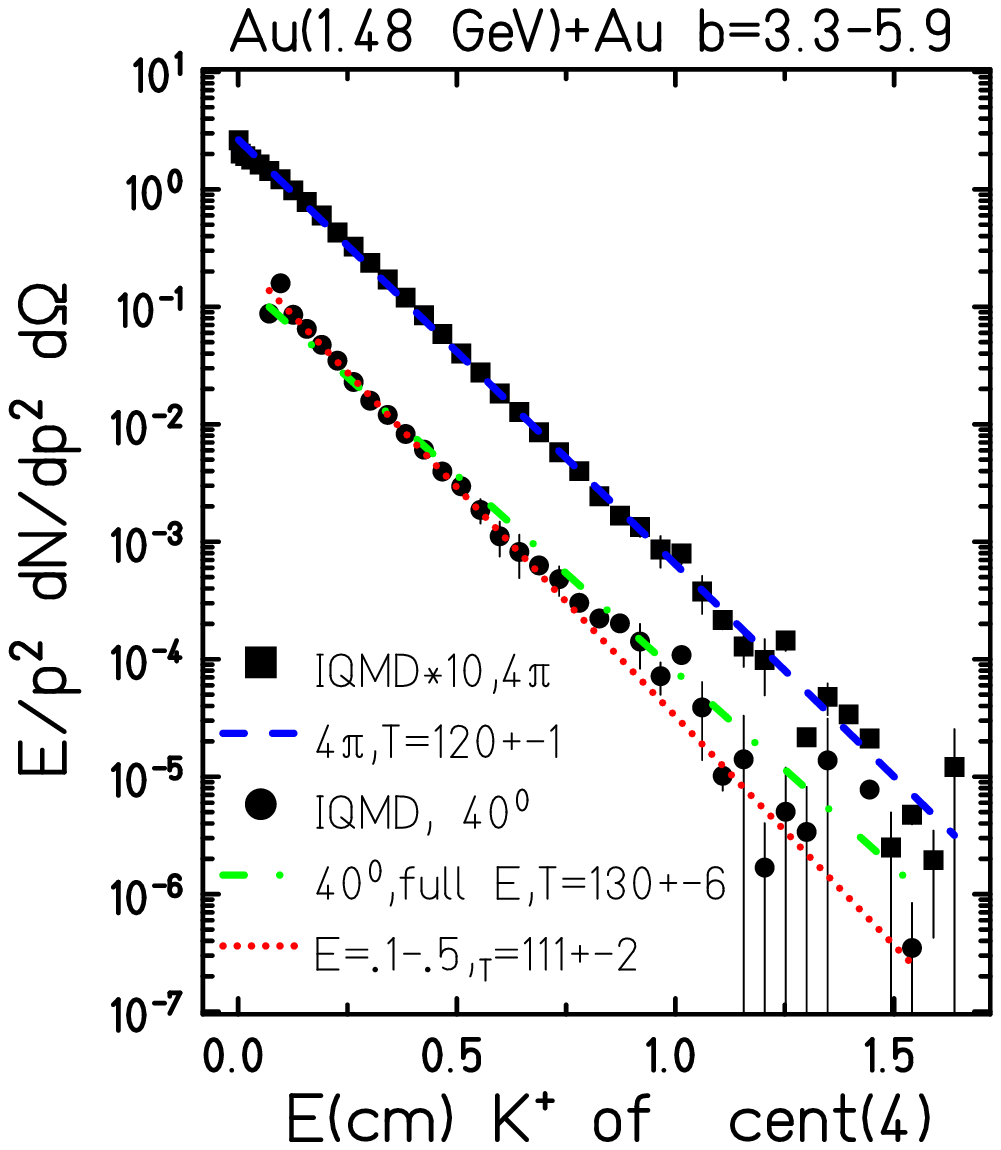,width=0.4\textwidth}
\epsfig{file=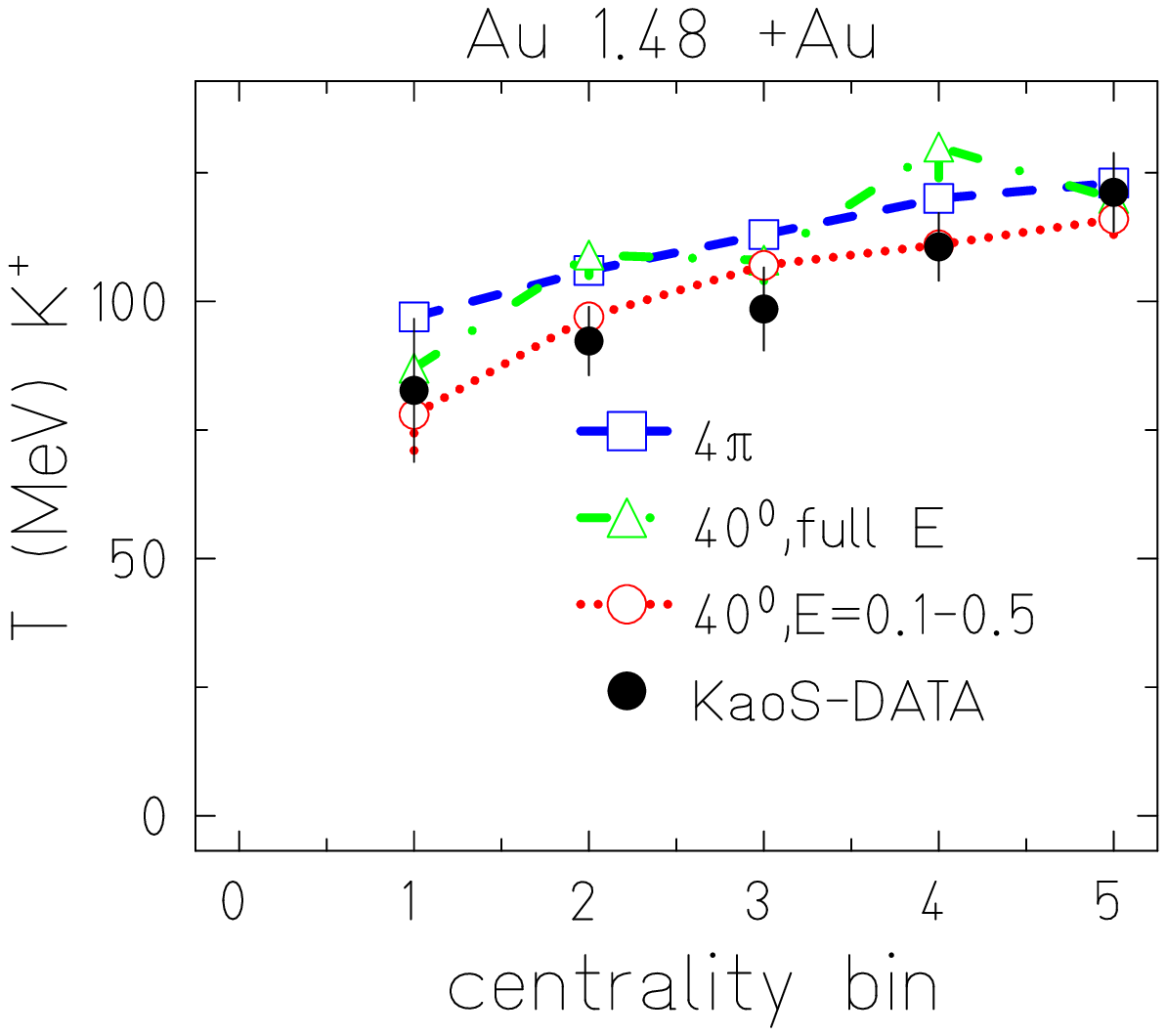,width=0.4\textwidth}
\caption{Determination of temperatures
}
\Label{au15-t-def}
\end{figure}

\Figref{au15-t-def} shows the effect of these constraints to the
determination of the temperatures. Using full $4\pi$ information (\bdl)
yields higher temperatures than observed experimentally. Reducing the
event to a $\vartheta=40^0 \pm 4^0$ region (\gml) and applying
additional cuts to the energy coverage (red dotted lines) yields low 
temperatures which are now compatible to those measured by experiment.


\begin{figure}
\epsfig{file=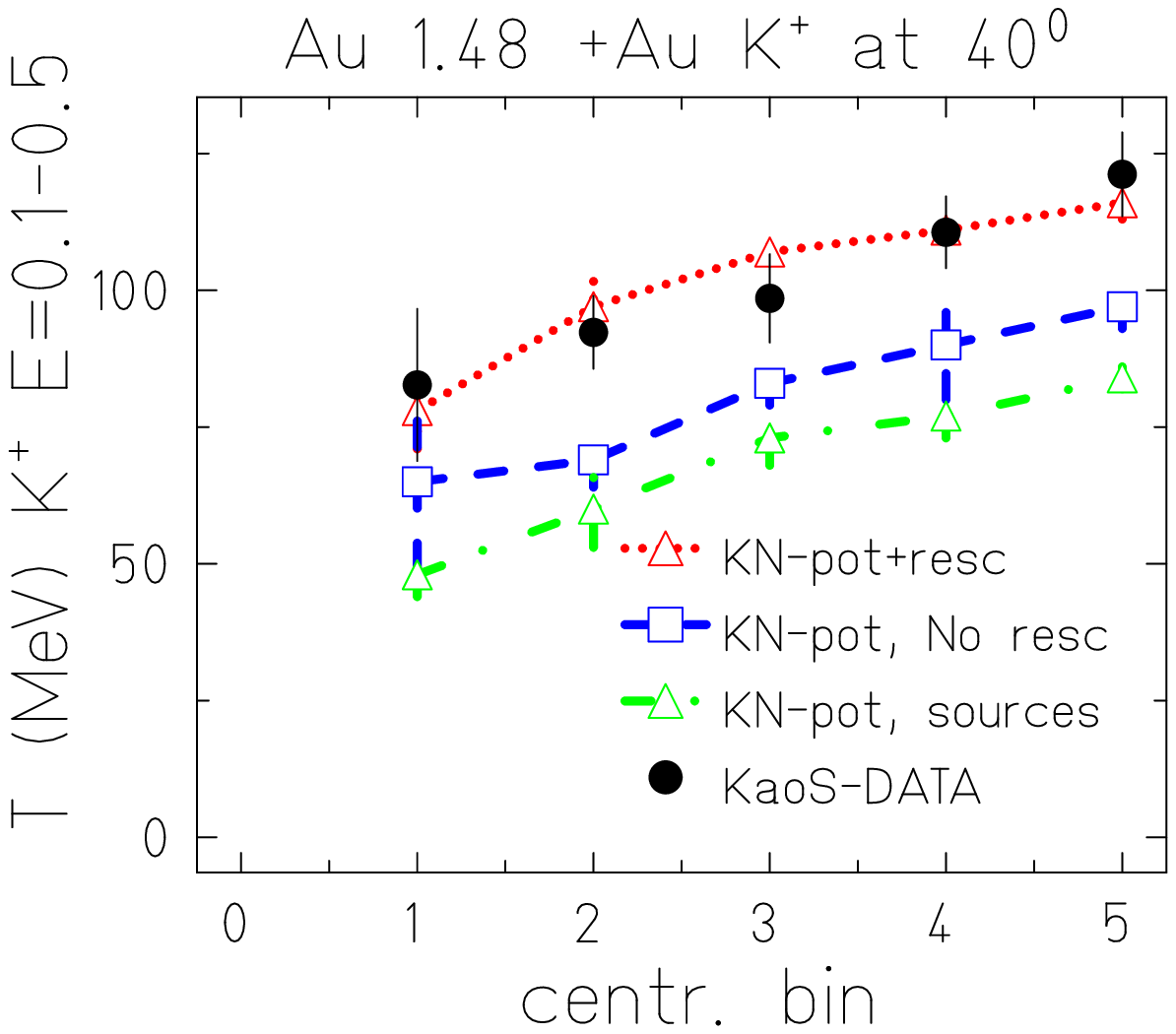,width=0.4\textwidth}
\epsfig{file=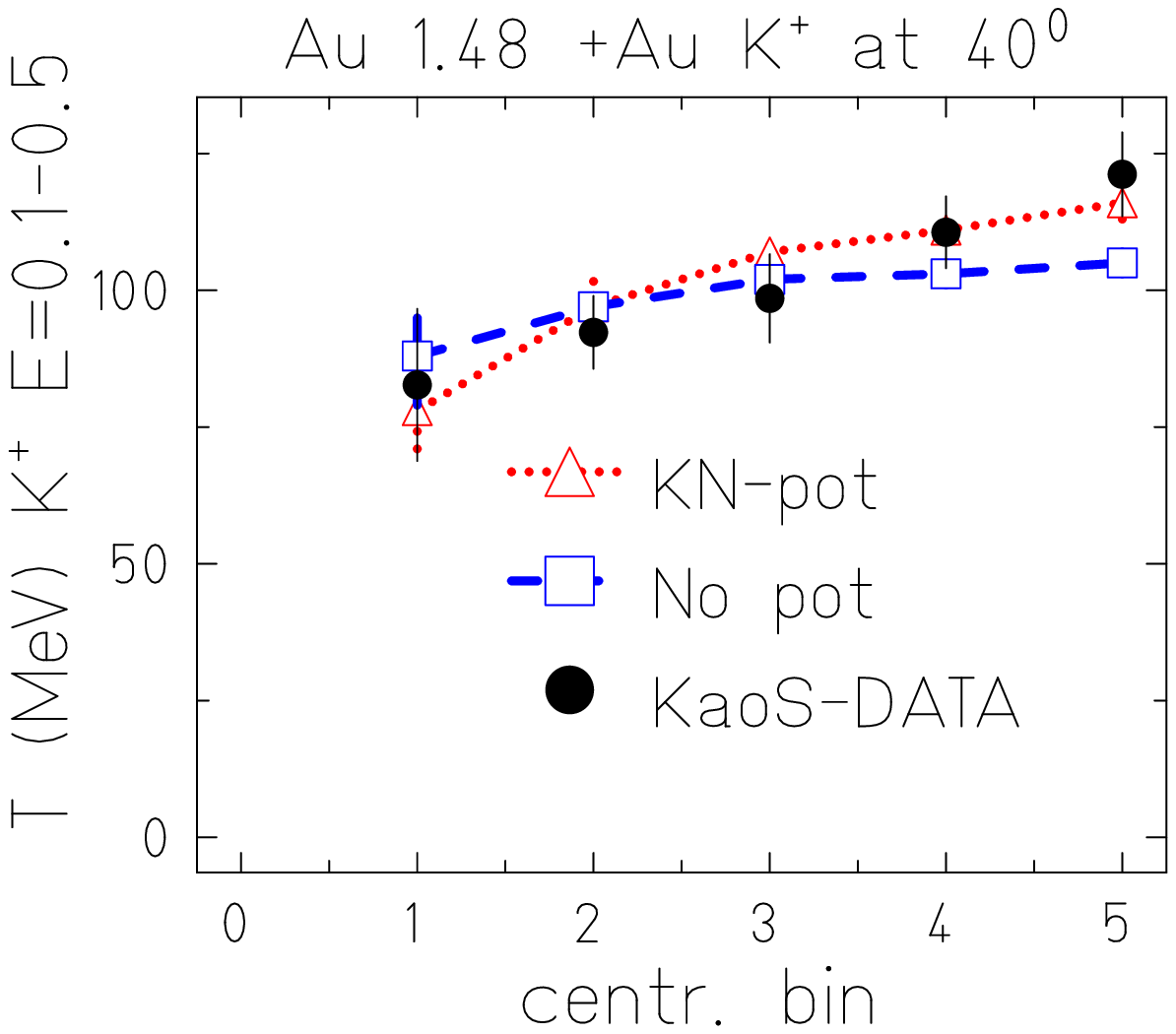,width=0.4\textwidth}
\caption{Comparison of temperatures to KaoS data
}
\Label{au15-t-comparison}
\end{figure}

Finally, \figref{au15-t-comparison} shows a comparison of $K^+$
temperatures (using optical potential but switching the
rescattering on and off, \lhs) and 
in a calculation with (red dotted lines with triangles) and without (blue dashed
line with squares) optical potential to the KaoS data \cite{foerster}.
The \lhs{} shows that the temperature of the initially produced kaons is much
too low to explain the data and that the potentials shift up the temperatures
but not sufficiently.  On the other hand if we calculate with full rescattering
(\rhs) but play on the potentials we end up with similar temperatures but a slightly
different centrality dependence.
We see that the calculations with potential show a nice agreement to the
data. 
However, the calculations without potential are not contrary to the
data. 

\subsection{Energy and system size dependence}
Let us shortly discuss the dependence of the inclusive
spectra (i.e. taking all impact parameters) on the
incident energy and on the system size

\begin{figure}
\epsfig{file=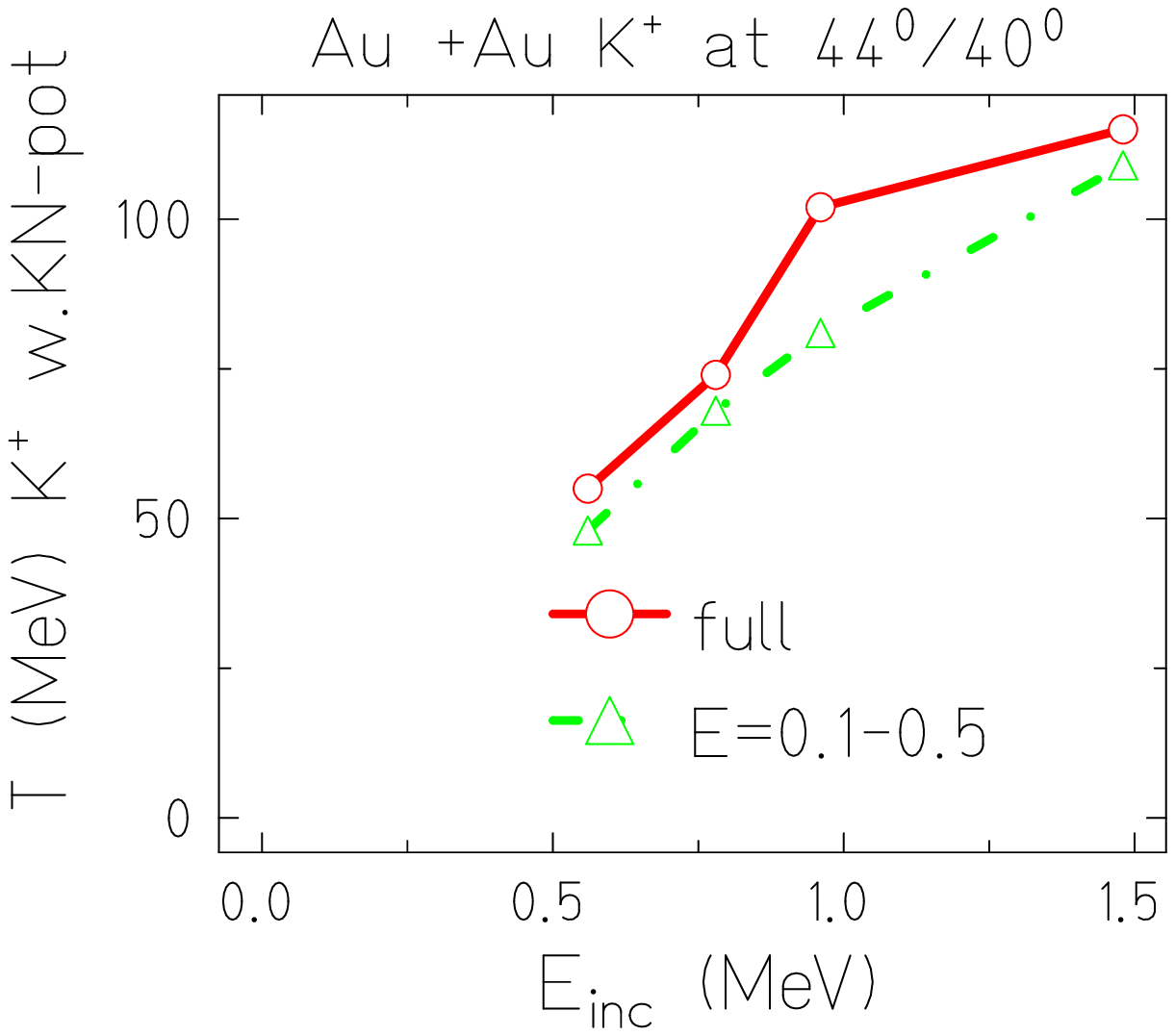,width=0.4\textwidth}
\epsfig{file=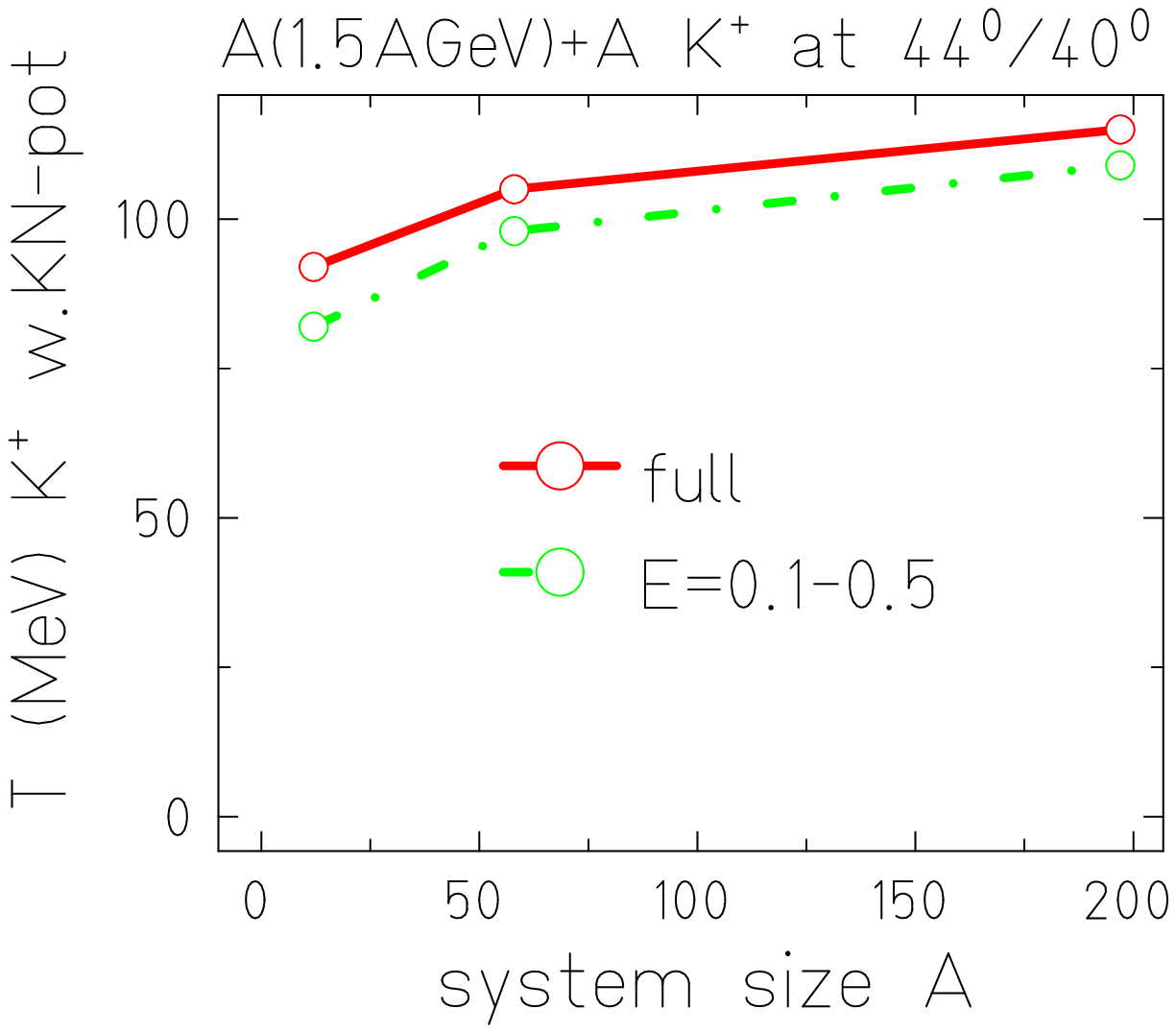,width=0.4\textwidth}
\caption{Excitation function and system size dependence of 
the temperatures at laboratory angles of $40^0$ or $44^0$. 
}
\Label{t-e-a}
\end{figure}

The \lhsref{t-e-a} shows the temperatures taken at laboratory angles
of $44^0$ (0.56, 0.78 and 0.96 GeV) or $40^0$ (1.48 GeV).
The choice  of the lab angles is guided by the angles 
taken by the experiment. All choices correspond to kaons at mid-rapidity
when going to higher laboratory momenta. We see an increase of the
temperatures with the incident energy. However, there is a slight 
difference between the temperatures fitted to the spectra at the
full energy scale and the spectra taken using an energy cut.
The system size dependence on the \rhsref{t-e-a} shows a slight
increase of the temperatures when going to larger systems.

\begin{figure}
\epsfig{file=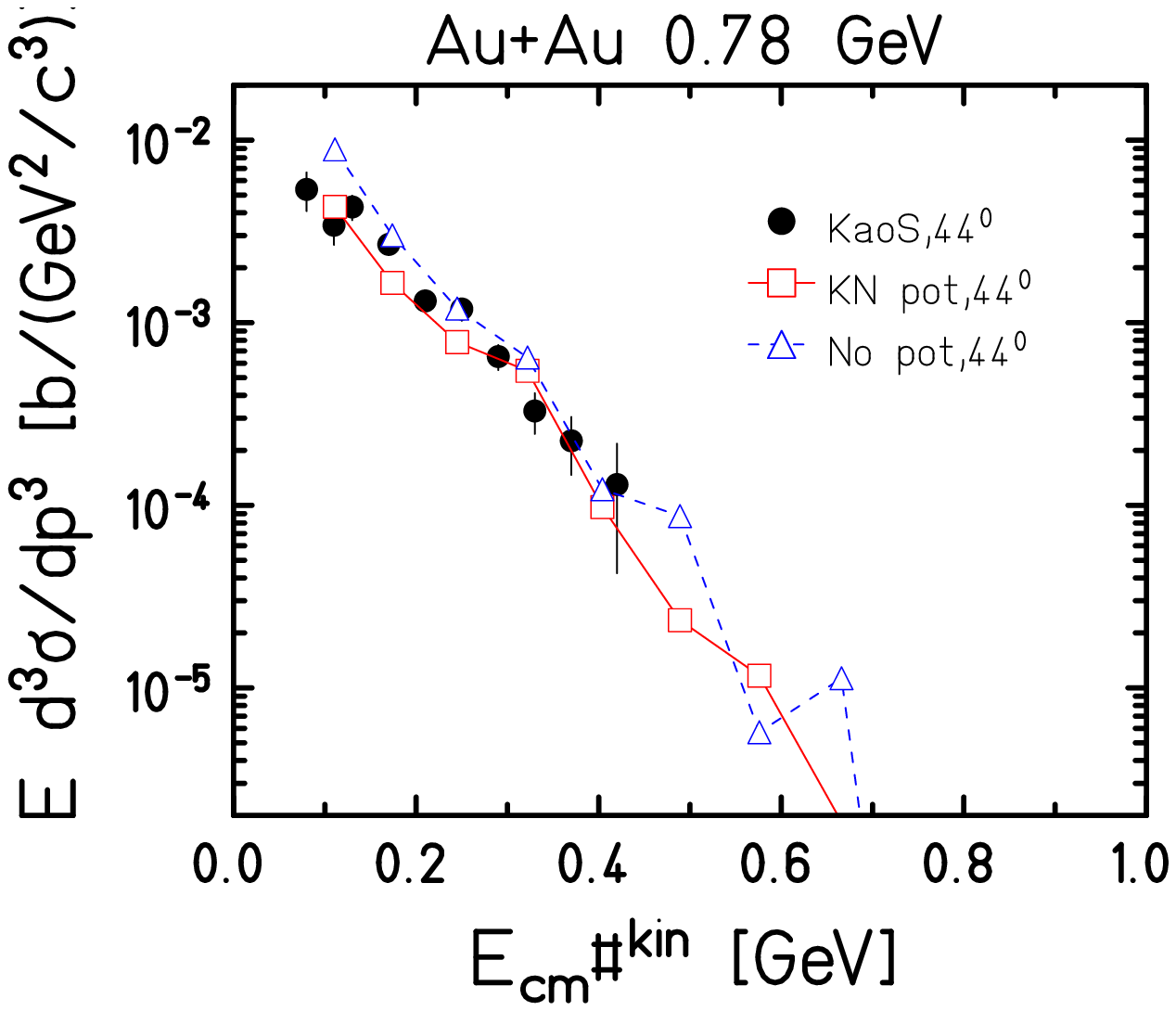,width=0.4\textwidth}
\epsfig{file=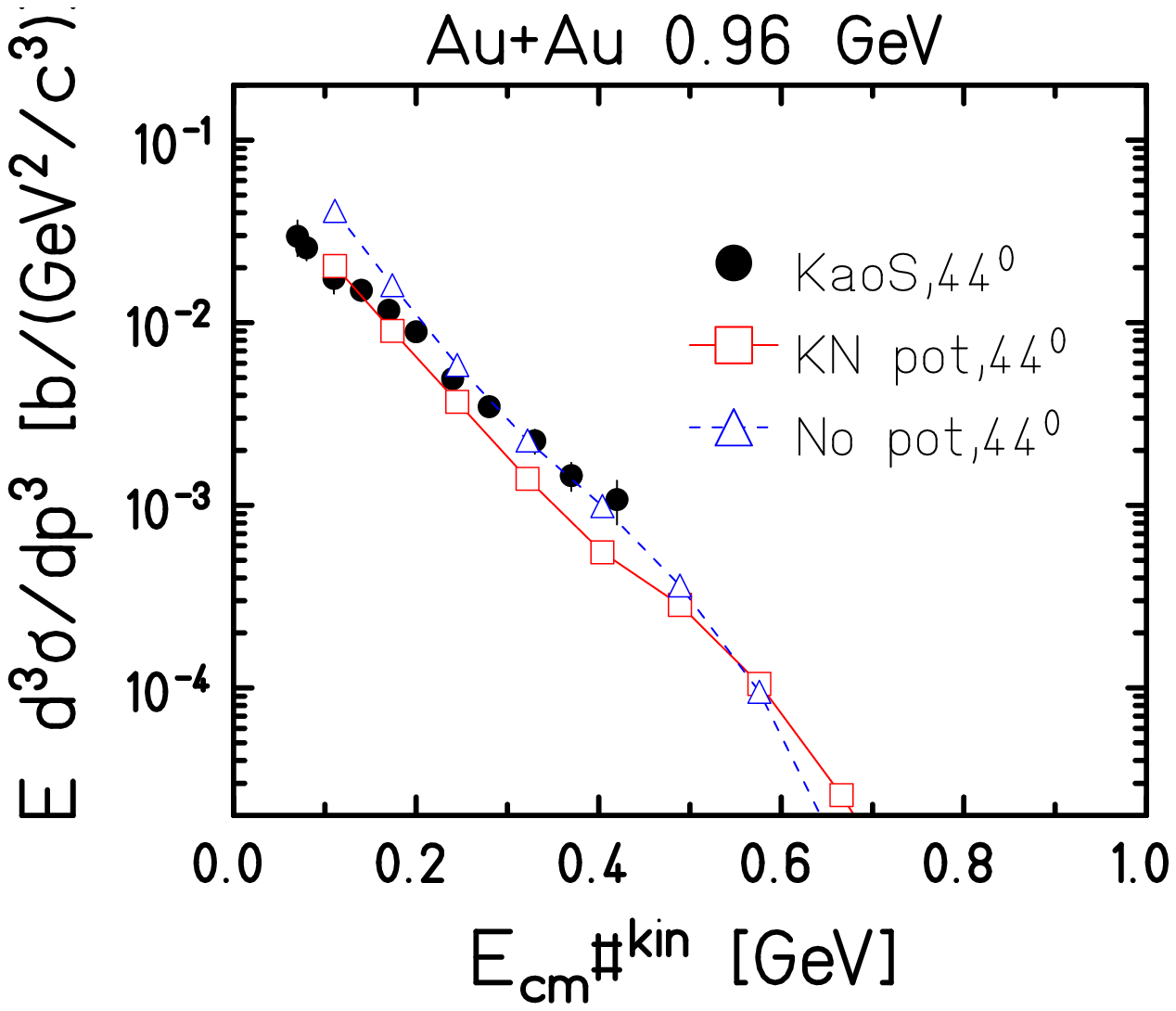,width=0.4\textwidth}
\caption{Comparison to KaoS data on Au+Au at lower energies
}
\Label{au-e-compa}
\end{figure}

In order to estimate the fidelity of this measurement we compare
the energy spectra of Au+Au collision at energies 0.78 and 0.96 AGeV.
\Figref{au-e-compa} shows comparison of IQMD data with KN potential
(\rfl) and without potential (\bdl) to data taken by the KaoS collaboration
\cite{Laue,Sturm} We see a nice agreement to the calculations with optical
potential. However the experimental temperature for the 0.96 case might
be higher.


\begin{figure}
\epsfig{file=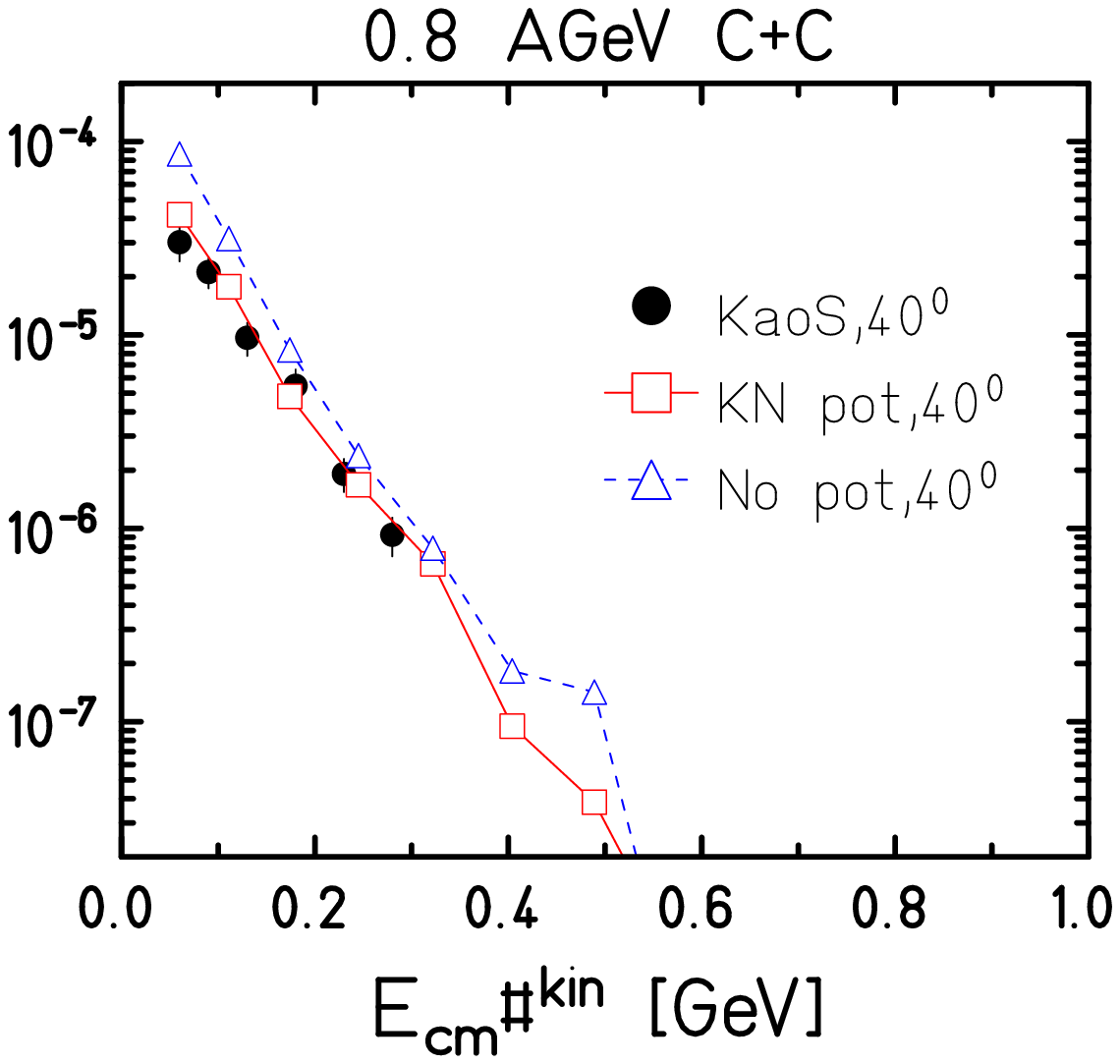,width=0.4\textwidth}
\epsfig{file=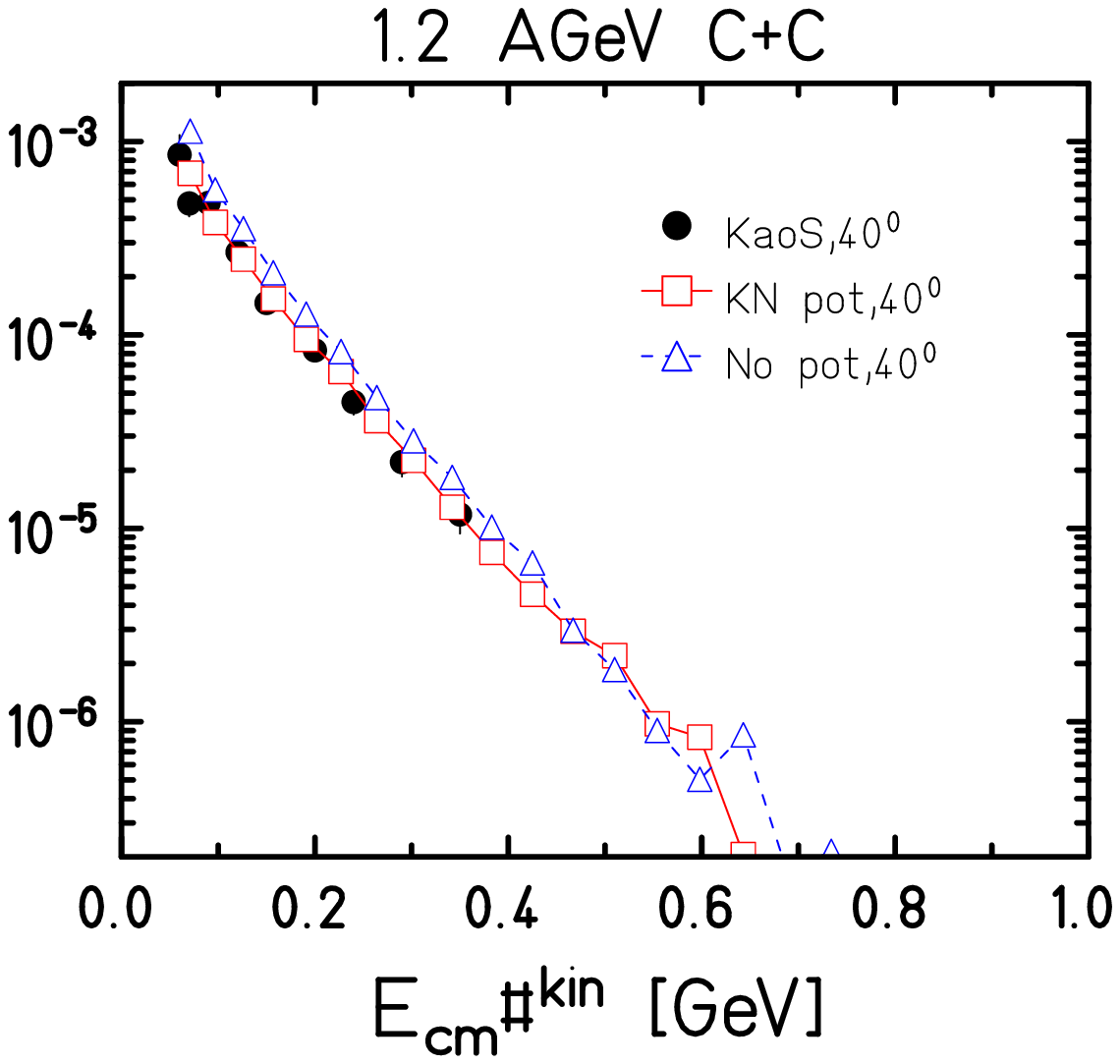,width=0.4\textwidth}
\caption{Comparison to KaoS data for C+C
}
\Label{cc-compa}
\end{figure}

\Figref{cc-compa} shows some comparisons for C+C at lower energies.
Again the calculation with potential (\rfl) seems to agree better
than the calculation without potential (\bdl).



\subsection{Spectra at different laboratory angles }
For the systems of C+C and Au at about 1.5 AGeV incident energies
several laboratory angles have been measured.

\begin{figure}[hbt]
\epsfig{file=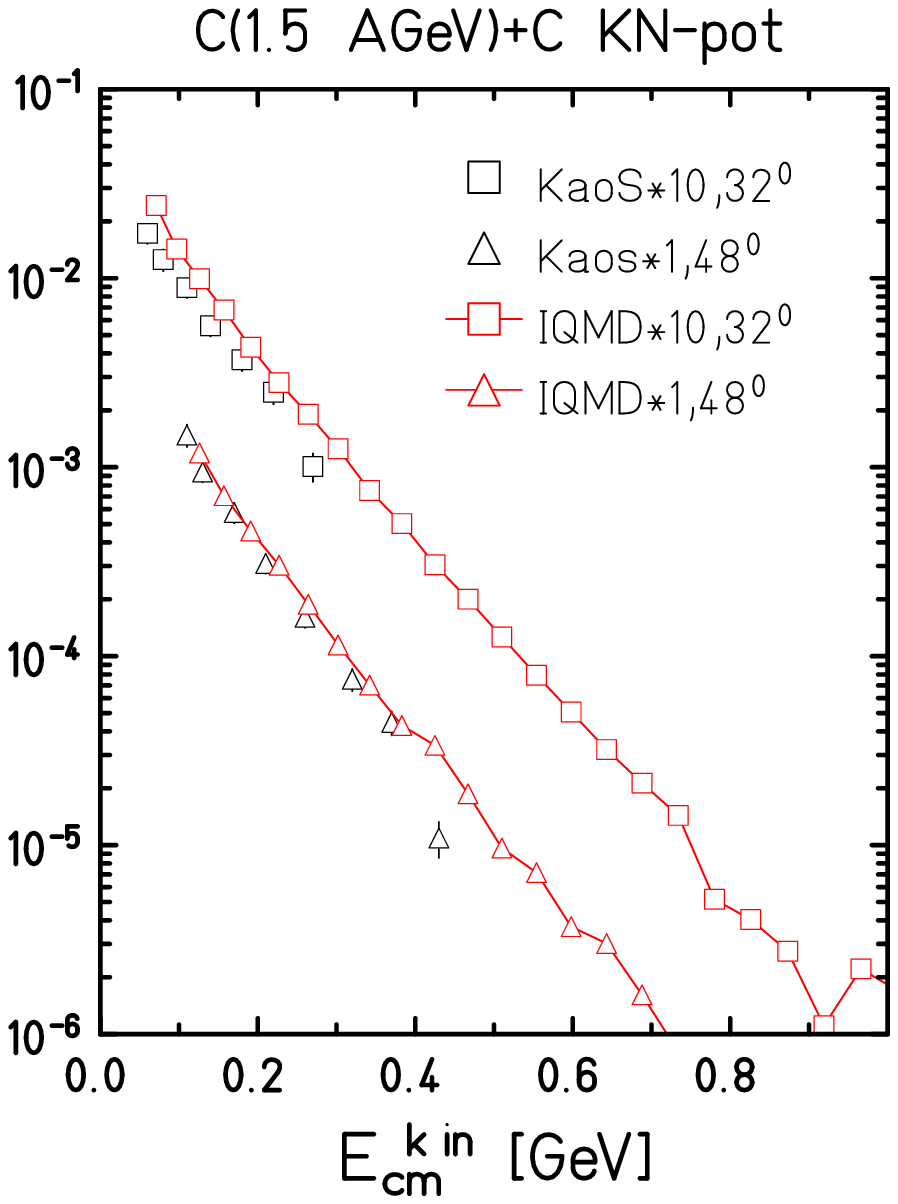,width=0.4\textwidth}
\epsfig{file=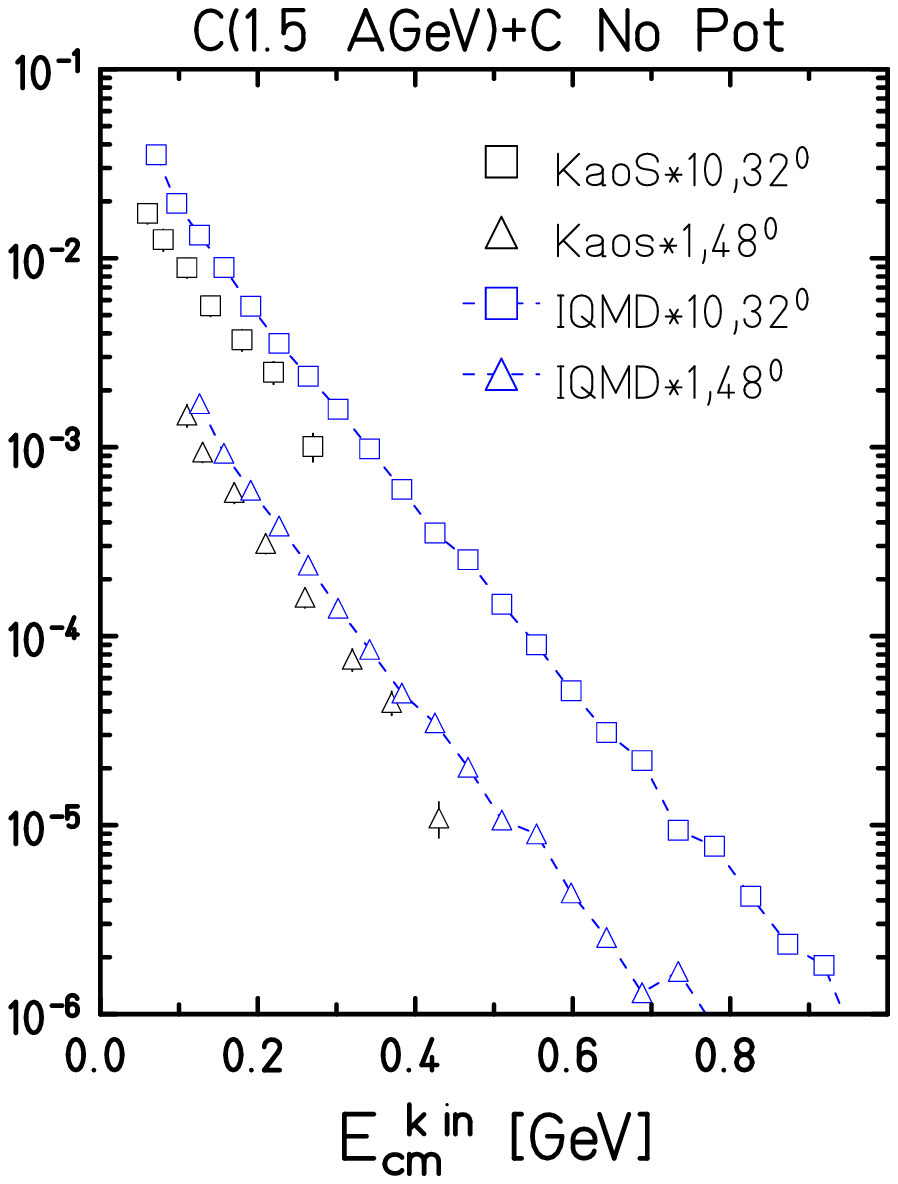,width=0.4\textwidth}
\caption{Invariant spectra measured at different laboratory angles
for C+C at 1.5 AGeV calculated with (left) and without (right) 
KN potentials compared to KaoS-data
}
\Label{angle-c}

\end{figure}
\Figref{angle-c} compares the experimental data (black symbols \cite{Sturm}) taken for
C+C at $\vartheta=32^0$ and $48^0$ to calculations of IQMD with (\lhs) and
without (\rhs) KN potential. The calculation with KN potential is nearer
to the data than the calculation without potential. 


\begin{figure}[hbt]
\epsfig{file=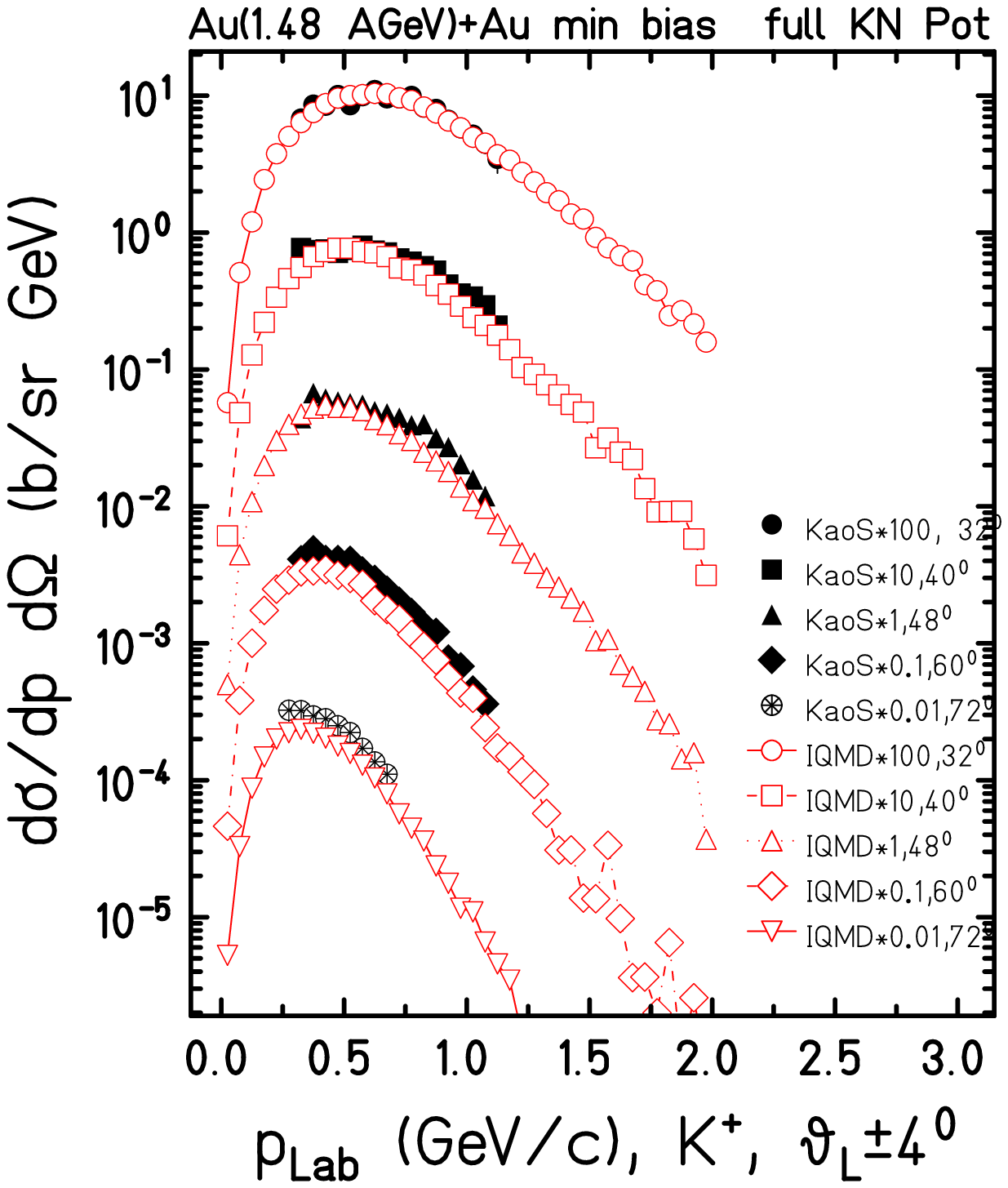,width=0.4\textwidth}
\epsfig{file=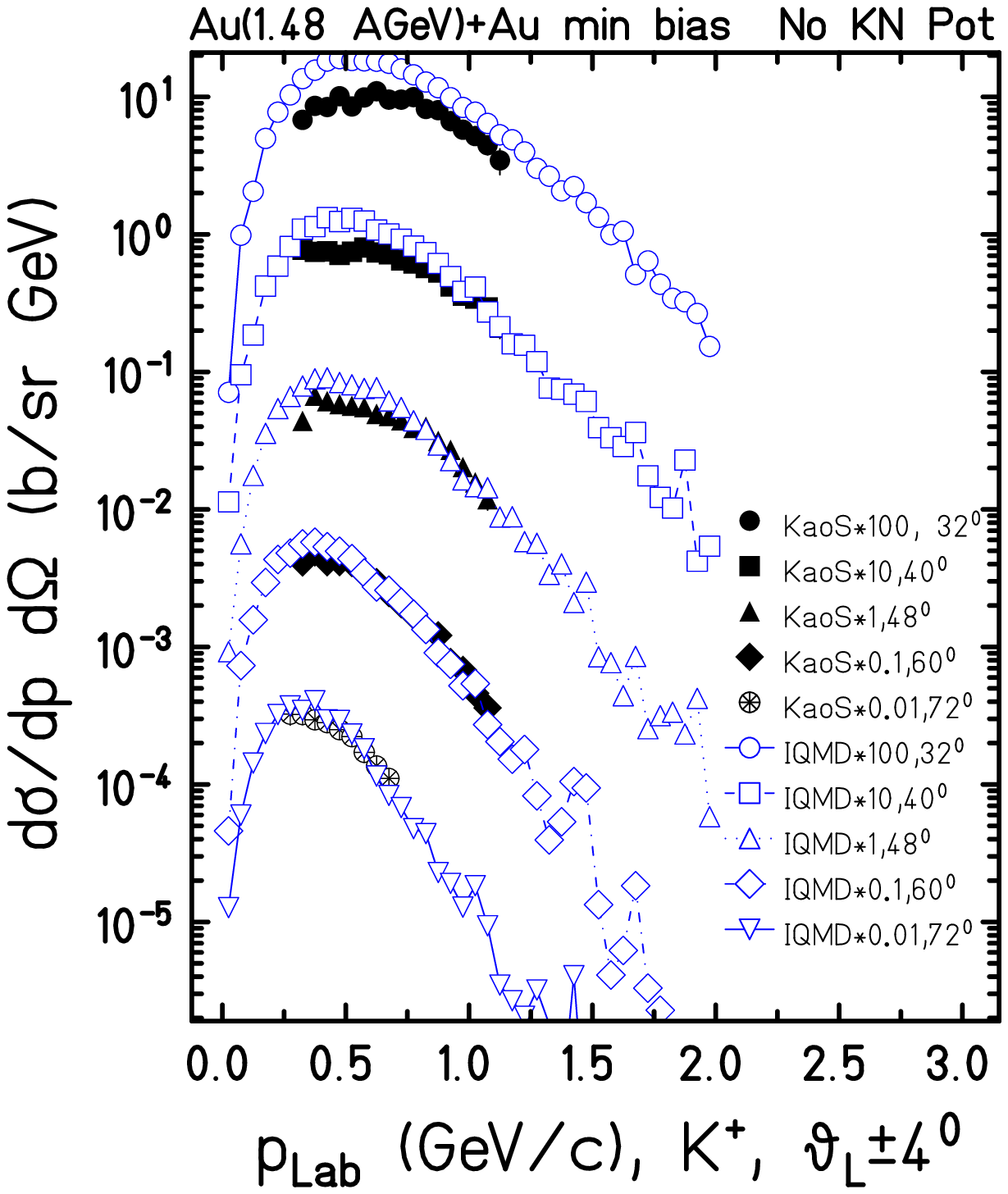,width=0.4\textwidth}
\caption{Momentum spectra measured at different laboratory angles
for C+C at 1.5 AGeV calculated with (left) and without (right) 
KN potentials compared to KaoS-data
}
\Label{angle-au}
\end{figure}

For the Au+Au data we directly compare in \figref{angle-au} the 
laboratory momentum spectra taken at different angles to IQMD
calculations. The data are taken by the KaoS collaboration \cite{foerster}.
 The \lhs\ shows calculations with optical potential, the \rhs\ calculations
without such a potential. We see that it is hard to decide which calculation
fits better since the calculation with potential can better describe
the data at smaller angles while the data at larger angles are better
described by the calculation without potential. Overall IQMD has some problems
in describing the full angular dependence. 

\begin{figure}[hbt]
\epsfig{file=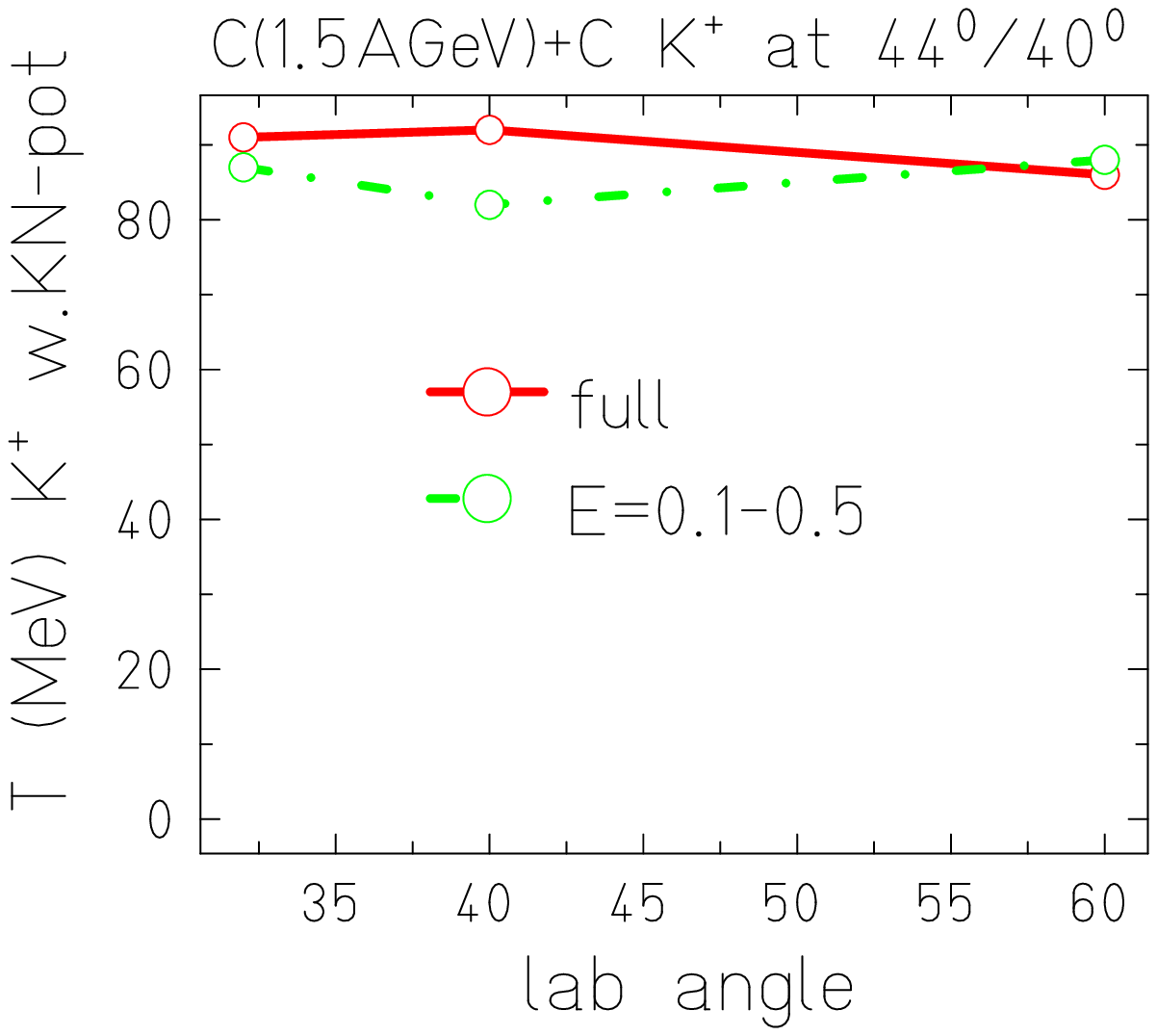,width=0.4\textwidth}
\epsfig{file=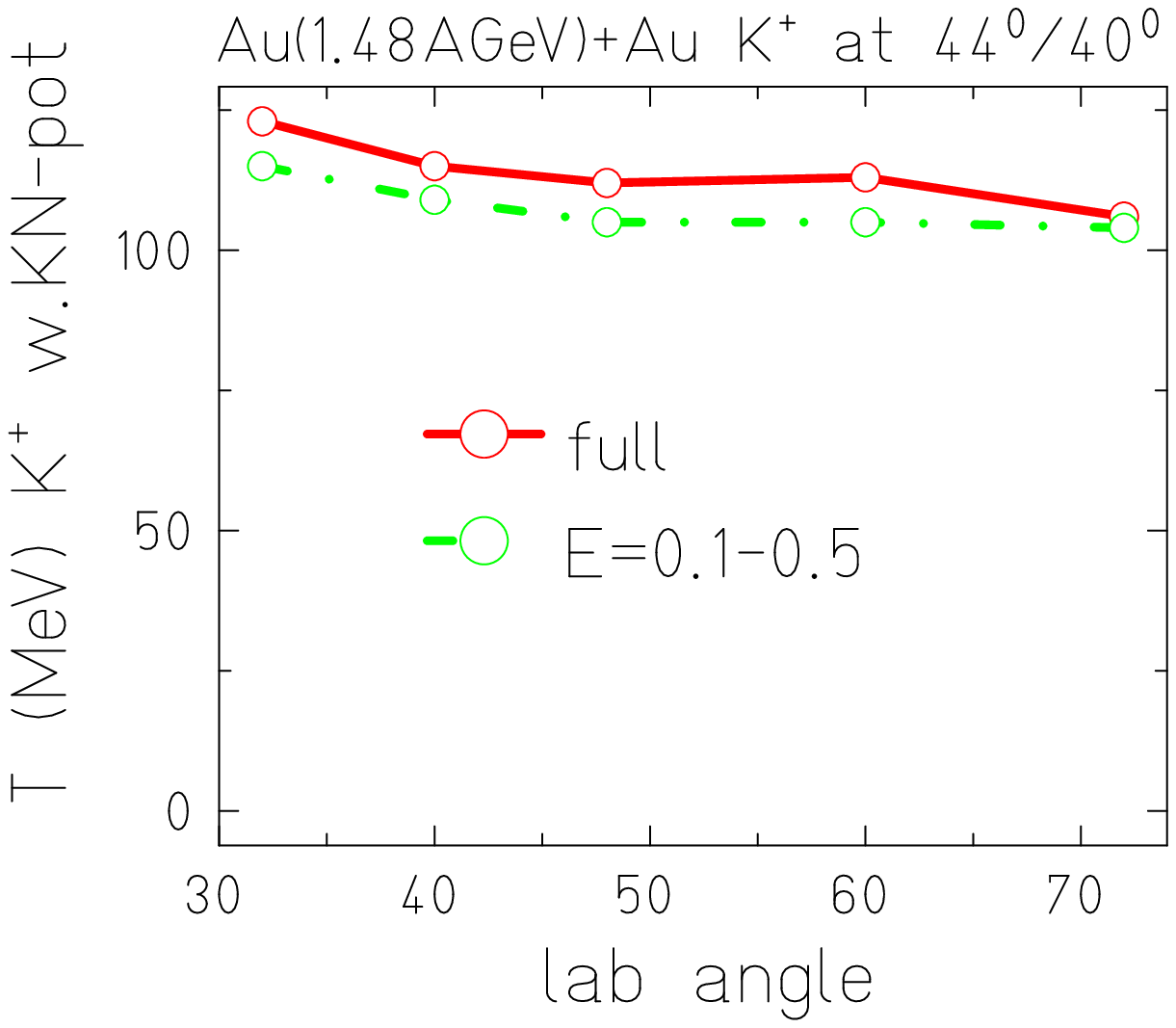,width=0.4\textwidth}
\caption{Temperatures of spectra measured at different laboratory angles
for C+C and Au+Au at 1.5 AGeV
}
\Label{t-theta}
\end{figure}

\Figref{t-theta} shows the temperatures deduced from the spectra taken
at different angles for C+C (\lhs) and Au+Au (\rhs). For the latter we see a slight 
dependence of the spectra on the lab angle. This should not be the case for
an ideal isotropic thermal source. 
Therefore, an analysis of polar distribution would be interesting in order to 
see whether the source is isotropic or not.


\section{Dynamical observables and anisotropic distributions}

\subsection{Polar distributions}
We now investigate the polar distributions $dN/d\cos\vartheta$, 
where $\vartheta=p_Z/p$ is the angle of the momentum vector to the beam-axis
and $d\cos\vartheta = - \sin\vartheta d\vartheta$ includes the Jacobian of
a polar distribution in spherical coordinates. 
For a better comparison of the anisotropies of the kaons in IQMD calculations
to experiment \cite{foerster} we divide the minimum bias (i.e. all $b$) events into two
centrality bins $b<5.9$fm and $b>5.9$fm and fit the polar distributions
by
\begin{equation}
\frac{dN}{d\cos\vartheta} = a_0 \cdot \bigl ( 1 + a_2 \cos^2 \vartheta \bigr )
\end{equation}

\begin{figure}[hbt]
\epsfig{file=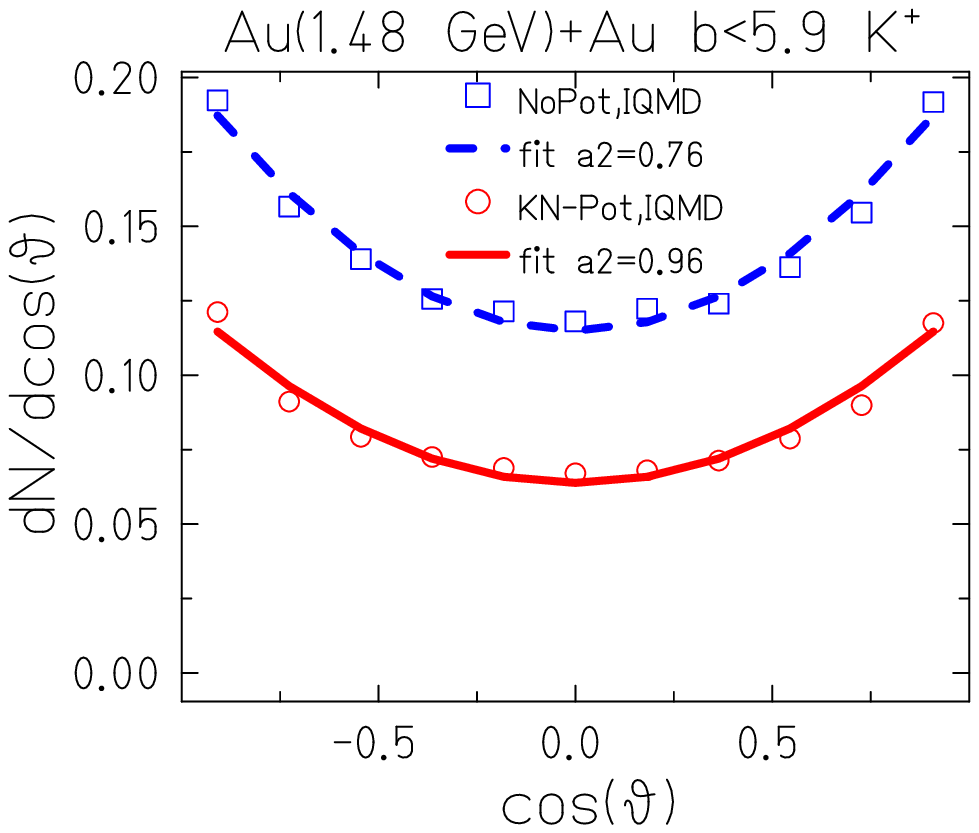,width=0.4\textwidth}
\epsfig{file=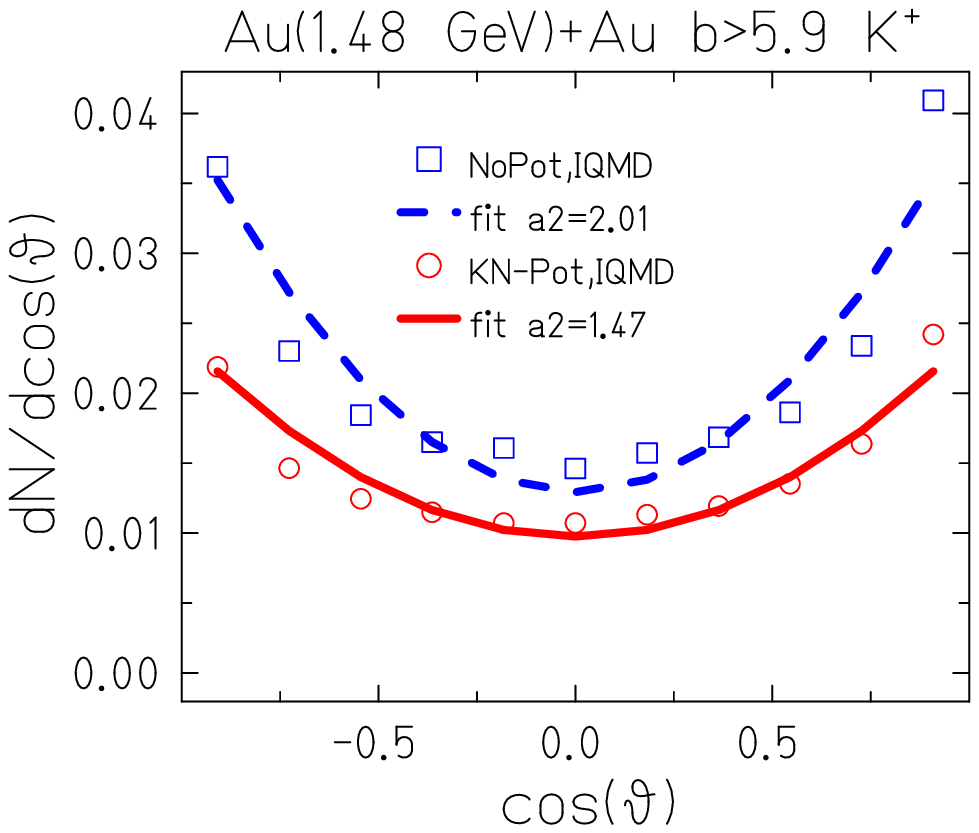,width=0.4\textwidth}
\caption{Polar distributions
}
\Label{au15-a2}
\end{figure}
\Figref{au15-a2} shows the corresponding distributions (symbols) and the
fits (lines) for the semi-central bin ($b<5.9$fm, \lhs) and for the
peripheral bin ($b>5.9$fm, \rhs).
The calculations with potential are represented by red circles and their
corresponding fit functions by red solid lines. 
The results of the fits are summarized in table \ref{a2-list}.

\begin{table}[hbt]
\begin{tabular}{|c|c|c|}
\hline 
centrality & $b<5.9$ fm & $b>5,9$ fm \\
\hline
with KN pot & 0.96 & 1.47 \\
without pot & 0.76 &2.01 \\
\hline
KaoS data & 1.1 & 1.9 \\
\hline
\end{tabular}
\caption{Comparison of the fit parameters to the angular distributions.}
\label{a2-list}
\end{table}

For semi-central collisions the calculations with the KN potential is closer
to the data, while it is the opposite case for peripheral collisions.
However, the IQMD-fits for peripheral collisions are dominated by the extreme
bins at $\cos\vartheta \approx \pm 1$ while the other bins comply with even
lower values. 
Thus, the polar distribution in peripheral collisions seems not be described
well neither with nor without potential.

We shall therefore focus on the semi-central collisions and investigate the
question, which effects are responsible for the anisotropy.

\begin{figure}[hbt]
\epsfig{file=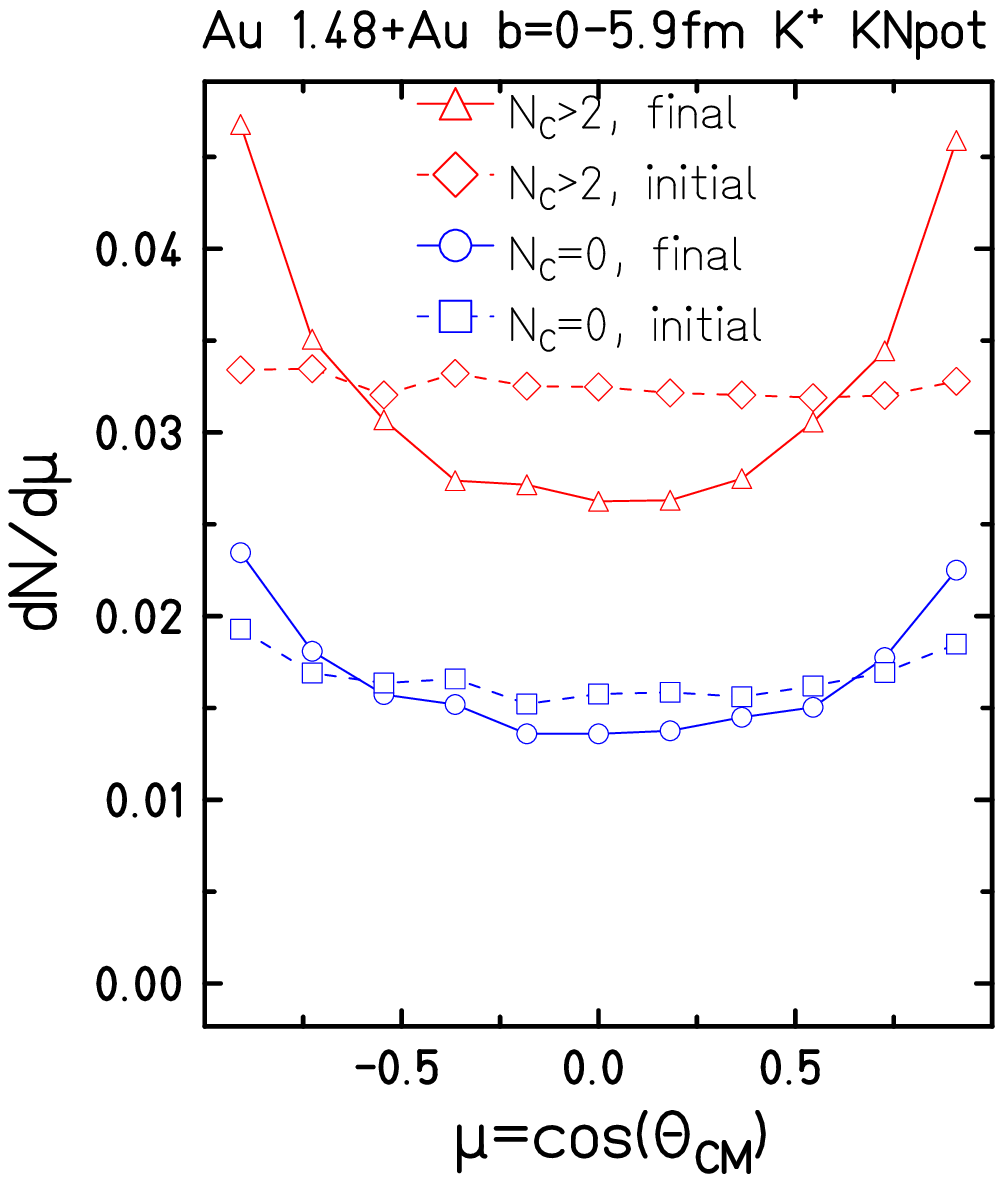,width=0.4\textwidth}
\epsfig{file=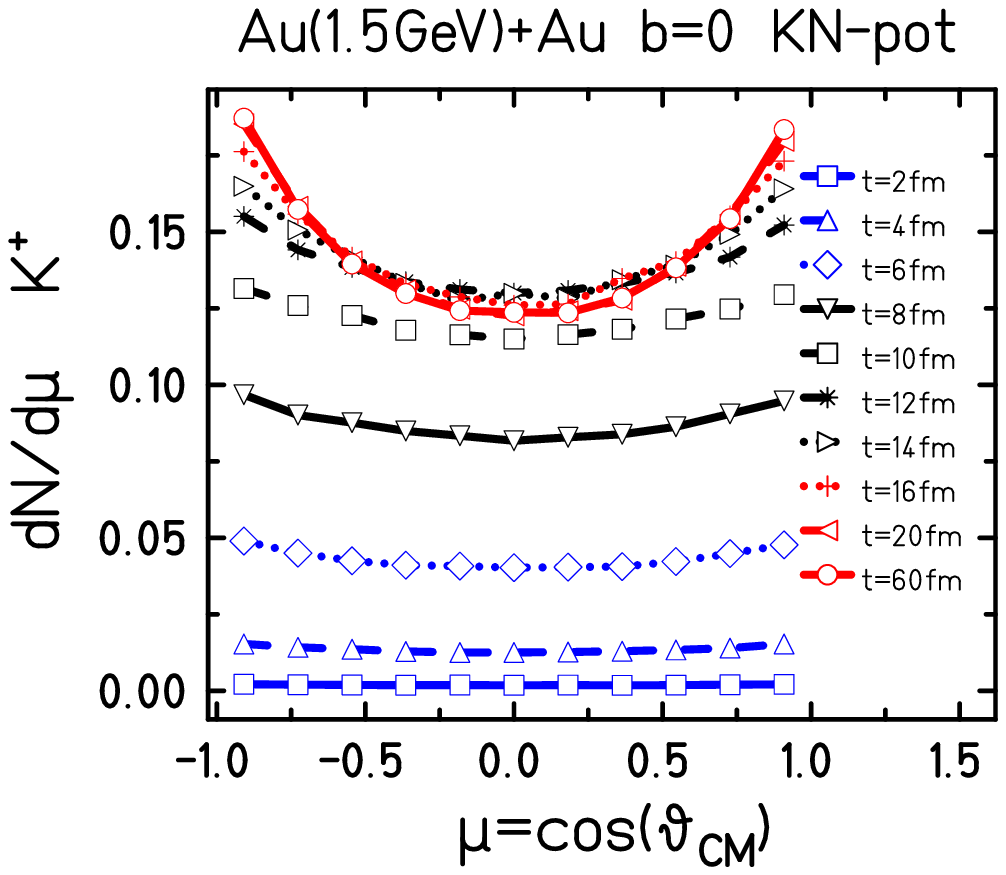,width=0.4\textwidth}
\caption{Time evolution of polar distributions
}
\Label{au15-tim-ct0}
\end{figure}
The \lhsref{au15-tim-ct0} shows the polar distributions of kaons
directly after their production (initial, dashed lines) and in the
final state (final, solid lines). We selected kaons which did not
collide (blue lines) and kaons which collided frequently.
The initial distributions are rather flat for both types of kaons.
For kaons without collisions there is a slight forward-backward enhancement
due to preselection effects. Kaons have some more chance to come out in forward-backward
direction without having a collision than at $\vartheta=90^0$. 
These kaons show a stronger forward-backward enhancement in the final
state. This enhancement can only be true due to the optical potential
since the kaons did not undergo any collision. Thus there is a visible
effect of the potential to the anisotropy, which may explain the different
fit parameters of the calculations with and without potential.

The multicolliding kaons (red lines) show a higher total yield and thus a
higher total contribution to the kaon distribution. 
Furthermore they show a much stronger forward-backward peaking.We will
explain this effect a few lines later.

The \rhsref{au15-tim-ct0} shows snapshots of the actual polar kaon distribution
at different time steps. We see that the total yield enhances to about
$t=12$ fm/c without showing a very strong anisotropy. Only after the production
of all kaons at times $t=12-20$ fm/c the distribution is getting its
strong anisotropy. This is the time when all kaons are already produced 
but still collide with the nuclear matter.


\begin{figure}[hbt]
\epsfig{file=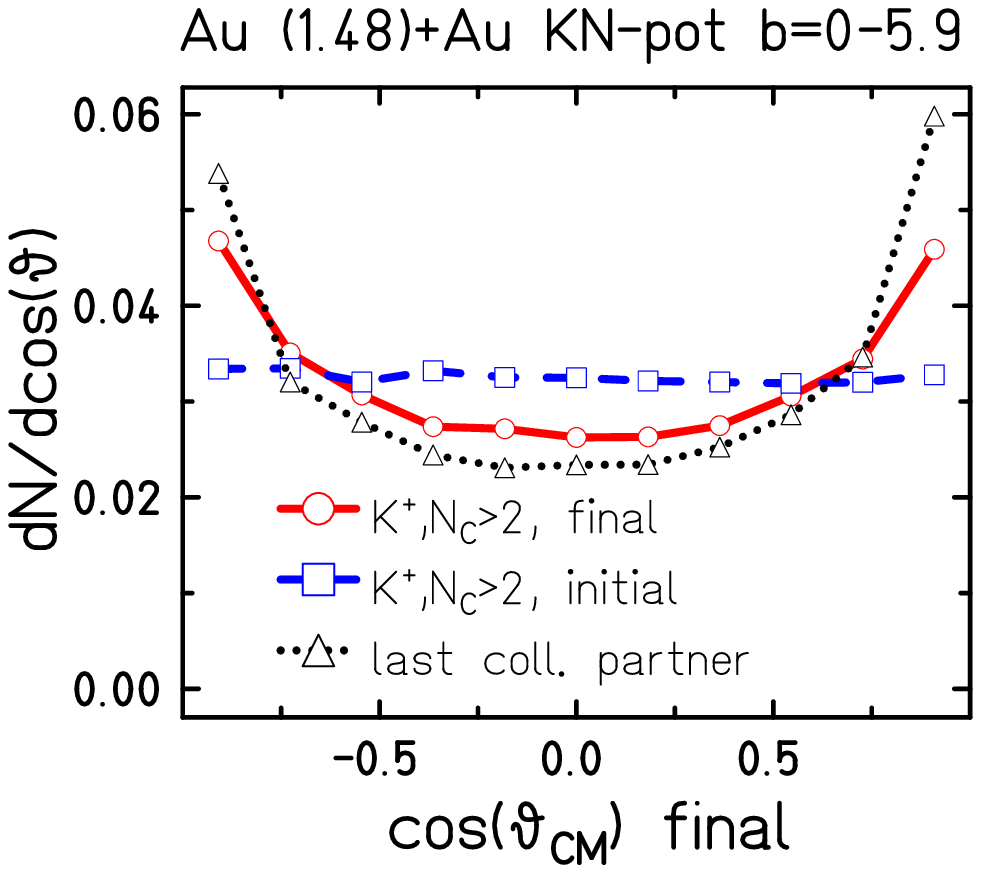,width=0.4\textwidth}
\epsfig{file=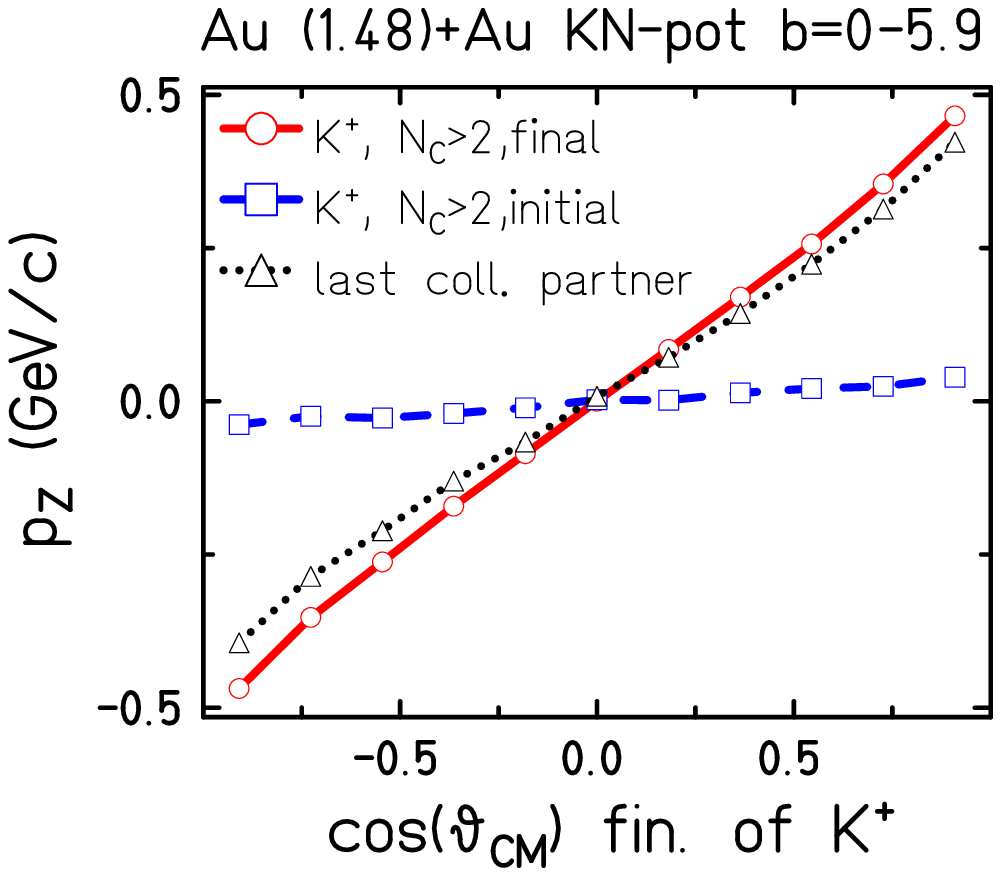,width=0.4\textwidth}
\caption{Contribution of the last collision partner
to the the polar azimuthal anisotropy
}
\Label{au15-col-ct0}
\end{figure}

\Figref{au15-col-ct0} addresses the question which effect provokes
the strong anisotropy for the multicollisionner particles.
On the \lhs\ we compare the (normalized) polar distributions of the kaons
directly after their production (\bdl), in the final state (\rfl)
and the distribution of the last collision partners of those particles
(\bpl). We see that the latter have a very strong forward-backward peaking,
even stronger than those of the kaons in the final state. 
The \rhsref{au15-col-ct0} show the mean longitudinal momentum of
the final kaons (\rfl), the kaons directly after their production (\bdl)
and of the last collision partners (\bpl) as a function of the final
polar angle of the kaon. 
The mean $p_Z$ of the final state (\rfl) is easy to understand, since it is 
directly correlated to the polar angle by $\cos\vartheta=p_Z/p$.
The mean $p_Z$ of the initial state after the production (\bdl) is zero. This means
that there is no preselection from the production where the particle would
go finally. Thus, particles ending in forward direction may have been produced
with any longitudinal momentum initially.
The mean $p_Z$ of the last collision partners (\bpl) is nearly the same
than the mean final $p_Z$. This supports the assumption that their   
longitudinal momentum is responsible for the final direction of the kaon.
We see that the kaons have collided with a nuclear medium, which is very
anisotropic in momentum space.
The anisotropy is thus majorly due to the rescattering.

The influence of rescattering may also be a key for explaining the
dependence of the polar distribution on the centre-of mass momentum
as depicted on the \lhsref{au15-ct-aa}.
We know that the contribution of kaons having undergone many collisions is 
higher for high energy kaons than for low energy kaons. Therefore they might
yield a slight enhancement of the anisotropy at high momenta.

\begin{figure}[hbt]
\epsfig{file=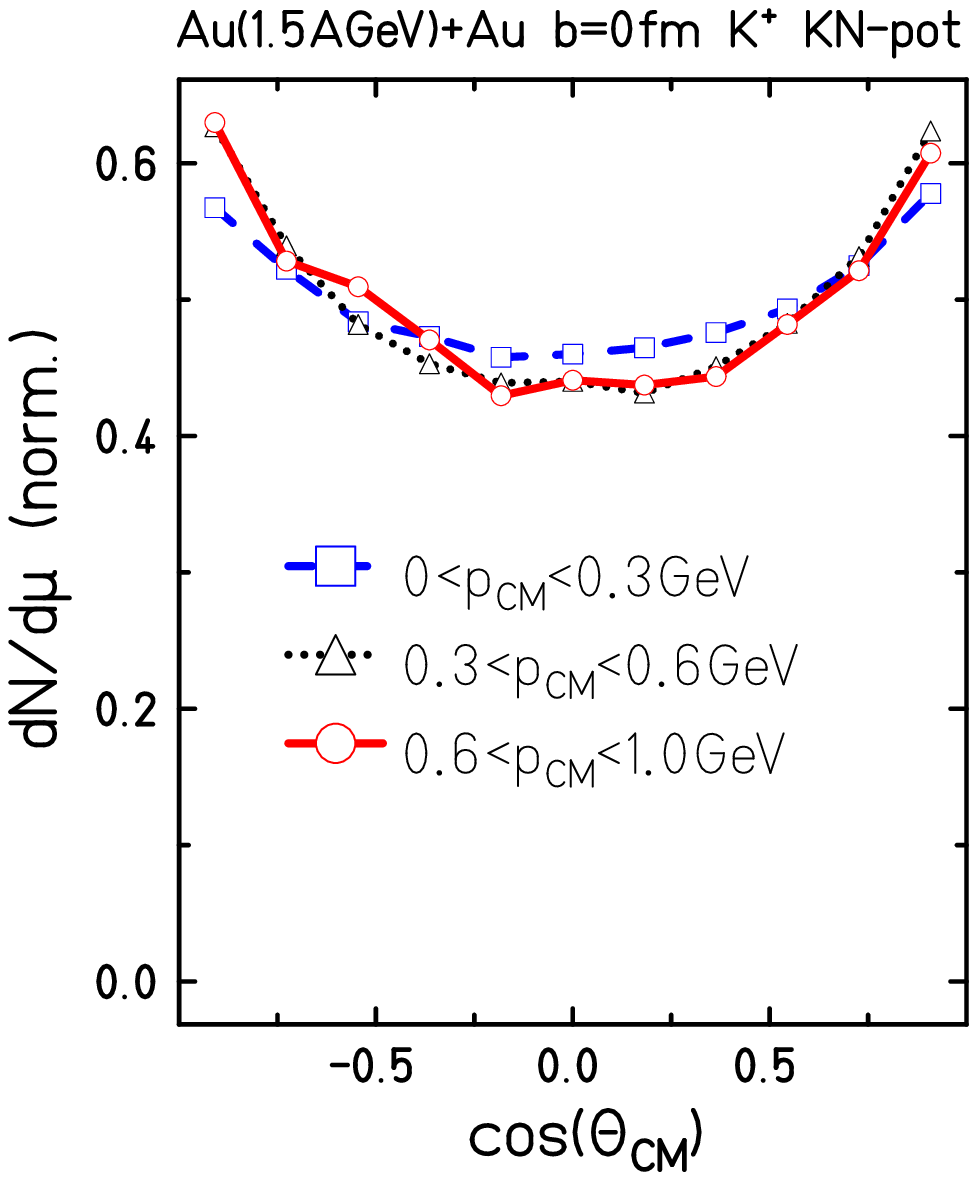,width=0.4\textwidth}
\epsfig{file=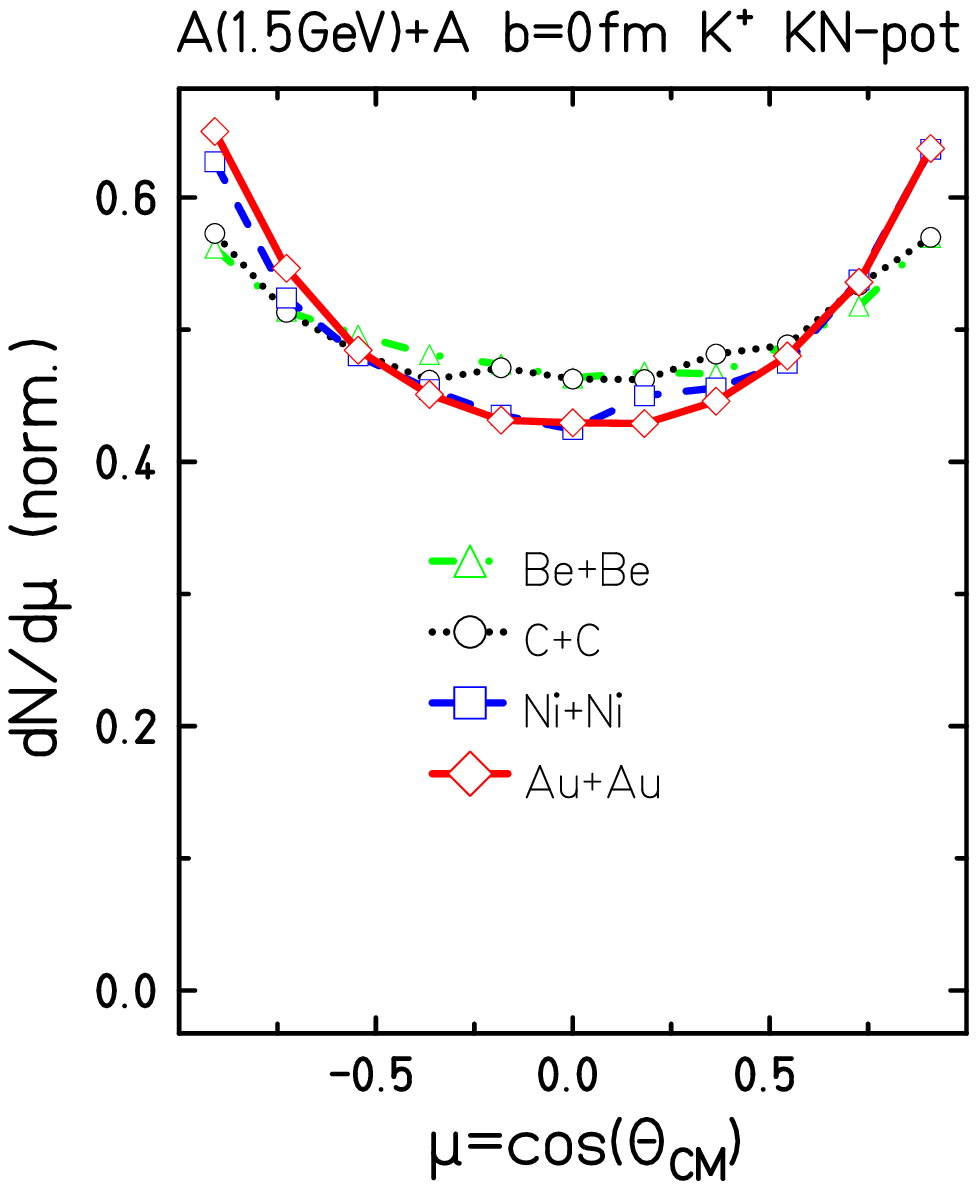,width=0.4\textwidth}
\caption{Polar distributions of Au+Au for different cm spectra (left)
and polar distributions for different system sizes
}
\Label{au15-ct-aa}
\end{figure}

\subsection{Energy and system size dependence of polar distributions}

Let us shortly discuss the energy and system size dependence 
of polar distributions. The \rhsref{au15-ct-aa} 
shows polar distributions of $A+A$ collisions at different system
sizes. We see an enhancement of the anisotropy for heavier systems.
This can again be explained by rescattering. In heavier systems
there is much more rescattering of the kaons. 
A similar argument holds for the energy dependence of the polar distributions
shown in \figref{au15-ct-e} for C+C (left) and Au+Au (right). 
At higher incident energies the nucleons show a stronger anisotropy especially
for light systems. In the rescattering of the kaons they transfer some
part of that anisotropy to the kaons.

\begin{figure}[hbt]
\epsfig{file=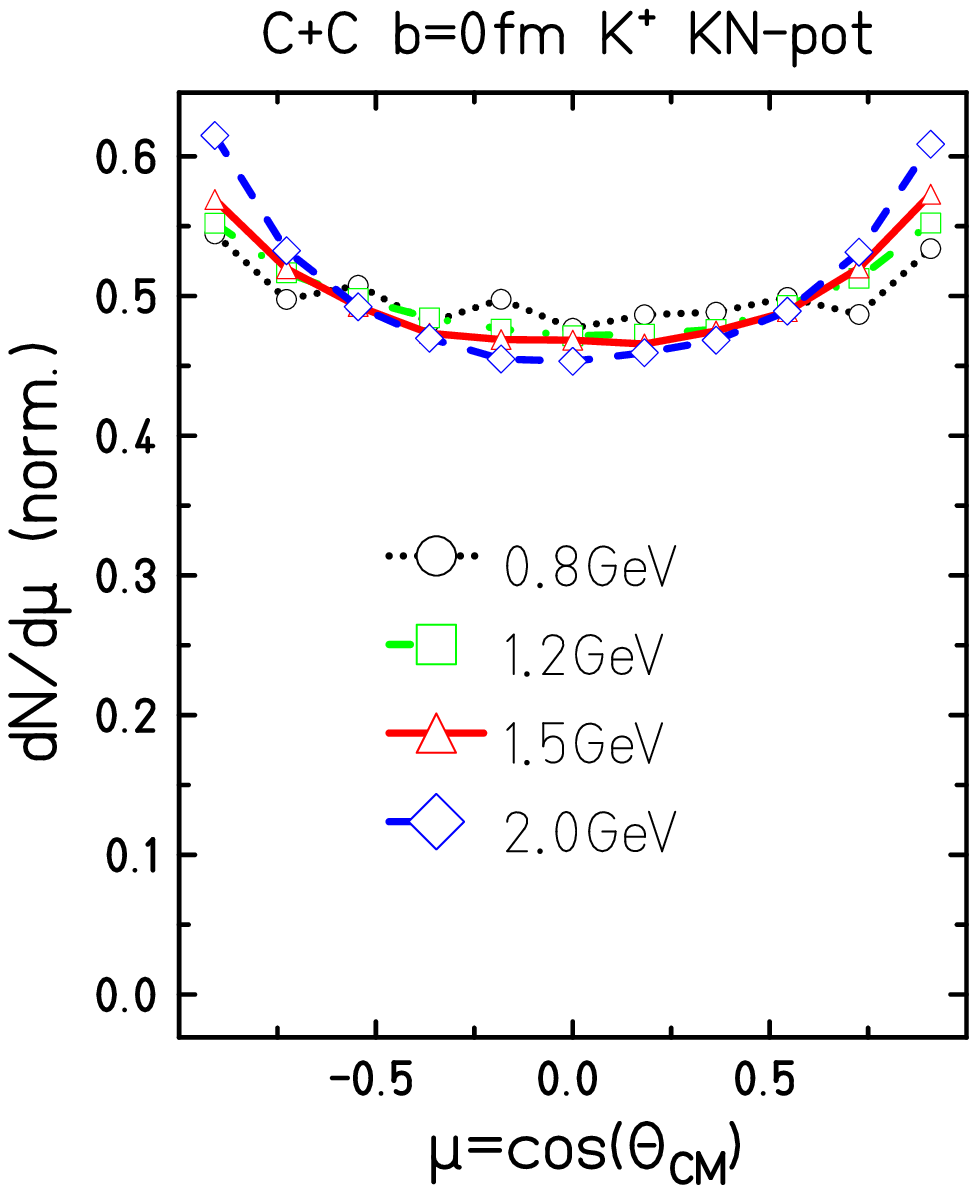,width=0.4\textwidth}
\epsfig{file=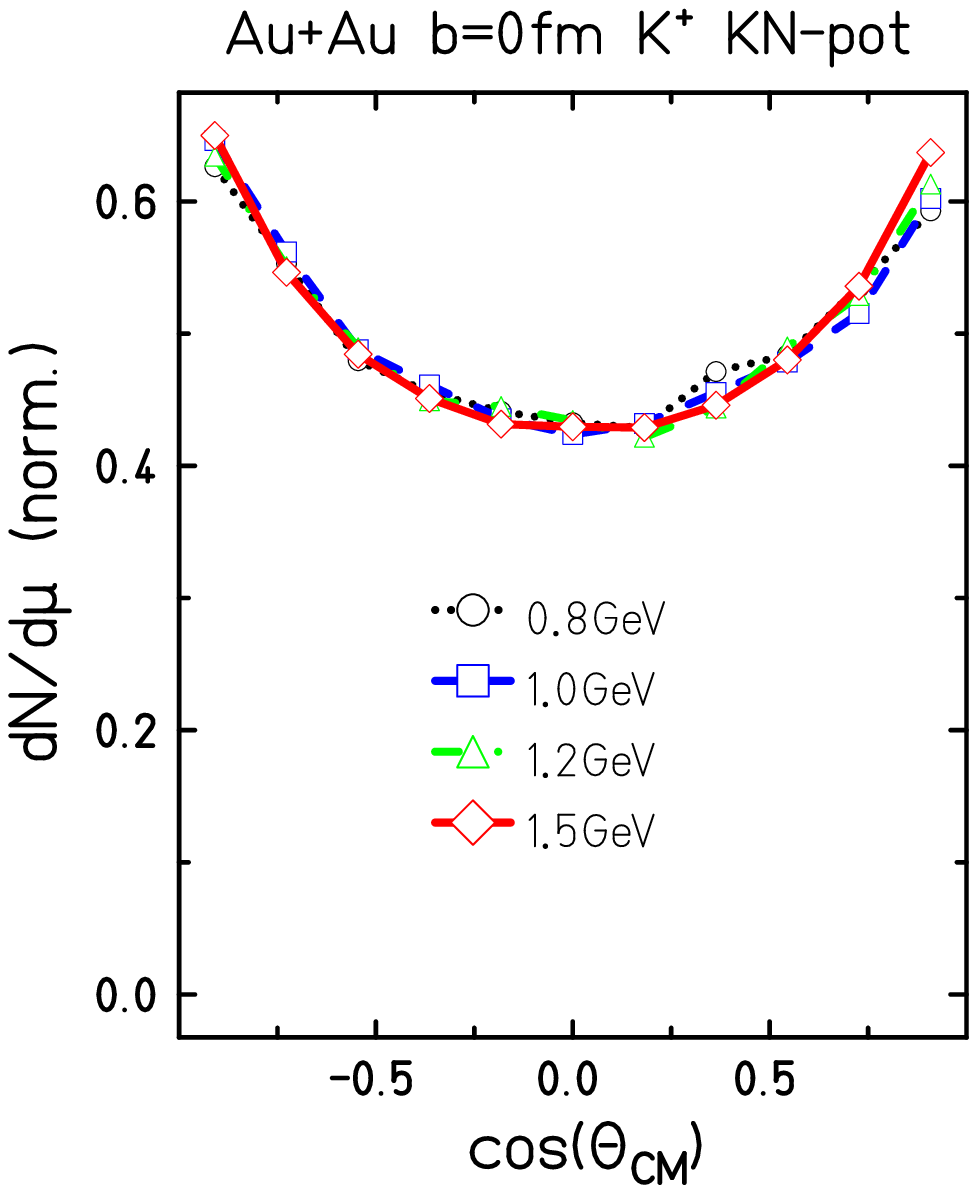,width=0.4\textwidth}
\caption{Polar distributions of C+C (left) and  Au+Au (right)
at different energies} 
\Label{au15-ct-e}
\end{figure}

\subsection{Azimuthal distributions}
After the investigation of the polar angle distributions let us finally study
the azimuthal angular distribution at mid-rapidity.
$\varphi$ is the angle between the transverse momentum $\vec{p_T}$ and
the x-axis, where the x-axis is the direction of the impact parameter
between projectile and target. 

\begin{figure}
\epsfig{file=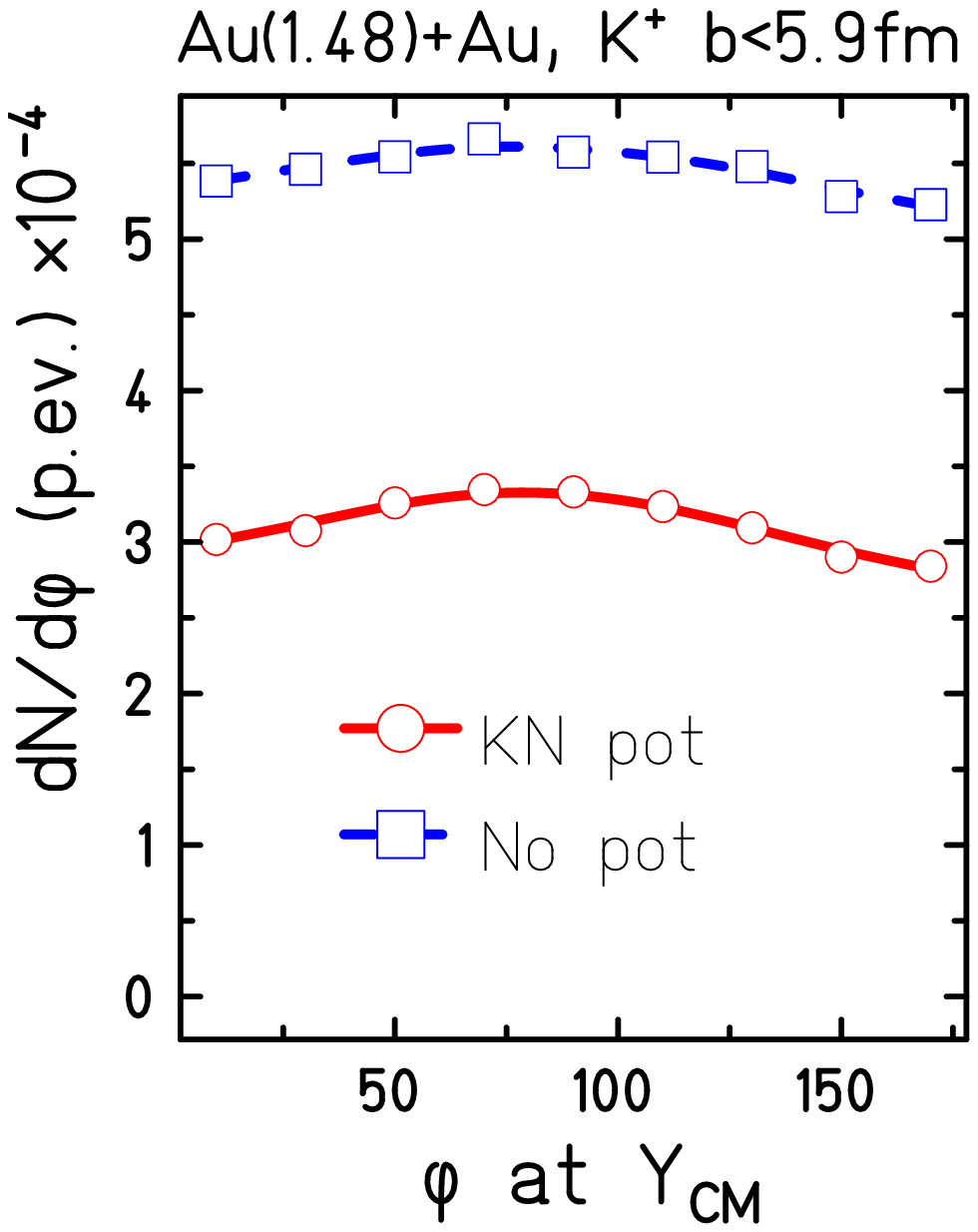,width=0.4\textwidth}
\epsfig{file=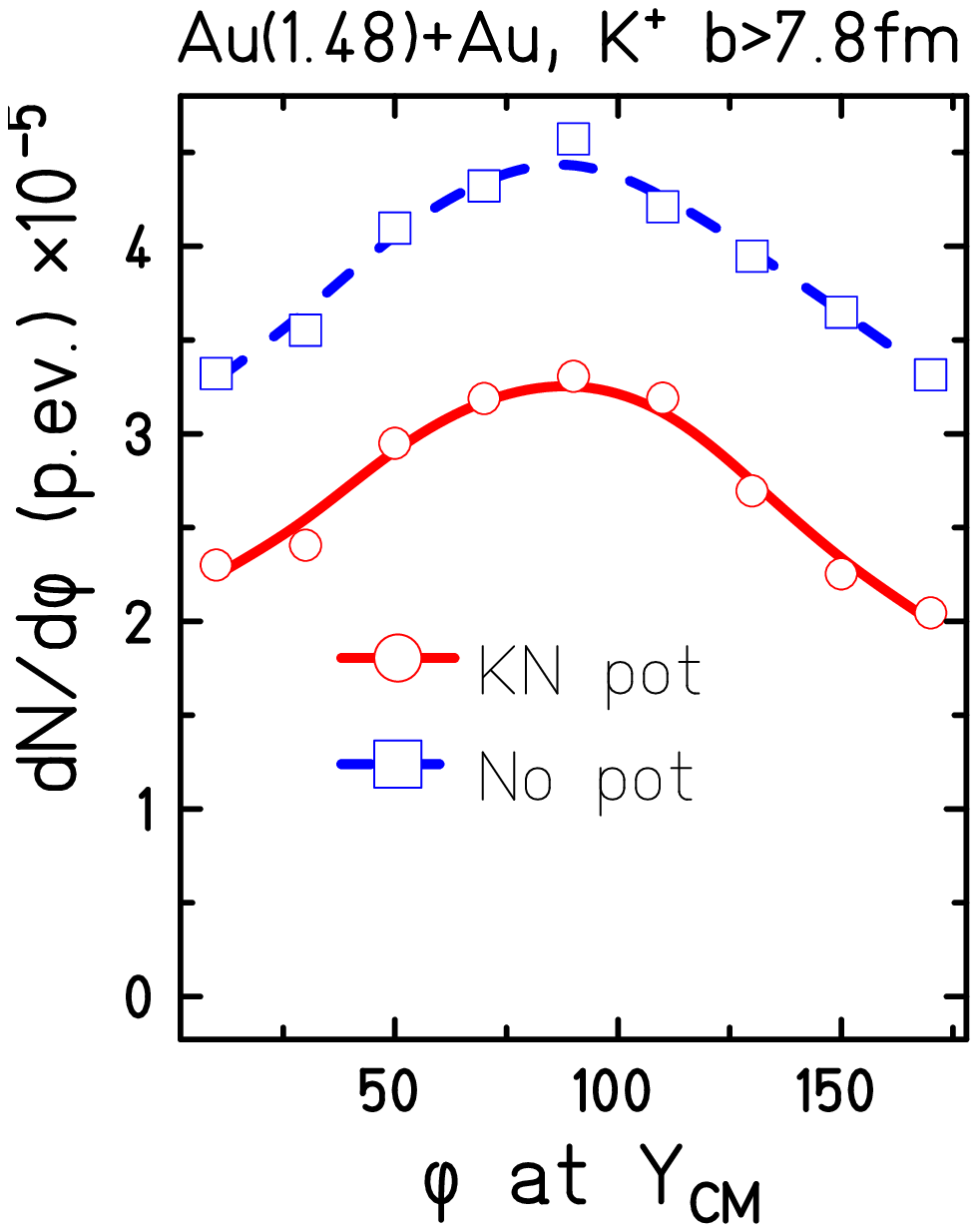,width=0.4\textwidth}
\caption{Azimuthal distribution for central and peripheral
collisions.
}
\Label{au15-phi-b}
\end{figure}
\Figref{au15-phi-b} shows the azimuthal distribution of the kaons at mid-rapidity
for a semi-peripheral centrality ($b<5.9$fm, \lhs) and for a peripheral
centrality bin ($b>7.8$, \rhs). We see that calculations with an optical
potential (\rfl) yield less absolute yields than calculations without potential
(\bdl) and that the distributions for semi-central events are much less
pronounced than those of the peripheral events. The latter finding is also
supported by experiment \cite{foerster}. 
We will thus concentrate on the peripheral events and analyze 
only normalized distributions to investigate the significance of the
out-of plane enhancement.
The strength of the effect is in agreement with experimental data \cite{sqm2003}.


\begin{figure}[hbt]
\epsfig{file=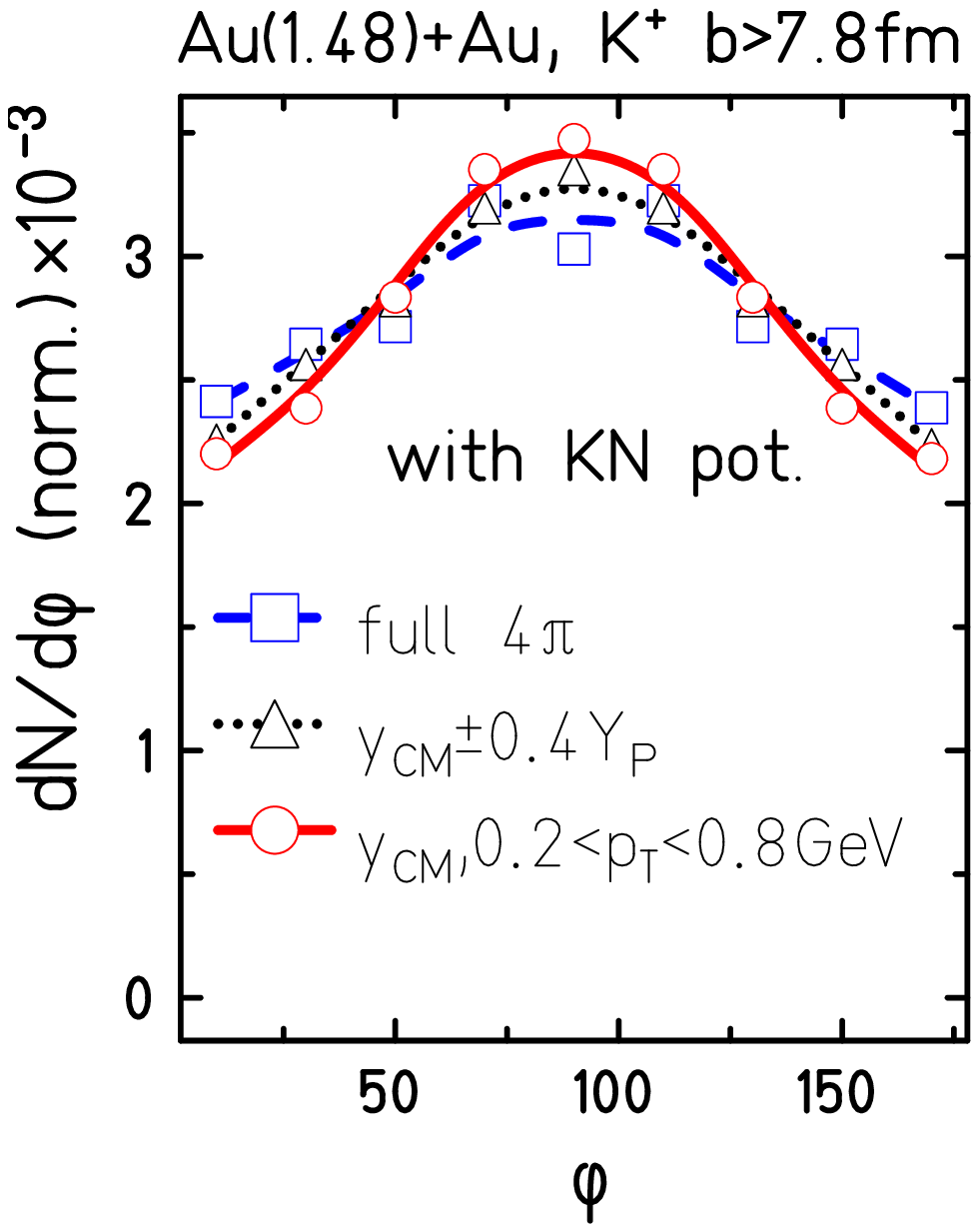,width=0.4\textwidth}
\epsfig{file=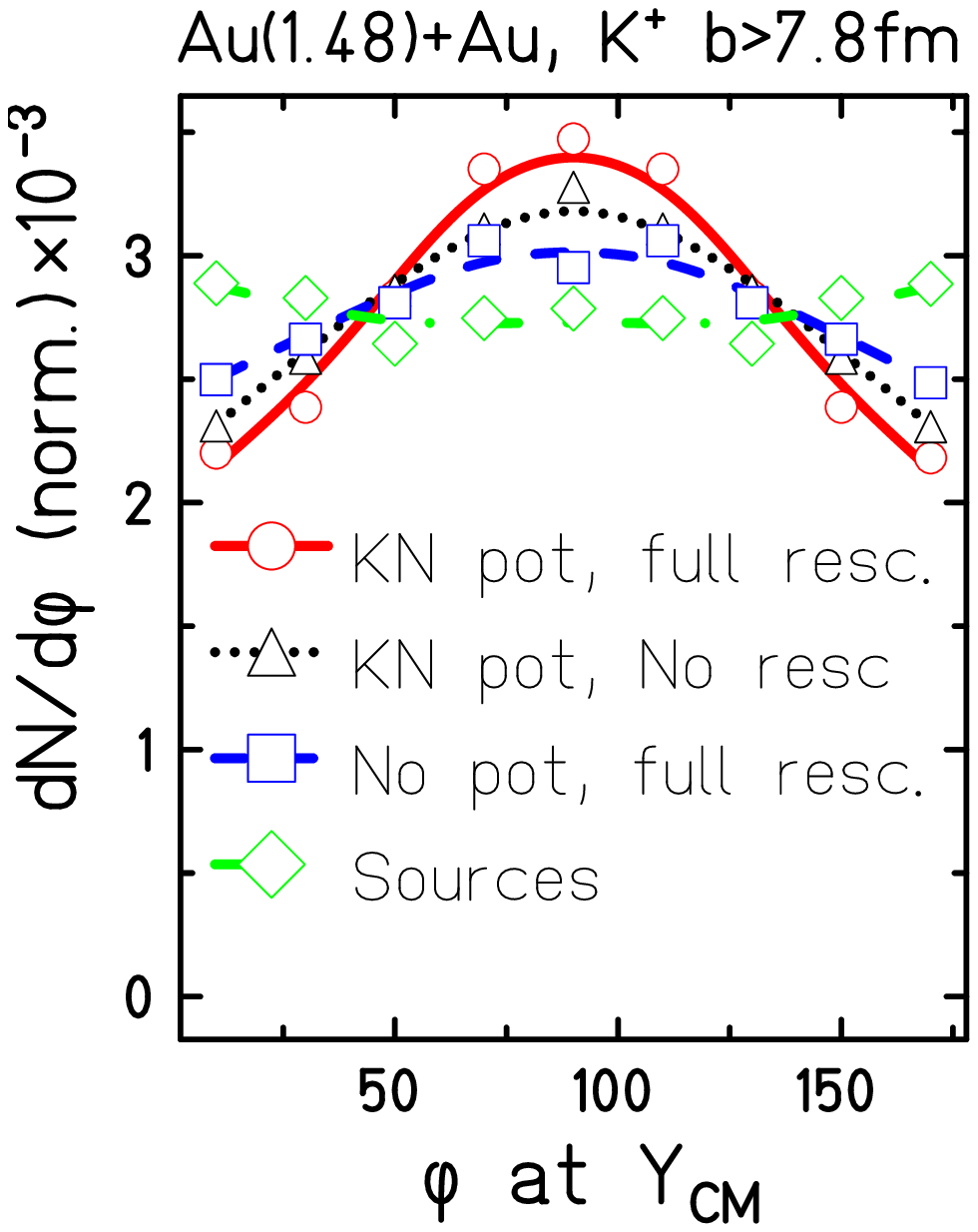,width=0.4\textwidth}
\caption{Contribution of cuts and of collisions
}
\Label{au15-phi-c}
\end{figure}
In order to understand the effect we analyze on the \lhsref{au15-phi-c}
the influence of the experimental cuts. 
An analysis of all kaons without any constraints on the rapidity 
($4\pi$, \bdl) already shows an out-of-plane peak. If we now 
reduce the rapidity window to the same size than the experiment 
(\bpl) we enhance the peak since we go more to the mid-rapidity region.
If we finally apply $p_T$ cuts (\rfl) we enhance the peak still more
since we cut out rather isotope low energy kaons.

The \rhsref{au15-phi-c} compares the distribution taking all experimental
cuts but varying on the potentials and rescattering.
If we forbid all potentials and rescattering (sources, \gml) the distribution
is quite flat. If we now allow rescattering but forbid the optical potential
(No pot, full resc., \bdl) we get an enhancement out-of-plane.
If on the other hand we allow the optical potential but forbid the
rescattering (KN pot, No resc., \bpl) the enhancement is stronger.
But the maximum final enhancement can only be got by allowing potential
and rescattering (KN pot, full resc. \rfl).

\begin{figure}[hbt]
\epsfig{file=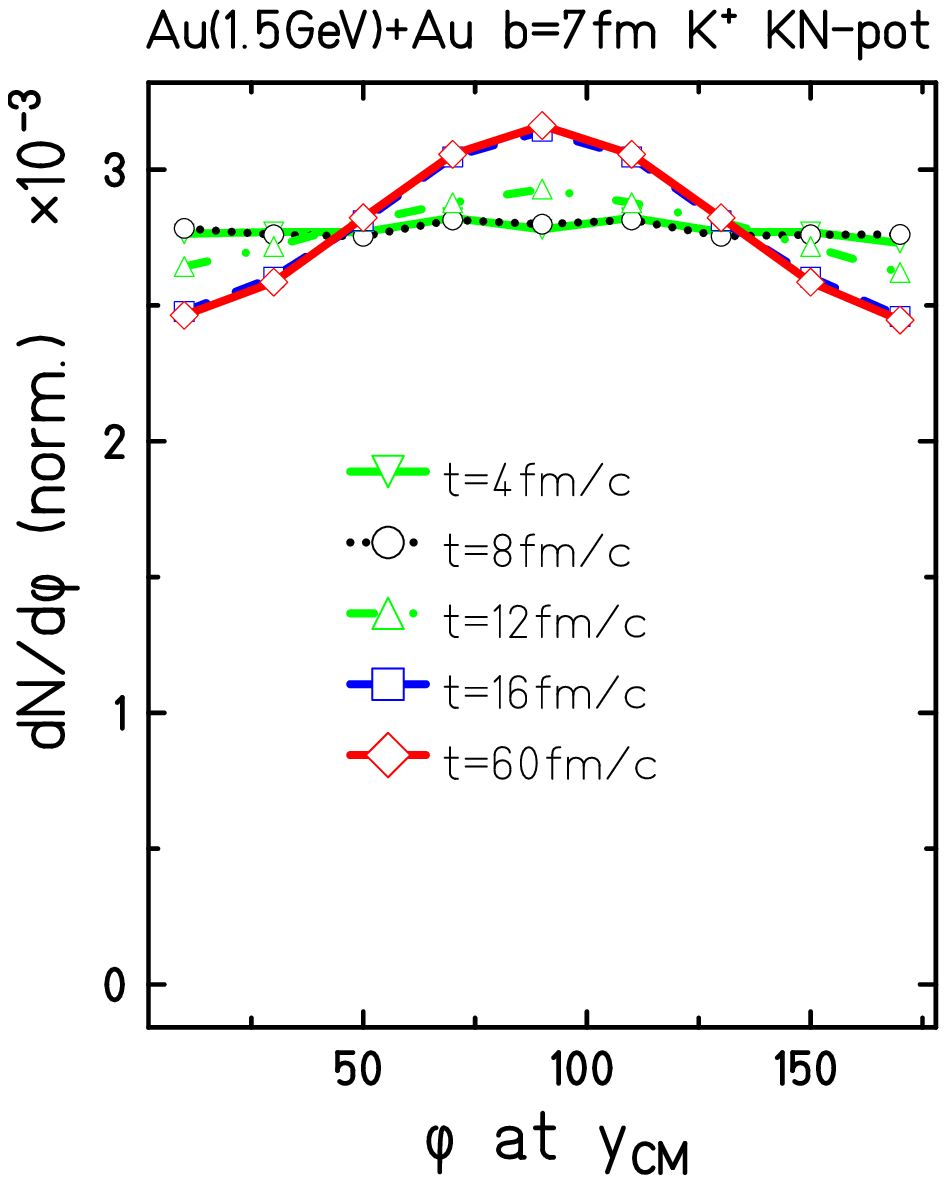,width=0.4\textwidth}
\epsfig{file=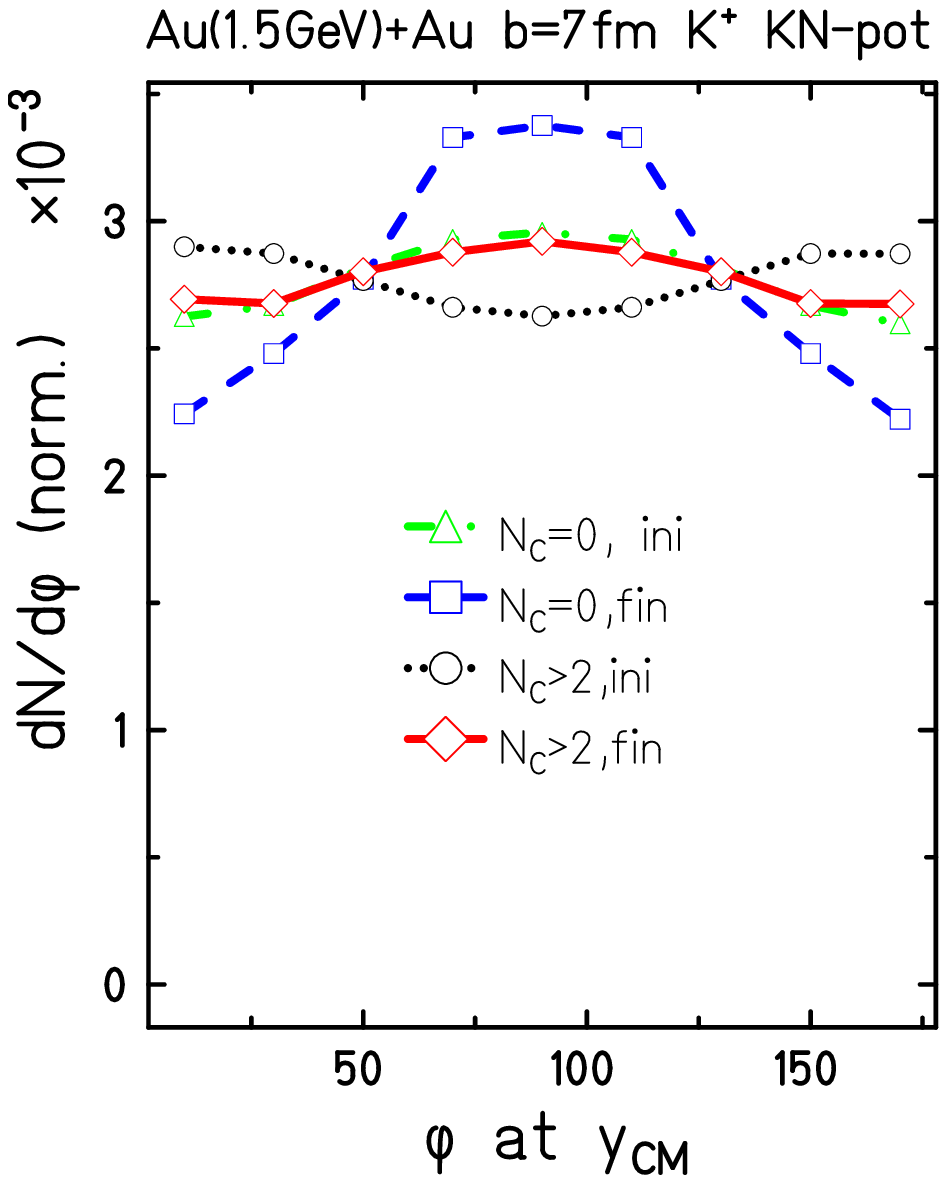,width=0.4\textwidth}
\caption{Time evolution of the azimuthal distribution (left) and
initial and final distributions of the kaons selected according to their
collision number}
\Label{au15-phi-t}
\end{figure}
Let us now look on the time evolution of the kaon squeeze as it is shown
on th \lhsref{au15-phi-t}. The curves are normalized, thus the effect of the
increasing number of kaons is not visible. We see that the kaons are produced
rather isotropically since up to about 8  fm/c there is a flat distribution
which changes slightly at about 12 fm/c (when the kaon production is majorly
finished) to obtain its final form at about 16 fm/c. At this time the collisions
are still active. If we decompose the kaon distributions according to their
number of collisions (\rhsref{au15-phi-t}) we find an effect of preselection.
The initial curve of kaons that won't collide during the reaction (\gml) shows
already a weak peaking out-of plane, while the initial curve of kaons which
will collide frequently (\bpl) has its maximum in plane. This is due to the 
effect that the kaons have more chance to collide when passing upward or
downward into the spectator matter while the chance to escape without collisions
is higher when moving out of plane. The final distributions show both
an out-of-plane peak where the peak of the particles which did not collide
(\bdl) is stronger that that of the particles which collided frequently
(\rfl).
We see that the out-of-plane squeeze is due to rescattering and potential,
where the potential seems to play the more important role.

\begin{figure}[hbt]
\epsfig{file=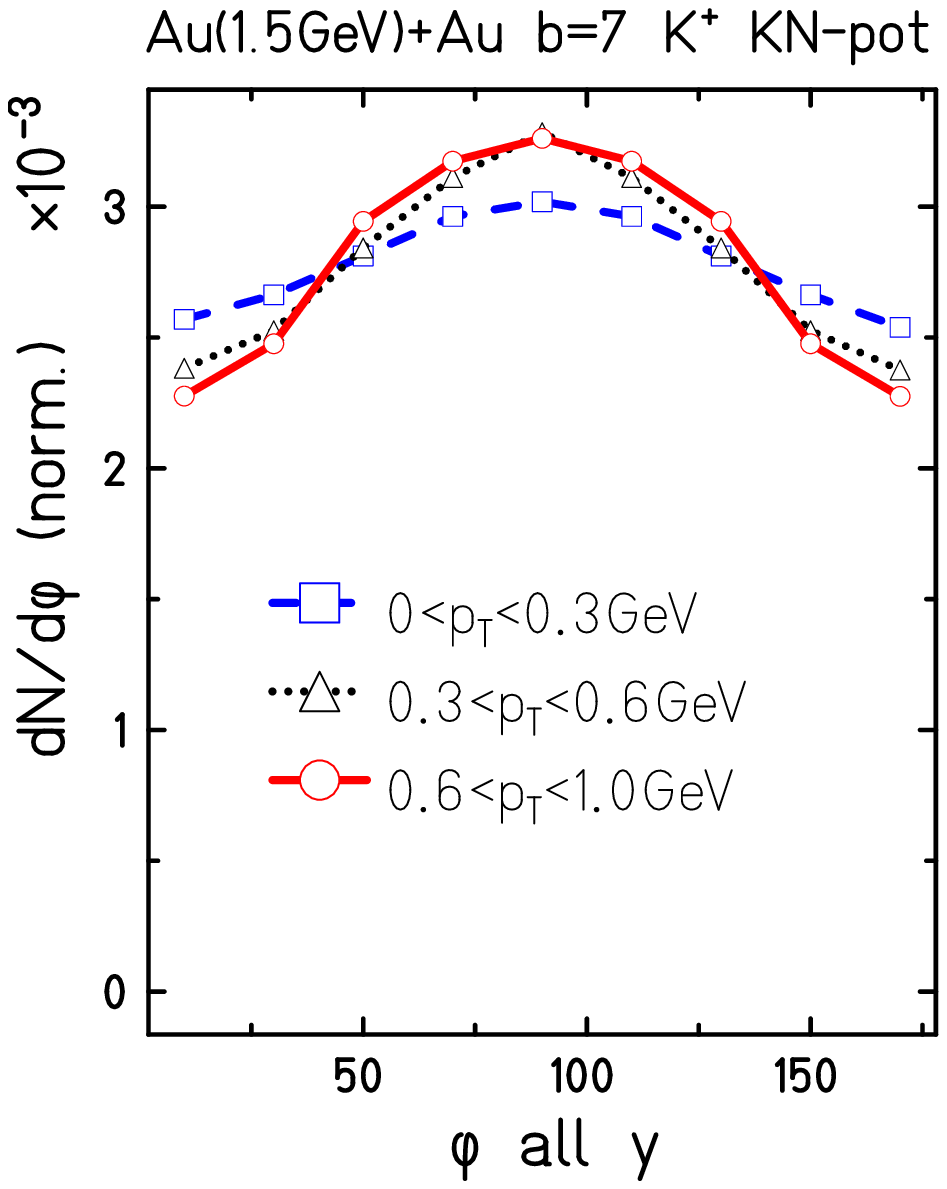,width=0.4\textwidth}
\epsfig{file=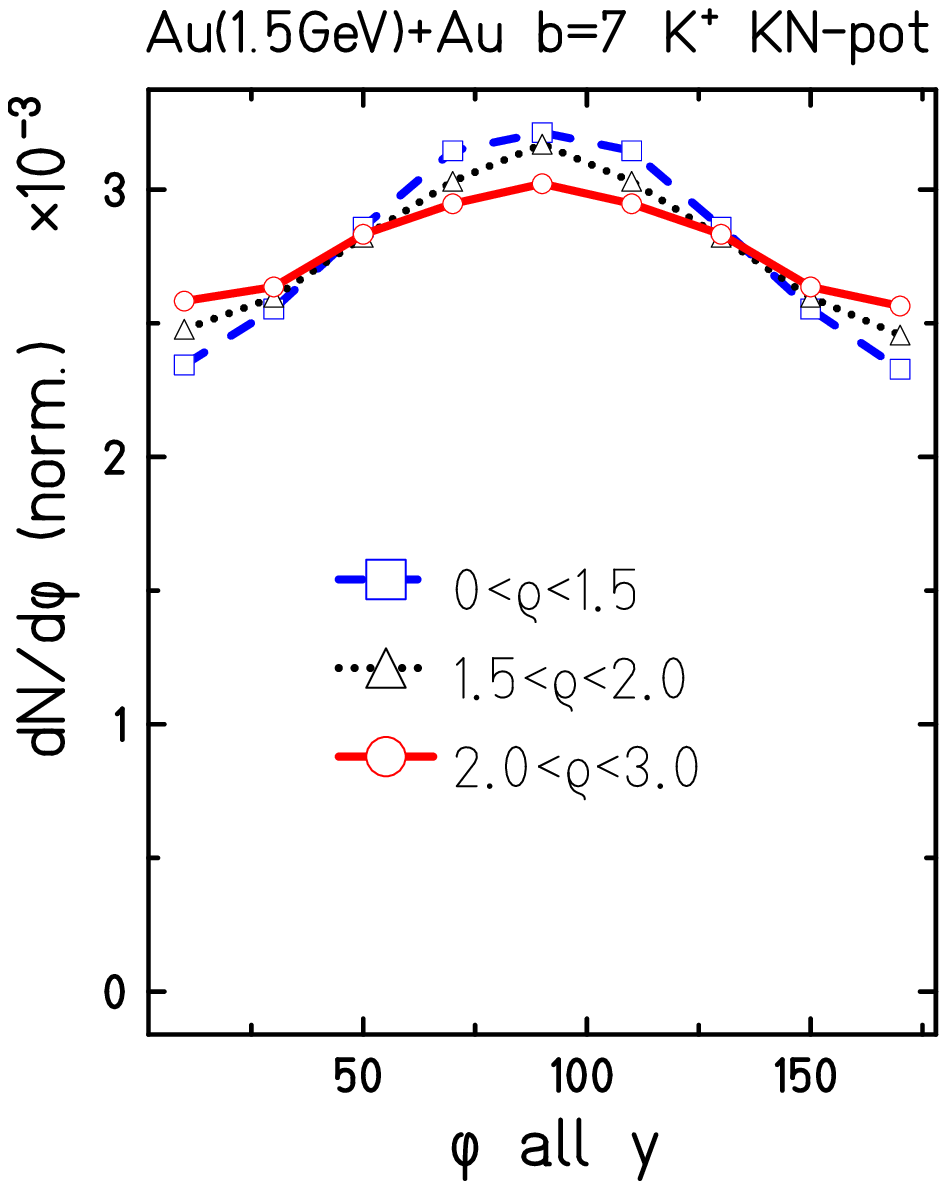,width=0.4\textwidth}
\caption{Azimuthal distributions averaged over all rapidities selected
according to the transverse momentum (left) and the production density (right)}
\Label{au15-phi-rho}
\end{figure}
Let us now decompose the spectra according to the transverse momentum(\lhsref{au15-phi-rho}). 
For a better statistics all rapidities are taken, which causes a slight reduction of
the squeeze signal (see \lhsref{au15-phi-c}). We see that the squeeze is stronger
for particles with high transverse momenta. This is in agreement with experimental
data \cite{foerster} who also noted this effect. We also
decompose the squeeze according to the production density of the kaons
(\rhsref{au15-phi-rho}). Kaons stemming from less high densities (\bdl) show
a stronger signal than those stemming from high densities (\rfl). This is in
analogy to what one finds for the nucleon squeeze \cite{hart}.

\subsection{Energy and system size dependence of the azimuthal distributions}

Let us finally take a glance at the energy and system size dependence as
shown in \figref{au15-phi-a}. We see a slight increase of the squeeze effect
with increasing system size and a slight decrease of the squeeze with 
increasing incident energy. This effect is similar but stronger for the
nucleon signal which also shows an increase with system size and an
decrease with incident energy for energies higher than 400 MeV.
Furthermore the kaon squeeze at energies of about 1 GeV and below 
is dominated by the KN-potentials. Calculations including KN potentials
give similar resulats with and without KN rescattering. Switching off
the KN potentials reduces the kaon squeeze significantly when rescattering
is allowed and yields a flat distribution without KN rescattering.
This again gives some hint to the strong correlation of nucleonic matter
to the dynamics of kaons.
\begin{figure}[hbt]
\epsfig{file=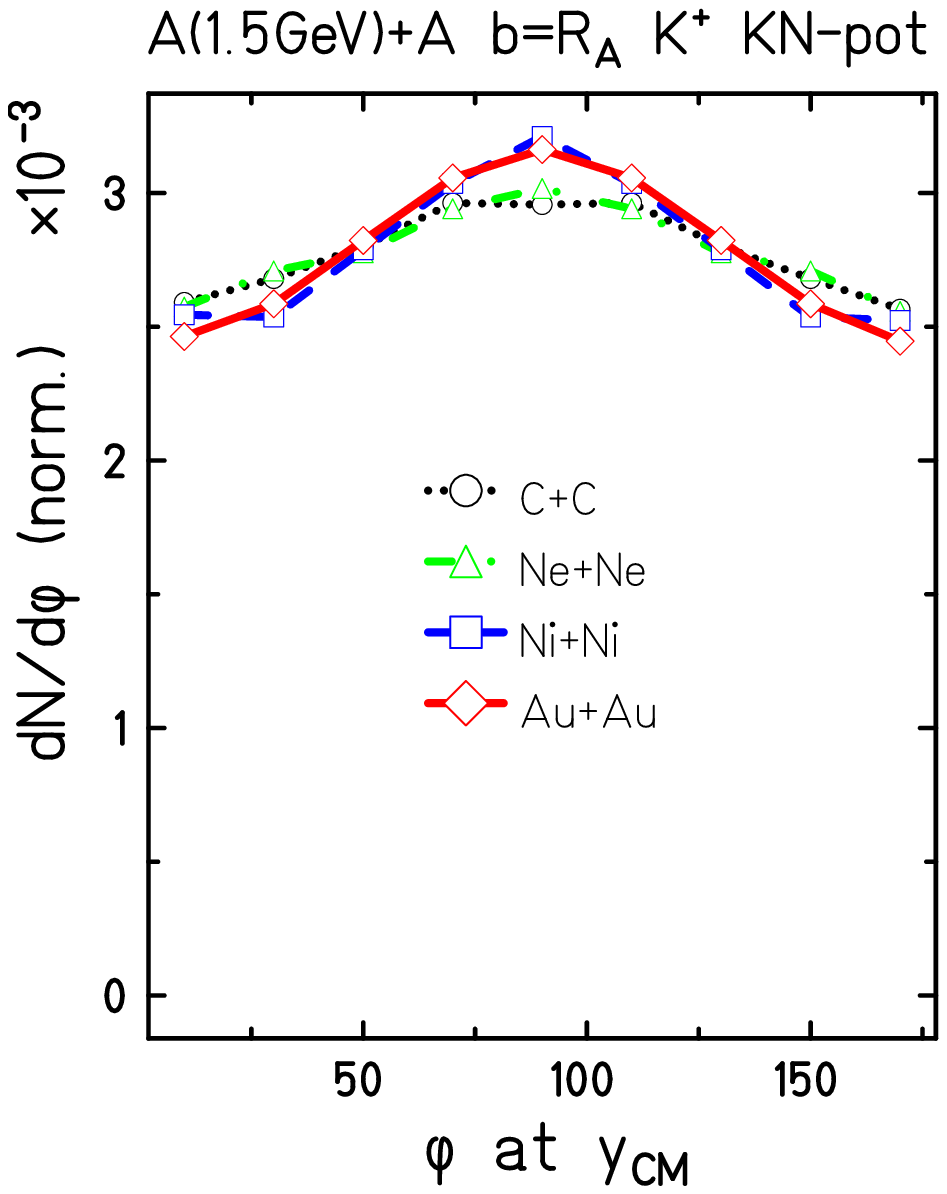,width=0.4\textwidth}
\epsfig{file=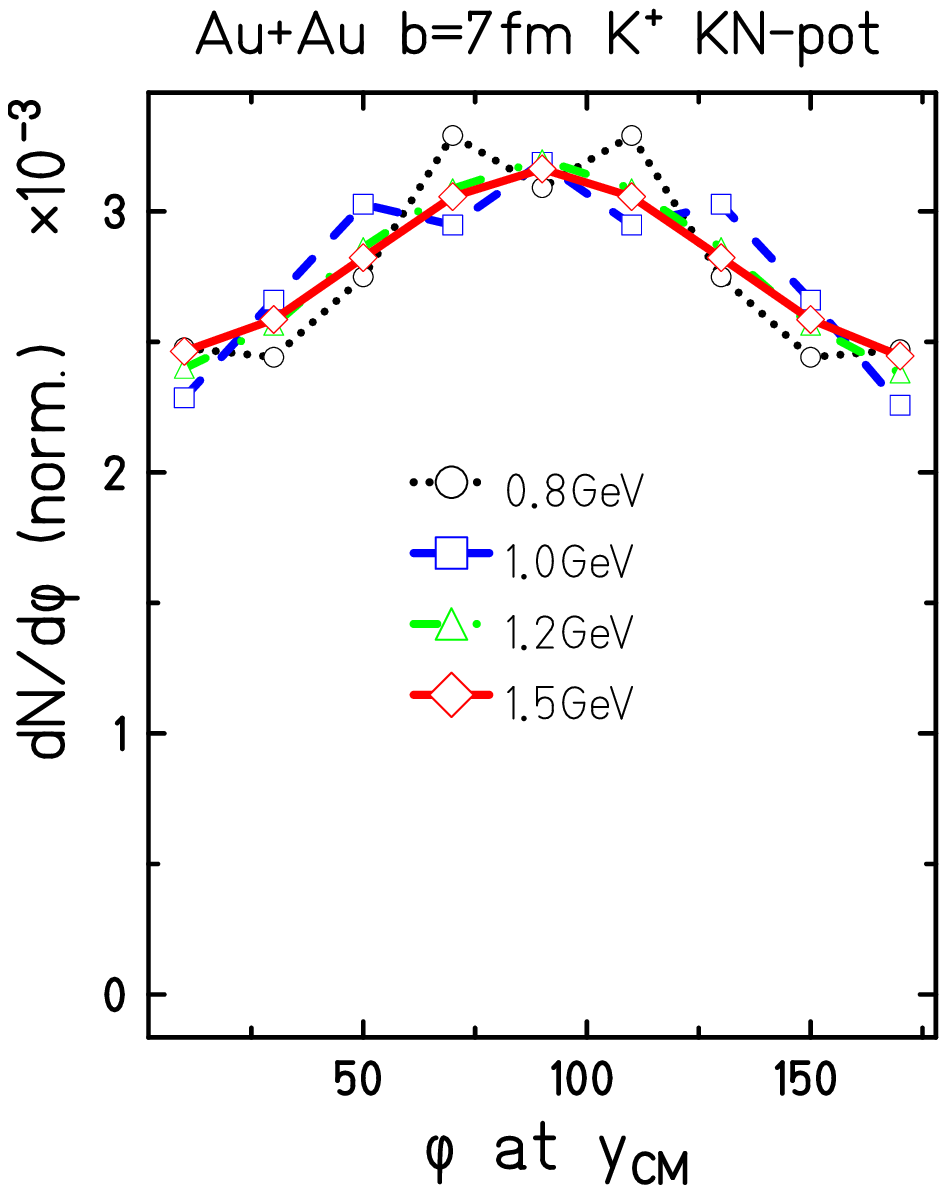,width=0.4\textwidth}
\caption{System size dependence (left) and dependence on the incident energy (right)}
\Label{au15-phi-a}
\end{figure}

\section{Conclusion}

We have analyzed the production of kaons in heavy ion collisions 
at incident energies below the elementary threshold. 
We see that the production of the kaons is rather collective phenomenon
requiring high densities and multi-step processes. 
Nevertheless the kaons do not take a unique signature of this high density zone
since they rescatter afterward.

The optical potential of the kaon in the nuclear medium influences strongly
the absolute number of the kaon yield and the low energy part ob the kaon
spectra. There is also a strong influence on the azimuthal distributions but 
only a weak influence on the polar distributions.

The rescattering influences the high energy part of the spectra and the
temperatures. There is a strong influence on the polar distributions but
a less strong influence on the azimuthal distributions.

The comparison of absolute kaon yield between experiment and calculations 
is still problematic due to the uncertainties relying in the parametrization
of unknown cross sections. 
Comparison of the spectra in p+A collisions nearby the threshold support rather
the calculations with an optical potential.
The use of ratios Au/C allow to cancel
at least the major parts of that uncertainties and support the assumptions of
a soft nuclear equation of state. 
This equation of state is also supported by the analysis of the dependence 
of the kaon yield on the participant number.

All in all we can conclude that the study of different observables on kaon 
production is a very helpful tool to view inside the dynamics of heavy ion
collisions. IQMD is able to describe the kaon data reasonably well. In general
the calculations using an optical potential and rescattering show the
best agreement. This can be seen as a hint to the importance of medium effects
for the kaon production close to the threshold.

\section*{Acknowledgments}
The author acknowledges the fruitful collaboration with J\"org Aichelin 
and the vivid and very inspiring exchange with Helmut Oeschler 
for his work on strangeness production.
Many fruitful discussions with Elena Bratkowskaja, Wolfgang Cassing,
Christian Fuchs, Burkhard K\"ampfer and Horst St\"ocker on the
theory side and with Andreas F\"orster, Yvonne Leifels, Christian Sturm 
and Florian Uhlig from the FOPI and KaoS collaboration are gratefully 
acknowledged. This work was in part supported by the GSI-IN2P3 
convention.
The calculations were run on the GSI Linux-cluster (Darmstadt), 
the Centre de Calcul de l'IN2P3 (Lyon) and the local SUBATECH
cluster.  






\begin{thebibliography}{99}


\bibitem{dowa} 
C.B. Dover and G.E. Walker, Phys. Rep. {\bf 89}, 1 (1982)

\bibitem{aik} J. Aichelin and C.M. Ko Phys. Rev. Lett. 55 (1985) 2661

\bibitem{RanKo}
J.~Randrup and C.M.~Ko, 
\newblock Nucl.~Phys. A 343, 519 (1980).

\bibitem{Barth} R. Barth et al., (KaoS  Collaboration),
Phys. Rev. Lett. {\bf 78} (1997) 4007.

\bibitem{Ahle}
L.~Ahle et al., (E802 Collaboration), Phys. Rev. {\bf C58} (1998)
3523.

\bibitem{schaffi} J. Schaffner et al, Phys. Lett. {\bf B334} (1994) 268; \\
J. Schaffner-Bielich et al., Nucl. Phys. {\bf A625};\\
(1997) 325 
J. Schaffner-Bielich et al., Nucl Phys {\bf A669} (2000) 153

\bibitem{cassing} W.~Cassing et al., Nucl.~Phys.~A{\bf 614} (1997) 415;\\
W. Cassing and E. Bratkovskaya, Phys. Rep. {\bf 308} (1999) 65.

\bibitem{koli} C.M. Ko, Phys. Lett.~{ B138} (1984) 361; \\
G. Q. Li and C. M. Ko, Nucl. Phys. {\bf A} (1995) 460 \\
G. Q. Li et al., Nucl. Phys.{\bf 625} (1997) 415,




 
\bibitem{ha89}
C.~Hartnack, L.~Zhuxia, L.~Neise, G.~Peilert, A.~Rosenhauer, H.~Sorge,
  J.~Aichelin, H.~St\"ocker, and W.~Greiner.
\newblock Nucl.~Phys.~{\bf A495}, 303 (1989).

\bibitem{hart}
Ch. Hartnack.
\newblock PhD thesis, GSI-Report 93-5 (1993).


\bibitem{baprc}
S.~A.~Bass, C.~Hartnack, H.~St\"ocker and W.~Greiner.
\newblock Phys.~Rev.~{\bf C 51 }, 3343 (1994).

\bibitem{iqmd}C. Hartnack et al., Eur.Phys.J. {\bf A1} (1998) 151.


\bibitem{st86}
H.~St\"ocker and W.~Greiner.
\newblock Phys.~Reports~{\bf 137}, 277 (1986).

\bibitem{cas90}
W.~Cassing, V.~Metag, U.~Mosel and K.~Niita.
\newblock Phys.~Reports~{\bf 188}, 361 (1990).


\bibitem{ai91}
J.~Aichelin.
\newblock  Phys.~Reports~{\bf 202}, 233 (1991).

\bibitem{urqmd}
S. A. Bass J. Phys. G {\bf 28} (2002) 1543

\bibitem{ar82}
L.~G.~Arnold et al.
\newblock Phys.~Rev.~{\bf C25}, 936 (1982).

\bibitem{pa67}
G. Passatore.
\newblock Nucl.~Phys. {\bf A95}, 694 (1967).

\bibitem{bert88b}
G.~F.~Bertsch and S.~Das~Gupta,
\newblock Phys.~Rep.~{\bf 160}, 189 (1988).

\bibitem{ai87b}
J.~Aichelin, A.~Rosenhauer, G.~Peilert, H.~St\"ocker, W.~Greiner.
\newblock  Phys.~Rev.~Lett.~{\bf 58}, 1926 (1987).

\bibitem{hama90}
S.~Hama et al. 
\newblock Phys.~Rev.~{\bf C41}, 2737 (1990).

\bibitem{newopt}
Ch.~Hartnack and J.~Aichelin
\newblock Phys.~Rev.~{\bf C49}, 2801 (1994).

\bibitem{bleicher}
M. Bleicher, UFTP Frankfurt, private communication.


\bibitem{kaon94}
C. Hartnack, J. Jaenicke, L. Sehn, H. St\"ocker, J. Aichelin,
\newblock Nucl. Phys. {\bf A 580}, 643 (1994). 

\bibitem{dnsiso}
Ch.~Hartnack, J.~Aichelin, H.~St\"ocker and W.~Greiner,
\newblock Phys.~Rev.~Lett. 72, 3767 (1994).   

%
\bibitem{hogan} W.J. Hogan, Phys. Rev. 166 (1968) 1472
\bibitem{sibirtsev} A. Sibirtsev et al. Phys. Lett. {\bf B 359} (1995) 29
\bibitem{tsu} K. Tsushima et al., Phys. Lett. {\bf B337} (1994) 245\\ 
Phys. Rev. {C 59} (1999) 369 (nucl-th/9801063) 

\bibitem{david} C. David et al, Nucl. Phys {\bf A 650} (1999) 358


\bibitem{cosy98} J.T. Balewski et al. nucl-ex/9803003

\bibitem{hong}
 B. Hong et al., Phys. Rev. {\bf C 66} (2002)034901
\bibitem{hartmann}
O.N. Hartmann, PhD thesis, TU Darmstadt 2003,

\bibitem{anke} 
M. B\"uscher et al. Phys. Rev. {\bf C 65} (2002) 14603 \\
M. B\"uscher and N. Nekipelov, proceedings of the Meson 2002, p.179.

\bibitem{scheinast} W. Scheinast, PhD thesis, University of Dresden (2004). 

\bibitem{menzel} M.~Menzel et al., (KaoS  Collaboration),
Phys. Lett. {\bf B 495} (2000) 26; M. Menzel, Dissertation,
Universit\"at Marburg, 2000.
\bibitem{CLE00} J. Cleymans, H. Oeschler and K. Redlich,
Phys. Lett.~{\bf B485} (2000) 27.
\bibitem{ho_s2000} H.~Oeschler, proceedings of ``Strangeness 2000'', Berkeley,
USA, July 2000; J.~Phys.~G: Nucl.~Part.~Phys. {\bf 27} (2001) 1.

\bibitem{lutz} M.F.M. Lutz et al., nucl-th/0112053
\bibitem{sqm2001} C. Hartnack and J. Aichelin, J. Phys. G {\bf 28} (2002) 1649
\bibitem{sqm2003} C. Hartnack and J. Aichelin, J. Phys. G {\bf 30} (2004) 531
\bibitem{pal} S. Pal, C.M. Ko and Z. Lin, Phys. Rev. {\bf C 64} (2001) 042201
\bibitem{foerster} A.~F\"orster (KaoS  Collaboration), PhD thesis, TH Darmstadt; 
A. F\"orster et al. Phys. Rev. Lett. {\bf 31} (2003) 152301; 
A. F\"orster et al. J. Phys. G. {\bf 30} (2004) 393 

\bibitem{Laue} F. Laue 
et al., (KaoS  Collaboration), Phys. Rev. Lett. {\bf 82} (1999)
1640.
\bibitem{rit} J.L. Ritman et al.,{\it  Z. Phys. }{\bf A352} (1995) 355

\bibitem{sturm} C. Sturm et al (KaoS Collaboration), Phys. Rev. Lett. {\bf 86}
(2001) 39 
\bibitem{Sturm} C. Sturm, PhD thesis, TH Darmstadt \\
C. Sturm et al (KaoS Collaboration), J. Phys. G {\bf 28} (2002) 1895
\bibitem{fuchs} 
C. Fuchs et al. Phys. Lett {\bf B 434} (1998) 245 ;
C. Fuchs et al. Phys. Rev. Lett {\bf 86} (2001) 1794;
C. Fuchs et al. J. Phys. G {\bf 28} (2002) 1615
\end{thebibliography}
\end{document}